\documentclass[fleqn,usenatbib]{mnras}

\usepackage{newtxtext,newtxmath}
\usepackage{comment}
\newcommand{\ergcm}[1]{$\times 10^{#1}\, {\rm erg \,s}^{-1} {\rm cm}^{-2}$}

\usepackage{multirow}
\usepackage{xspace} 

\newcommand{\xmm}{\textit{XMM-Newton}\xspace}
\newcommand{\nus}{{\it NuSTAR}\xspace}
\usepackage[T1]{fontenc}
\usepackage{threeparttable}

\DeclareRobustCommand{\VAN}[3]{#2}
\let\VANthebibliography\thebibliography
\def\thebibliography{\DeclareRobustCommand{\VAN}[3]{##3}\VANthebibliography}

\usepackage{graphicx}	
\usepackage{amsmath}	

\title[X-ray secrets of $\gamma$-NLSy1s]{Unveiling the X-ray Secrets of Fermi-detected Narrow-Line Seyfert 1 Galaxies with \xmm\ Observations}

\author[SUVAS et al.]{
Suvas Chandra Chaudhary,$^{1}$\thanks{E-mail: suvas0101phy@gmail.com}
Raj Prince,$^{2}$\thanks{E-mail: priraj@bhu.ac.in}
Brian van Soelen$^{1}$\thanks{E-mail: VanSoelenB@ufs.ac.za}
and I.P.\, van der Westhuizen$^{1}$
\\
$^{1}$Department of Physics, University of the Free State, 205 Nelson Mandela Dr., Bloemfontein, 9300, South Africa.\\
$^{2}$Department of Physics, Institute of Science, Banaras Hindu University, Varanasi-221005, India.}

\pubyear{\the\year{}}

\begin{document}
\label{firstpage}
\pagerange{\pageref{firstpage}--\pageref{lastpage}}
\maketitle

\begin{abstract}
In the innermost regions of active galactic nuclei, where the accretion disk, corona, and jet processes are closely coupled, X-ray observations offer a direct probe to study the physics of disk–jet coupling and the mechanisms driving relativistic outflows. We present a comprehensive analysis of the X-ray timing and spectral variability of 16 Narrow Line Seyfert 1 galaxies detected by {\it Fermi}-LAT, based on 29 epochs of \xmm observations. A moderate intraday flux variability is observed throughout the sample, with fractional variability ranging from 5 to 16\%. The temporal study of 1H\,0323+342 reveals a distinctive turnover timescale in structure functions, along with notable variations in flux and power spectral density slopes. The hardness ratio in some epochs demonstrates a clear trend of softer-when-brighter. The X-ray spectra of 1H\,0323+342, PMN\, J0948+0022, RGB\, J1644+263, PKS\, 1502+036, and J1246+0238 are well fitted by a power-law + blackbody model, suggesting a bright disk along with a jet, and J1222+0413 is fitted by broken power-law, while the remaining sources are well fitted by a power-law model revealing the non-thermal domination. The X-ray luminosity exhibits a strong correlation with $\gamma$-ray and disk luminosity, and a strong correlation with the jet power, suggesting a close coupling of disk and jet. Additionally, we have found an anti-correlation between the X-ray spectral index and the X-ray luminosity, as well as with the FWHM of H$\beta$ line, indicating a complex interaction between the central engine, jet activity, and the accretion disk in these sources.
\end{abstract}




\section{INTRODUCTION}
Narrow-Line Seyfert 1 galaxies (NLSy1) are a subclass of active galactic nuclei (AGN) that exhibit weak forbidden [O III] lines with line ratio [O III]$\lambda$5007/H$\beta < 3$, strong Iron emission lines (high $\mathrm{Fe\,II}$/H$\beta$, \citealt{1985ApJ...297..166O, 1996A&A...305...53B}), and narrow permitted emission lines (FWHM H$\beta < $ 2000 km s$^{-1}$, \citealt{1989ApJ...342..224G}). According to a study reported in \cite{2006AJ....132..531K}, nearly 2.5\% of the NLS1 galaxies have a radio loudness $R > 100$, which is approximately one-third of the radio-loud galaxy population. In optical, the narrow Balmer emission lines are attributed to the low rotational velocity of the broad line region (BLR) clouds around a low or intermediate mass black hole ($<$10$^8$\,M$_\odot$; \citealt{2011arXiv1109.4181P}). Many studies suggest the black hole mass ranges between 10$^6$ - 10$^{8}$ M$_\odot$ \citep{2001ApJ...558..578O, 2007ApJ...658L..13Z, 2017ApJS..229...39R, 2013MNRAS.431..210C, 2015AJ....150...23Y, 2016ApJ...824..149W}. \citet{Peterson:2011sM} also argued that most of the NLSy1, which have a high Eddington ratio \citep{Komossa_2006, 2015A&A...575A..13F}, are similar to bright quasars. The reverberation studies conducted on many NLSy1s support the idea of the presence of a low-mass black hole \citep{2021A&A...654A.125B, Du_2014}. Considering these are low-mass AGN, many studies suggest they might be in the early stage of their life, and over time, they can grow to become supermassive ($>10^{9-10}$ M$_\odot$) AGN \citep{10.1046/j.1365-8711.2000.03530.x, Sulentic_2000}. It is now unequivocally known that some of the AGNs are massive enough to launch powerful relativistic jets, and in some cases, non-collimated outflows and winds. The calculated jet power distribution, which is normally lower than that of FSRQs but partly overlaps with the BL Lac objects, varies from around $10^{42.6}$ to about $10^{45.6}$ erg,"s$^{-1}$. It is also argued that the jet power of the different jetted AGNs is consistent with one another after being normalized by the core black hole masses, suggesting that the observed variations are caused by mass scaling factors and that the jets are identical in nature \citep{2015A&A...575A..13F}. The central engine of radio-loud NLSy1s appears to be somewhat comparable to that of blazars, despite the observed dissimilarities in their timing and spectral properties. Radio-loud NLSy1 galaxies have historically been challenging to identify within the jetted AGN sequence because of their low power and irregular jet activity. \citep{2015A&A...575A..13F}.


The concept that NLSy1 galaxies could produce relativistic jets gained attraction after their detection at gamma-ray energies by {\it Fermi}-LAT, suggesting a blazar-like emission process, structure, and rapid flux variability \citep{2013MNRAS.436..191D, Abdo_2009, 2015A&A...575A..13F}. The gamma-ray detected NLSy1 are referred to as $\gamma-$NLSy1 in the entire text. Previous studies have shown that the jetted NLSy1 galaxies are very similar to FSRQ-type blazars \citep{2019A&A...632A.120B, 2020Univ....6..136F}, especially at lower gamma-ray luminosities \citep{galaxies7040087}. As discussed in e.g.\ \cite{2002ApJ...564...86B}, radio loud AGN (i.e.\ jetted AGN) typically contain central black holes with masses exceeding $\sim 3\times 10^{8}$\,M$_\odot$ \citep{2000ApJ...543L.111L}; 
However, the detection of gamma-rays from NLSy1 galaxies poses various questions on the nature of these sources and the jet launching mechanisms in various types of galaxies, making them an interesting candidates for disk-jet coupling, and high-energy emission processes. 
To date, more than two dozen NLSy1 galaxies have been detected in gamma rays with {\it Fermi}-LAT, and with its continuation of operation, this number is expected to rise in the coming years \citep{2018MNRAS.481.5046R, 2022Univ....8..587F, 2025arXiv251115814D}. These sources are identified as candidate VHE emitters as well as possible neutrino candidates, though to date only upper-limits have been reported -- see, for example, VERITAS observations of PMN J0948+0022 \citep{2015MNRAS.446.2456D}, and IceCube observations of 1H\,0323+342 \citep{2025ATel17423....1M}.
Therefore, these are promising candidates for future observations using the Cherenkov Telescope Array \citep[CTA;][]{2018MNRAS.481.5046R}. Further, constraints on the evolution of AGNs, jet dynamics, and the connection between relativistic outflows and accretion in various galactic environments will be obtained by increasing the number of known and observed sources. New gamma-ray bright NLSy1 candidates may be discovered among the unclassified AGN listed in the {\it Fermi}-LAT catalogues \citep{2023arXiv230712546B, 2022ApJS..263...24A}, by undertaking a multi-wavelength examination of their radio, optical, and X-ray properties \citep{2021MNRAS.501.1384Y, 2025arXiv251115814D}\\
In comparison to Broad-line Seyfert-1 galaxies (BLSy1s), NLSy1s typically exhibit steeper spectral indices \citep{1999ApJS..125..317L} and substantial flux fluctuation at X-ray energies \citep{1999ApJ...526...52T, 1999ApJS..125..317L}. A strong soft X-ray excess has also been observed from these sources, the origin of which is still unknown \citep{1996A&A...305...53B, 1999ApJS..125..317L, 2006MNRAS.370..245G, 2006ApJS..166..128Z, 2020MNRAS.491..532G, 2023MNRAS.522.5456Y}. On top of that, some of them also show a clear hard-excess emission above 35 \text{keV} \citep{2020MNRAS.496.2922M, 2018MNRAS.479.2464G}, making them an interesting class for X-ray studies. Additionally, the X-ray spectral slope $\Gamma_X$ and FWHM (H\,$\beta$) show an anti-correlation. The X-ray spectra of sources with smaller FWHMs (H\,$\beta$) are often steeper \citep{1996A&A...305...53B, 2018BSRSL..87..379R}. Therefore, the X-ray properties of these sources can help to understand their origin and the interplay between the disk-corona or disk-jet dynamics. In our first paper \citep{2025arXiv250404492C}, we discussed the hard X-ray properties of gamma-ray-detected NLSy1 and found that these objects are more similar to jet-dominated, LBL and IBL type blazars. In this work, we explore the soft X-ray properties of {\it Fermi}-LAT-detected NLSy1 galaxies utilizing their temporal and spectral studies and compare them with their hard X-ray and gamma-ray properties. Comparing this information can reveal the interaction between the disk and jet, which is the primary goal of the paper. \\
The paper is structured as follows: in Section \ref{sample}, we outline the sample selection, followed by the {\it XMM-Newton} data reduction in Section \ref{xmmdata}; in Section \ref{analysis}, we discuss the various methodologies used to examine the X-ray properties and results, followed by the discussion and interpretation in Section \ref{discussion}, and a conclusion in Section \ref{conclusion}.


\section{SAMPLE SOURCES}  \label{sample}
This paper presents a study of NLSy1 galaxies that have been detected  using the {\it Fermi}-LAT telescope \cite[see e.g.][]{2018MNRAS.481.5046R, 2018rnls.confE..15K, 2020yCat..22470033A, 2022Univ....8..587F, 2024MNRAS.527.7055P, 2025arXiv251115814D} and have been observed with {\it XMM-Newton}. We cross-matched the {\it Fermi} detected NLSy1 objects with the {\it XMM-Newton} archive and found that 16 sources had {\it XMM-Newton} observations from 29 epochs. Tables \ref{TAB1} and \ref{TAB2} list the sample sources and their {\it XMM-Newton} observations, respectively.

\begin{table*}
    \caption{Summary of the basic information about the sources in this work.}
    \centering
    \begin{tabular}{lccccccc}\hline
        Name & 4FGL Name & RA (J2000) & Dec. (J2000)  & z  & $L_{{\rm H}\beta}$  &  $L_{\gamma}$ & $\log(M_{\rm BH})$ \\
        &  &  &  & & erg\,s$^{-1}$ & erg\,$s^{-1}$ & M$_{\odot}$  \\ \hline
        1H\,0323+342 & J0324.8+3412 & 51.1715  & +34.1794 & 0.06 &  42.02  & $2.1 \times 10^{44}$ & 7.30  \\
        SDSS J1641+3454 & - & 250.2504 & +34.9146  &  0.164   &  41.37   &  $7.2 \times 10^{42}$   & 6.35 \\
        3C 286 & J1331.0+3032  &  202.7845 & +30.5091  & 0.86  &  43.67   &  $8.7 \times 10^{45}$  & 7.79 \\
        FBQS J1102+2239 & -  & 165.5974  & +22.6557  & 0.45  &   42.59  &  $7.3 \times 10^{43}$ & 7.25\\
       SDSS J0946+1017 &   J0946.6+1016 &  146.6461 & +10.2850  &  1.00 & 43.48 &   $4.5 \times 10^{46}$   & 8.28\\
       PMN J0948+0022 & J0948.9+0022  & 147.2388 & +00.3737 & 0.58 &  42.91 & $7.5 \times 10^{46}$ & 7.16 \\
CGRaBS J1222+0413  & J1222.5+0414  & 185.5939 & +04.2210 & 0.97 & 43.60 & $2.3 \times 10^{47}$ &  8.85  \\
SDSS J124634.64+023809.0 & -   &   191.6443 & +02.6358   &  0.36 &  42.51   &  $7.1 \times 10^{43}$ & 7.11 \\
       TXS 1419+391 & J1421.1+3859  &  215.2751 & +38.9230   &  0.49 &  -   &  $1.5 \times 10^{45}$   & 7.36 \\
       PKS 1502+036  & J1505.0+0326 & 226.2769 & +03.4418 &  0.41 & 41.78 & $1.0 \times 10^{46}$ &7.60   \\
       RGB J1644+263 & J1644.9+2620  & 251.1772 & +26.3203  & 0.14 & 41.93 & $2.7 \times 10^{44}$ & 6.99\\
       TXS 2116-077 &  J2118.8-0723 &  319.7206 & -07.5409   & 0.26  &   42.07  &  $7.4 \times 10^{44}$   & 6.87 \\

       PKS 2004-447  & J2007.9-4432 & 301.9799 & -44.5789 & 0.24 & 41.74  & $1.7 \times 10^{45}$ &  6.70 \\

       TXS 0103+189  & J0105+1912 & 16.4801 & 19.2077 &  0.88 & 41.51  & - &  6.70 \\
       PKS 1244-255  & J1246-2548 & 191.6950 & -25.7970 &  0.64 & 42.95  &  $1.2 \times 10^{47}$ & 8.22 \\
       TXS 1308+554  & J1310+5514 & 197.7633 & +55.2317 &  0.93 & -  & $1.6 \times 10^{46}$ &  7.84 \\
 \hline
    \end{tabular}
    \par\smallskip
    \noindent\textbf{Table Note.} Col. 1: Source Name; Col. 2: 4FGL Name; Col. 3: RA (in degree); Col. 4: Dec (in degree); Col. 5: redshift; Col. 6: H$\beta$ luminosity; Col. 7: Gamma-ray luminosity; Col. 8: Black hole mass. Various source parameters such as BH mass \citep{2007ApJ...658L..13Z, 2017ApJS..229...39R, 2013MNRAS.431..210C, 2015AJ....150...23Y, 2001ApJ...558..578O, 2011MNRAS.413.1671F,2013MNRAS.431..210C, 2015AJ....150...23Y, 2015A&A...573A..76J}, $L_{\gamma}$ \citep{2023ApJS..265...31A, 2020MNRAS.496.2213D, 2022ApJS..260...53A}, FWHM(H$\beta$), and L(H$\beta$) of NLSy1 galaxies are taken from \citep{1996A&A...305...53B, 1996A&A...309...81W, 2006ApJS..166..128Z, 2015A&A...575A..13F}. Some of the sources are still not included in the recent {\it Fermi}-LAT {\it 4FGL-DR4} catalogue; hence, their 4FGL names are missing from this Table. 
    \label{TAB1}
\end{table*}

\section{{\it XMM-Newton} OBSERVATIONS AND DATA REDUCTION} \label{xmmdata}
The X-ray Multi-Mirror Mission ({\it XMM-Newton}) was launched by an Ariane 504 on December 10, 1999 \citep{2001A&A...365L...1J}. The Optical Monitor (OM), Reflection Grating Spectrometer (RGS), and European Photon Imaging Camera (EPIC) are the three onboard instruments in the {\it XMM-Newton} observatory \citep[see e.g.][]{2000SPIE.4012..154B, 2001A&A...365L...7D}. It features three X-ray cameras (EPIC-PN, EPIC-MOS1, and EPIC-MOS2) with an exceptionally large effective area. Observations are extremely sensitive due to the wide collecting area and the ability to conduct long, continuous exposures. In this work, we have used only EPIC-PN data due to its high signal-to-noise ratio, high quantum efficiency, and high effective area, particularly in comparison to EPIC-MOS \citep{2001A&A...365L..18S}. Light curves, and source and background spectra were extracted with the {\it XMM-Newton} Science Analysis System (SAS) v22.0.0\footnote{\url{https://www.cosmos.esa.int/web/xmm-newton/sas-threads}}. Calibration Files and summary files for the Observation Data Files (ODFs) were created from Updated Calibration Files (CCFs), under the guidelines presented in the Users' Guide to the {\it XMM-Newton} Science Analysis System (SAS). Event files were processed with the {\it EPPROC} pipeline. To correct for pile-up effects, we performed iterative corrections with the {\it EPATPLOT} task in SAS, determining optimal source extraction areas. For {\it imaging mode}, the extraction region is an {\it annulus}, and for {\it timing mode}, it is a rectangular box. Before extracting light curves and spectra, we applied the filter \textit{FLAG} == 0 \texttt{\&\&} \textit{PATTERN}$\leq$4 to all PN data. Background regions were selected from source-free regions within the same chip. To ensure data quality, we picked good time intervals (GTIs) in which the background count rate was less than the threshold {\it "RATE$\leq$0.4"}. Three distinct energy bands were used to obtain light curves: the soft band (0.3–2 keV), the hard band (2–10 keV), and the entire {\it XMM-Newton} band (0.3–10 keV) with a constant time bin of 100 seconds. For spectral analysis, we employed a typical energy range of 0.3–10 keV. Response matrix files (RMFs) and ancillary response files (ARFs) were created using the {\it RMFGEN} and {\it ARFGEN} tasks. Extracted spectra were also binned to have a minimum of {\it 25 counts} in each spectral channel. To fit the X-ray spectra of our sample sources, we used models built in {\it XSPEC V12.15.0}\footnote{\url{https://heasarc.gsfc.nasa.gov/xanadu/xspec/manual/Models.html}} \citep{1996ASPC..101...17A} and obtained the best fitting parameters for X-ray spectra presented in Table \ref{TAB3}.

\begin{table*}
\caption{Summary of the {\it XMM-Newton} Observations of the $\gamma$-NLSy1 sources selected in this work.} 
    \centering
    \begin{tabular}{ccccccccccc} \hline
       Object  & Obs. ID & Obs. Date & Time (ks) & $\langle F \rangle$ (count\,$s^{-1}$) & $F_{\rm var} (\%)$ & VA & $\tau_{\rm var}$ (ks)& $\beta_p$ & $\tau_{SF}$ (ks) &\\ \hline
       1H\,0323+342  & 0764670101 & 2015-08-23 & 80.9 & 4.97$\pm$0.25 & 11.56$\pm$0.07 & 1.18$\pm$0.16 & 0.39$\pm$0.11 & 1.40$\pm$0.15 & 5.0 \\
         & 0823780201 & 2018-08-14  & 54.2 & 4.43$\pm$0.23 & 9.89$\pm$0.10 & 0.92$\pm$0.15 & 0.49$\pm$0.18 & 1.20$\pm$0.20 & 5.0\\
         & 0823780301 & 2018-08-18 & 49.3 & 3.90$\pm$0.22 & 15.60$\pm$0.10 & 1.29$\pm$0.19 & 0.37$\pm$0.12 & 1.52$\pm$0.19 & 5.0\\
         & 0823780401 & 2018-08-20 & 49.1 & 6.64$\pm$0.29 & 10.67$\pm$0.07 & 0.68$\pm$0.11 & 0.52$\pm$0.16 & 0.93$\pm$0.18 & 5.0\\
         & 0823780501 & 2018-08-24 & 49.6 & 4.39$\pm$0.23 & 12.06$\pm$0.10 & 1.67$\pm$1.34 & 0.15$\pm$0.11 & 1.40$\pm$0.18 & 6.0\\
         & 0823780601 & 2018-09-05 & 51.9 & 4.14$\pm$0.23  & 11.80$\pm$0.12 & 0.94$\pm$0.16 & 0.34$\pm$0.10 & 1.12$\pm$0.26 & -\\
         & 0823780701 & 2018-09-09 & 50.6 & 2.98$\pm$0.20 & 14.67$\pm$0.14 & 1.35$\pm$0.24 & 0.27$\pm$0.07 & 1.19$\pm$0.26 & 12.5\\  \\
      SDSS J1641+3454 & 0860640201 & 2022-03-06 & 19.8 & 0.17$\pm$0.05  & - & 3.80$\pm$2.91 & 0.08$\pm$0.04 & 0.64$\pm$0.30 & -\\
         & 0860640301 & 2022-03-24 & 22.5 & 0.09$\pm$0.04 & 12.23$\pm$31.60 & 15.12$\pm$27.08 & 0.04$\pm$0.01 & - & -& \\  \\
        3C 286 & 0881760101 & 2021-12-17 & 49.4 & 0.26$\pm$0.06 & 8.33$\pm$7.15 & 4.64$\pm$2.84 & 0.07$\pm$0.03 & 1.27$\pm$0.17 & - \\  \\
     FBQS J1102+2239 & 0690090301 & 2012-06-11 & 13.9 & 0.07$\pm$0.03 & - & - & 0.03$\pm$0.04 & 0.77$\pm$0.15 & -\\  \\
     SDSS J0946+1017 & 0800040101 & 2017-11-03 & 65.0 & 0.12$\pm$0.04  & - & - & 0.05$\pm$0.02 & 0.23$\pm$0.06 & 5.0 \\  \\
     PMN J0948+0022 & 0502061001 & 2008-04-29 & 24.3 & 2.84$\pm$0.41  & 7.24$\pm$1.56 & 1.50$\pm$0.77 & 0.14$\pm$0.04 & 0.25$\pm$0.11 & -\\  
         & 0673730101 & 2011-05-28 & 92.9 & 0.04$\pm$0.03  & - & - & 0.03$\pm$0.09 & 0.51$\pm$0.16 & - \\
         & 0790860101 & 2016-11-04 & 93.9 & 0.75$\pm$0.12 & 6.55$\pm$1.08 & 2.01$\pm$0.74 & 0.11$\pm$0.03 & 0.25$\pm$0.10 & 10.0\\ \\
     CGRaBS J1222+0413 & 0401790601 & 2006-07-12 & 11.9 & 1.08$\pm$0.13 & 5.36$\pm$2.20 & 1.61$\pm$ 0.74 & 0.19$\pm$0.10 & 0.66$\pm$0.11 & -\\  \\
     SDSS J1246+0238   & 0690090201 & 2012-12-14 & 16.9 & 0.05$\pm$0.03  & - & - & 0.10$\pm$0.02 & 0.23$\pm$0.11 & -\\  \\
     TXS 1419+391  & 0845040601 & 2020-06-11 & 38.0  & 0.22$\pm$0.06 & - &4.02$\pm$2.77 & 0.08$\pm$0.03 & 0.27$\pm$0.07 & -\\  \\
     PKS 1502+036 & 0690090101 & 2012-08-07 & 17.3 & 0.14$\pm$0.04 & 11.64$\pm$4.56 & 8.09$\pm$6.71 & 0.07$\pm$0.03 & - & -\\  \\
      RGB J1644+263 & 0783230101 & 2017-03-03 & 90.0 & 1.10$\pm$0.05  & 5.57$\pm$0.81 & 1.49$\pm$0.43 & 0.20$\pm$0.07 & 1.14$\pm$0.09 & -\\  \\
      TXS 2116-077  & 0784090201 & 2016-05-10 & 33.0 & 0.02$\pm$0.01  & - & - & 0.03$\pm$0.04 & 0.30$\pm$0.10 & -\\
         & 0784090301 & 2016-10-27 & 33.9 & 0.09$\pm$0.04 & - & 17.24$\pm$37.41 & 0.05$\pm$0.03 & 0.56$\pm$0.07 & -\\   \\
      PKS 2004-447 & 0200360201 & 2004-04-11 & 41.9 & 0.48$\pm$0.08 & 10.50$\pm$2.88 & 3.16$\pm$1.17 & 0.16$\pm$0.06 & 0.98$\pm$0.10 & - \\ 
         & 0694530101 & 2012-05-01 & 37.9 &  0.17$\pm$0.05 & - & 4.41$\pm$2.96 & 0.07$\pm$0.02 & 0.53$\pm$0.15 & - \\
         & 0694530201 & 2012-10-18 & 39.8 & 0.26$\pm$0.06 & 11.48$\pm$5.17 & 7.36$\pm$8.50 & 0.09$\pm$0.03 & 0.69$\pm$0.15 & - \\
         & 0790630101 & 2016-05-05 & 53.4 & 0.33$\pm$0.07  & - & 2.58$\pm$1.23 & 0.41$\pm$0.04 & 0.42$\pm$0.14 & -\\
         & 0853980701 & 2019-10-31 & 13.0 & 0.39$\pm$0.06 & 8.95$\pm$4.29 & 2.47$\pm$0.09 & 0.14$\pm$0.05 & 0.44$\pm$0.30 & -\\ \\
    TXS 0103+189   & 0932790201 & 2023-12-26 & 38.4 & 0.21$\pm$0.06 & - & 7.41$\pm$8.52 & 0.03$\pm$0.05 & - & -\\  \\
    PKS 1244-255   & 0650382301 &  2010-07-13 & 7.4 & 1.15$\pm$0.18 & 7.00$\pm$3.79 & 1.79$\pm$1.12 & 0.12$\pm$0.06 & 0.52$\pm$0.28 & -\\  \\
    TXS 1308+554  & 0741031601 & 2014-12-18 & 8.0 & 0.13$\pm$0.04 & - & 6.37$\pm$6.13 & 0.05$\pm$0.02 & 0.10$\pm$0.24 & -\\  \\
          \hline
    \end{tabular}
    \par\smallskip
    \noindent\textbf{Table Note.} Col. 1: Source Name; Col. 2: Observations ID; Col. 3: Start Time of Observations; Col. 4: Observation length; Col. 5: Mean Count-rate; Col. 6: Fractional Variability; Col. 7: Variability Amplitude; Col. 8: Minimum Variability Timescale; Col. 9: PSD slope; Col. 10: SF turnover time scales.
    \label{TAB2}
\end{table*}

\section{ANALYSIS METHODS: VARIABILITY PROPERTIES} \label{analysis}
To investigate X-ray timing and spectral properties of $\gamma$-ray emitting NLSy1 galaxies ($\gamma$-NLSy1), we have used several methods to quantify the X-ray variability properties of these sources, which are discussed in this section.

\subsection{X-ray timing analysis}
One of the most distinctive features of AGNs is X-ray flux variability, which allows us to map the size, geometry, and energetics of the innermost regions by observing rapid variations in disc-jet systems. We have used Fractional variability, Variability Amplitude, Structure Function, Power Spectral Density, and Flux histograms to estimate the intrinsic source variability of all selected $\gamma$-NLSy1.

 An example of an X-ray light curve is shown in Figure \ref{LCplot}, and the remaining light curves are presented in the appendix.

\subsubsection{Flux Variability}
The fractional root mean square variability amplitude ($F_{\rm var}$) is used to measure the average source variability \citep{1990ApJ...359...86E, 1997ApJS..110....9R, 2003MNRAS.345.1271V}. For a given light curve of $N$ data points, having a source variance $S^2$, a mean square error in the flux of 
\begin{equation}
    \overline{\sigma^{2}_{\rm err}} = \frac{1}{N}\sum_{i=1}^{N} \sigma_{{\rm err}, i}^{2},
\end{equation}
and a mean flux $\bar F$, the $F_{\rm var}$ is given by, 
\begin{equation} \label{FVs}
    F_{\rm var} = \sqrt{\frac{S^{2} - \overline{\sigma^2_{\rm err}}}{\bar{F}^2}},
\end{equation}
 where the uncertainty in $F_{\rm var}$ is estimated as \citep{2003MNRAS.345.1271V},
\begin{equation}
   \sigma_{F_{\mathrm{var}}} =  \sqrt{\left(\frac{1}{\sqrt{2N}}\frac{\overline{\sigma^{2}_{\rm err}}}{F_{\rm var}}\frac{1}{\bar{F}^2}\right)^2 + \left(\sqrt{\frac{\overline{\sigma^{2}_{\rm err}}}{N}}\frac{1}{\bar{F}^2}\right)^2}. 
\end{equation}
We can also quantify peak-to-peak light curve variability using the Variability amplitude (VA) defined as \citep[e.g,][]{2022MNRAS.510.5280M, 2023ApJ...955..121D, 2025ApJ...981..118B},
\begin{equation}
    VA = \frac{F_{\mathrm{max}}-F_{\mathrm{min}}}{F_{\mathrm{min}}}, \label{eqn:VA}
\end{equation}
where $F_{\mathrm{max}}$ and $F_{\mathrm{min}}$ are the maximum and minimum flux in counts\,s$^{-1}$, respectively. In contrast to $F_{\rm var}$, the VA typically detects extreme variability. The associated uncertainty in the VA was estimated using the error propagation formulae (see Eq. (3.14) in \citealt{2003drea.book.....B}), which is given as follows:

\begin{equation}
    \sigma_{\mathrm{VA}} = (VA+1)\cdot\sqrt{\left(\frac{\sigma_{F_{\mathrm{max}}}}{F_{\mathrm{max}}}\right)^2+\left(\frac{\sigma_{F_{\mathrm{min}}}}{F_{\mathrm{min}}}\right)^2},
\end{equation}

Since AGNs exhibit high variability over a diverse timescale across the full electromagnetic spectrum, we can also quantify the minimum variability timescale \citep[see][]{1974ApJ...193...43B} for a given light curve,
\begin{equation} \label{tvar}
    \tau_{\mathrm{var}} = \left|\frac{\Delta t}{\Delta \ln F}\right|,
\end{equation}
where $\Delta t$, is the time interval between flux measurements \citep{2008ApJ...672...40H}. The uncertainty in the variability timescale is given by a generalized error propagation formula akin to Eq. (3.14) in \citet{2003drea.book.....B}:
\begin{equation} 
    \Delta_{\tau_{\mathrm{var}}} \approx \sqrt{\frac{F_1^2\Delta F_2^2+F_2^2\Delta F_1^2}{F_1^2 F_2^2 (\ln[F_1/F_2])^4}}\cdot\Delta t,
\end{equation}
where the flux uncertainties used to calculate the minimal variability timescales for the count-rate $F_1$ and $F_2$ are denoted by $\Delta F_1$ and $\Delta F_2$.

\begin{figure}
    \centering
    \includegraphics[angle=0,width=0.99\linewidth]{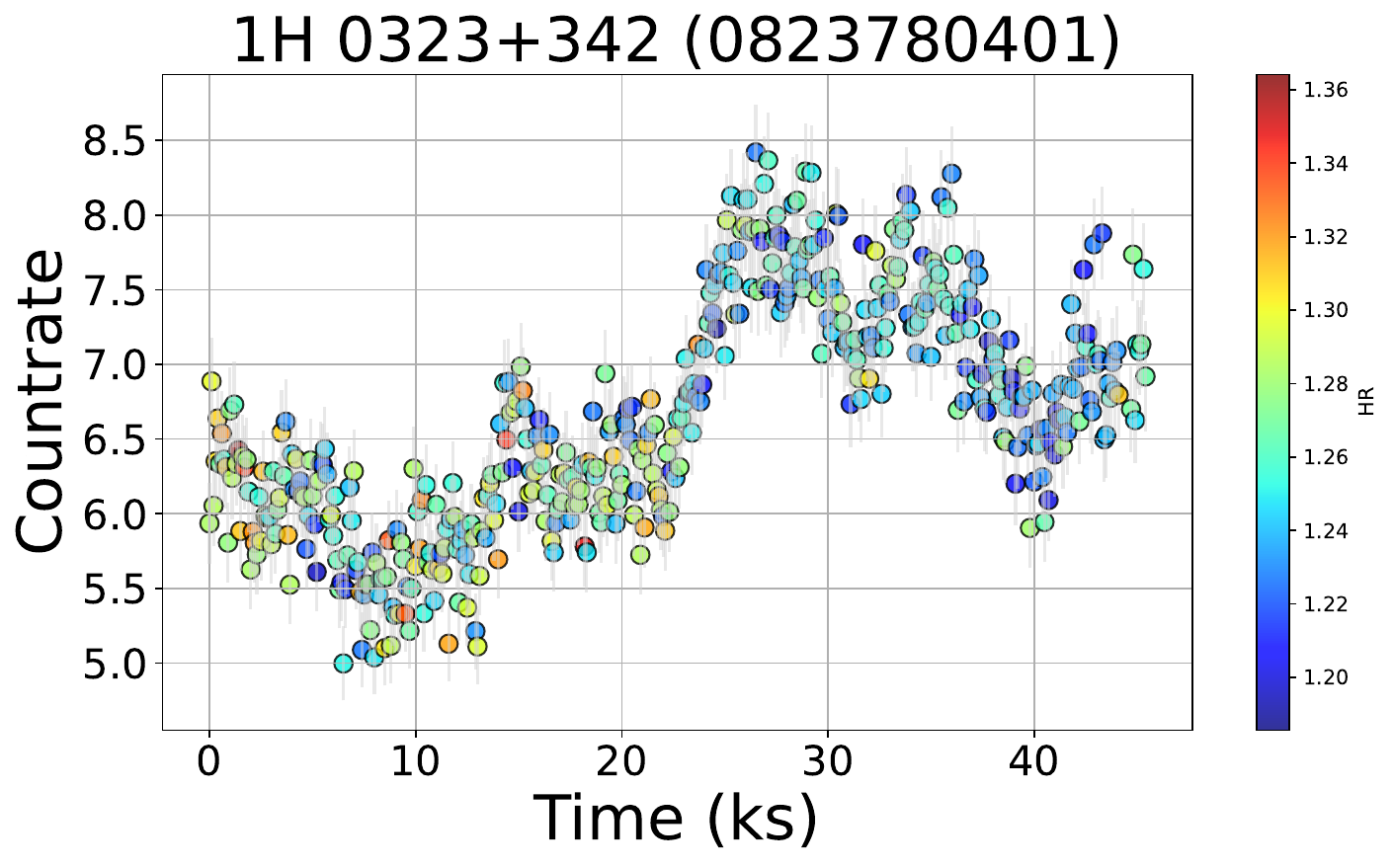}
    \caption{The X-ray (0.3-10 keV) light curve of NLSy1 1H\,0323+342 obtained from OBS-ID: 0823780401.}
    \label{LCplot}
\end{figure}

In Table \ref{TAB2}, Columns (6) and (7) provide the VA and $F_{\rm var}$ for each source. In summary, most objects in our sample exhibit $F_{\rm var}$ values above 5\% and reach a maximum of 16\% (although some lack quality data for variability studies), which is significantly higher than that of some blazars. As an example, comparing this variability with FSRQs and BL Lacs, as presented in literature, we see that for FSRQ 3C\,273, the $F_{\rm var}$, and \text{VA} estimated from the {\it XMM-Newton} observation lie between 0.7 -- 6\% and 0.40 -- 1.33, respectively \citep{Dinesh_2023}. Similarly, for a sample of TeV blazars, it is estimated to be between 0.2 -- 3\% \citep{Devanand_2022}. \cite{Noel_2022}, however, estimated the $F_{\rm var}$ between 0.6 -- 11\% for all the available {\it XMM-Newton} observation for Mkn 421. A mixed sample of FSRQs and BL Lacs has been studied in \cite{10.1093/mnras/stw1667} using {\it XMM-Newton} observations, and the study showed that the FSRQs have lower $F_{\rm var}$ most of the time compared to BL Lac objects. Another study done on blazar PG 1553+113 (LBL) showed the $F_{\rm var}$ between 0.6 -- 18\% using the {\it XMM-Newton} observation \citep{10.1093/mnras/stab1743}. According to the \cite{2022MNRAS.510.5280M} soft X-ray analysis of four TeV blazars (RBS 2070, OJ 287, Mrk 501, and S4 0716+714), all sources exhibit significantly reduced variability $F_{\rm var}<6\%$, except S4 0716+714 ($F_{\rm var}\sim23-33\%$), while \text{VA} varies from 0.51 to 4.42. These results suggest that, based solely on the {\it XMM-Newton} variability, we cannot separate the sources into different classes. \citet{2017MNRAS.466.3309R} compiled a large sample of jetted and non-jetted AGNs and computed the $F_{\rm var}$ using NuSTAR observations in three different energy bands. The NLSy1s exhibit the highest duty cycle variability (87


For 1H\,0323+342, $F_{\rm var}$ ranges from $\sim$10\% to $\sim$16\%, which is also consistent with its hard X-ray $F_{\rm var}$ estimates in \cite{2025arXiv250404492C}, and its VA ranges from 0.68 to 1.67 throughout seven epochs. PMN J0948+0022, $F_{\rm var}$ changes from 5 - 7\%, while VA goes 1.5 to 2.01 over three epochs, PKS 2004-447 shows $F_{\rm var}\sim$9-11\%, while $VA\sim$2-7 in five epochs, and 3C 286 reveals $F_{\rm var}\sim$8\%, while VA=4.6 in a single epoch variable sources. Moreover, sources J1222+0413 and J1644+263 exhibit $F_{\rm var}$, and  $VA$ values of $\sim$5\%, 1.61 and 6\%, 1.49, respectively. On the other hand, sources such as J1246+0238, J1641+3454, J1102+2239, and TXS 2116-077 do not exhibit considerable variability. This is primarily due to high error bars and poor photon statistics, which result in negative excess variance and are therefore excluded from Table \ref{TAB2}. Long-term Swift-XRT studies of a few $\gamma$-NLSy1 \citep{2020MNRAS.496.2213D}, in comparison, report higher $F_{\rm var}$ values (e.g., 31\% for 1H\,0323+342, 44\% for J0948+0022, 51\% for PKS 1502+036, 27\% for J1644+263, and 37\% for PKS 2004-447) on year-long timescales. This underscores the significance of long-term {\it XMM-Newton} observations for a more accurate analysis of X-ray variability in these sources. It also suggests that these objects are more variable on longer timescales compared to the intraday timescale. 
Like blazars, $\gamma$-NLSy1s exhibit significant X-ray variability, which might offer important information on the physical processes in their inner regions constrained by the rapid flux fluctuation.

We investigated the size and dynamics of the emission zones of all sources via the minimum variability timescale (equation~\ref{tvar}; Column~(8) in Table~\ref{TAB2}). The $\tau_{\mathrm{var}}$ can be used to quantify the upper limit for the minimum emitting region size ($R_H$), following the causality argument. The expression for the $R_H$ is given as:
\begin{equation}
    R_H \ge \frac{\delta}{1+z}\, c\, \tau_{\mathrm{var}},
\end{equation}
where $\delta$ is the Doppler factor, $z$ is the source redshift, $c$ is the speed of light.
For 1H\,0323+342, $\tau_{\mathrm{var}}$ varied between 0.15 and 0.52 ks, with an 
average value of 0.36 ks, suggesting a variable emission region of size $R_H \sim 10^{13-14}$ \text{cm}, near the central supermassive black hole. Across five epochs, PKS 2004$-$447 exhibits $\tau_{\mathrm{var}}$ values ranging from 0.07 to 0.41 ks, corresponding to a region size of $R_H \sim 10^{13-14}$ \text{cm}. With $\tau_{\mathrm{var}}$ spanning 0.03 to 0.14 ks, PMN~J0948+0022 exhibits some of the quickest variability, suggesting extremely confined and active zones ($R_H \sim 10^{12-13}$ cm). Similar short timescales (0.13 to 0.47 ks) are also displayed by TXS 2116-077. Across the full sample, $\tau_{\mathrm{var}}$ has a mean value of 0.2 ks and ranges from 0.03 to 0.52 ks. This indicates that the short-term scale variability in the X-ray emission of $\gamma$-NLSy1s comes from very compact regions with a size of $R_H \sim 10^{12-14}$ \text{cm} close to the base of the jet. Based on the yearlong {\it Swift-XRT} light curves, \cite{2020MNRAS.496.2213D} estimated the variability timescale, $\tau_{\mathrm{var}}$, and found it to be on the order of $\sim10$ \text{ks}, which corresponds to $R_H \sim 10^{15}$ cm. This value is obviously much longer than our intraday variability estimates. On the other hand, the X-ray investigations of blazars presented in \cite{2022MNRAS.510.5280M, 2025ApJ...981..118B} indicate an intermediate range of $\tau_{\mathrm{var}} \sim$ a few \text{ks}, which corresponds to an emitting region size of $R_H \sim 10^{14-15}$ \text{cm}.

\subsubsection{STRUCTURE FUNCTION}
In the time domain, the structure function (SF) is used to quantify AGN variability \citep{1994ApJ...433..494T, 2009ApJ...696.1241B, 2016A&A...593A..55V}. The break time scale appears in the SF can reveals the characteristic time scale present in the source.  
The SF for a given light curve is calculated as:
\begin{equation}
\mathrm{SF}(\tau) \equiv \sqrt{\frac{\pi}{2} \left\langle \left| \log f_X(t + \tau) - \log f_X(t) \right|^2 \right\rangle - \sigma^2_{\text{noise}}}.
\end{equation}
Here, $f_X(t)$ and $f_X(t + \tau)$ are the measured X-ray fluxes at time $t$ and $(\tau+t)$, and, $
\sigma^2_{\text{noise}} = \left\langle \sigma^2_n(t) + \sigma^2_n(t + \tau) \right\rangle $. 

\begin{figure}
    \centering
    \includegraphics[angle=0,width=0.99\linewidth]{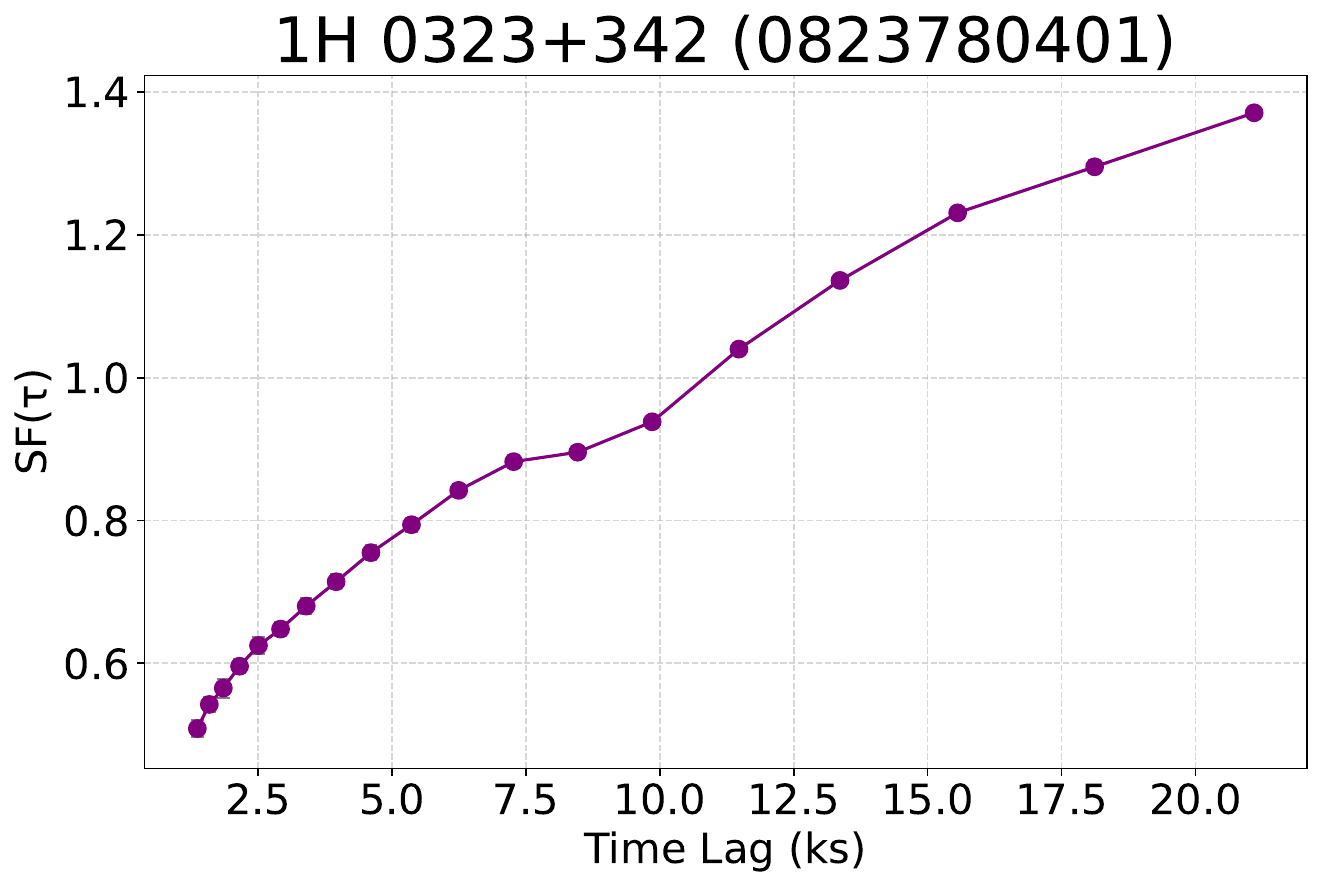}
    \caption{Structure function of one of the {\it XMM-Newton} observations of 1H\,0323+342.}
    \label{SFplot}
\end{figure}
Figure \ref{SFplot} shows the SF plot for one observation of 1H\,0323+342, while for the remaining sources, the plots are shown in the appendix. In this particular case, we do not see any break in the SF, suggesting the presence of a longer timescale in the system. In the full sample, only 8 out of 29 epochs show an indication of a break (Table~\ref {TAB2}), which shows that most sources do not show short-term turnover frequencies. The details for each individual object are discussed in its respective subsection. It should be noted that \cite{2010MNRAS.404..931E} argued that the SF can exhibit artificial breaks, which are determined by the length of the data and gaps, as well as the PSD shape, rather than any real physical timescales.



\subsubsection{POWER SPECTRAL DENSITY}
Finding the underlying physical mechanisms of AGN variability can be aided by employing techniques that encompass both the temporal and frequency domains. Measuring the ``variability power'' at a specific temporal frequency (or timescale), the Power Spectral Density (PSD) is a crucial analytical tool in the frequency domain.  This helps us understand the characteristics of variability and provides clues about its potential source.  For discrete sets of observations, the Discrete Fourier Transform (DFT) is a commonly used PSD measure \citep[see][]{2003MNRAS.345.1271V,2012A&A...544A..80G,2025ApJ...981..118B}. For a given lightcurve sampled at times  $t_j$ with $j = 1, 2, ..., n$,  which occurs at frequencies $\nu_{\rm min} = \frac{1}{T}$, $2 \nu_{\rm min}$ ... , $\nu_{\rm max} = \frac{1}{2\Delta t}$, it can be expressed as,
\begin{equation}\label{eq8}
   P(\nu) = \frac{T}{\bar{x}^2n^2}\left|\sum_{j=1}^n x(t_j)e^{-i2\pi\nu t_j}\right|^2,
\end{equation}
where $T$ is the length of the light curve, $\Delta t$ is the mean sampling step and $\bar{x}$ is the mean flux (counts\,s$^{-1}$) of the lightcurve. Statistical uncertainty in the variability power associated with the detector is commonly referred to as Poisson noise and is given by, 
\begin{equation}
P_{\rm Poisson} =
   \frac{T \sigma_{\rm stat}^{2}}{n \bar{x}^2} \quad    \text{and} \quad
\sigma_{\rm statistical}^2 =
    \sum_{j=0}^{n-1} \frac{(\Delta x_{\rm j})^2 }{n} 
\end{equation}
where $\Delta x_{\rm j}$ represents the error in the observed flux at a given time $t_j$. A PSD study of AGNs has been conducted, assuming a power-law variation of temporal frequency in both phenomenological and physical models, utilizing light curves across different energy regimes. The blazar PSD can be expressed using the PL form of the temporal frequency
\begin{equation}
    P(\nu)=A\cdot \nu^{-\beta_{\rm{P}}}+C,
\end{equation}
where $C$ is the Poisson noise level, $\beta_{\rm{P}}$ is the spectral power index, and {\it A} is a normalization constant, as explained in \citep{2002MNRAS.332..231U, 2018A&A...620A.185N}. We estimated the PSD for all light curves and fitted it with a simple power-law model, after binning the PSD data in frequency \citep[see for example the discussion in][]{2003MNRAS.345.1271V}.

A power-law fitted PSD is presented in Figure \ref{PSDplot} along with the remaining ones in the appendix. The power-law's slope  PSD can offer crucial hints regarding the physical mechanisms causing the blazer's variability. These fluctuating PSD slopes correspond to various stochastic processes, such as red noise ($\beta_p \sim$2), flicker noise ($\beta_p \sim$1), and other noise processes with a steeper PL index \citep[i.e., $\beta_p >$ 2;][]{1978ComAp...7..103P, 1997MNRAS.292..679L, 2025A&A...693A.319K}. Additionally, $\beta_p \sim 0$ is sometimes observed, which is indicative of an uncorrelated white-noise-type stochastic process.\citep{1978ComAp...7..103P}. 
The MWL studies conducted in the past suggest a variable and frequency-dependent PSD slope, such as in the optical R-band analysis of 31 blazars presented in \cite{2018A&A...620A.185N}, which shows that $\beta_P$ varies from 0.8 to 2.8. Another R-band work by \cite{2021ApJ...909...39G} also shows a similar result with a $\beta_P$ value lying in 1.4-4.0, and for {\it TESS} blazars, it shows a range of 1.7 to 3.2 \citep{2023MNRAS.518.1459P}. A radio monitoring of blazars at 15 \text{GHz} frequency suggests that there is a clustering of $\beta_P$ around 2.3 \citep{2014MNRAS.445..428M, 2015IAUS..313...17M}. The X-ray analysis conducted in \cite[see, e.g.,][]{2022MNRAS.510.5280M, 2025ApJ...981..118B} also shows variable slopes, with slopes varying from 0.02 to 4.5. A sample of {\it Fermi}-LAT blazars studied in \cite{2020ApJ...891..120B} shows the range of $\beta_P$ values from 0.03 to 0.82.

The power-law fitted PSD slope ($\beta_p$) in Table \ref{TAB2} Column (9) sheds light on the source temporal variability behavior. The $\beta_P$ values in this work range from 0.1 to 1.52, exhibiting a significant scatter, which implies a complex transition from white-flicker to red noise. For 1H\,0323+342, $\beta_p$ exhibits a considerable variation from 0.93 to 1.52 over all seven epochs, indicating a flicker-noise dominated variability component throughout time. The PSD slopes of 3C 286 and J1644+263 are 1.27 and 1.14, respectively, which again suggests a flicker-noise-like behaviour. PKS 2004-447 exhibits a vast range of $\beta_p$ values, from 0.42 to 0.98, indicating a combination of random and flicker noise in different epochs, which displays moderate variability with a somewhat consistent temporal structure. In cases where the observed $\beta_p$ values are close to 1, it suggests that the light curves exhibit more power in the short-term variability  \citep{2003MNRAS.345.1271V}. Sources like J1641+3454, J1102+2239, J1222+0413, PKS 1244-255, and TXS 2116-077 show lower $\beta_P \sim$0.5 values. The $\beta_p$ values found for $\gamma$-NLSy1s in our analysis are generally in agreement with those found in blazars, as discussed above, suggesting that the X-ray variability characteristics of $\gamma$-NLSy1 galaxies might be similar to those of blazars.

\begin{figure}
    \centering
    \includegraphics[angle=0,width=0.99\linewidth]{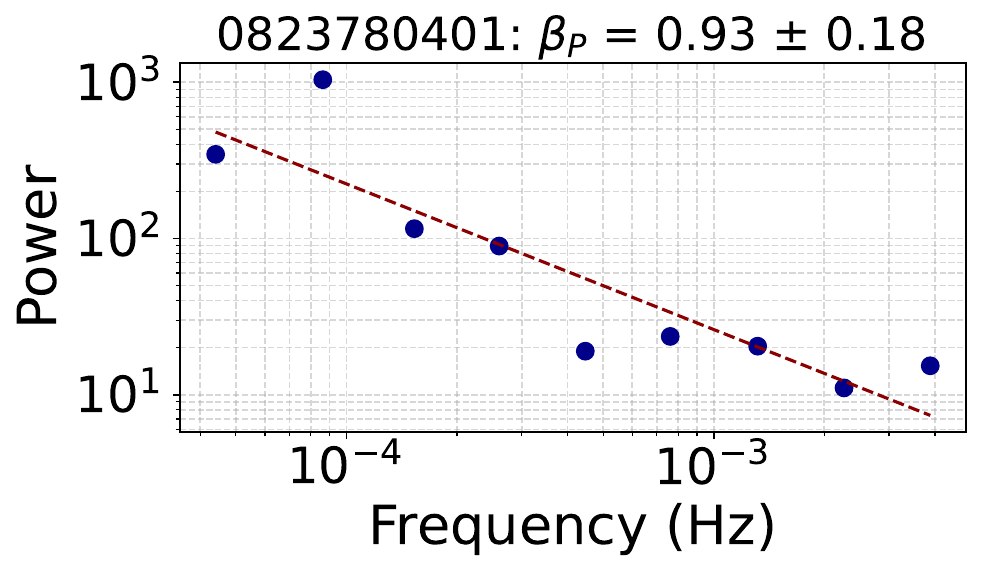}
    \caption{PSD produced for one of the observations of 1H\,0323+342 is shown in green data points, and the yellow line is the best-fit power-law model.}
    \label{PSDplot}
\end{figure}

\subsubsection{FLUX HISTOGRAMS}
AGN lightcurves often reveal flux variability across the electromagnetic spectrum; however, the details of the physical mechanisms responsible for this variability remain a topic of debate. Examining the count-rate distribution, which can be represented by a suitable probability density function (PDF), is an essential technique for understanding and controlling physical processes.  As discussed, for instance, for optical \citep{2024A&A...686A.228O}, X-ray \citep[see, e.g.,][]{2022MNRAS.510.5280M, 2025ApJ...981..118B} Gamma-ray observations \citep{2020ApJ...891..120B} and {2022MNRAS.510.5280M, 2025ApJ...981..118B}.

The normal flux distribution is given as:
\begin{equation}
        N(f) = \frac{1}{\sqrt{2\pi\sigma_1^2}} {\exp\left({-\ \frac{(F - \langle F_1 \rangle)^2}{2\sigma_1^2}}\right)}{\ \rm ,}
\end{equation}
where $\langle F_1 \rangle$ and $\sigma_1$ are the mean and standard deviations of the normal distribution.

A lognormal distribution function for a given mean flux $\langle F_{2} \rangle$ and standard deviation $\sigma_2$ is given as:
\begin{equation}
    Ln(f) =  \frac{1}{\sqrt{2\pi}\sigma_2F}\exp\left(-\frac{(\ln F - \langle F_2 \rangle)^2}{2\sigma_2^2}\right),
\end{equation}

\begin{figure}
    \centering
    \includegraphics[angle=0,width=0.99\linewidth]{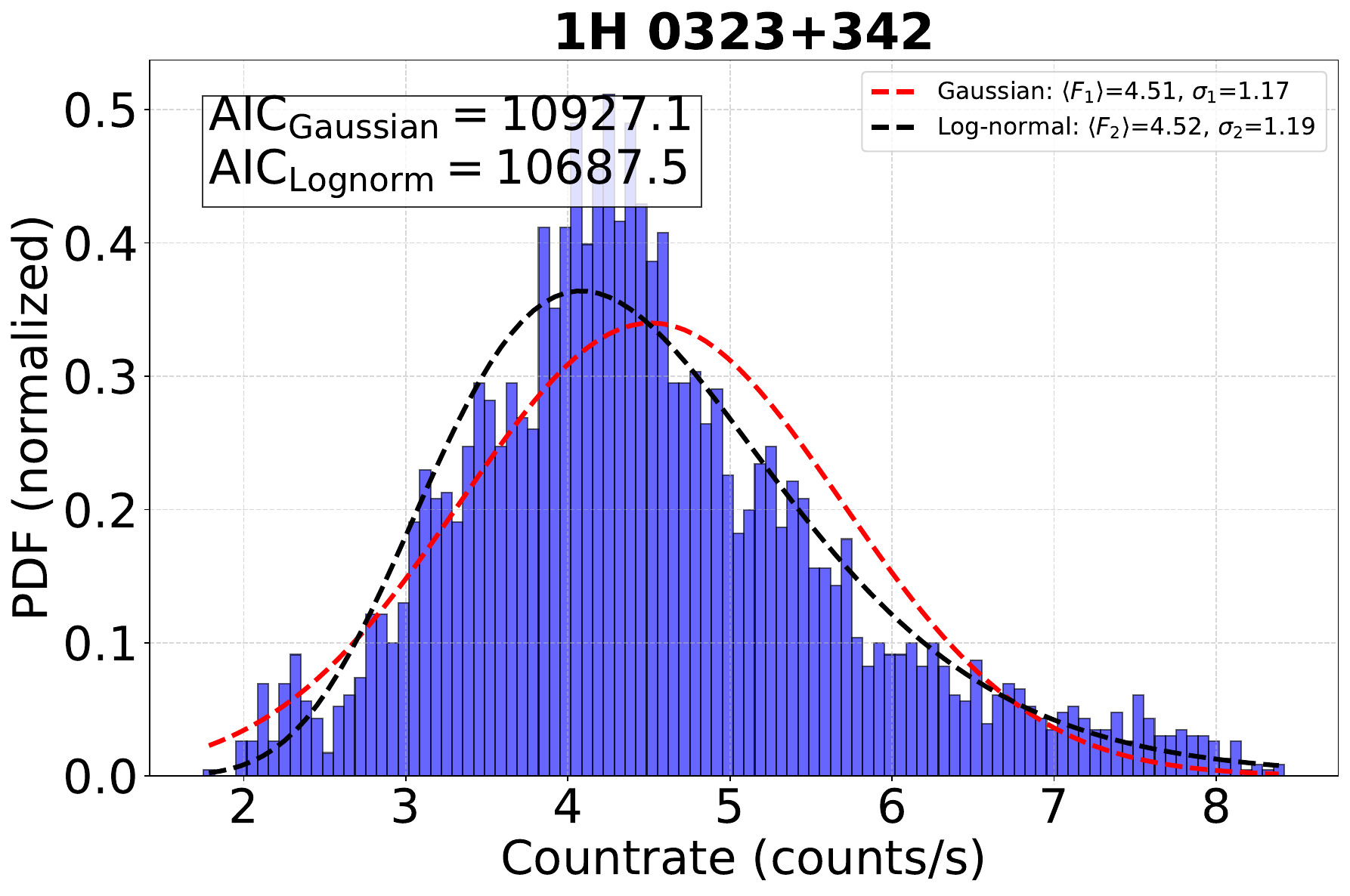}
    \includegraphics[angle=0,width=0.99\linewidth]{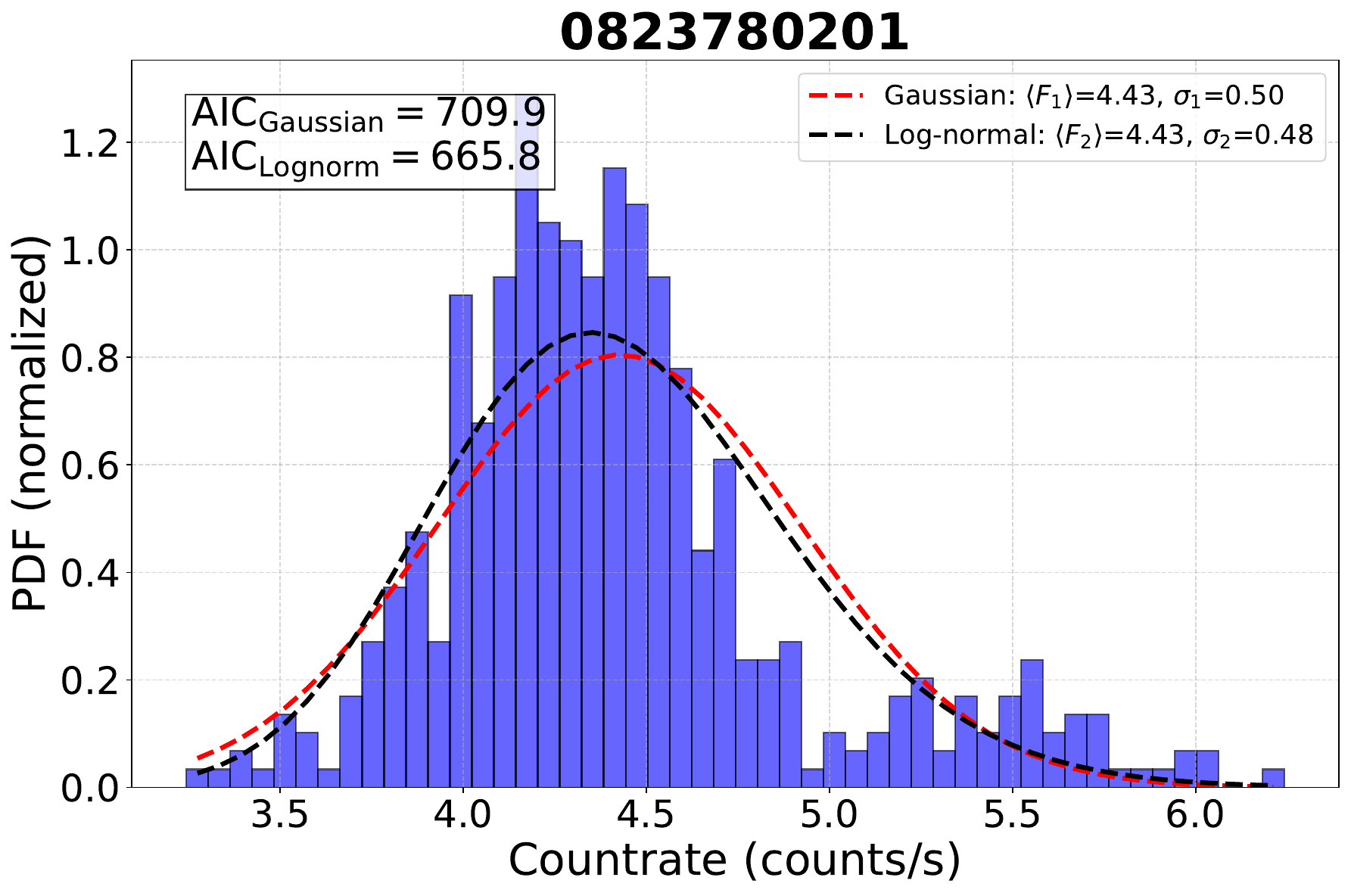}
    \includegraphics[angle=0,width=0.99\linewidth]{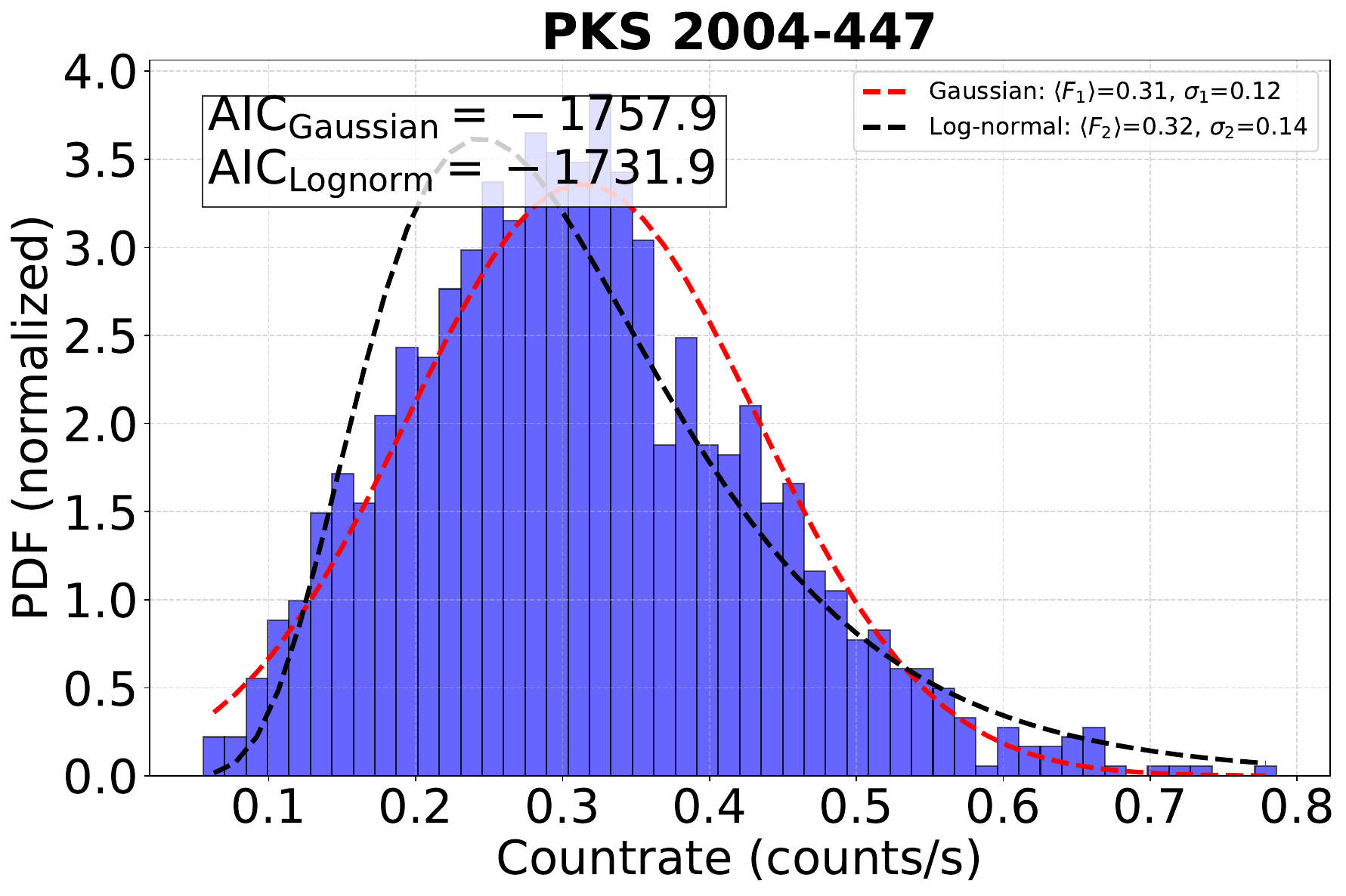}
    \caption{Top panel: The combined flux (counts\,s$^{-1}$) distribution of 1H\,0323+342 in all seven epochs, Middle panel: one epoch of 1H\,0323+342, Bottom panel: Combined flux distribution PKS 2004-447.}
    \label{PDFplot}
\end{figure}

Flux distribution studies are extensively conducted in X-ray binaries and AGNs, where disk emission is dominated by accretion activity. A normal flux distribution suggests a stochastic and linear variation mostly caused by the additive process \citep{2010LNP...794..203M}. A log-normal flux distributions are mostly associated with a stochastic non-linear variation, mostly seen in AGNs \citep{2005MNRAS.359..345U} and blazars \citep{2017ApJ...849..138K,2018RAA....18..141S,2021MNRAS.502.5245P}.
We have produced the flux distribution of all the available light curves and fitted them with both models. For the fitting, we have adopted the maximum likelihood method presented in \citep{2023MNRAS.518.1459P}. The PDFs of 1H\,0323+342 across display somewhat complex patterns,  with some indications of bi-modality found for some epochs  (e.g., middle panel of Figure \ref{PDFplot}), suggesting multiple emission sources or state transitions. However, when the fluxes from all epochs are combined (upper and bottom panel of Figure \ref{PDFplot}), the resulting distribution clearly takes on a log-normal, unimodal shape, suggesting the variability is caused by multiplicative processes as seen in radio-quiet AGN, where the emissions are dominated by the disk. On the other hand, as argued in \cite{2012A&A...548A.123B}, the log-normal distribution can also be produced in the jet as a result of disordered mini-jets within the primary relativistic jet. The remaining flux distribution for each epoch is presented in the appendix.

\subsection{SPECTRAL VARIABILITY}
\subsubsection{HARDNESS RATIO}
To quantify spectral variations over time and flux states in a model-independent manner, as a probe of the evolution of the X-ray-emitting electron population, we have utilized a Hardness Ratio (HR). A harder-when-brighter (HWB) trend suggests that the acceleration of high-energy electrons dominates, shifting the spectrum towards higher energies; a softer-when-brighter (SWB) pattern is indicative of intense radiative cooling or enhanced external seed-photon fields, causing inverse-Compton emission \citep[common in FSRQs; e.g.][]{2003A&A...402..929B, 2004A&A...424..841R}. The HR is defined as:
\begin{equation}
\label{hard}
    {\rm HR} = \frac{H}{S},
\end{equation}
where $H$ and $S$ are the flux (counts\,s$^{-1}$) in the hard (2--10 keV) and soft (0.3--2 keV) bands, respectively. The associated error in the HR ($\sigma_{\rm HR}$) is estimated as,
\begin{equation}
    \sigma_{\rm HR} = \frac{2}{S^2}\sqrt{H^{2}\sigma^{2}_{\rm S} + S^{2}\sigma^{2}_{\rm H}},
\end{equation}
where $\sigma_H$ and $\sigma_S$ are the errors in hard and soft bands, respectively. A HR ratio plot of one epoch is shown in the Figure \ref{HRplot}, while the remaining are shown in the appendix. A clear SWB tendency was observed in sources such as 1H\,0323+342 (Obs IDs: 0823780301, 0823780501) and PKS 2004-447 (Obs ID: 0200360201), while sources like 3C 286, 1H\,0323+342 (Obs ID: 0823780701), and J1644+263 (Obs ID: 0783230101), shows a mixed trend of SWB and HWB, in our investigation of $\gamma$-NLSy1. However, in the remaining sources, the behaviour remains inconclusive or does not exhibit a distinct spectral pattern due to weak photon statistics and high errors in their light curves and hence in HR measurements. This spectral behavior is frequently observed in certain blazars, especially in optical bands. On intraday timescales, BL Lacs exhibit a bluer-when-brighter (BWB) trend, while the FSRQ population shows a redder-when-brighter (RWB) trend \citep{2022MNRAS.510.1791N}. In the X-ray \citep{2015ApJ...807...79H, 2020MNRAS.491..858O, 2022MNRAS.510.5280M, 2025ApJ...981..118B} and the $\gamma$-ray \citep{2017ApJ...847....7B, 2025A&A...703A.162D} range, both blazars and $\gamma$-NLSy1s also show a similar spectral trend, where the spectra soften/harden with count-rate enhancements.

\begin{figure}
    \centering
    \includegraphics[angle=0,width=0.99\linewidth]{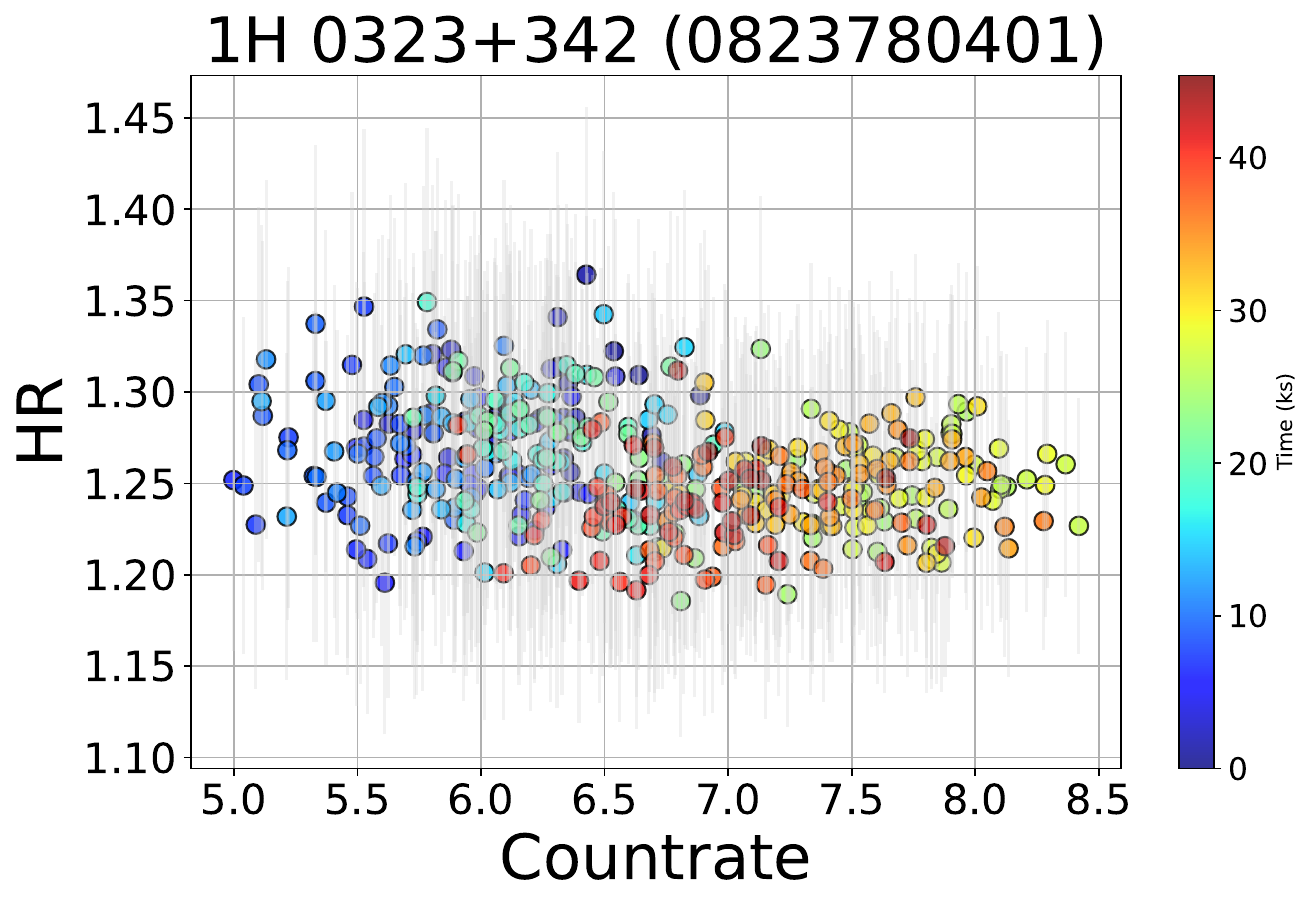}
    \caption{HR plot of one of the good light curves of 1H\,0323+342. The color bars represent the exposure time.}
    \label{HRplot}
\end{figure}

\begin{table*}
\caption{The Power-Law (PL) and Power-Law+Blackbody (PL+BB)  X-ray spectral fitting results. }
    \centering
    \begin{tabular}{cccccccccc} \hline
      Object & Obs. ID & model & $\Gamma$ & $kT$ (keV) & Flux ($10^{-13}$ erg s$^{-1}$\,cm$^{-2}$) &  $\chi^2_{r}$ & $N_H$ (cm$^{-2})$\\ \hline
      1H\,0323+342  & 0764670101  & PL+BB &  1.82$\pm$0.01 & 0.145$\pm$0.003 & 190.72$\pm$0.61  & 249.51/167  & 1.14$\pm$0.04 \\
       & 0823780201  & PL+BB &  1.88$\pm$0.01 & 0.149$\pm$0.037 & 165.79$\pm$0.38   & 204.83/163 & 1.08$\pm$0.05\\
        & 0823780301  & PL+BB & 1.98$\pm$0.01  & 0.135$\pm$0.028  & 149.48$\pm$0.43  & 221.78/159 & 1.23$\pm$0.05\\
         & 0823780401 & PL+BB & 1.89$\pm$0.01 & 0.140$\pm$0.003  & 265.02$\pm$0.67 & 176.23/165 & 1.24$\pm$0.04\\
     & 0823780501 & PL+BB & 1.95$\pm$0.01  & 0.133$\pm$0.028  &  174.97$\pm$0.37 &  189.44/161 & 1.27$\pm$0.05\\
       & 0823780601 & PL+BB & 1.91$\pm$0.01  & 0.139$\pm$0.03  &  160.08$\pm$0.57   & 221.72/162 & 1.31$\pm$0.06 \\
        & 0823780701 & PL+BB & 1.90$\pm$0.01  & 0.138$\pm$0.003 &   111.23$\pm$0.41   & 191.43/155 & 1.17$\pm$0.06\\ \\
        
         SDSS J1641+3454 & 0860640201    &  PL &  1.77$\pm$0.10 &  - & 9.03$\pm$0.34  & 40.56/39 & 2.48$\pm$0.31\\ 
     & 0860640301   & PL & 1.64$\pm$0.09  &  - & 4.45$\pm$0.18  &   37.70/46  & 2.22$\pm$0.88\\  \\
     
    3C 286 & 0881760101 & PL &  2.23$\pm$0.05 & -  &   6.45$\pm$0.17   & 73.03/58 & 0.23$\pm$0.09 \\  \\
        FBQS J1102+2239 & 0690090301    &  PL & 2.19$\pm$0.22  &  - & 2.15$\pm$0.02  & 2.57/7 & 0.12 \\ \\
         SDSS J0946+1017 & 0800040101  & PL &  1.47$\pm$0.03 &  - & 4.06$\pm$0.09  & 122.39/91  & 0.25 \\  \\
      PMN J0948+0022 & 0790860101  &  PL+BB   &  1.52$\pm$0.02 & 0.123$\pm$0.009  &  23.79$\pm$0.25 & 178.47/149 & 0.30$\pm$0.09 \\
       & 0673730101 & PL+BB & 1.59$\pm$0.02  & 0.114$\pm$0.008  & 45.10$\pm$0.41  &  148.24/141   & 0.37$\pm$0.08\\
        & 0502061001 & PL+BB & 1.58$\pm$0.05     &  0.105$\pm$0.008 & 68.52$\pm$0.35  & 135.17/149 & 0.33$\pm$0.06\\  \\
        
          CGRaBS J1222+0413 & 0401790601  & PL+BB & 1.29$\pm$0.04     & 0.164$\pm$0.012  & 36.08$\pm$0.64  & 142.57/112 & 0.16 \\  \\
         SDSS J1246+0238   & 0690090201  & PL+BB &  1.71$\pm$0.14 &      0.046$\pm$0.012 & 2.72$\pm$0.25  & 22.62/14 & 0.19 \\ \\
         TXS 1419+391  & 0845040601  & PL &  2.05$\pm$0.03 &  -    &  4.62$\pm$0.11 & 93.68/73 & 1.13 \\  \\
         PKS 1502+036 & 0690090101  & PL+BB & 1.54$\pm$0.09  &  0.11$\pm$0.02  & 4.08$\pm$0.09  & 52.07/37 & 0.20$\pm$0.03 \\   \\
         RGB J1644+263 & 0783230101  & PL+BB & 1.73$\pm$0.02 & 0.11$\pm$0.01 &  34.98$\pm$0.27 & 126.85/142  & 0.70$\pm$0.09 \\  \\
         TXS 2116-077  & 0784090201  & PL &  1.84$\pm$0.05 &   -   &  8.85$\pm$0.21 & 62.41/41 & 1.00$\pm$0.12  \\
         & 0784090301 & PL & 1.75$\pm$0.09  & -  &  3.36$\pm$0.14  & 47.63/44  & 1.28$\pm$0.22 \\   \\
         PKS 2004-447  & 0790630101 & PL & 1.62$\pm$0.02  &  -   & 10.02$\pm$0.16  & 114.04/111 & 0.26$\pm$0.06 \\ 
           & 0853980701 & PL & 1.49$\pm$0.02  &  -   & 21.74$\pm$0.35  & 91.47/77 & 0.04$\pm$0.08 \\ 
         & 0694530101 & PL &  1.74$\pm$0.05 &  -  & 4.77$\pm$0.19  & 51.57/56 & 0.35$\pm$0.12 \\
         & 0694530201 & PL & 1.69$\pm$0.03  &  -  & 7.71$\pm$0.20  & 114.04/90 & 0.32$\pm$0.07 \\ 
         &  0200360201 & PL+BB & 1.53$\pm$0.04 & 0.04$\pm$0.01 & 17.01$\pm$0.30 & 86.79/96 & 0.17$\pm$0.14\\ \\
         TXS 0103+189  & 0932790201  & PL &  1.73$\pm$0.01 &  -    &  3.56$\pm$0.20 & 37.75/35 & 0.03$\pm$0.01 \\  \\

        PKS 1244-255  & 0650382301  & PL &  1.82$\pm$0.06 &  -    &  40.03$\pm$1.10 & 56.25/54 & 0.09$\pm$0.01 \\  \\

        TXS 1308+554  & 0741031601  & PL &  1.79$\pm$0.15 &  -    &  3.20$\pm$0.31 & 12.78/13 & 0.01$\pm$0.01 \\  \\
         \hline    
    \end{tabular}
    \par\smallskip
    \noindent\textbf{Table Note.} Col 1: Source name; Col 2: Obs. ID; Col 3: Model; Col 4: X-ray best-fitted model parameters $\Gamma$ for PL model; Col 5: blackbody temperature in (KeV); Col 6: Unabsorbed X-ray flux in 0.2-10 keV energy range in units of \ergcm{-13}; Col 7: Reduced chi-square value.
    \label{TAB3}
    \end{table*}

\begin{table*}
\caption{Broken Power Law (BPL) spectral fitting results.}
    \centering
    \begin{tabular}{ccccccccc} \hline
      Name   & Obs. ID & Model & $\Gamma_1$ & $\Gamma_2$ & $E_{\rm break}$ (keV) & Flux ($10^{-13}$ erg s$^{-1}$\,cm$^{-2}$) & $\chi^2_{\rm red}$  & $N_H$\\ \hline
      1H\,0323+342  & 0764670101 & BPL & 2.71$\pm$0.03 & 1.84$\pm$0.08 & 1.58$\pm$0.02 & 222.56$\pm$0.34 &345.88/167 & 1.71$\pm$0.04\\
         & 0823780201 & BPL  & 2.65$\pm$0.04 & 1.89$\pm$0.01 & 1.64$\pm$0.03 & 242.05$\pm$0.27 & 235.09/163 & 1.62$\pm$0.05\\
         & 0823780301 & BPL  & 2.97$\pm$0.04 & 1.98$\pm$0.01 & 1.58$\pm$0.02 & 264.21$\pm$0.65 & 272.01/159 & 1.83$\pm$0.05\\
         & 0823780401 & BPL  & 2.62$\pm$0.04 & 1.92$\pm$0.01 & 1.55$\pm$0.02 & 364.60$\pm$0.70 & 237.02/165 & 1.69$\pm$0.05 \\
         & 0823780501 & BPL  &  2.90$\pm$0.05 & 1.98$\pm$0.01 & 1.48$\pm$0.02 & 285.25$\pm$0.48 & 250.93/161 & 1.83$\pm$0.06 \\
         & 0823780601 & BPL  & 2.87$\pm$0.04 & 1.94$\pm$0.01 & 1.53$\pm$0.02 & 273.17$\pm$0.10 & 291.90/162 & 1.91$\pm$0.06\\
         & 0823780701 & BPL  & 2.96$\pm$0.06  & 1.93$\pm$0.01 & 1.53$\pm$0.02 & 200.68$\pm$0.39 & 217.47/155 & 1.84$\pm$0.07\\ \\
         
      CGRaBS J1222+0413  &  0401790601 &BPL  & 1.62$\pm$0.09 & 1.11$\pm$0.08  & 2.48$\pm$0.37 & 37.90$\pm$0.11  & 127.26/111 & 0.18$\pm$0.10\\     

 PKS 1502+036  &  0690090101 &BPL  & 2.20$\pm$0.54 & 1.50$\pm$0.10  & 1.24$\pm$0.22 & 4.32$\pm$0.32  & 48.19/36 & 0.19$\pm$0.52\\ \\
 RGB J1644+263 & 0783230101 &BPL  & 2.25$\pm$0.10 & 1.74$\pm$0.02  & 1.19$\pm$0.05 & 4.32$\pm$0.32  & 130.12/142 & 0.83$\pm$0.10\\ \\
      
       PMN J0948+0022  & 0790860101 & BPL  &1.92$\pm$0.09 & 1.53$\pm$0.02  & 1.25$\pm$0.07 & 24.94$\pm$0.24  & 183.33/149 & 0.41$\pm$0.09  \\ 
                       & 0673730101 &BPL   & 2.30$\pm$0.09  & 1.59$\pm$0.02 &  1.28$\pm$0.04 & 50.58$\pm$0.48 & 152.01/141 & 0.54$\pm$0.85\\
                       & 0502061001 & BPL  & 2.26$\pm$0.18  & 1.64$\pm$0.07 & 1.22$\pm$0.14 & 7.05$\pm$0.30 & 127.02/122 & 0.61$\pm$0.14 \\ \hline
    \end{tabular}
    \par\smallskip
    \noindent\textbf{Table Note.} Col. 1: Source Name; Col. 2: Observation ID; Col. 3: Model; Col. 4: low-energy photon index; Col. 5: high-energy photon index; Col. 6: break energy (keV); Col. 7: 0.2-10 keV unabsorbed flux \ergcm{-13}; Col. 8: $\chi^2/dof$; Col. 9: Column density ($10^{21}$,$cm^{-2}$).
    \label{TAB4}
\end{table*}

\subsubsection{SPECTRAL FITTING}
In this study, we used \emph{XSPEC} models in the 0.3-10 keV energy range to process the X-ray spectrum analysis of 16 $\gamma$-ray detected NLSy1 galaxies. We used a power-law (PL) model to fit all X-ray spectra of all observations in \emph{XSPEC}. The mathematical formulation of the model is given as follows:
\begin{equation}\label{EQPL}
    \frac{dN}{dE} = N_0\cdot E^{-\Gamma},
\end{equation} 
where $N_0$ (photons \(\mathrm{keV^{-1}\, cm^{-2}\, s^{-1}}\) at 1 keV) and $\Gamma$ stand for the normalization constant and the X-ray photon index, respectively. A PL injection of high-energy particles in the turbulent jet often produces a non-thermal emission spectrum that resembles the particle distribution. 
We have found that 10 sources exhibit a PL spectral behaviour, while six sources do not yield an acceptable fit to the PL, resulting in high residuals below 2 keV and Chi-square ($\chi^2$) values. Consequently, to improve this, we have employed alternative models, such as the Broken Power-law (BPL), which also implies the existence of a spectral break, with two PL indexes and is expressed as follows:
\begin{equation}\label{bpl}
A(E)=\left\{\begin{array}{ll}
K E^{-\Gamma_{1}}, & \text { if } E \leq E_{\rm break}, \\
K E_{\rm break}^{\Gamma_{2}-\Gamma_{1}}(E / 1 k e V)^{-\Gamma_{2}}, & \text { if } E>E_{\rm break},
\end{array}\right.
\end{equation}
where $K$ is the normalization constant, the spectral indices before and after the spectral break are $\Gamma_1$ and $\Gamma_2$, and $E_{\text{break}}$ is the break energy. The BPL spectral index suggests that relativistic electrons in the jet cool more rapidly due to radiation loss, leading to a spectral break. The BPL improves the $\chi^2$ values for five sources (1H\,0323+342, PMN J0948+0022, CGRaBS J1222+0413, PKS 1502+036, RGB J1644+263), compared to the PL fitting. The spectral fitting results for the BPL model are presented in Table \ref{TAB4}. The distribution of the BPL parameters is presented in Figure \ref{1parhist}. 
For 1H\,0323+342, the spectral parameters are given as $\langle \Gamma_1\rangle=2.81$, $\langle \Gamma_2\rangle=1.88$, and $\langle E_{b}\rangle=1.56$ keV, suggesting a steeper spectrum and higher $E_{\text{break}}$ in the whole sample (except for J1222+0413, which shows the highest $E_{\text{break}}$ of 2.48 keV and the hardest spectrum). The $\Gamma_2$ values show a jet domination above the break energy. 

NLSy1 galaxies are known for their distinctive soft-excess feature below 2 keV \citep{2006MNRAS.370..245G, 2020MNRAS.496.2213D}, and this is also evident in the residuals of the PL fit. 
Hence, to address this soft excess, we added a blackbody (BB) model in conjunction with a PL, indicating a simultaneous coexistence of thermal and non-thermal emission. This is possible either because the sources are observed at a large viewing angle or have a weak jet, which allows disk radiation to couple with the jet. The BB model is defined as:
\begin{equation}\label{EQBB}
    \frac{dN}{dE} = N_{0}\frac{E^2}{(kT)^4 .(\exp(E/kT) - 1)},
\end{equation} 
where the normalization is denoted by $N_0$ and $kT$ is the temperature (in keV). Again, all the sources that we were unable to fit with the PL model show a significant improvement with the PL+BB model. The spectral fit with this model is presented in Table \ref{TAB3}, and the distribution of parameters is shown in Figure \ref{parhist}. Compared to the BPL, the PL+BB gives a better fit for 1H\,0323+342, J1246+0238, and one epoch of PKS 2004-447, while PMN J0948+0022 and PKS 1502+036 show similar $\chi^2_{\text{red}}$ values, while a BPL fit give a better  $\chi^2_{\text{red}}$ value for  CGRaBS J1222+0413.  
An example of an X-ray spectrum fitted with a PL+BB is presented in Figure \ref{spectraplot}, while the other 28 spectra are shown in the appendix. The NLSy1 1H\,0323+342 has seven {\it XMM-Newton} observations, and we have found that all these are well-fitted by a PL+BB, with improved $\chi^2_{\text{red}}$ compared to a BPL. We have not noticed any significant spectral change; the $\Gamma$ varies from 1.79 to 1.96 with a mean value of $\langle \Gamma \rangle=1.88$, and $kT$ changes from 0.133 to 0.145\,keV, with a mean value of $\langle kT \rangle=0.14$\,keV in all seven epochs. The NuSTAR spectral study of 1H\,0323+342 presented in \cite{2025arXiv250404492C} also exhibits a consistent photon index with a mean value of $\langle \Gamma \rangle =1.81$. The spectra in all three epochs of PMN J0948+0022 were fit by a 
BB+PL with an average $\langle \Gamma \rangle $=1.57 and $\langle kT \rangle $=0.115 keV. 
A BB+PL model was also found to be an acceptable fit to the X-ray spectra of 
CGRaBS J1222+0413 ($\Gamma=1.29$, $kT=0.164$ keV), 
SDSS J1246+0238 ($\Gamma=1.71$, $kT=0.046$ keV), 
PKS 1502+036 ($\Gamma=1.54$, $kT=0.11$ keV), 
and RGB J1644+263 ($\Gamma=1.75$, $kT=0.11$ keV).

\citet{2006MNRAS.370..245G} have analysed the X-ray spectra of PKS 2004-447 and have reported a soft-excess below 1 keV in during the 2004 epoch, which we also recovered, however we have analysed additional four epochs and have not found any signature of a soft-excess, which is also consistent with the studies presented in \cite{2016A&A...585A..91K} and \citet{2021A&A...654A.125B}. The two-component spectral fitting suggests that a luminous accretion disk produces thermal radiation, whereas the power-law component represents jet-related non-thermal emission, most likely from synchrotron and inverse-Compton processes. This dual emission model further confirms that some NLSy1s are a transitional class of AGN and supports the idea that they host relativistic jets similar to blazars. It also offers valuable insights into jet-disk interaction and AGN evolution. It suggests that these are the best candidates to study the disk-jet coupling.

\begin{figure}
    \centering
     \includegraphics[angle=-90,width=0.99\linewidth]{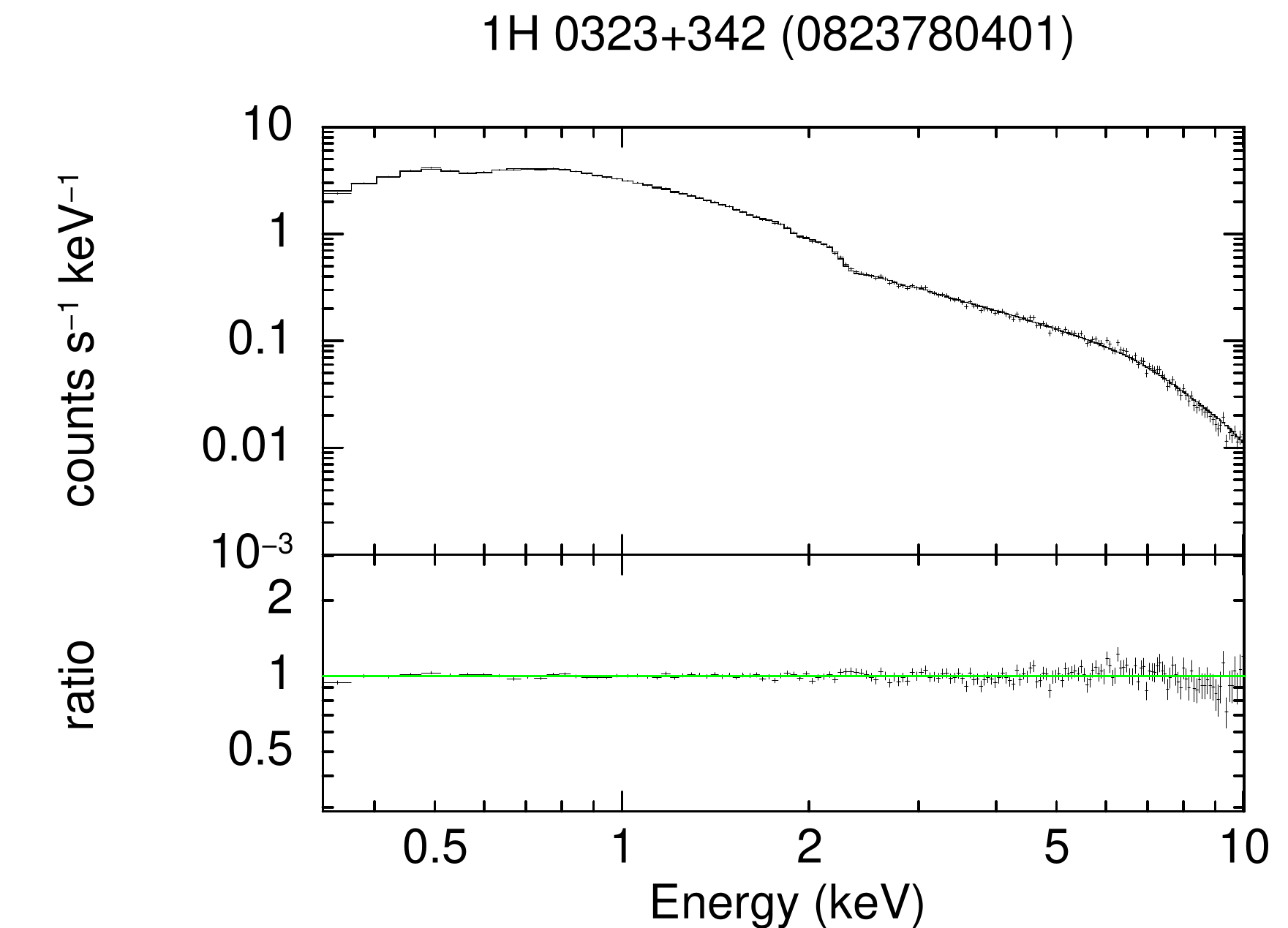}
    \caption{{\it XMM-Newton} spectrum of one of the observations of 1H\,0323+342 fitted with a PL+BB.}
    \label{spectraplot}
\end{figure}

\begin{figure*}
    \centering
    \includegraphics[width=18.0cm]{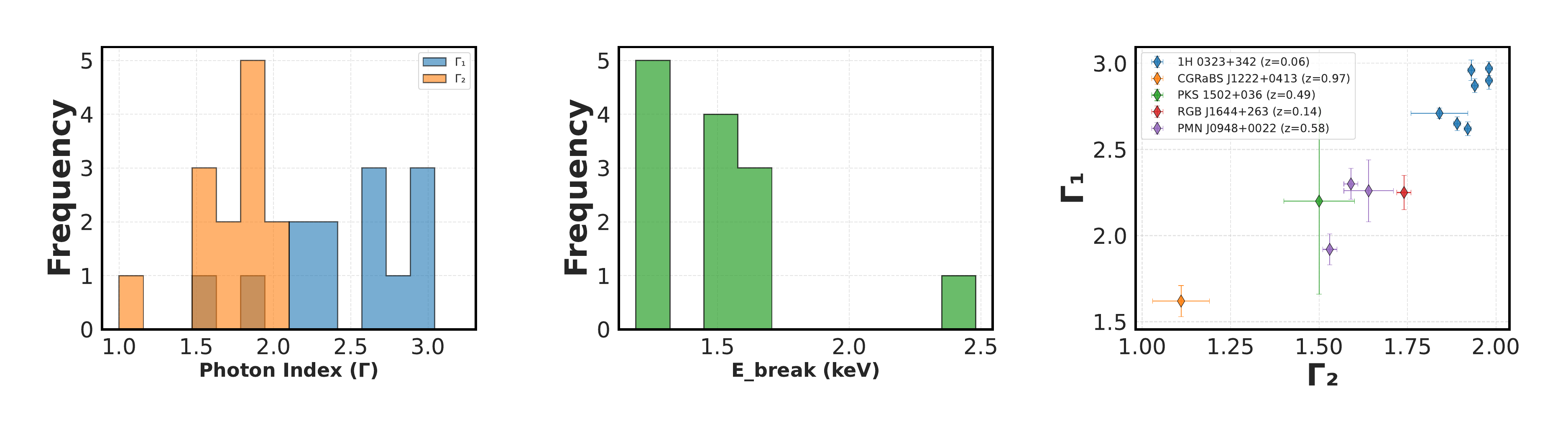}
    \caption{BPL spectral parameters; Left: Distribution of BPL $\Gamma$ in 0.3-10 keV; Centre: Break energy $E_b$ (keV); Right: $\Gamma_1$ Vs $\Gamma_2$}
    \label{1parhist} 
\end{figure*}

\begin{figure*}
    \centering
    \includegraphics[width=17.0cm]{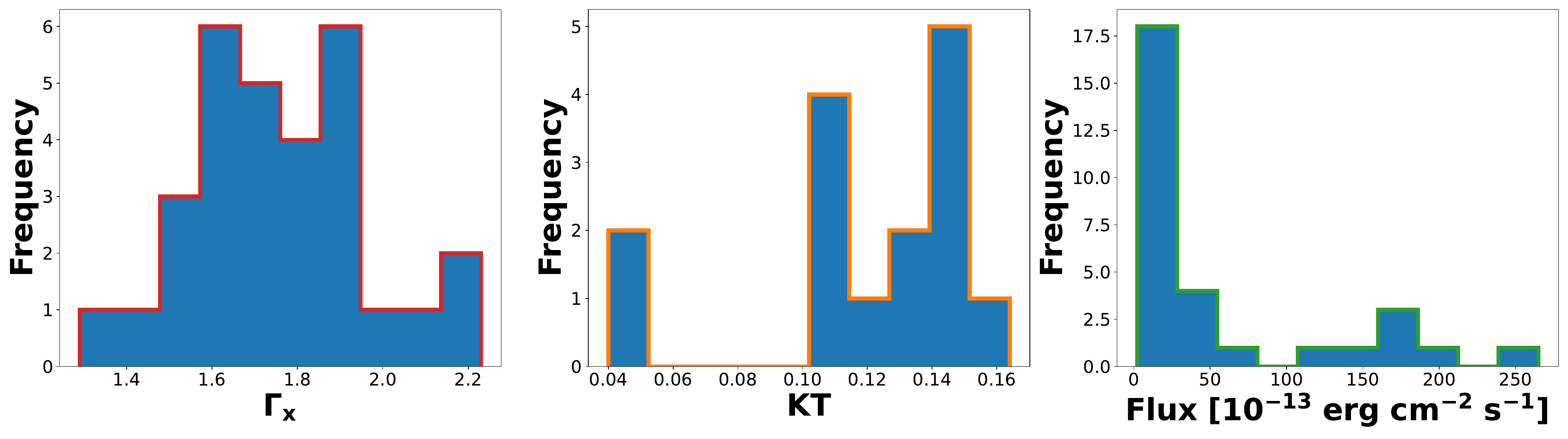}
    \caption{Distribution of PL+BB spectral parameters; Left: $\Gamma_X$ in 0.3-10 keV; Centre: Temperature distribution; Right: histogram of X-ray flux in the energy range 0.3-10 keV.}
    \label{parhist} 
\end{figure*}


\begin{figure}
    \centering
     \includegraphics[width=\columnwidth]{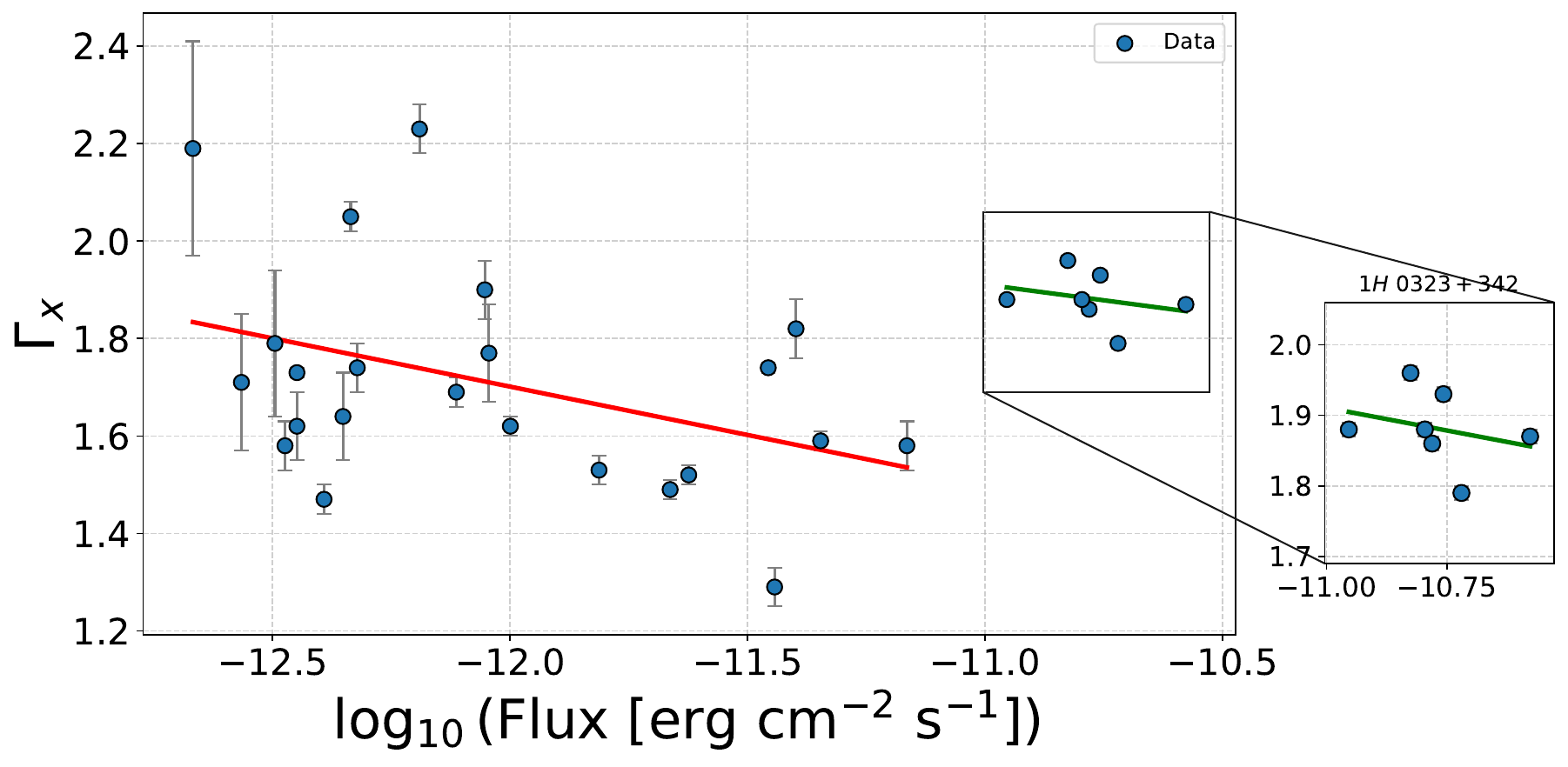}
     \includegraphics[width=\columnwidth]{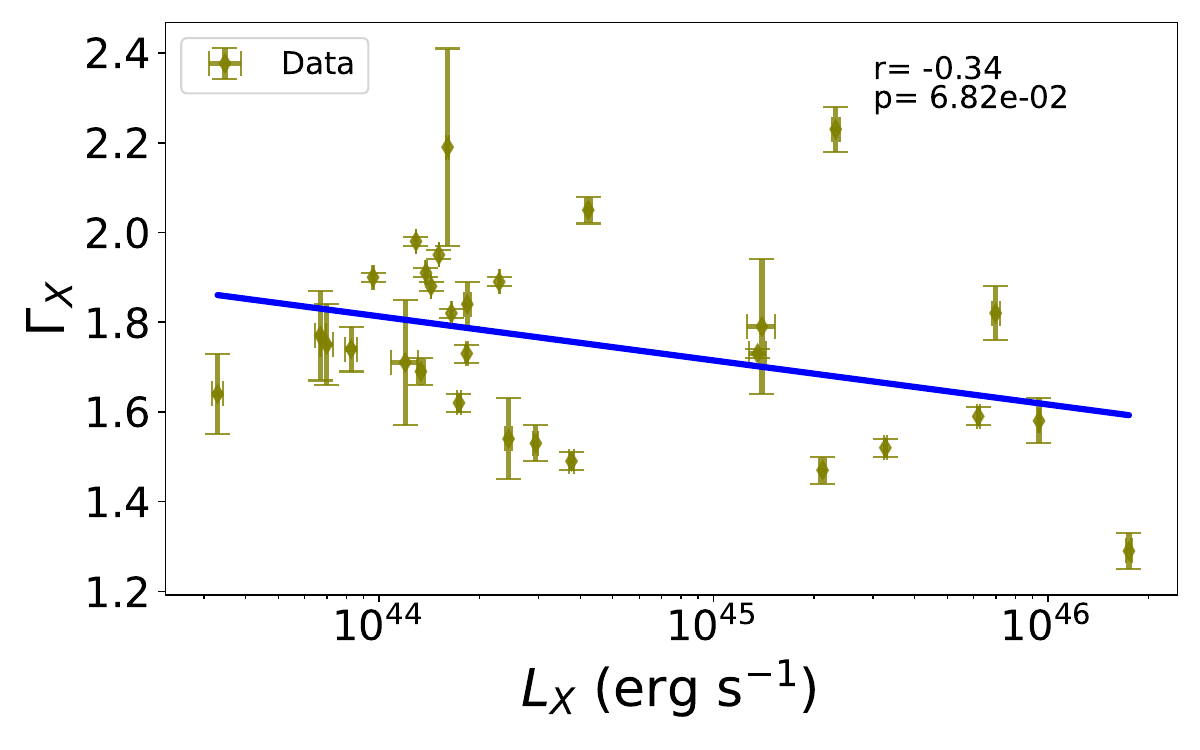}
    \caption{Flux vs $\Gamma_X$: X-ray spectral index variation as a function of unabsorbed X-ray flux and luminosity (0.3-10 keV)}
    \label{ind_F} 
\end{figure}

\begin{figure*}
    \centering
     \includegraphics[width=5.80cm]{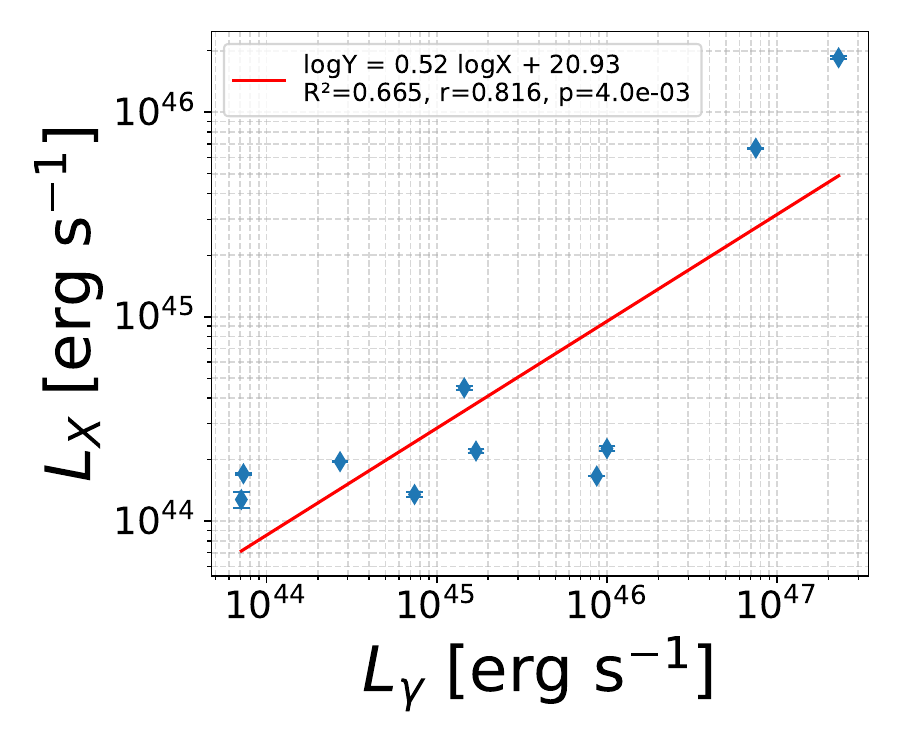}
     \includegraphics[width=5.80cm]{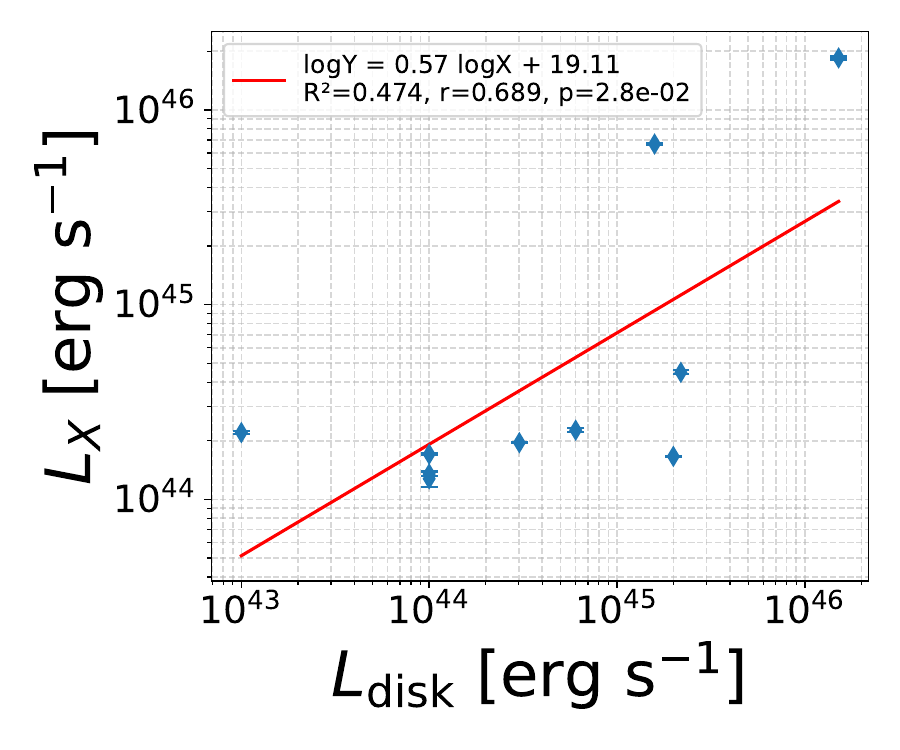}
     \includegraphics[width=5.80cm]{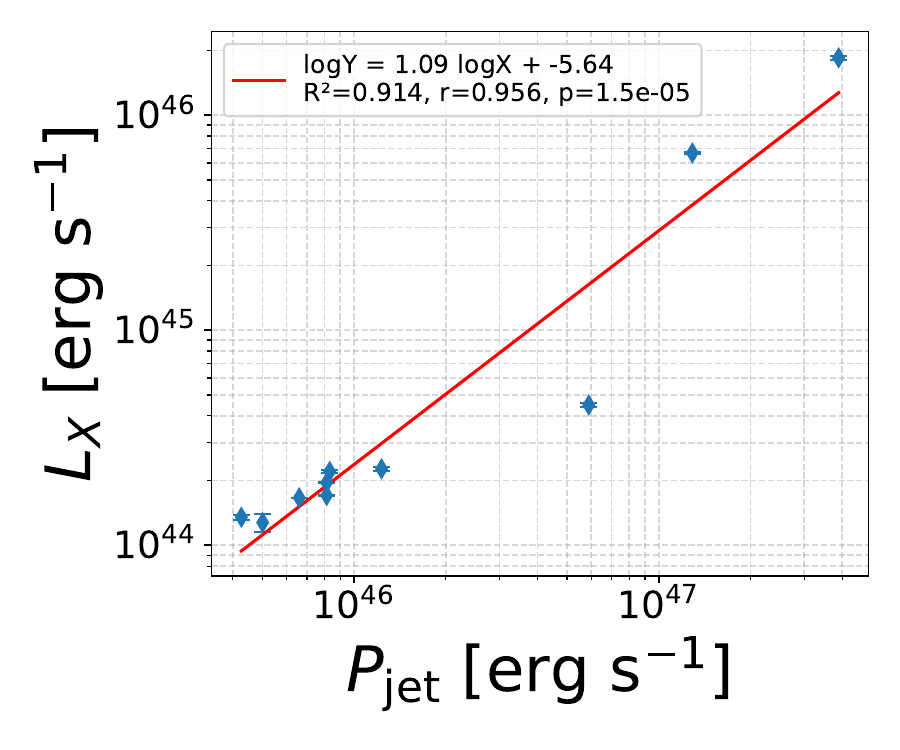}
    \caption{X-ray Luminosity vs $\gamma$-ray Luminosity, Disk Luminosity, and Jet Power. These parameters were taken from literature \citep{2024MNRAS.532.3729L, 2019ApJ...872..169P, 2023ApJS..265...31A, 2020MNRAS.496.2213D}}
    \label{Lum} 
\end{figure*}


\begin{figure}
    \centering
     \includegraphics[width=\columnwidth]{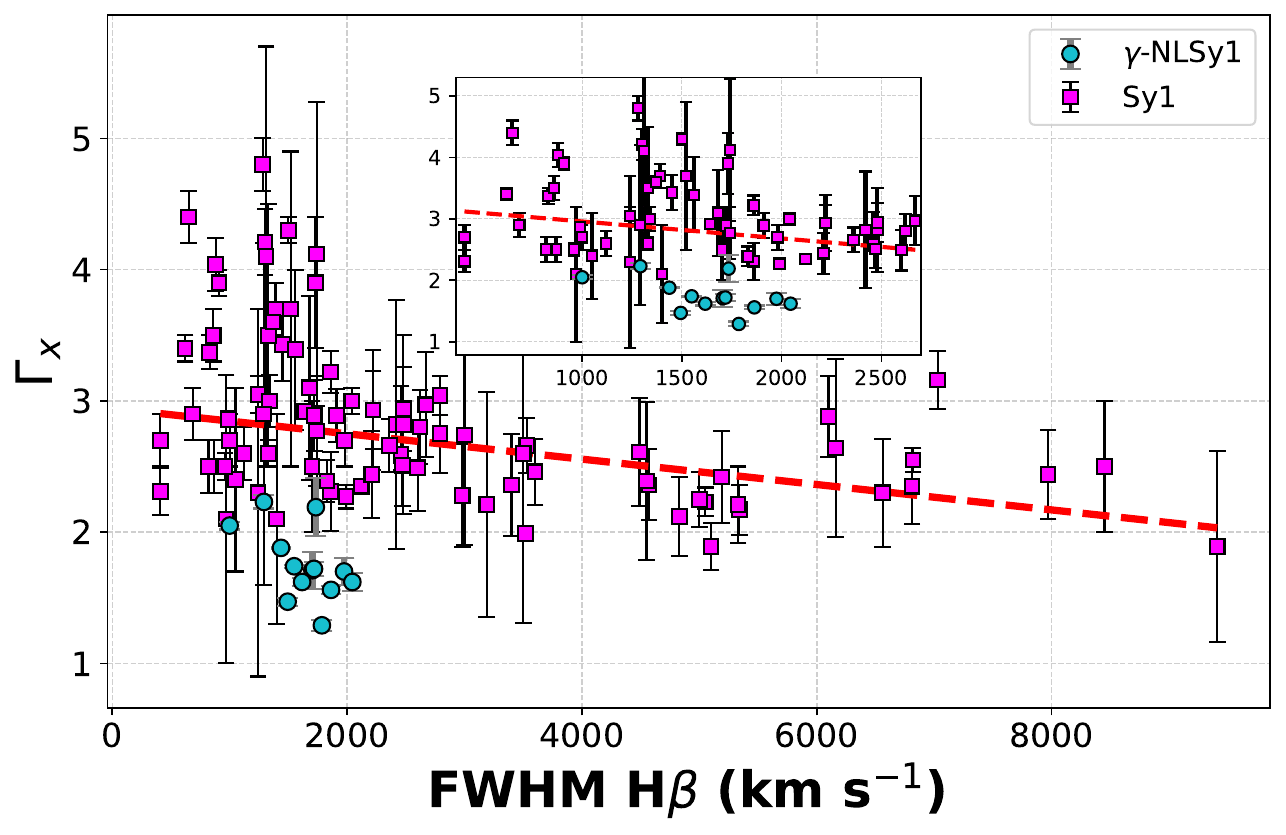}
    \caption{The X-ray photon index $\Gamma_X$ versus FWHM of H$\beta$ for both $\gamma$-NLSy1 \citep[from][]{2001ApJ...558..578O, 2015A&A...573A..76J, 2015A&A...575A..13F, 2015MNRAS.454L..16Y, 2017ApJS..229...39R, 2017MNRAS.464.2565L, 2019MNRAS.487L..40Y, 2021A&A...654A.125B}) and non-jetted Seyfert 1 galaxies \citep[from][]{1996A&A...305...53B, 1996A&A...309...81W, 2006ApJS..166..128Z}.}
    \label{ind_Hb} 
\end{figure}

\section{DISCUSSION AND INTERPRETATION} \label{discussion}
This work presents a homogeneous X-ray analysis of $\gamma$-NLSy1 using archival {\it XMM-Newton} observations, investigating their similarity to non-jetted Seyfert galaxies and blazars. The sample shows a variable flux, with $F_{\rm var}$ as high as $\sim$16\%, while the VA amplitude varies from 0.92 to 4.6, and the $\tau_{\rm var}$ changes from 0.03-0.49 ks across all the selected objects and observations. These values suggest similar X-ray variability to that of blazars. Sources like 1H\,0323+342, PKS 2004-447, RGB J1644+263, PKS 1502+036, and PMN J0948+0022 exhibit rapid variability, which is also consistent with the Swift-XRT and NuSTAR analysis presented in \cite{2020MNRAS.496.2213D, 2025arXiv250404492C}. The $\tau_{\rm var}$ values imply a compact emitting region with a size of $R_H \sim 10^{12-14}$\text{cm}, in the turbulent jet, although the spectral analysis strongly suggests these features may also have a disk origin due to a change in the accretion rates or disc-jet coupling. The recent $\gamma$-ray flare that occurred in late 2025 in 1H 0232+342, showed an enhanced flux state observed by {\it Fermi}-LAT \citep[20 times greater than the average flux;][]{2025ATel17407....1L}, and in hard X-ray  \citep[$\sim3-4$ times the mean flux;][]{2025arXiv250404492C}, and for which a neutrino upper limit was reported by {\it{IceCube}} \citep{2025ATel17423....1M}.
Based on the variability present in the {\it XMM-Newton} light curve, we argued that the 1H\,0323+342 object is highly variable, and we observe a turnover in the SF. The turnover time does not vary significantly for 1H\,0323+342 across the observations, and it is found to be close to 5 ksec in most observations, except for observation 0823780701, where the turnover time scale is estimated to be 12.5 ksec. Except for 1H\,0323+342, SDSS J0946+1017, and one observation in PMN J0948+0022, all other objects and observations do not show a clear turnover time scale in SF, either due to a lack of variability in the light curve or due to variability on time scales longer than the observational data. In fact, in some cases, we observe a rising trend in SF, suggesting the presence of a longer timescale. 

The PSD slopes of NLSy1s show a range of slopes (0.4-1.5), indicating that the jet and the accretion disk may be combined. A flatter slope is indicative of chaotic or turbulent processes, while a steeper slope suggests significant low-frequency variability.  Understanding the complex processes that distinguish $\gamma$--NLSy1s from blazars in the context of their accretion and jets is facilitated by their variability. The flux histograms of several epochs are well-fitted by log-normal or normal distributions, or occasionally by more complex patterns with multiple peaks, similar to blazars \citep{2022MNRAS.510.5280M, 2025ApJ...981..118B}, suggesting a delicate interaction between the jet and the disk/corona.

Similar to blazars, the $\gamma$-NLSy1 in our sample typically exhibits a SWB trend and occasionally displays mixed characteristics, as indicated in HR investigations. These spectral patterns can be explained in light of synchrotron emission; the detected pattern can be attributed to the inverse relationship between the cooling timescale and electron energy ($t_{\rm cool} \sim 1/E_{\rm e}$). There is an excess of lower-energy (soft) photons because high-energy electrons cool faster. Depending on the spectral states of the source, the countrate-HR diagram may often result in more complex trends, like a hysteresis loop \citep{2009A&A...501..879T, 2017ApJ...834....2A}, according to \cite{2016MNRAS.458...56W}.

Such features can be explained within the framework of synchrotron emission; the observed trend can be attributed to the cooling timescale being inversely related to photon energy $(t_{\rm cool} \sim 1/E_{\rm ph})$. Energetic high-energy photons cool more quickly, leaving an excess of lower-energy (soft) photons that cool more slowly. According to \cite{2016MNRAS.458...56W}, the countrate-HR diagram may generally produce more complicated trends, such as a hysteresis loop \citep{2009A&A...501..879T, 2017ApJ...834....2A}, depending on the source's activity states.

The spectrum of all observations for all sources was initially fitted with a single PL. We observed that in some cases, the PL model fits the spectrum well, but in others, the reduced $\chi^2$ values were 
quite high, which led us to choose 
a BPL or PL+BB model. 
The hydrogen column density was optimized, but in some cases, it was fixed to its nominal value. 
The best-fit parameters are summarized in Tables~\ref{TAB3}  and \ref{TAB4}, and the distribution of the parameters are plotted in Figures~\ref{1parhist} and \ref{parhist}. For the PL+BB fits, the 
spectral index, $\Gamma_X$, ranges between 1.3 and 2.2, with most observations showing indices between $\Gamma_X = 1.6 $ and 1.9. The temperature, $kT$, lies between 0.04 to 0.16\,keV, which is expected in AGN and well corroborates with the findings of \cite{2020MNRAS.496.2213D}. The spectral index and temperature values are consistent with those of NLSy1 galaxies, which exhibit a strong soft excess, as presented in \cite{2020MNRAS.491..532G}.
The dominance of the beamed jet radiation may be weakened by a relatively large jet viewing angle \citep[$\theta\sim12^{\circ}$;][]{2018Natur.563..657G}, which could be the source of the significant thermal disk component combined with jet emission. Some FSRQs, such as 3C~273, have also shown similar disk features \citep{2009MNRAS.400.1521J, Dinesh_2023}. Nevertheless, the spectral fit shows that $\gamma$-NLSy1 has a softer spectral index than FSRQs \citep[see e.g.\ discussion in][]{2020MNRAS.496.2213D}, suggesting that the X-ray spectra of these jetted systems might still feature contributions from the corona and the accretion disc, which is again supported by the presence of a significant soft-excess. The BPL fitting of some of the sources (see Table \ref{TAB4}) suggests that $\Gamma_1$ varies from 1.62 (for J1222+0413) to 2.97 (for 1H\,0323+342), $\Gamma_2$ changes from 1.11 to 1.98 (for 1H\,0323+342), and $E_b$ from  $\sim$1.19\,keV (for J1644+263) to 2.48 keV (for J1222+0413). The BPL fitting results are well aligned with the findings of \cite{2020MNRAS.496.2213D}. The unabsorbed X-ray flux shows that while most of the observations have low flux values, ranging between $F_X = 10^{-13}$ -- $10^{-12}$ erg cm$^{-2}$ s$^{-1}$, some observations shows flux values $F_X >10^{-11}$ erg cm$^{-2}$ s$^{-1}$. 

Figure \ref{ind_F} shows the anti-correlation between the X-ray spectral index $\Gamma_X$ and X-ray flux/luminosity, when considering the full sample of sources, which suggests that as the X-ray flux rises, the X-ray spectrum hardens (i.e., $\Gamma_X$ falls), inferring a brighter-when-harder trend which is generally seen in blazars. The flux and spectral index for 1H\,0323+342 are shown in the box on the same plot, also indicating a mild trend of brighter flux when the spectral index is harder. Spectral analysis of a sample of blazars and $\gamma$-NLSy1, presented in \cite{2016A&A...585A..91K} (see their fig. 7), also demonstrates a comparable trend. Similar to 1H\,0323+342, this tendency suggests that higher flux levels are associated with more energetic or efficient emission, most likely due to changes in jet or corona activity. It is usually assumed that this kind of behavior indicates a change in the system's cooling or acceleration processes of high-energy particles. In contrast to $\gamma$-NLSy1 \citep{2020MNRAS.496.2213D}, the study of non-jetted NLSy1 galaxies using different X-ray data from Chandra, {\it XMM-Newton}, Swift-XRT presented in \cite{2015A&A...575A..13F, 2017A&A...603C...1F} indicates that they are weak X-ray emitters with a flux range of 5.1$\times 10^{-15}$ to 4.6$\times 10^{-12}\, {\rm erg \,s}^{-1} {\rm cm}^{-2}$. The unabsorbed X-ray fluxes of our sample shown in Table \ref{TAB3}, \ref{TAB4} and Figure \ref{1parhist} and \ref{parhist} instead indicate that $\gamma$-NLSy1 have a higher flux range of 2.65$\times 10^{-13}$ to 2.15$\times 10^{-11}\, {\rm erg \,s}^{-1} {\rm cm}^{-2}$, (and the photon index which falls between 1.29 and 2.23 ) in Figure \ref{1parhist} and \ref{parhist}. These results are in good agreement with \cite{2020MNRAS.496.2213D}. 

We have found that $\gamma$-NLSy1 galaxies have an increasing trend between their X-ray, $L_X$, and gamma-ray, $L_\gamma$,  luminosities (Pearson correlation coefficient r=0.816, and p-value=$4.0\times10^{-3}$), between the disk luminosity, $L_{\rm disk}$, and $L_X$ (r=0.69, p-value=$2.8\times10^{-2}$) and the between the jet power, $P_{\rm jet}$ and $L_X$ (r=0.95, p-value=$1.5\times10^{-5}$), as shown in the Figure \ref{Lum}. The $L_{\rm disk}$, $P_{\rm jet}$, and $L_\gamma$ values are taken from the \citet{2024MNRAS.532.3729L, 2019ApJ...872..169P, 2020MNRAS.496.2213D}. However, these values are not simultaneous with the X-ray observations, which may affect these correlations. This suggests that the X-ray emission is closely linked to both relativistic jet activity and accretion processes. This is somewhat comparable to blazars, particularly FSRQs, where the high-energy output, jet power, and disk radiation are known to be strongly coupled. On the other hand, radio-quiet NLSy1 galaxies usually exhibit X-ray emission correlated mainly with the accretion disk and corona and without any contribution from jets. In contrast, BL Lacs with weak accretion disks do not exhibit such strong disk–X-ray correlations. Conversely, our \nus analysis presented in \cite{2025arXiv250404492C} exhibits the opposite pattern, suggesting that more energetic X-rays test different physical mechanisms where the correlation with disk or total jet power is reduced or inverted, most likely jet-dominated emission and inverse-Compton scattering. Thus, hard X-rays are produced by more jet-dominated processes, whereas soft X-rays are a combination of emission processes from both the disk and the jet.

In Figure~\ref{ind_Hb}, we also plot the X-ray spectral index of Seyfert-I galaxies (Sy1) and compare it with the $\gamma$--NLSy1 galaxies; this clearly shows that their spectral index differs. The Sy1 galaxies mostly have a softer index, suggesting the presence of a strong soft excess in these sources. As argued in \citet{1996A&A...305...53B, 1996A&A...309...81W}, the variation of $\Gamma_X$ with the FWHM of H\,$\beta$ line should show an anti-correlation. A similar study presented in \cite{2006ApJS..166..128Z} shows a turnover at FWHM (H\,$\beta$) $\sim$1000 km\,s$^{-1}$, above which a decreasing trend exists and below which an increasing trend exists. A change in the dominating accretion or coronal processes at extreme accretion rates is suggested by the turnover near 1000 km\,s$^{-1}$. \cite{2018BSRSL..87..379R} have studied a large sample of NLSy1 galaxies and reported a steeper spectral index which exhibits an anti-correlation with FWHM(H\,$\beta$). We plot the $\Gamma_X$ versus FWHM of H\,$\beta$, and we observed a mild anti-correlation between them. Although these measurements (FWHM of H\,$\beta$, and $\Gamma_X$) are not taken simultaneously, and H\,$\beta$ line profiles can change over time (e.g., NGC 4151, 3C 390.3), these variations do not affect the overall trend in population studies. However, a more detailed investigation is needed, although it is limited by the small size of the $\gamma$-NLSy1 population.
The observed correlation between the FWHM (H\,$\beta$) and the $\Gamma_X$ illustrates how the black hole mass and accretion rate affect the coronal emission in active galactic nuclei (AGN). Softer X-ray spectra result from stronger cooling of the corona by enhanced disk radiation, which is correlated with narrower H$\beta$ lines and lower black hole masses as well as higher Eddington ratios. On the other hand, broader H$\beta$ lines are linked by harder X-ray spectra and are indicative of more massive black holes and lower accretion rates. However, in jetted sources, the X-ray spectrum may be dominated or contaminated by non-thermal jet emission, which flattens $\Gamma_X$ and conceals or reduces the underlying disk–corona association.

\section{CONCLUSIONS} \label{conclusion}
In this section, we summarize the X-ray temporal and spectral variability properties of 16 $\gamma$-NLSy1 galaxies found using \xmm observations.

\begin{itemize}
    \item The soft X-ray variability of most sources is moderate, with $F_{\rm var}$ falling between 5 and 16\%, which is marginally lower than its hard X-ray variability (10 to 20\%). \\
    \item One of the most intriguing $\gamma$-NLSy1 sources is 1H\,0323+342, which displays interesting temporal features in all seven epochs, such as flux variability $F_{\rm var}\sim$10-15\%, variable PSD slope $\beta_P\sim$0.80-1.06, VA$\sim$0.68-1.67, $\tau_{\rm var}\sim$0.15-0.49, and characteristic turnover in its structure function.\\
    \item Normal or log-normal distributions fit the flux histograms of all sources; nevertheless, because of the low count rate in the flux, it is not possible to definitively distinguish between additive and multiplicative emission mechanisms.\\
    \item The bright sources in the sample have a PSD slope ($\beta_P$) that indicates flicker noise dominance, but the remaining sources do not exhibit any clear trends.\\
    \item Most HR plots show no discernible trends, although in some observations, the trend is softer when brighter (e.g.\ for 1H\,0323+342; obsid 0823780301 \&  0823780501), whilst other observations (obsid 0823780701) exhibit a mixed trend that is consistent with its optical colour-magnitude patterns.\\
    \item X-ray spectral analysis shows that these sources are significantly brighter than typical NLSy1 galaxies, implying that most of the excess X-ray emission comes from a jet.\\
    
    \item In this work, half of the targets (1H\,0323+342, PMN J0948+0022, CGRaBS J1222+0413, SDSS J1246+0238, PKS 1502+036, and RGB J1644+263) show complicated or degenerate X-ray spectra that were well fit with either a PL+BB model, and $kT = 0.04$–$0.15$ keV, $\Gamma = 1.29$–$1.98$, or a BPL model with $\Gamma_1 = 1.62$–$2.97$, $\Gamma_2 = 1.11$–$1.98$, and break energies $E_b = 1.22$–$2.48$ keV. The $\chi^2_{red}$ values favour the PL+BB model (except CGRaBS J1222+0413, which is well fit with a BPL), which is generally more consistent with a disk–jet interaction. A conventional power-law model does a good fit of simulating the remaining sources, which are jet-dominated. \\
    
    \item We have noticed a moderate correlation of $L_X$ with $L_{\gamma}$, $L_{\rm disk}$, and $P_{\rm jet}$. Such a correlation implies that both the jet radiation and the high-energy emission grow with the jet power and accretion power, which supports the notion that a disk–jet connection regulates the energy output of such sources.\\
    \item The X-ray flux and the FWHM of the H$\beta$ emission line demonstrate a moderate anti-correlation with the $\Gamma$, suggesting a complex conjunction of disk-corona-jet system. \\ \\
In conclusion, $\gamma$-NLSy1 galaxies exhibit properties that are shared by both blazars and Seyfert galaxies, making them an ideal class of AGNs to study the disk-jet coupling. They also represent the intriguing sources for current and future missions, such as CTA and {\it ATHENA}.

\end{itemize}

\section*{Acknowledgements}
 RP acknowledges the support from the BHU-IoE seed grant. BvS acknowledges this work is based on the research supported in part by the National Research Foundation of South Africa (Ref Numbers 119430 and CSRP23041894484).

\section*{Data Availability}
The data utilized in this study are accessible via the HEASARC database. Upon reasonable request, the data products can be provided.



\bibliographystyle{mnras}
\bibliography{Refs} 

\begin{thebibliography}{}
\makeatletter
\relax
\def\mn@urlcharsother{\let\do\@makeother \do\$\do\&\do\#\do\^\do\_\do\%\do\~}
\def\mn@doi{\begingroup\mn@urlcharsother \@ifnextchar [ {\mn@doi@}
  {\mn@doi@[]}}
\def\mn@doi@[#1]#2{\def\@tempa{#1}\ifx\@tempa\@empty \href
  {http://dx.doi.org/#2} {doi:#2}\else \href {http://dx.doi.org/#2} {#1}\fi
  \endgroup}
\def\mn@eprint#1#2{\mn@eprint@#1:#2::\@nil}
\def\mn@eprint@arXiv#1{\href {http://arxiv.org/abs/#1} {{\tt arXiv:#1}}}
\def\mn@eprint@dblp#1{\href {http://dblp.uni-trier.de/rec/bibtex/#1.xml}
  {dblp:#1}}
\def\mn@eprint@#1:#2:#3:#4\@nil{\def\@tempa {#1}\def\@tempb {#2}\def\@tempc
  {#3}\ifx \@tempc \@empty \let \@tempc \@tempb \let \@tempb \@tempa \fi \ifx
  \@tempb \@empty \def\@tempb {arXiv}\fi \@ifundefined
  {mn@eprint@\@tempb}{\@tempb:\@tempc}{\expandafter \expandafter \csname
  mn@eprint@\@tempb\endcsname \expandafter{\@tempc}}}

\bibitem[\protect\citeauthoryear{Abdo et~al.,}{Abdo et~al.}{2009}]{Abdo_2009}
Abdo A.~A.,  et~al., 2009, \mn@doi [The Astrophysical Journal]
  {10.1088/0004-637X/699/2/976}, 699, 976

\bibitem[\protect\citeauthoryear{{Abdollahi} et~al.,}{{Abdollahi}
  et~al.}{2020}]{2020yCat..22470033A}
{Abdollahi} S.,  et~al., 2020, {VizieR Online Data Catalog: The Fermi LAT
  fourth source catalog (4FGL) (Abdollahi+, 2020)}, VizieR On-line Data
  Catalog: J/ApJS/247/33. Originally published in: 2020ApJS..247...33A,
  \mn@doi{10.26093/cds/vizier.22470033}

\bibitem[\protect\citeauthoryear{{Abdollahi} et~al.,}{{Abdollahi}
  et~al.}{2022}]{2022ApJS..260...53A}
{Abdollahi} S.,  et~al., 2022, \mn@doi [\apjs] {10.3847/1538-4365/ac6751},
  \href {https://ui.adsabs.harvard.edu/abs/2022ApJS..260...53A} {260, 53}

\bibitem[\protect\citeauthoryear{{Abdollahi} et~al.,}{{Abdollahi}
  et~al.}{2023}]{2023ApJS..265...31A}
{Abdollahi} S.,  et~al., 2023, \mn@doi [\apjs] {10.3847/1538-4365/acbb6a},
  \href {https://ui.adsabs.harvard.edu/abs/2023ApJS..265...31A} {265, 31}

\bibitem[\protect\citeauthoryear{{Abeysekara} et~al.,}{{Abeysekara}
  et~al.}{2017}]{2017ApJ...834....2A}
{Abeysekara} A.~U.,  et~al., 2017, \mn@doi [\apj] {10.3847/1538-4357/834/1/2},
  \href {https://ui.adsabs.harvard.edu/abs/2017ApJ...834....2A} {834, 2}

\bibitem[\protect\citeauthoryear{{Ajello} et~al.,}{{Ajello}
  et~al.}{2022}]{2022ApJS..263...24A}
{Ajello} M.,  et~al., 2022, \mn@doi [\apjs] {10.3847/1538-4365/ac9523}, \href
  {https://ui.adsabs.harvard.edu/abs/2022ApJS..263...24A} {263, 24}

\bibitem[\protect\citeauthoryear{{Arnaud}}{{Arnaud}}{1996}]{1996ASPC..101...17A}
{Arnaud} K.~A.,  1996, in {Jacoby} G.~H.,  {Barnes} J.,  eds,  Astronomical
  Society of the Pacific Conference Series Vol. 101, Astronomical Data Analysis
  Software and Systems V. p.~17

\bibitem[\protect\citeauthoryear{{Ballet}, {Bruel}, {Burnett}, {Lott}  \& {The
  Fermi-LAT collaboration}}{{Ballet} et~al.}{2023}]{2023arXiv230712546B}
{Ballet} J.,  {Bruel} P.,  {Burnett} T.~H.,  {Lott} B.,   {The Fermi-LAT
  collaboration} 2023, \mn@doi [arXiv e-prints] {10.48550/arXiv.2307.12546},
  \href {https://ui.adsabs.harvard.edu/abs/2023arXiv230712546B} {p.
  arXiv:2307.12546}

\bibitem[\protect\citeauthoryear{{Bauer}, {Baltay}, {Coppi}, {Ellman}, {Jerke},
  {Rabinowitz}  \& {Scalzo}}{{Bauer} et~al.}{2009}]{2009ApJ...696.1241B}
{Bauer} A.,  {Baltay} C.,  {Coppi} P.,  {Ellman} N.,  {Jerke} J.,  {Rabinowitz}
  D.,   {Scalzo} R.,  2009, \mn@doi [\apj] {10.1088/0004-637X/696/2/1241},
  \href {https://ui.adsabs.harvard.edu/abs/2009ApJ...696.1241B} {696, 1241}

\bibitem[\protect\citeauthoryear{{Berton}, {Braito}, {Mathur}, {Foschini},
  {Piconcelli}, {Chen}  \& {Pogge}}{{Berton}
  et~al.}{2019}]{2019A&A...632A.120B}
{Berton} M.,  {Braito} V.,  {Mathur} S.,  {Foschini} L.,  {Piconcelli} E.,
  {Chen} S.,   {Pogge} R.~W.,  2019, \mn@doi [\aap]
  {10.1051/0004-6361/201935929}, \href
  {https://ui.adsabs.harvard.edu/abs/2019A&A...632A.120B} {632, A120}

\bibitem[\protect\citeauthoryear{{Berton} et~al.,}{{Berton}
  et~al.}{2021}]{2021A&A...654A.125B}
{Berton} M.,  et~al., 2021, \mn@doi [\aap] {10.1051/0004-6361/202141409}, \href
  {https://ui.adsabs.harvard.edu/abs/2021A&A...654A.125B} {654, A125}

\bibitem[\protect\citeauthoryear{{Bevington} \& {Robinson}}{{Bevington} \&
  {Robinson}}{2003}]{2003drea.book.....B}
{Bevington} P.~R.,  {Robinson} D.~K.,  2003, {Data reduction and error analysis
  for the physical sciences}

\bibitem[\protect\citeauthoryear{{Bhatta}}{{Bhatta}}{2017}]{2017ApJ...847....7B}
{Bhatta} G.,  2017, \mn@doi [\apj] {10.3847/1538-4357/aa86ed}, \href
  {https://ui.adsabs.harvard.edu/abs/2017ApJ...847....7B} {847, 7}

\bibitem[\protect\citeauthoryear{{Bhatta} \& {Dhital}}{{Bhatta} \&
  {Dhital}}{2020}]{2020ApJ...891..120B}
{Bhatta} G.,  {Dhital} N.,  2020, \mn@doi [\apj] {10.3847/1538-4357/ab7455},
  \href {https://ui.adsabs.harvard.edu/abs/2020ApJ...891..120B} {891, 120}

\bibitem[\protect\citeauthoryear{{Bhatta} et~al.,}{{Bhatta}
  et~al.}{2025}]{2025ApJ...981..118B}
{Bhatta} G.,  et~al., 2025, \mn@doi [\apj] {10.3847/1538-4357/adb0c9}, \href
  {https://ui.adsabs.harvard.edu/abs/2025ApJ...981..118B} {981, 118}

\bibitem[\protect\citeauthoryear{{Bhattacharyya}, {Bhatt}, {Bhatt}  \&
  {Singh}}{{Bhattacharyya} et~al.}{2014}]{2014MNRAS.440..106B}
{Bhattacharyya} S.,  {Bhatt} H.,  {Bhatt} N.,   {Singh} K.~K.,  2014, \mn@doi
  [\mnras] {10.1093/mnras/stu239}, \href
  {https://ui.adsabs.harvard.edu/abs/2014MNRAS.440..106B} {440, 106}

\bibitem[\protect\citeauthoryear{{Biteau} \& {Giebels}}{{Biteau} \&
  {Giebels}}{2012}]{2012A&A...548A.123B}
{Biteau} J.,  {Giebels} B.,  2012, \mn@doi [\aap]
  {10.1051/0004-6361/201220056}, \href
  {https://ui.adsabs.harvard.edu/abs/2012A&A...548A.123B} {548, A123}

\bibitem[\protect\citeauthoryear{{Boller}, {Brandt}  \& {Fink}}{{Boller}
  et~al.}{1996}]{1996A&A...305...53B}
{Boller} T.,  {Brandt} W.~N.,   {Fink} H.,  1996, \mn@doi [\aap]
  {10.48550/arXiv.astro-ph/9504093}, \href
  {https://ui.adsabs.harvard.edu/abs/1996A&A...305...53B} {305, 53}

\bibitem[\protect\citeauthoryear{{B{\"o}ttcher} \& {Dermer}}{{B{\"o}ttcher} \&
  {Dermer}}{2002}]{2002ApJ...564...86B}
{B{\"o}ttcher} M.,  {Dermer} C.~D.,  2002, \mn@doi [\apj] {10.1086/324134},
  \href {https://ui.adsabs.harvard.edu/abs/2002ApJ...564...86B} {564, 86}

\bibitem[\protect\citeauthoryear{{Briel} et~al.,}{{Briel}
  et~al.}{2000}]{2000SPIE.4012..154B}
{Briel} U.~G.,  et~al., 2000, in {Truemper} J.~E.,  {Aschenbach} B.,  eds,
  Society of Photo-Optical Instrumentation Engineers (SPIE) Conference Series
  Vol. 4012, X-Ray Optics, Instruments, and Missions III. pp 154--164,
  \mn@doi{10.1117/12.391613}

\bibitem[\protect\citeauthoryear{{Brinkmann}, {Papadakis}, {den Herder}  \&
  {Haberl}}{{Brinkmann} et~al.}{2003}]{2003A&A...402..929B}
{Brinkmann} W.,  {Papadakis} I.~E.,  {den Herder} J.~W.~A.,   {Haberl} F.,
  2003, \mn@doi [\aap] {10.1051/0004-6361:20030264}, \href
  {https://ui.adsabs.harvard.edu/abs/2003A&A...402..929B} {402, 929}

\bibitem[\protect\citeauthoryear{{Burbidge}, {Jones}  \& {O'Dell}}{{Burbidge}
  et~al.}{1974}]{1974ApJ...193...43B}
{Burbidge} G.~R.,  {Jones} T.~W.,   {O'Dell} S.~L.,  1974, \mn@doi [\apj]
  {10.1086/153125}, \href
  {https://ui.adsabs.harvard.edu/abs/1974ApJ...193...43B} {193, 43}

\bibitem[\protect\citeauthoryear{{Calderone}, {Ghisellini}, {Colpi}  \&
  {Dotti}}{{Calderone} et~al.}{2013}]{2013MNRAS.431..210C}
{Calderone} G.,  {Ghisellini} G.,  {Colpi} M.,   {Dotti} M.,  2013, \mn@doi
  [\mnras] {10.1093/mnras/stt157}, \href
  {https://ui.adsabs.harvard.edu/abs/2013MNRAS.431..210C} {431, 210}

\bibitem[\protect\citeauthoryear{{Chaudhary} \& {Prince}}{{Chaudhary} \&
  {Prince}}{2025}]{2025arXiv250404492C}
{Chaudhary} S.~C.,  {Prince} R.,  2025, arXiv e-prints, \href
  {https://ui.adsabs.harvard.edu/abs/2025arXiv250404492C} {p. arXiv:2504.04492}

\bibitem[\protect\citeauthoryear{{D'Ammando}}{{D'Ammando}}{2020}]{2020MNRAS.496.2213D}
{D'Ammando} F.,  2020, \mn@doi [\mnras] {10.1093/mnras/staa1580}, \href
  {https://ui.adsabs.harvard.edu/abs/2020MNRAS.496.2213D} {496, 2213}

\bibitem[\protect\citeauthoryear{{D'Ammando} et~al.,}{{D'Ammando}
  et~al.}{2013}]{2013MNRAS.436..191D}
{D'Ammando} F.,  et~al., 2013, \mn@doi [\mnras] {10.1093/mnras/stt1560}, \href
  {https://ui.adsabs.harvard.edu/abs/2013MNRAS.436..191D} {436, 191}

\bibitem[\protect\citeauthoryear{{D'Ammando} et~al.,}{{D'Ammando}
  et~al.}{2015}]{2015MNRAS.446.2456D}
{D'Ammando} F.,  et~al., 2015, \mn@doi [\mnras] {10.1093/mnras/stu2251}, \href
  {https://ui.adsabs.harvard.edu/abs/2015MNRAS.446.2456D} {446, 2456}

\bibitem[\protect\citeauthoryear{{Dalla Barba} et~al.,}{{Dalla Barba}
  et~al.}{2025}]{2025arXiv251115814D}
{Dalla Barba} B.,  et~al., 2025, \mn@doi [arXiv e-prints]
  {10.48550/arXiv.2511.15814}, \href
  {https://ui.adsabs.harvard.edu/abs/2025arXiv251115814D} {p. arXiv:2511.15814}

\bibitem[\protect\citeauthoryear{Devanand, Gupta, Jithesh  \& Wiita}{Devanand
  et~al.}{2022}]{Devanand_2022}
Devanand P.~U.,  Gupta A.~C.,  Jithesh V.,   Wiita P.~J.,  2022, \mn@doi [The
  Astrophysical Journal] {10.3847/1538-4357/ac9064}, 939, 80

\bibitem[\protect\citeauthoryear{Dhiman, Gupta, Gaur  \& Wiita}{Dhiman
  et~al.}{2021}]{10.1093/mnras/stab1743}
Dhiman V.,  Gupta A.~C.,  Gaur H.,   Wiita P.~J.,  2021, \mn@doi [Monthly
  Notices of the Royal Astronomical Society] {10.1093/mnras/stab1743}, 506,
  1198

\bibitem[\protect\citeauthoryear{{Dinesh}, {Bhatta}, {Adhikari}, {Mohorian},
  {Dhital}, {Chaudhary}, {P{\'a}nis}  \& {G{\'o}ra}}{{Dinesh}
  et~al.}{2023a}]{2023ApJ...955..121D}
{Dinesh} A.,  {Bhatta} G.,  {Adhikari} T.~P.,  {Mohorian} M.,  {Dhital} N.,
  {Chaudhary} S.~C.,  {P{\'a}nis} R.,   {G{\'o}ra} D.,  2023a, \mn@doi [\apj]
  {10.3847/1538-4357/acf316}, \href
  {https://ui.adsabs.harvard.edu/abs/2023ApJ...955..121D} {955, 121}

\bibitem[\protect\citeauthoryear{Dinesh, Bhatta, Adhikari, Mohorian, Dhital,
  Chaudhary, Pánis  \& Góra}{Dinesh et~al.}{2023b}]{Dinesh_2023}
Dinesh A.,  Bhatta G.,  Adhikari T.~P.,  Mohorian M.,  Dhital N.,  Chaudhary
  S.~C.,  Pánis R.,   Góra D.,  2023b, \mn@doi [The Astrophysical Journal]
  {10.3847/1538-4357/acf316}, 955, 121

\bibitem[\protect\citeauthoryear{{Dinesh}, {Dominguez}, {Paliya}, {Contreras},
  {Buson}  \& {Ajello}}{{Dinesh} et~al.}{2025}]{2025A&A...703A.162D}
{Dinesh} A.,  {Dominguez} A.,  {Paliya} V.,  {Contreras} J.~L.,  {Buson} S.,
  {Ajello} M.,  2025, \mn@doi [\aap] {10.1051/0004-6361/202556241}, \href
  {https://ui.adsabs.harvard.edu/abs/2025A&A...703A.162D} {703, A162}

\bibitem[\protect\citeauthoryear{Du et~al.,}{Du et~al.}{2014}]{Du_2014}
Du P.,  et~al., 2014, \mn@doi [The Astrophysical Journal]
  {10.1088/0004-637X/782/1/45}, 782, 45

\bibitem[\protect\citeauthoryear{D’Ammando}{D’Ammando}{2019}]{galaxies7040087}
D’Ammando F.,  2019, Galaxies, 7

\bibitem[\protect\citeauthoryear{{Edelson}, {Krolik}  \& {Pike}}{{Edelson}
  et~al.}{1990}]{1990ApJ...359...86E}
{Edelson} R.~A.,  {Krolik} J.~H.,   {Pike} G.~F.,  1990, \mn@doi [\apj]
  {10.1086/169036}, \href
  {https://ui.adsabs.harvard.edu/abs/1990ApJ...359...86E} {359, 86}

\bibitem[\protect\citeauthoryear{{Emmanoulopoulos}, {McHardy}  \&
  {Uttley}}{{Emmanoulopoulos} et~al.}{2010}]{2010MNRAS.404..931E}
{Emmanoulopoulos} D.,  {McHardy} I.~M.,   {Uttley} P.,  2010, \mn@doi [\mnras]
  {10.1111/j.1365-2966.2010.16328.x}, \href
  {https://ui.adsabs.harvard.edu/abs/2010MNRAS.404..931E} {404, 931}

\bibitem[\protect\citeauthoryear{{Foschini}}{{Foschini}}{2011}]{2011nlsg.confE..24F}
{Foschini} L.,  2011, in {Foschini} L.,  {Colpi} M.,  {Gallo} L.,  {Grupe} D.,
  {Komossa} S.,  {Leighly} K.,   {Mathur} S.,  eds, Narrow-Line Seyfert 1
  Galaxies and their Place in the Universe. p.~24 (\mn@eprint {arXiv}
  {1105.0772}), \mn@doi{10.22323/1.126.0024}

\bibitem[\protect\citeauthoryear{{Foschini}}{{Foschini}}{2020}]{2020Univ....6..136F}
{Foschini} L.,  2020, \mn@doi [Universe] {10.3390/universe6090136}, \href
  {https://ui.adsabs.harvard.edu/abs/2020Univ....6..136F} {6, 136}

\bibitem[\protect\citeauthoryear{{Foschini} et~al.,}{{Foschini}
  et~al.}{2011}]{2011MNRAS.413.1671F}
{Foschini} L.,  et~al., 2011, \mn@doi [\mnras]
  {10.1111/j.1365-2966.2011.18240.x}, \href
  {https://ui.adsabs.harvard.edu/abs/2011MNRAS.413.1671F} {413, 1671}

\bibitem[\protect\citeauthoryear{{Foschini} et~al.,}{{Foschini}
  et~al.}{2015}]{2015A&A...575A..13F}
{Foschini} L.,  et~al., 2015, \mn@doi [\aap] {10.1051/0004-6361/201424972},
  \href {https://ui.adsabs.harvard.edu/abs/2015A&A...575A..13F} {575, A13}

\bibitem[\protect\citeauthoryear{{Foschini} et~al.,}{{Foschini}
  et~al.}{2017}]{2017A&A...603C...1F}
{Foschini} L.,  et~al., 2017, \mn@doi [\aap] {10.1051/0004-6361/201424972e},
  \href {https://ui.adsabs.harvard.edu/abs/2017A&A...603C...1F} {603, C1}

\bibitem[\protect\citeauthoryear{{Foschini} et~al.,}{{Foschini}
  et~al.}{2022}]{2022Univ....8..587F}
{Foschini} L.,  et~al., 2022, \mn@doi [Universe] {10.3390/universe8110587},
  \href {https://ui.adsabs.harvard.edu/abs/2022Univ....8..587F} {8, 587}

\bibitem[\protect\citeauthoryear{{Gallo} et~al.,}{{Gallo}
  et~al.}{2006}]{2006MNRAS.370..245G}
{Gallo} L.~C.,  et~al., 2006, \mn@doi [\mnras]
  {10.1111/j.1365-2966.2006.10482.x}, \href
  {https://ui.adsabs.harvard.edu/abs/2006MNRAS.370..245G} {370, 245}

\bibitem[\protect\citeauthoryear{{Ghosh}, {Dewangan}, {Mallick}  \&
  {Raychaudhuri}}{{Ghosh} et~al.}{2018}]{2018MNRAS.479.2464G}
{Ghosh} R.,  {Dewangan} G.~C.,  {Mallick} L.,   {Raychaudhuri} B.,  2018,
  \mn@doi [\mnras] {10.1093/mnras/sty1571}, \href
  {https://ui.adsabs.harvard.edu/abs/2018MNRAS.479.2464G} {479, 2464}

\bibitem[\protect\citeauthoryear{{Gliozzi} \& {Williams}}{{Gliozzi} \&
  {Williams}}{2020}]{2020MNRAS.491..532G}
{Gliozzi} M.,  {Williams} J.~K.,  2020, \mn@doi [\mnras]
  {10.1093/mnras/stz3005}, \href
  {https://ui.adsabs.harvard.edu/abs/2020MNRAS.491..532G} {491, 532}

\bibitem[\protect\citeauthoryear{{Gonz{\'a}lez-Mart{\'\i}n} \&
  {Vaughan}}{{Gonz{\'a}lez-Mart{\'\i}n} \&
  {Vaughan}}{2012}]{2012A&A...544A..80G}
{Gonz{\'a}lez-Mart{\'\i}n} O.,  {Vaughan} S.,  2012, \mn@doi [\aap]
  {10.1051/0004-6361/201219008}, \href
  {https://ui.adsabs.harvard.edu/abs/2012A&A...544A..80G} {544, A80}

\bibitem[\protect\citeauthoryear{{Goodrich}}{{Goodrich}}{1989}]{1989ApJ...342..224G}
{Goodrich} R.~W.,  1989, \mn@doi [\apj] {10.1086/167586}, \href
  {https://ui.adsabs.harvard.edu/abs/1989ApJ...342..224G} {342, 224}

\bibitem[\protect\citeauthoryear{{Goyal}}{{Goyal}}{2021}]{2021ApJ...909...39G}
{Goyal} A.,  2021, \mn@doi [\apj] {10.3847/1538-4357/abd7fb}, \href
  {https://ui.adsabs.harvard.edu/abs/2021ApJ...909...39G} {909, 39}

\bibitem[\protect\citeauthoryear{{Gravity Collaboration} et~al.,}{{Gravity
  Collaboration} et~al.}{2018}]{2018Natur.563..657G}
{Gravity Collaboration} et~al., 2018, \mn@doi [\nat]
  {10.1038/s41586-018-0731-9}, \href
  {https://ui.adsabs.harvard.edu/abs/2018Natur.563..657G} {563, 657}

\bibitem[\protect\citeauthoryear{Gupta, Kalita, Gaur  \& Duorah}{Gupta
  et~al.}{2016}]{10.1093/mnras/stw1667}
Gupta A.~C.,  Kalita N.,  Gaur H.,   Duorah K.,  2016, \mn@doi [Monthly Notices
  of the Royal Astronomical Society] {10.1093/mnras/stw1667}, 462, 1508

\bibitem[\protect\citeauthoryear{{Hagen-Thorn}, {Larionov}, {Jorstad},
  {Arkharov}, {Hagen-Thorn}, {Efimova}, {Larionova}  \&
  {Marscher}}{{Hagen-Thorn} et~al.}{2008}]{2008ApJ...672...40H}
{Hagen-Thorn} V.~A.,  {Larionov} V.~M.,  {Jorstad} S.~G.,  {Arkharov} A.~A.,
  {Hagen-Thorn} E.~I.,  {Efimova} N.~V.,  {Larionova} L.~V.,   {Marscher}
  A.~P.,  2008, \mn@doi [\apj] {10.1086/523841}, \href
  {https://ui.adsabs.harvard.edu/abs/2008ApJ...672...40H} {672, 40}

\bibitem[\protect\citeauthoryear{{Hayashida} et~al.,}{{Hayashida}
  et~al.}{2015}]{2015ApJ...807...79H}
{Hayashida} M.,  et~al., 2015, \mn@doi [\apj] {10.1088/0004-637X/807/1/79},
  \href {https://ui.adsabs.harvard.edu/abs/2015ApJ...807...79H} {807, 79}

\bibitem[\protect\citeauthoryear{{Jansen} et~al.,}{{Jansen}
  et~al.}{2001}]{2001A&A...365L...1J}
{Jansen} F.,  et~al., 2001, \mn@doi [\aap] {10.1051/0004-6361:20000036}, \href
  {https://ui.adsabs.harvard.edu/abs/2001A&A...365L...1J} {365, L1}

\bibitem[\protect\citeauthoryear{{J{\"a}rvel{\"a}}, {L{\"a}hteenm{\"a}ki}  \&
  {Le{\'o}n-Tavares}}{{J{\"a}rvel{\"a}} et~al.}{2015}]{2015A&A...573A..76J}
{J{\"a}rvel{\"a}} E.,  {L{\"a}hteenm{\"a}ki} A.,   {Le{\'o}n-Tavares} J.,
  2015, \mn@doi [\aap] {10.1051/0004-6361/201424694}, \href
  {https://ui.adsabs.harvard.edu/abs/2015A&A...573A..76J} {573, A76}

\bibitem[\protect\citeauthoryear{{Jolley}, {Kuncic}, {Bicknell}  \&
  {Wagner}}{{Jolley} et~al.}{2009}]{2009MNRAS.400.1521J}
{Jolley} E.~J.~D.,  {Kuncic} Z.,  {Bicknell} G.~V.,   {Wagner} S.,  2009,
  \mn@doi [\mnras] {10.1111/j.1365-2966.2009.15554.x}, \href
  {https://ui.adsabs.harvard.edu/abs/2009MNRAS.400.1521J} {400, 1521}

\bibitem[\protect\citeauthoryear{{Kankkunen}, {Tornikoski}  \&
  {Hovatta}}{{Kankkunen} et~al.}{2025}]{2025A&A...693A.319K}
{Kankkunen} S.,  {Tornikoski} M.,   {Hovatta} T.,  2025, \mn@doi [\aap]
  {10.1051/0004-6361/202450562}, \href
  {https://ui.adsabs.harvard.edu/abs/2025A&A...693A.319K} {693, A319}

\bibitem[\protect\citeauthoryear{{Komossa}}{{Komossa}}{2018}]{2018rnls.confE..15K}
{Komossa} S.,  2018, in Revisiting Narrow-Line Seyfert 1 Galaxies and their
  Place in the Universe. p.~15 (\mn@eprint {arXiv} {1807.03666}),
  \mn@doi{10.22323/1.328.0015}

\bibitem[\protect\citeauthoryear{{Komossa}, {Voges}, {Xu}, {Mathur}, {Adorf},
  {Lemson}, {Duschl}  \& {Grupe}}{{Komossa}
  et~al.}{2006a}]{2006AJ....132..531K}
{Komossa} S.,  {Voges} W.,  {Xu} D.,  {Mathur} S.,  {Adorf} H.-M.,  {Lemson}
  G.,  {Duschl} W.~J.,   {Grupe} D.,  2006a, \mn@doi [\aj] {10.1086/505043},
  \href {https://ui.adsabs.harvard.edu/abs/2006AJ....132..531K} {132, 531}

\bibitem[\protect\citeauthoryear{Komossa, Voges, Xu, Mathur, Adorf, Lemson,
  Duschl  \& Grupe}{Komossa et~al.}{2006b}]{Komossa_2006}
Komossa S.,  Voges W.,  Xu D.,  Mathur S.,  Adorf H.-M.,  Lemson G.,  Duschl
  W.~J.,   Grupe D.,  2006b, \mn@doi [The Astronomical Journal]
  {10.1086/505043}, 132, 531

\bibitem[\protect\citeauthoryear{{Komossa}, {Yao}, {Grupe}  \&
  {Kraus}}{{Komossa} et~al.}{2024}]{2024Univ...10..289K}
{Komossa} S.,  {Yao} S.,  {Grupe} D.,   {Kraus} A.,  2024, \mn@doi [Universe]
  {10.3390/universe10070289}, \href
  {https://ui.adsabs.harvard.edu/abs/2024Univ...10..289K} {10, 289}

\bibitem[\protect\citeauthoryear{{Kreikenbohm} et~al.,}{{Kreikenbohm}
  et~al.}{2016}]{2016A&A...585A..91K}
{Kreikenbohm} A.,  et~al., 2016, \mn@doi [\aap] {10.1051/0004-6361/201424818},
  \href {https://ui.adsabs.harvard.edu/abs/2016A&A...585A..91K} {585, A91}

\bibitem[\protect\citeauthoryear{{Kushwaha}, {Sinha}, {Misra}, {Singh}  \& {de
  Gouveia Dal Pino}}{{Kushwaha} et~al.}{2017}]{2017ApJ...849..138K}
{Kushwaha} P.,  {Sinha} A.,  {Misra} R.,  {Singh} K.~P.,   {de Gouveia Dal
  Pino} E.~M.,  2017, \mn@doi [\apj] {10.3847/1538-4357/aa8ef5}, \href
  {https://ui.adsabs.harvard.edu/abs/2017ApJ...849..138K} {849, 138}

\bibitem[\protect\citeauthoryear{{Landt} et~al.,}{{Landt}
  et~al.}{2017}]{2017MNRAS.464.2565L}
{Landt} H.,  et~al., 2017, \mn@doi [\mnras] {10.1093/mnras/stw2447}, \href
  {https://ui.adsabs.harvard.edu/abs/2017MNRAS.464.2565L} {464, 2565}

\bibitem[\protect\citeauthoryear{{Laor}}{{Laor}}{2000}]{2000ApJ...543L.111L}
{Laor} A.,  2000, \mn@doi [\apjl] {10.1086/317280}, \href
  {https://ui.adsabs.harvard.edu/abs/2000ApJ...543L.111L} {543, L111}

\bibitem[\protect\citeauthoryear{{Leighly}}{{Leighly}}{1999}]{1999ApJS..125..317L}
{Leighly} K.~M.,  1999, \mn@doi [\apjs] {10.1086/313287}, \href
  {https://ui.adsabs.harvard.edu/abs/1999ApJS..125..317L} {125, 317}

\bibitem[\protect\citeauthoryear{{Longo}, {Holzmann Airasca}  \& {La
  Mura}}{{Longo} et~al.}{2025}]{2025ATel17407....1L}
{Longo} F.,  {Holzmann Airasca} A.,   {La Mura} G.,  2025, The Astronomer's
  Telegram, \href {https://ui.adsabs.harvard.edu/abs/2025ATel17407....1L}
  {17407, 1}

\bibitem[\protect\citeauthoryear{{Luna-Cervantes}, {Tramacere}  \&
  {Ben{\'\i}tez}}{{Luna-Cervantes} et~al.}{2024}]{2024MNRAS.532.3729L}
{Luna-Cervantes} J.,  {Tramacere} A.,   {Ben{\'\i}tez} E.,  2024, \mn@doi
  [\mnras] {10.1093/mnras/stae1687}, \href
  {https://ui.adsabs.harvard.edu/abs/2024MNRAS.532.3729L} {532, 3729}

\bibitem[\protect\citeauthoryear{{Lyubarskii}}{{Lyubarskii}}{1997}]{1997MNRAS.292..679L}
{Lyubarskii} Y.~E.,  1997, \mn@doi [\mnras] {10.1093/mnras/292.3.679}, \href
  {https://ui.adsabs.harvard.edu/abs/1997MNRAS.292..679L} {292, 679}

\bibitem[\protect\citeauthoryear{{Mand} \& {Blaufuss}}{{Mand} \&
  {Blaufuss}}{2025}]{2025ATel17423....1M}
{Mand} A.,  {Blaufuss} E.,  2025, The Astronomer's Telegram, \href
  {https://ui.adsabs.harvard.edu/abs/2025ATel17423....1M} {17423, 1}

\bibitem[\protect\citeauthoryear{Mathur}{Mathur}{2000}]{10.1046/j.1365-8711.2000.03530.x}
Mathur S.,  2000, \mn@doi [Monthly Notices of the Royal Astronomical Society]
  {10.1046/j.1365-8711.2000.03530.x}, 314, L17

\bibitem[\protect\citeauthoryear{{Max-Moerbeck} et~al.,}{{Max-Moerbeck}
  et~al.}{2014}]{2014MNRAS.445..428M}
{Max-Moerbeck} W.,  et~al., 2014, \mn@doi [\mnras] {10.1093/mnras/stu1749},
  \href {https://ui.adsabs.harvard.edu/abs/2014MNRAS.445..428M} {445, 428}

\bibitem[\protect\citeauthoryear{{Max-Moerbeck}, {Richards}, {Hovatta},
  {Pavlidou}, {Pearson}, {Readhead}, {King}  \& {Reeves}}{{Max-Moerbeck}
  et~al.}{2015}]{2015IAUS..313...17M}
{Max-Moerbeck} W.,  {Richards} J.~L.,  {Hovatta} T.,  {Pavlidou} V.,  {Pearson}
  T.~J.,  {Readhead} A.~C.~S.,  {King} O.~G.,   {Reeves} R.,  2015, in
  {Massaro} F.,  {Cheung} C.~C.,  {Lopez} E.,   {Siemiginowska} A.,  eds,  IAU
  Symposium Vol. 313, Extragalactic Jets from Every Angle. pp 17--20
  (\mn@eprint {arXiv} {1411.0334}), \mn@doi{10.1017/S1743921315001799}

\bibitem[\protect\citeauthoryear{{McHardy}}{{McHardy}}{2010}]{2010LNP...794..203M}
{McHardy} I.,  2010, in {Belloni} T.,  ed., , Vol.~794, Lecture Notes in
  Physics, Berlin Springer Verlag.
p.~203, \mn@doi{10.1007/978-3-540-76937-8_8}

\bibitem[\protect\citeauthoryear{{Mohorian} et~al.,}{{Mohorian}
  et~al.}{2022}]{2022MNRAS.510.5280M}
{Mohorian} M.,  et~al., 2022, \mn@doi [\mnras] {10.1093/mnras/stab3738}, \href
  {https://ui.adsabs.harvard.edu/abs/2022MNRAS.510.5280M} {510, 5280}

\bibitem[\protect\citeauthoryear{{Mundo} et~al.,}{{Mundo}
  et~al.}{2020}]{2020MNRAS.496.2922M}
{Mundo} S.~A.,  et~al., 2020, \mn@doi [\mnras] {10.1093/mnras/staa1744}, \href
  {https://ui.adsabs.harvard.edu/abs/2020MNRAS.496.2922M} {496, 2922}

\bibitem[\protect\citeauthoryear{{Negi}, {Joshi}, {Chand}, {Chand}, {Wiita},
  {Ho}  \& {Singh}}{{Negi} et~al.}{2022}]{2022MNRAS.510.1791N}
{Negi} V.,  {Joshi} R.,  {Chand} K.,  {Chand} H.,  {Wiita} P.,  {Ho} L.~C.,
  {Singh} R.~S.,  2022, \mn@doi [\mnras] {10.1093/mnras/stab3591}, \href
  {https://ui.adsabs.harvard.edu/abs/2022MNRAS.510.1791N} {510, 1791}

\bibitem[\protect\citeauthoryear{{Nilsson} et~al.,}{{Nilsson}
  et~al.}{2018}]{2018A&A...620A.185N}
{Nilsson} K.,  et~al., 2018, \mn@doi [\aap] {10.1051/0004-6361/201833621},
  \href {https://ui.adsabs.harvard.edu/abs/2018A&A...620A.185N} {620, A185}

\bibitem[\protect\citeauthoryear{Noel, Gaur, Gupta, Wierzcholska, Ostrowski,
  Dhiman  \& Bhatta}{Noel et~al.}{2022}]{Noel_2022}
Noel A.~P.,  Gaur H.,  Gupta A.~C.,  Wierzcholska A.,  Ostrowski M.,  Dhiman
  V.,   Bhatta G.,  2022, \mn@doi [The Astrophysical Journal Supplement Series]
  {10.3847/1538-4365/ac7799}, 262, 4

\bibitem[\protect\citeauthoryear{{Orienti}, {D'Ammando}, {Giroletti},
  {Dallacasa}, {Giovannini}  \& {Ciprini}}{{Orienti}
  et~al.}{2020}]{2020MNRAS.491..858O}
{Orienti} M.,  {D'Ammando} F.,  {Giroletti} M.,  {Dallacasa} D.,  {Giovannini}
  G.,   {Ciprini} S.,  2020, \mn@doi [\mnras] {10.1093/mnras/stz2949}, \href
  {https://ui.adsabs.harvard.edu/abs/2020MNRAS.491..858O} {491, 858}

\bibitem[\protect\citeauthoryear{{Oshlack}, {Webster}  \& {Whiting}}{{Oshlack}
  et~al.}{2001}]{2001ApJ...558..578O}
{Oshlack} A.~Y.~K.~N.,  {Webster} R.~L.,   {Whiting} M.~T.,  2001, \mn@doi
  [\apj] {10.1086/322299}, \href
  {https://ui.adsabs.harvard.edu/abs/2001ApJ...558..578O} {558, 578}

\bibitem[\protect\citeauthoryear{{Osterbrock} \& {Pogge}}{{Osterbrock} \&
  {Pogge}}{1985}]{1985ApJ...297..166O}
{Osterbrock} D.~E.,  {Pogge} R.~W.,  1985, \mn@doi [\apj] {10.1086/163513},
  \href {https://ui.adsabs.harvard.edu/abs/1985ApJ...297..166O} {297, 166}

\bibitem[\protect\citeauthoryear{{Otero-Santos} et~al.,}{{Otero-Santos}
  et~al.}{2024}]{2024A&A...686A.228O}
{Otero-Santos} J.,  et~al., 2024, \mn@doi [\aap] {10.1051/0004-6361/202449647},
  \href {https://ui.adsabs.harvard.edu/abs/2024A&A...686A.228O} {686, A228}

\bibitem[\protect\citeauthoryear{{Paliya}, {Parker}, {Jiang}, {Fabian},
  {Brenneman}, {Ajello}  \& {Hartmann}}{{Paliya}
  et~al.}{2019}]{2019ApJ...872..169P}
{Paliya} V.~S.,  {Parker} M.~L.,  {Jiang} J.,  {Fabian} A.~C.,  {Brenneman} L.,
   {Ajello} M.,   {Hartmann} D.,  2019, \mn@doi [\apj]
  {10.3847/1538-4357/ab01ce}, \href
  {https://ui.adsabs.harvard.edu/abs/2019ApJ...872..169P} {872, 169}

\bibitem[\protect\citeauthoryear{{Paliya} et~al.,}{{Paliya}
  et~al.}{2020}]{2020ApJ...892..133P}
{Paliya} V.~S.,  et~al., 2020, \mn@doi [\apj] {10.3847/1538-4357/ab754f}, \href
  {https://ui.adsabs.harvard.edu/abs/2020ApJ...892..133P} {892, 133}

\bibitem[\protect\citeauthoryear{{Paliya}, {Stalin}, {Dom{\'\i}nguez}  \&
  {Saikia}}{{Paliya} et~al.}{2024}]{2024MNRAS.527.7055P}
{Paliya} V.~S.,  {Stalin} C.~S.,  {Dom{\'\i}nguez} A.,   {Saikia} D.~J.,  2024,
  \mn@doi [\mnras] {10.1093/mnras/stad3650}, \href
  {https://ui.adsabs.harvard.edu/abs/2024MNRAS.527.7055P} {527, 7055}

\bibitem[\protect\citeauthoryear{{Peterson}}{{Peterson}}{2011a}]{2011arXiv1109.4181P}
{Peterson} B.~M.,  2011a, \mn@doi [arXiv e-prints] {10.48550/arXiv.1109.4181},
  \href {https://ui.adsabs.harvard.edu/abs/2011arXiv1109.4181P} {p.
  arXiv:1109.4181}

\bibitem[\protect\citeauthoryear{Peterson}{Peterson}{2011b}]{Peterson:2011sM}
Peterson B.,  2011b, \mn@doi [PoS] {10.22323/1.126.0032}, NLS1, 032

\bibitem[\protect\citeauthoryear{{Pininti}, {Bhatta}, {Paul}, {Kumar},
  {Rajgor}, {Barnwal}  \& {Gharat}}{{Pininti}
  et~al.}{2023}]{2023MNRAS.518.1459P}
{Pininti} V.~R.,  {Bhatta} G.,  {Paul} S.,  {Kumar} A.,  {Rajgor} A.,
  {Barnwal} R.,   {Gharat} S.,  2023, \mn@doi [\mnras]
  {10.1093/mnras/stac3125}, \href
  {https://ui.adsabs.harvard.edu/abs/2023MNRAS.518.1459P} {518, 1459}

\bibitem[\protect\citeauthoryear{{Press}}{{Press}}{1978}]{1978ComAp...7..103P}
{Press} W.~H.,  1978, Comments on Astrophysics, \href
  {https://ui.adsabs.harvard.edu/abs/1978ComAp...7..103P} {7, 103}

\bibitem[\protect\citeauthoryear{{Prince}, {Khatoon}  \& {Stalin}}{{Prince}
  et~al.}{2021}]{2021MNRAS.502.5245P}
{Prince} R.,  {Khatoon} R.,   {Stalin} C.~S.,  2021, \mn@doi [\mnras]
  {10.1093/mnras/stab369}, \href
  {https://ui.adsabs.harvard.edu/abs/2021MNRAS.502.5245P} {502, 5245}

\bibitem[\protect\citeauthoryear{{Rakshit}, {Stalin}, {Chand}  \&
  {Zhang}}{{Rakshit} et~al.}{2017}]{2017ApJS..229...39R}
{Rakshit} S.,  {Stalin} C.~S.,  {Chand} H.,   {Zhang} X.-G.,  2017, \mn@doi
  [\apjs] {10.3847/1538-4365/aa6971}, \href
  {https://ui.adsabs.harvard.edu/abs/2017ApJS..229...39R} {229, 39}

\bibitem[\protect\citeauthoryear{{Rakshit}, {Stalin}, {Chand}  \&
  {Zhang}}{{Rakshit} et~al.}{2018}]{2018BSRSL..87..379R}
{Rakshit} S.,  {Stalin} C.~S.,  {Chand} H.,   {Zhang} X.-G.,  2018, \mn@doi
  [Bulletin de la Societe Royale des Sciences de Liege]
  {10.48550/arXiv.1706.00797}, \href
  {https://ui.adsabs.harvard.edu/abs/2018BSRSL..87..379R} {87, 379}

\bibitem[\protect\citeauthoryear{{Rani}, {Stalin}  \& {Rakshit}}{{Rani}
  et~al.}{2017}]{2017MNRAS.466.3309R}
{Rani} P.,  {Stalin} C.~S.,   {Rakshit} S.,  2017, \mn@doi [\mnras]
  {10.1093/mnras/stw3228}, \href
  {https://ui.adsabs.harvard.edu/abs/2017MNRAS.466.3309R} {466, 3309}

\bibitem[\protect\citeauthoryear{{Ravasio}, {Tagliaferri}, {Ghisellini}  \&
  {Tavecchio}}{{Ravasio} et~al.}{2004}]{2004A&A...424..841R}
{Ravasio} M.,  {Tagliaferri} G.,  {Ghisellini} G.,   {Tavecchio} F.,  2004,
  \mn@doi [\aap] {10.1051/0004-6361:20034545}, \href
  {https://ui.adsabs.harvard.edu/abs/2004A&A...424..841R} {424, 841}

\bibitem[\protect\citeauthoryear{{Rodr{\'\i}guez-Pascual}
  et~al.,}{{Rodr{\'\i}guez-Pascual} et~al.}{1997}]{1997ApJS..110....9R}
{Rodr{\'\i}guez-Pascual} P.~M.,  et~al., 1997, \mn@doi [\apjs]
  {10.1086/312996}, \href
  {https://ui.adsabs.harvard.edu/abs/1997ApJS..110....9R} {110, 9}

\bibitem[\protect\citeauthoryear{{Romano}, {Vercellone}, {Foschini},
  {Tavecchio}, {Landoni}  \& {Kn{\"o}dlseder}}{{Romano}
  et~al.}{2018}]{2018MNRAS.481.5046R}
{Romano} P.,  {Vercellone} S.,  {Foschini} L.,  {Tavecchio} F.,  {Landoni} M.,
   {Kn{\"o}dlseder} J.,  2018, \mn@doi [\mnras] {10.1093/mnras/sty2484}, \href
  {https://ui.adsabs.harvard.edu/abs/2018MNRAS.481.5046R} {481, 5046}

\bibitem[\protect\citeauthoryear{{Shah}, {Mankuzhiyil}, {Sinha}, {Misra},
  {Sahayanathan}  \& {Iqbal}}{{Shah} et~al.}{2018}]{2018RAA....18..141S}
{Shah} Z.,  {Mankuzhiyil} N.,  {Sinha} A.,  {Misra} R.,  {Sahayanathan} S.,
  {Iqbal} N.,  2018, \mn@doi [Research in Astronomy and Astrophysics]
  {10.1088/1674-4527/18/11/141}, \href
  {https://ui.adsabs.harvard.edu/abs/2018RAA....18..141S} {18, 141}

\bibitem[\protect\citeauthoryear{{Str{\"u}der} et~al.,}{{Str{\"u}der}
  et~al.}{2001}]{2001A&A...365L..18S}
{Str{\"u}der} L.,  et~al., 2001, \mn@doi [\aap] {10.1051/0004-6361:20000066},
  \href {https://ui.adsabs.harvard.edu/abs/2001A&A...365L..18S} {365, L18}

\bibitem[\protect\citeauthoryear{Sulentic, Zwitter, Marziani  \&
  Dultzin-Hacyan}{Sulentic et~al.}{2000}]{Sulentic_2000}
Sulentic J.~W.,  Zwitter T.,  Marziani P.,   Dultzin-Hacyan D.,  2000, \mn@doi
  [The Astrophysical Journal] {10.1086/312717}, 536, L5

\bibitem[\protect\citeauthoryear{{Tramacere}, {Giommi}, {Perri}, {Verrecchia}
  \& {Tosti}}{{Tramacere} et~al.}{2009}]{2009A&A...501..879T}
{Tramacere} A.,  {Giommi} P.,  {Perri} M.,  {Verrecchia} F.,   {Tosti} G.,
  2009, \mn@doi [\aap] {10.1051/0004-6361/200810865}, \href
  {https://ui.adsabs.harvard.edu/abs/2009A&A...501..879T} {501, 879}

\bibitem[\protect\citeauthoryear{{Trevese}, {Kron}, {Majewski}, {Bershady}  \&
  {Koo}}{{Trevese} et~al.}{1994}]{1994ApJ...433..494T}
{Trevese} D.,  {Kron} R.~G.,  {Majewski} S.~R.,  {Bershady} M.~A.,   {Koo}
  D.~C.,  1994, \mn@doi [\apj] {10.1086/174661}, \href
  {https://ui.adsabs.harvard.edu/abs/1994ApJ...433..494T} {433, 494}

\bibitem[\protect\citeauthoryear{{Turner}, {George}  \& {Netzer}}{{Turner}
  et~al.}{1999}]{1999ApJ...526...52T}
{Turner} T.~J.,  {George} I.~M.,   {Netzer} H.,  1999, \mn@doi [\apj]
  {10.1086/307995}, \href
  {https://ui.adsabs.harvard.edu/abs/1999ApJ...526...52T} {526, 52}

\bibitem[\protect\citeauthoryear{{Uttley}, {McHardy}  \& {Papadakis}}{{Uttley}
  et~al.}{2002}]{2002MNRAS.332..231U}
{Uttley} P.,  {McHardy} I.~M.,   {Papadakis} I.~E.,  2002, \mn@doi [\mnras]
  {10.1046/j.1365-8711.2002.05298.x}, \href
  {https://ui.adsabs.harvard.edu/abs/2002MNRAS.332..231U} {332, 231}

\bibitem[\protect\citeauthoryear{{Uttley}, {McHardy}  \& {Vaughan}}{{Uttley}
  et~al.}{2005}]{2005MNRAS.359..345U}
{Uttley} P.,  {McHardy} I.~M.,   {Vaughan} S.,  2005, \mn@doi [\mnras]
  {10.1111/j.1365-2966.2005.08886.x}, \href
  {https://ui.adsabs.harvard.edu/abs/2005MNRAS.359..345U} {359, 345}

\bibitem[\protect\citeauthoryear{{Vagnetti}, {Middei}, {Antonucci}, {Paolillo}
  \& {Serafinelli}}{{Vagnetti} et~al.}{2016}]{2016A&A...593A..55V}
{Vagnetti} F.,  {Middei} R.,  {Antonucci} M.,  {Paolillo} M.,   {Serafinelli}
  R.,  2016, \mn@doi [\aap] {10.1051/0004-6361/201629057}, \href
  {https://ui.adsabs.harvard.edu/abs/2016A&A...593A..55V} {593, A55}

\bibitem[\protect\citeauthoryear{{Vaughan}, {Edelson}, {Warwick}  \&
  {Uttley}}{{Vaughan} et~al.}{2003}]{2003MNRAS.345.1271V}
{Vaughan} S.,  {Edelson} R.,  {Warwick} R.~S.,   {Uttley} P.,  2003, \mn@doi
  [\mnras] {10.1046/j.1365-2966.2003.07042.x}, \href
  {https://ui.adsabs.harvard.edu/abs/2003MNRAS.345.1271V} {345, 1271}

\bibitem[\protect\citeauthoryear{{Wang}, {Brinkmann}  \& {Bergeron}}{{Wang}
  et~al.}{1996}]{1996A&A...309...81W}
{Wang} T.,  {Brinkmann} W.,   {Bergeron} J.,  1996, \aap, \href
  {https://ui.adsabs.harvard.edu/abs/1996A&A...309...81W} {309, 81}

\bibitem[\protect\citeauthoryear{{Wang} et~al.,}{{Wang}
  et~al.}{2016}]{2016ApJ...824..149W}
{Wang} F.,  et~al., 2016, \mn@doi [\apj] {10.3847/0004-637X/824/2/149}, \href
  {https://ui.adsabs.harvard.edu/abs/2016ApJ...824..149W} {824, 149}

\bibitem[\protect\citeauthoryear{{Wierzcholska} \& {Wagner}}{{Wierzcholska} \&
  {Wagner}}{2016}]{2016MNRAS.458...56W}
{Wierzcholska} A.,  {Wagner} S.~J.,  2016, \mn@doi [\mnras]
  {10.1093/mnras/stw095}, \href
  {https://ui.adsabs.harvard.edu/abs/2016MNRAS.458...56W} {458, 56}

\bibitem[\protect\citeauthoryear{{Yao} \& {Komossa}}{{Yao} \&
  {Komossa}}{2021}]{2021MNRAS.501.1384Y}
{Yao} S.,  {Komossa} S.,  2021, \mn@doi [\mnras] {10.1093/mnras/staa3708},
  \href {https://ui.adsabs.harvard.edu/abs/2021MNRAS.501.1384Y} {501, 1384}

\bibitem[\protect\citeauthoryear{{Yao} \& {Komossa}}{{Yao} \&
  {Komossa}}{2023}]{2023MNRAS.523..441Y}
{Yao} S.,  {Komossa} S.,  2023, \mn@doi [\mnras] {10.1093/mnras/stad1415},
  \href {https://ui.adsabs.harvard.edu/abs/2023MNRAS.523..441Y} {523, 441}

\bibitem[\protect\citeauthoryear{{Yao}, {Yuan}, {Komossa}, {Grupe}, {Fuhrmann}
  \& {Liu}}{{Yao} et~al.}{2015a}]{2015AJ....150...23Y}
{Yao} S.,  {Yuan} W.,  {Komossa} S.,  {Grupe} D.,  {Fuhrmann} L.,   {Liu} B.,
  2015a, \mn@doi [\aj] {10.1088/0004-6256/150/1/23}, \href
  {https://ui.adsabs.harvard.edu/abs/2015AJ....150...23Y} {150, 23}

\bibitem[\protect\citeauthoryear{{Yao}, {Yuan}, {Zhou}, {Komossa}, {Zhang},
  {Qiao}  \& {Liu}}{{Yao} et~al.}{2015b}]{2015MNRAS.454L..16Y}
{Yao} S.,  {Yuan} W.,  {Zhou} H.,  {Komossa} S.,  {Zhang} J.,  {Qiao} E.,
  {Liu} B.,  2015b, \mn@doi [\mnras] {10.1093/mnrasl/slv119}, \href
  {https://ui.adsabs.harvard.edu/abs/2015MNRAS.454L..16Y} {454, L16}

\bibitem[\protect\citeauthoryear{{Yao}, {Komossa}, {Liu}, {Yi}, {Yuan}, {Zhou}
  \& {Wu}}{{Yao} et~al.}{2019}]{2019MNRAS.487L..40Y}
{Yao} S.,  {Komossa} S.,  {Liu} W.-J.,  {Yi} W.,  {Yuan} W.,  {Zhou} H.,   {Wu}
  X.-B.,  2019, \mn@doi [\mnras] {10.1093/mnrasl/slz071}, \href
  {https://ui.adsabs.harvard.edu/abs/2019MNRAS.487L..40Y} {487, L40}

\bibitem[\protect\citeauthoryear{{Yao}, {Komossa}, {Kraus}  \& {Grupe}}{{Yao}
  et~al.}{2024}]{2024MNRAS.533.1281Y}
{Yao} S.,  {Komossa} S.,  {Kraus} A.,   {Grupe} D.,  2024, \mn@doi [\mnras]
  {10.1093/mnras/stae1827}, \href
  {https://ui.adsabs.harvard.edu/abs/2024MNRAS.533.1281Y} {533, 1281}

\bibitem[\protect\citeauthoryear{{Yu}, {Jiang}, {Bambi}, {Gallo}, {Grupe},
  {Fabian}, {Reynolds}  \& {Brandt}}{{Yu} et~al.}{2023}]{2023MNRAS.522.5456Y}
{Yu} Z.,  {Jiang} J.,  {Bambi} C.,  {Gallo} L.~C.,  {Grupe} D.,  {Fabian}
  A.~C.,  {Reynolds} C.~S.,   {Brandt} W.~N.,  2023, \mn@doi [\mnras]
  {10.1093/mnras/stad1327}, \href
  {https://ui.adsabs.harvard.edu/abs/2023MNRAS.522.5456Y} {522, 5456}

\bibitem[\protect\citeauthoryear{{Zhou}, {Wang}, {Yuan}, {Lu}, {Dong}, {Wang}
  \& {Lu}}{{Zhou} et~al.}{2006}]{2006ApJS..166..128Z}
{Zhou} H.,  {Wang} T.,  {Yuan} W.,  {Lu} H.,  {Dong} X.,  {Wang} J.,   {Lu} Y.,
   2006, \mn@doi [\apjs] {10.1086/504869}, \href
  {https://ui.adsabs.harvard.edu/abs/2006ApJS..166..128Z} {166, 128}

\bibitem[\protect\citeauthoryear{{Zhou} et~al.,}{{Zhou}
  et~al.}{2007}]{2007ApJ...658L..13Z}
{Zhou} H.,  et~al., 2007, \mn@doi [\apjl] {10.1086/513604}, \href
  {https://ui.adsabs.harvard.edu/abs/2007ApJ...658L..13Z} {658, L13}

\bibitem[\protect\citeauthoryear{{den Herder} et~al.,}{{den Herder}
  et~al.}{2001}]{2001A&A...365L...7D}
{den Herder} J.~W.,  et~al., 2001, \mn@doi [\aap] {10.1051/0004-6361:20000058},
  \href {https://ui.adsabs.harvard.edu/abs/2001A&A...365L...7D} {365, L7}

\makeatother
\end{thebibliography}



\clearpage
\appendix

\section{Results on Individual Sources} \label{results}
In this section, we will summarize the timing and spectral properties of the individual sources. 


\subsection{1H\,0323+342}
Between 2015 and 2018, XMM-Newton observed the Narrow-Line Seyfert 1 galaxy 1H\,0323+342 over seven epochs, revealing notable variation in its X-ray emission.  The mean count rate varies between 2.98 and 6.64 counts\,s$^{-1}$, reflecting variations in the brightness of the source. Across all observations, the $F_{\rm var}$ varied from 10\% to 16\% (highest among all sources), suggesting moderate-to-high intrinsic variability. The maximum amplitude was observed in the 2018-08-24 observation, and the VA values ranged from 0.68 to 1.67. The minimum variability timescale $\tau_{\rm var}$ varied from 0.15 to 0.52 ks, indicating that fluxes could shift quickly on timescales as short as a few hundred seconds. Analyzing the HR versus count-rate during seven epochs shows a variety of spectral variability patterns. Three observations exhibit a distinct softening trend with increasing brightness (0823780301, 0823780401, 0823780701), indicating greater soft X-ray emission at higher flux levels, which are often associated with disk-corona coupling or thermal Comptonization. On the other hand, two epochs (0764670101, 0823780601) show a softer-when-brighter trend, suggesting a major contribution from a harder component, which might be the jet or a changeable corona.  There is no discernible spectral trend in the other two epochs, indicating more intricate or steady spectral behaviour during those times.  This complexity is further supported by the count-rate distributions, which show unimodal distributions in the other five epochs, indicating relatively consistent flux behaviour, and bimodal patterns in two epochs (0823780701, 0823780301), suggesting distinct flux states or transitions. Finally, the power spectral density (PSD) slope ($\beta_p$) fluctuated somewhat between epochs, ranging from 0.79 to 1.06, suggesting a red-noise-driven variability process that was reasonably constant. All things considered, the temporal analysis reveals that 1H\,0323+342 is a highly variable X-ray source exhibiting intricate and changing variability features over both short and long timescales. 

The X-ray spectrum of the source exhibits a soft excess below 2 keV and a hard excess above 35 keV, as reported in \citep{2020MNRAS.496.2922M,2018MNRAS.479.2464G}. Our X-ray spectral study of 1H\,0323+342 shows moderate spectral evolution and soft excess over time when fitted with a power-law plus blackbody (PL+BB) model spanning seven epochs. The source exhibits considerable flux fluctuation, as evidenced by the wide variation in the unabsorbed flux in the 0.3-10 keV range from a minimum of 1.1$\times 10^{-13}\, {\rm erg \,s}^{-1} {\rm cm}^{-2}$ in 0823780701 to a maximum of 2.6$\times 10^{-13}\, {\rm erg \,s}^{-1} {\rm cm}^{-2}$ in 0823780401. Reduced $\chi^{2}$ values are typically at or slightly over 1.4, indicating that the selected spectral model is adequate for all epochs. The photon index ($\Gamma$) varies between 1.79 and 1.96, with epoch 0823780301 having the softest spectrum ($\Gamma$ = 1.96), which is consistent with the HR analysis's softer-when-brighter pattern. A variable contribution from a steep, potentially disk-related component is suggested by this softening pattern. Over all epochs, the blackbody temperature ($kT$) stays relatively constant within the small range of $\sim$0.134-0.151 keV, suggesting a continuous soft excess that most likely originates from the warm Comptonizing area or inner accretion disk. The X-ray spectra in all epochs also show an acceptable fit with a BPL model (relatively bit higher $\chi^2_{red}$) with fit parameters in the range $2.62<\Gamma_1<2.97$,  $1.84<\Gamma_2<1.98$, $1.48<E_b<1.64$ keV, while the $N_H$ stays consistent throughout all observations.

\subsection{3C 286}
{\it XMM-Newton} observations of 3C 286 reveal low-level X-ray variability with a mean count rate of 0.26 ± 0.06 counts\,s$^{-1}$ and $F_{\rm var}$ = 8.33 ± 7.15\%. However, interpretation is limited by the high uncertainties and poor statistics. The hardness ratio exhibits a complex, non-monotonic trend, indicating weak or mixed spectral variability, while the structure function shows an unreasonable trend. Red-noise-dominated behavior is indicated by the PSD slope ($\beta_p$ = 0.85). The source's spectrum is well represented by a power-law model with 
$\Gamma$=2.23, which is compatible with a soft, non-thermal continuum and a flux of 6.4\ergcm{-13}. A deep X-ray investigation of 3C 286, conducted by \cite{2024Univ...10..289K, 2024MNRAS.533.1281Y}, explains its X-ray emission through a prominent jet feature, which is very similar to the one observed in this work.

\subsection{SDSS J0946+1017}
The timing analysis of this source is limited by its low mean count rate of 0.12 $\pm$0.04 counts per second. The SF, HR, and PSD thus exhibit no discernible patterns. The $\tau_{\rm var}$ is 0.05 $\pm$0.02 ks. The source's spectrum is well fit 
with a power-law model ($\Gamma$=1.47), indicating a relatively hard spectrum with a reasonable fit statistic ($\chi^{2}_{red}$=122.39/91) and a unabsorbed flux of 4.06$\times 10^{-13}\, {\rm erg \,s}^{-1} {\rm cm}^{-2}$, which well collaborates with the findings of \cite{2019MNRAS.487L..40Y}.

\subsection{PMN J0948+0022}
Three {\it XMM-Newton} observations (2008–2016) of PMN\,J0948+0022's X-ray timing and spectrum show steady spectral features with little change. Flux variability over epochs was reflected in the mean count rate, which fluctuated significantly, ranging from a low of 0.04 counts\,s$^{-1}$ in 2011 to a high of 2.84 counts\,s$^{-1}$ in 2008. Only two epochs (2008 and 2016) have quantifiable $F_{\rm var}$, with moderate values of 6–7\% and VA within the range of.50–2.01. The $\tau_{\rm var}$ was consistently short (0.11–0.14 ks), suggesting fast variability. Nevertheless, no noteworthy patterns were discernible in the SF, HR, or PSD slopes due to the low statistics in all three observations. The source exhibits a soft excess below 2 keV along with a power-law component above that. \cite{2014MNRAS.440..106B} analyzed two epochs of \xmm along with the {\textit{SWIFT-XRT}} data and fitted the soft excess with a disk temperature of 0.15±0.01 0.173±0.003 keV and $\Gamma \sim 1.59-1.6$. We have also modeled the soft excess and found that $kT$ fluctuates between 0.105 and 0.123 keV, whereas the $\Gamma$ stays comparatively constant between 1.52 and 1.59, suggesting a continuous soft excess component. Gradually, the unabsorbed flux rose from 23.79 in 2016 to 68.52\ergcm{-13} in 2008. In general, PMN J0948+0022 exhibits stable spectral features and modest X-ray fluctuation; however, in certain epochs, this is limited by poor count statistics. All three epochs show an acceptable fit with a BPL with $1.92<\Gamma_1<2.30$,  $1.53<\Gamma_2<1.64$, and the break energy and column density stays nearly constant $1.22<E_b<1.28$ keV, and $5.0\times10^{20}$ cm$^{-2}$.  However, a PL+BB,  was found to give a more acceptable $\chi^{2}_{red}$ value. 

\subsection{FBQS J1102+2239}
The {\it XMM-Newton} observation of the Narrow-Line Seyfert 1 galaxy FBQS J1102+2239 on 2012-06-11 lasting 13.9 ks produced a low mean count rate of 0.07±0.03 counts\,s$^{-1}$. The SF, HR, and PSD did not show any notable patterns due to insufficient photon statistics, which made evaluating their variable characteristics challenging. A rather soft spectrum is suggested by the  $\Gamma$=2.19, flux = 2.15\ergcm{-13} obtained from spectral analysis using a straightforward PL model.

\subsection{CGRaBS J1222+0413}
Modest X-ray variability is observed over a brief exposure of 11.9 ks in the {\it XMM-Newton} observation of CGRaBS J1222+0413 on 2006-07-12. With a VA of 1.61 ± 0.74, a $F_{\rm var}$ of 5.36\%, and a mean count rate of 1.08±0.13 counts\,s$^{-1}$, the source exhibits low-to-moderate flux variability. There are no obvious patterns in the SF, HR, or PSD because of the low statistics, but the $\tau_{\rm var}$ is calculated to be 0.19 ± 0.10 ks, and the PSD slope $\beta_P$ is 0.50.  The results of spectral fitting using a PL+BB model include a soft excess component with $kT = 0.164 \pm 0.012$ keV, a reasonably hard $\Gamma=1.29 \pm 0.04$, and a good fit ($\chi^{2}_{red}$ = 142.57/112) with a total flux of 36.08\ergcm{-13}. This hard X-ray spectrum suggests a jet dominates over the disk. The fit parameters $\Gamma_1$=1.62±0.09 $\Gamma_2$=1.11±0.08, $E_b$=2.48±0.37 keV, flux = 3.8$\times 10^{-11}\, {\rm erg \,s}^{-1} {\rm cm}^{-2}$, $\chi^{2}_{red}$ = 127.26/111, and $N_H$=0.18±0.10 are better for BPL, even though a couple of the epochs exhibit superior fit by PL+BB for BPL.

\subsection{SDSS J1246+0238}
Poor statistics and a low mean count rate of 0.05 ± 0.03 counts\,s$^{-1}$ limit the reliability of the timing analysis for the X-ray observation of SDSS J1246+0238 on 2012-12-14 (ObsID: 0690090201), which had a short exposure of 16.9 ks. Thus, the HR, SF, and PSD show no significant trends. A PL+BB model is used for spectral fitting, and the results show a statistically satisfactory fit ($\chi^{2}_{red}$=22.62/14), a $\Gamma$=1.71 ± 0.14, a $kT$ = 0.046 ± 0.012 keV, and a flux of 2.72\ergcm{-13}. Overall, only a simple spectrum characterization is possible for this epoch due to the low photon statistics.

\subsection{TXS 1419+391}
TXS 1419+391 was detected on 2020-06-11 with a mean count rate of 0.22 ± 0.06 counts\,s$^{-1}$ and a net exposure of 38 ks.  With a $F_{\rm var}\sim$14\%, a VA of 0.47 ± 59.1, and a $\tau_{\rm var}$ of 0.08 ± 0.03 ks, the source showed significant variability. However, there were no discernible changes in the SF, HR, or PSD because of the poor statistical quality. The spectral fitting showed a typical soft X-ray spectrum in a relatively low-flux state with a $\Gamma$=2.05 ± 0.03 and a flux of 4.62\ergcm{-13}. 

\subsection{PKS 1502+036}
The X-ray timing analysis for PKS 1502+036 shows a mean count rate of 0.14±0.04  counts\,s$^{-1}$, with $F_{\rm var} = 11.64±4.56\%$, VA =8.09±6.71, $\tau_{\rm var}$=0.07±0.03 ks, and $\beta_p = 0.04$.  No discernible trend is seen in PSD, SF, or HR because of the low S/N.  Using a PL model for spectral fitting yields an adequate fit with a photon index $\Gamma = 1.62\pm0.07$, unabsorbed flux=3.56\ergcm{-13}, and $\chi^{2}_{red}$=52.07/37, aligning with the findings of \cite{2019ApJ...872..169P,2023MNRAS.523..441Y}. This source also gives an acceptable fit with the BPL model and the model parameters $\Gamma_1=2.20\pm0.54$, $\Gamma_2=1.50\pm0.10$, and $E_b=1.24\pm0.22$ keV and $\chi^{2}_{red}$= 48.19/36. 

\subsection{SDSS J1641+3454}
Two {\it XMM-Newton} observations (Obs. IDs: 0860640201 and 0860640301) were available for this source. Timing analysis suggests a mean count rate of 0.17±0.05 and 0.09±0.04  counts\,s$^{-1}$, respectively. The estimates of $F_{\rm var}$, VA and $\tau_{\rm var}$ are again not reliable due to large error bars in the data and poor signal-to-noise. However, the X-ray spectra are well fitted by the PL model with a mean $\Gamma=1.70$ and unabsorbed flux of 9.03 to 4.45\ergcm{-13}.

\subsection{RGB J1644+263}
The NLSy1 RGB J1644+263 exhibits notable X-ray variability in the {\it XMM-Newton} observation of 90 ks. Moderate changes in flux are indicated by the mean count rate of 1.10±0.05 counts\,s$^{-1}$, $F_{\rm var}$ of 5.57±0.81\%, and VA of 1.49±0.43. Short-timescale variability is indicated by $\tau_{\rm var} = 0.20\pm0.07$\,ks. The PSD slope ($\beta_p\sim0.95$) indicates variability dominated by red noise.  However, the SF is inconclusive due to substantial uncertainties, and the HR plots do not exhibit any spectrum change. A jet-dominated X-ray emission, characteristic of radio-loud NLSy1s, is compatible with spectral fitting using a PL model, yielding $\Gamma=1.70$=1.74±0.01 and unabsorbed  flux$\sim35$\ergcm{-13}, matching with the work presented in \cite{2011nlsg.confE..24F}. The best fit BPL parameters are given as $\Gamma_1$=2.25±0.10, $\Gamma_2$=1.74±0.02, and $E_b$=1.19±0.05, although a BB+PL model provides a better fit.

\subsection{TXS 2116-077}
X-ray temporal investigations of this source in both {\it XMM-Newton} observations are significantly impacted by the large flux errors (mean flux = 0.09±0.04 and 0.02±0.01) and low photon statistics. Timing solutions are, therefore, unreliable. However, an X-ray spectral analyses show that the PL model with photon indices of 1.90 and 1.58 and a flux range of $\sim4-6$ \ergcm{-13} can adequately characterize both spectral epochs, following a jet-based emission process \citep{2020ApJ...892..133P}.

\subsection{PKS 2004-447}
Significant fluctuations are observed in the X-ray timing and spectral analysis of PKS 2004-447 across five {\it XMM-Newton} epochs. The mean count rate exhibits intrinsic variability, ranging from 0.17  (0694530101) to 0.48 counts\,s$^{-1}$ (0200360201), with a consistent $F_{\rm var} \sim 10\%$. Rapid flux changes are highlighted by the VA's range of 2.4–7.4 and the minimal variability timescale $\tau_{\rm var}$, which varies from 0.07 to 0.41 ks. Across epochs, the PSD slope ($\beta_p$) varies between 0.70 (0694530101), 1.12 (0200360201), and 0.63 (0694530201), indicating that red-noise-dominated variable processes are present; the other two exhibit substantial scatter. There is no clear trend in either SF or HR. The spectral properties of the source, including flux, vary from $\sim5$ to $\sim22$ \ergcm{-13}, and photon indices range from $\Gamma=1.49$ to 1.74 are consistent with the previous studies \citep{2006MNRAS.370..245G, 2016A&A...585A..91K, 2019ApJ...872..169P}. \\

\subsection{TXS 0103+189}
TXS 0103+189 was observed with \xmm once in 2023, with 38 \text{ks} exposure time. We have found the mean-rate count of 0.21(counts\,s$^{-1}$. The light curve does not exhibit any significant variability. The PSD, HR and SF also do not give any conclusive results. The flux distribution favors a normal distribution over a log-normal distribution, hence implying an additive process. However, the X-ray spectra were completely fitted by the PL model, with fit parameters of $\Gamma \sim1.7$, indicating a jet-dominated X-ray emission. 

\subsection{PKS 1244-255}
PKS 1244-255 has a single \xmm observation in 2010. The X-ray timing analysis suggests a mean count rate of $\sim$1.15 counts\,s$^{-1}$, the source shows a significant flux variability with $F_{\rm var} \sim 7\%$, and \text{VA}$\sim$1.79 and $\tau_{\rm var} \sim$0.12 \text{ks}, and PSD analysis also suggests $\beta_P\sim0.52$. The X-ray spectra are well fit by a PL model with $\Gamma \sim 1.8$. Overall, the source exhibits moderate intraday variability, driven primarily by the jet.

\subsection{TXS 1308+554}
The NLSy1, TXS 1308+554, has also only one \xmm observation taken in 2014. On the intraday timescale, the source does not exhibit any significant variability, primarily due to high error bars. The mean count rate was found to be $\sim0.13$ (count/s). We do not observe any trends in the SF, PSD, and HR plots. On the other hand, the X-ray spectra show a PL component with $\Gamma \sim 1.8$, indicating highly beamed jet emission.


\begin{figure*}\label{app:1H0323+342}
	\centering
	\caption{LCs, HR plots, Structure Function, PSD, PDF, and spectral fits derived for all epochs of 1H\,0323+342.}
	\begin{minipage}{.3\textwidth} 
		\centering 
		\includegraphics[width=.99\linewidth]{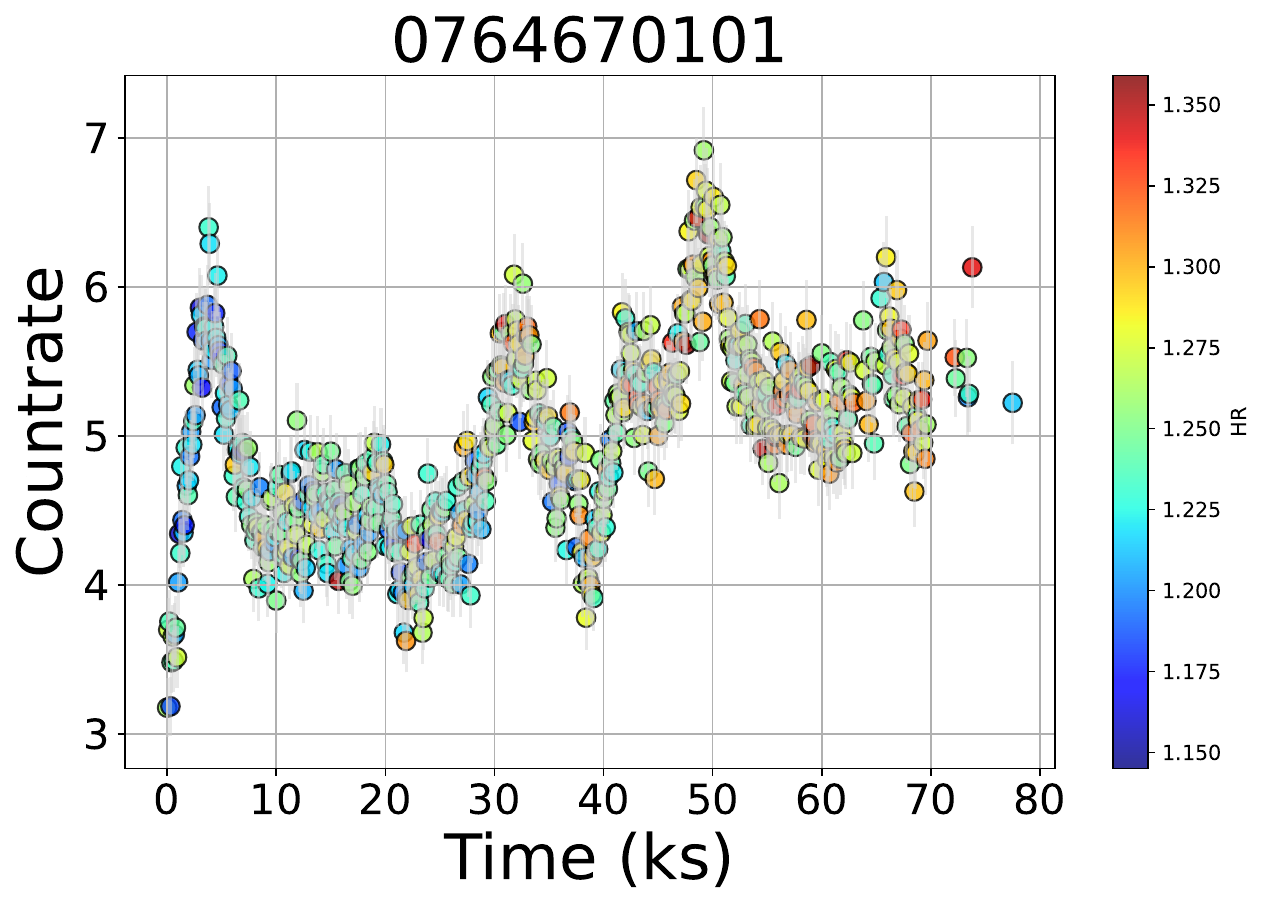}
	\end{minipage}
	\begin{minipage}{.3\textwidth} 
		\centering 
		\includegraphics[width=.99\linewidth]{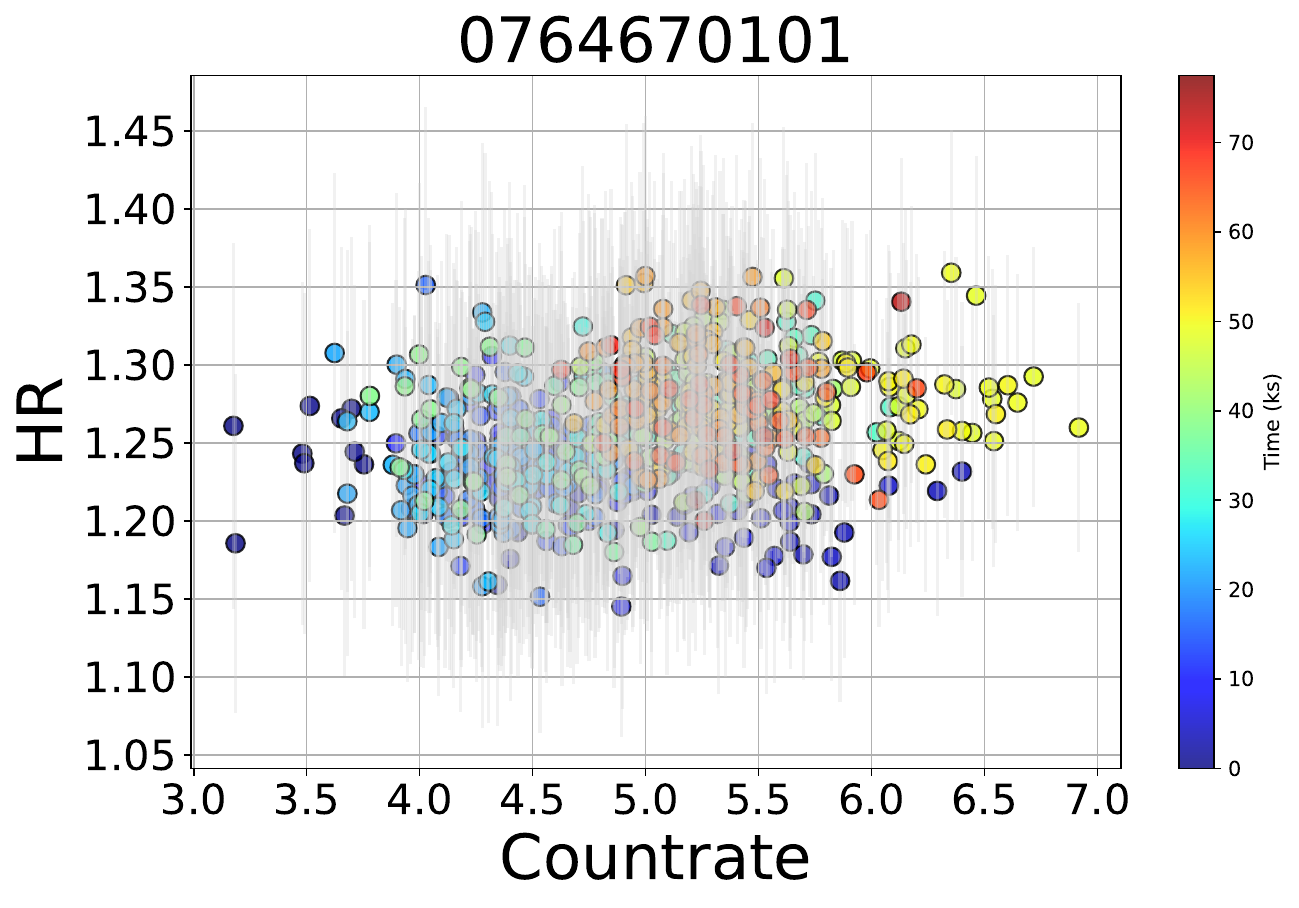}
	\end{minipage}
	\begin{minipage}{.3\textwidth} 
		\centering 
		\includegraphics[width=.99\linewidth, angle=0]{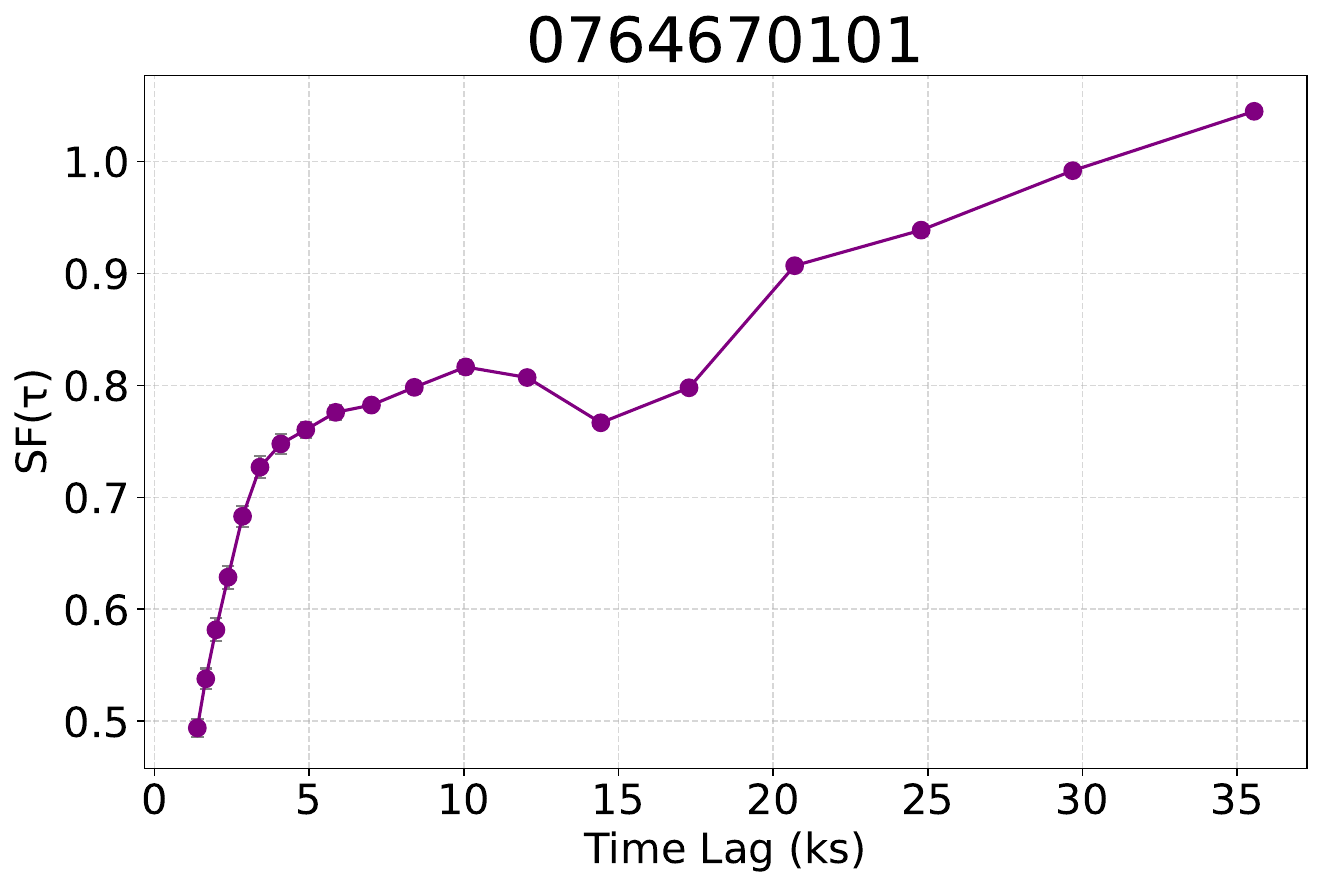}
	\end{minipage}
	\begin{minipage}{.3\textwidth} 
		\centering 
		\includegraphics[width=.99\linewidth]{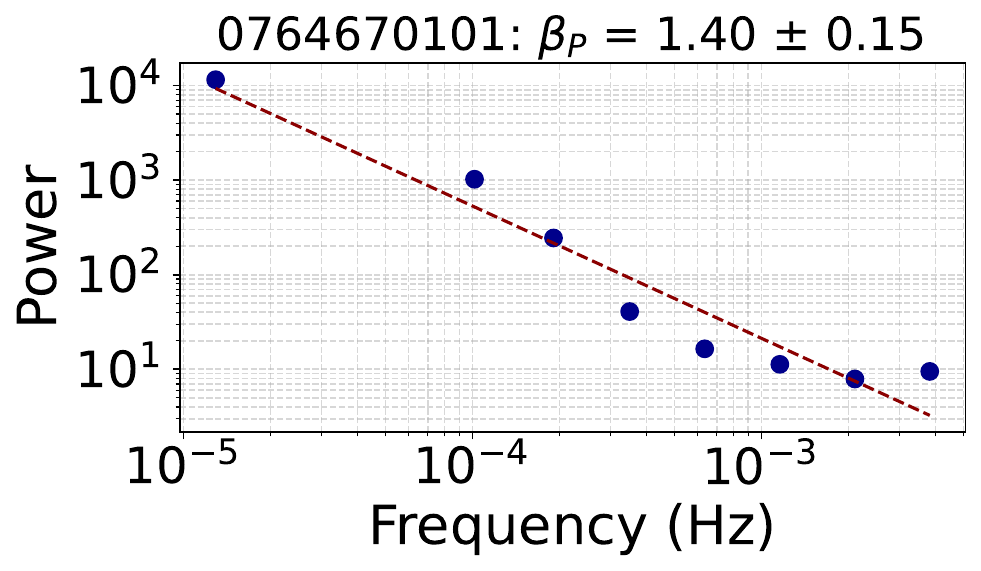}
	\end{minipage}
	\begin{minipage}{.3\textwidth} 
		\centering 
		\includegraphics[width=.99\linewidth]{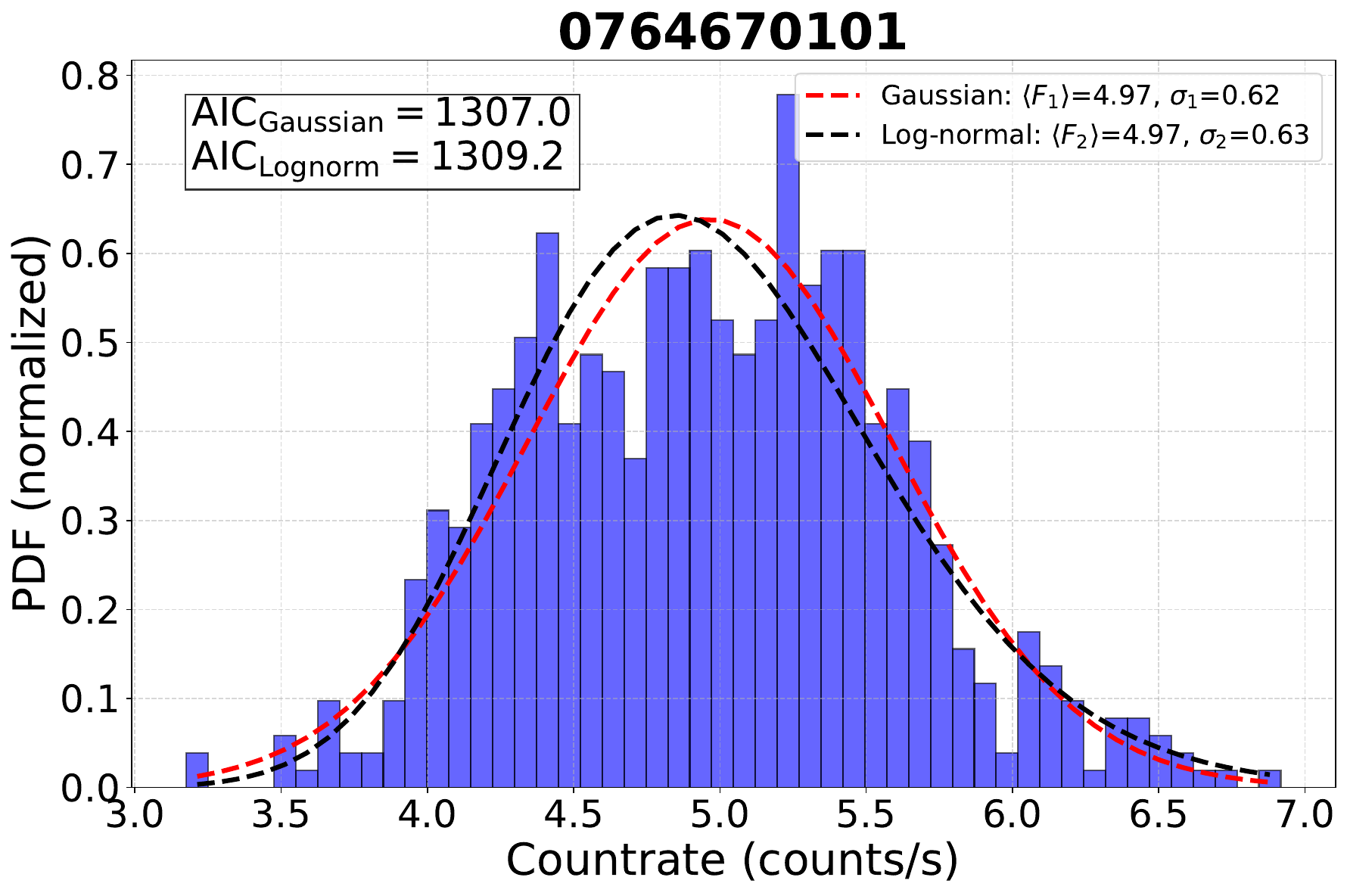}
	\end{minipage}
	\begin{minipage}{.3\textwidth} 
		\centering 
		\includegraphics[height=.99\linewidth, angle=-90]{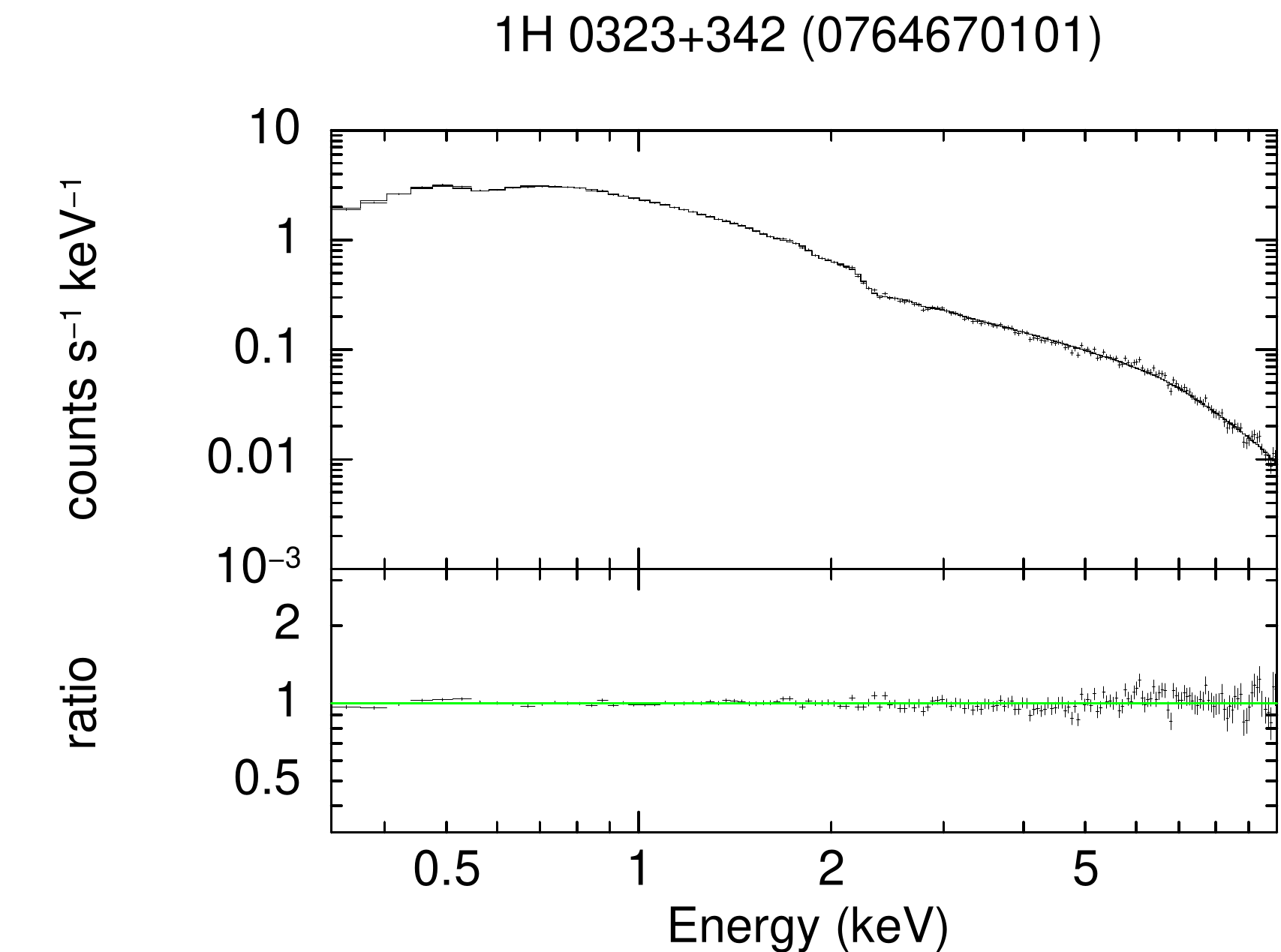}
	\end{minipage}
	\begin{minipage}{.3\textwidth} 
		\centering 
		\includegraphics[width=.99\linewidth]{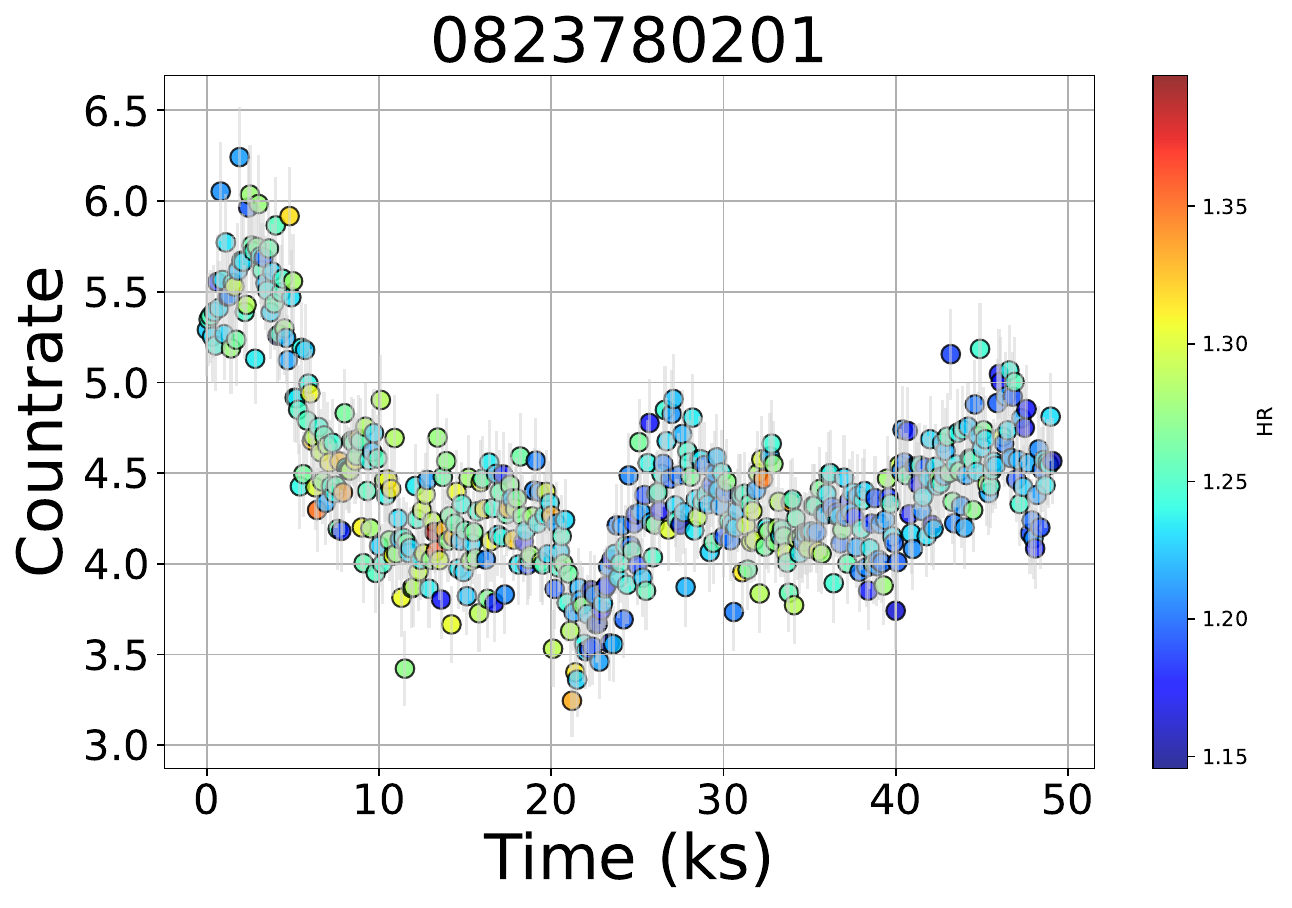}
	\end{minipage}
	\begin{minipage}{.3\textwidth} 
		\centering 
		\includegraphics[width=.99\linewidth]{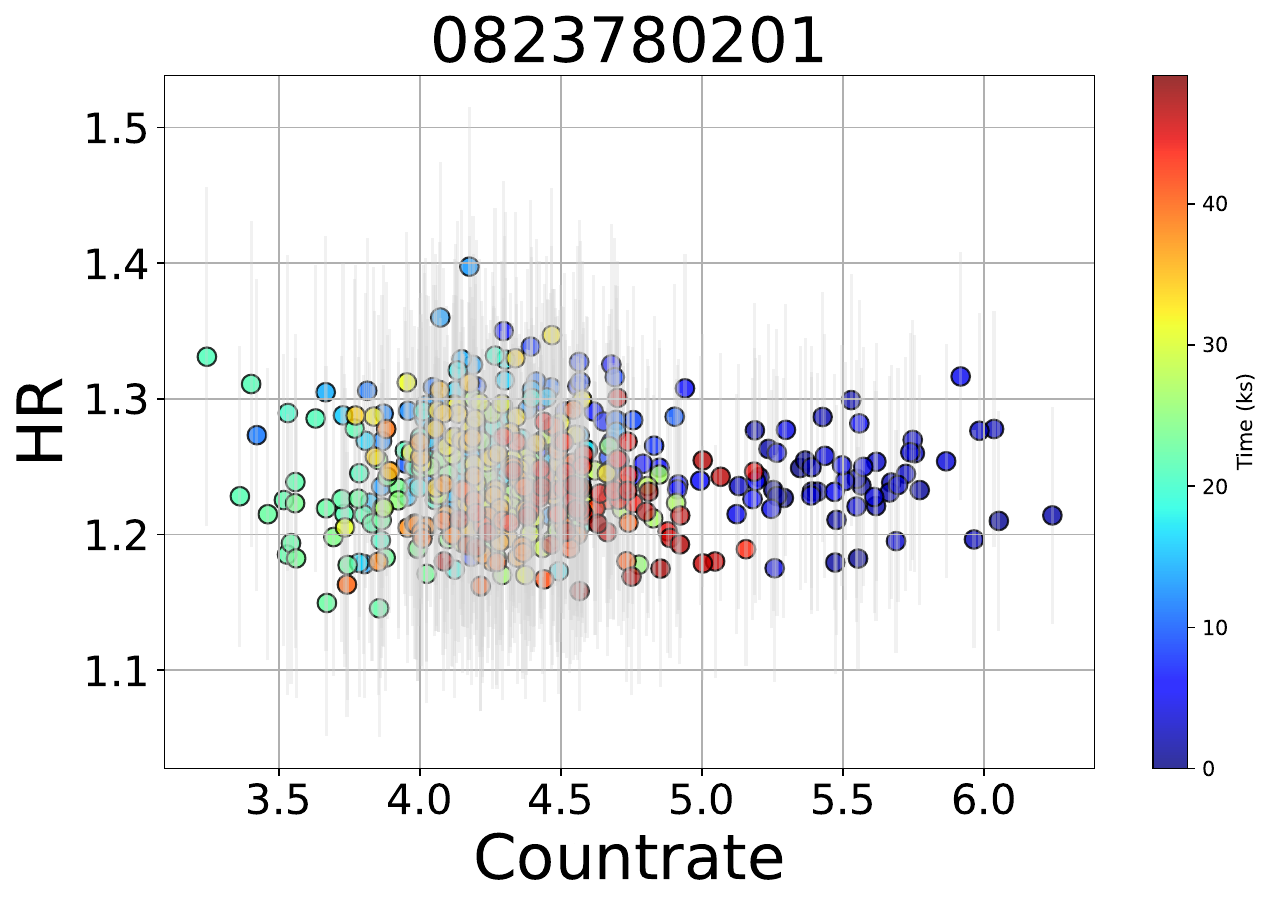}
	\end{minipage}
	\begin{minipage}{.3\textwidth} 
		\centering 
		\includegraphics[width=.99\linewidth, angle=0]{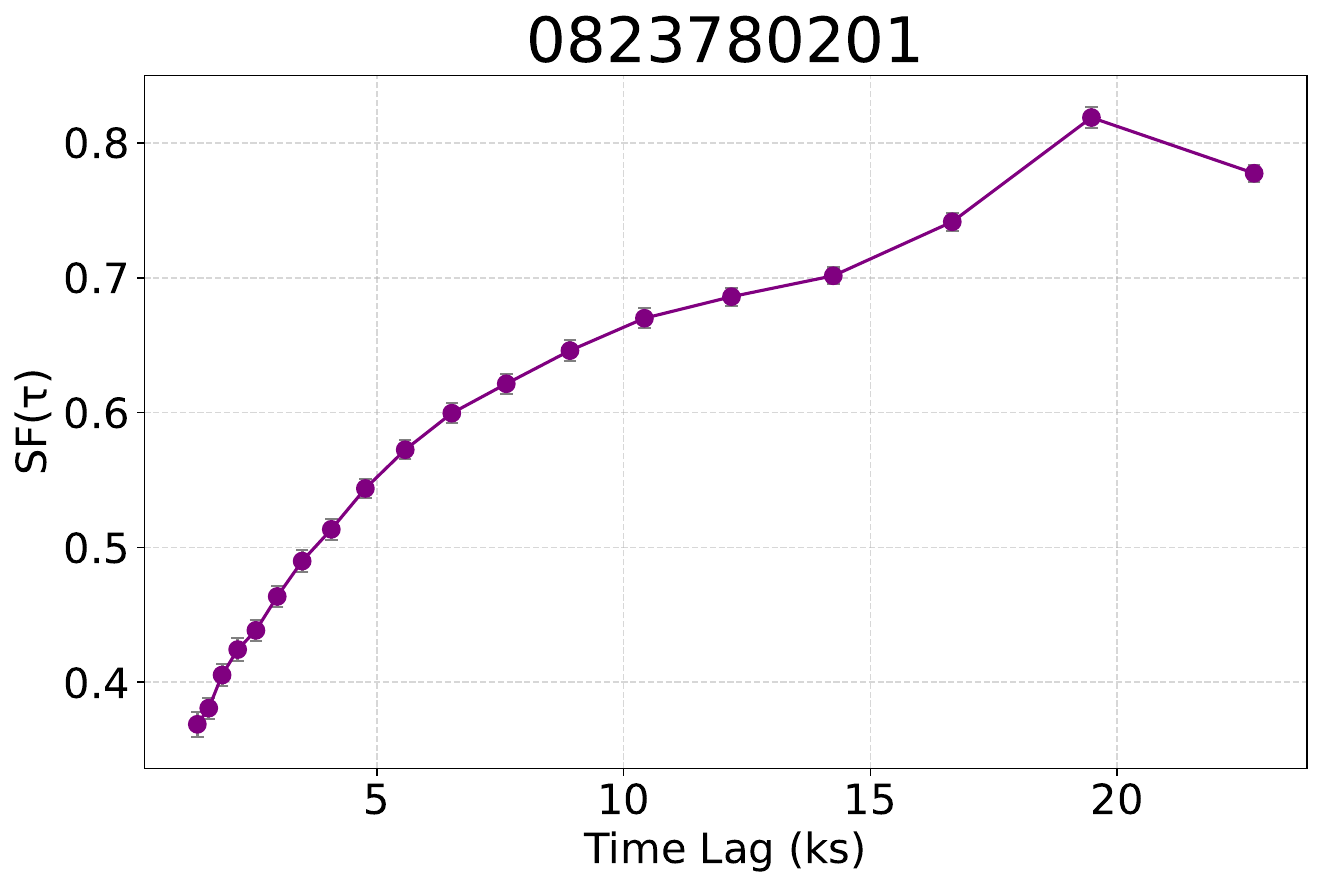}
	\end{minipage}
	\begin{minipage}{.3\textwidth} 
		\centering 
		\includegraphics[width=.99\linewidth]{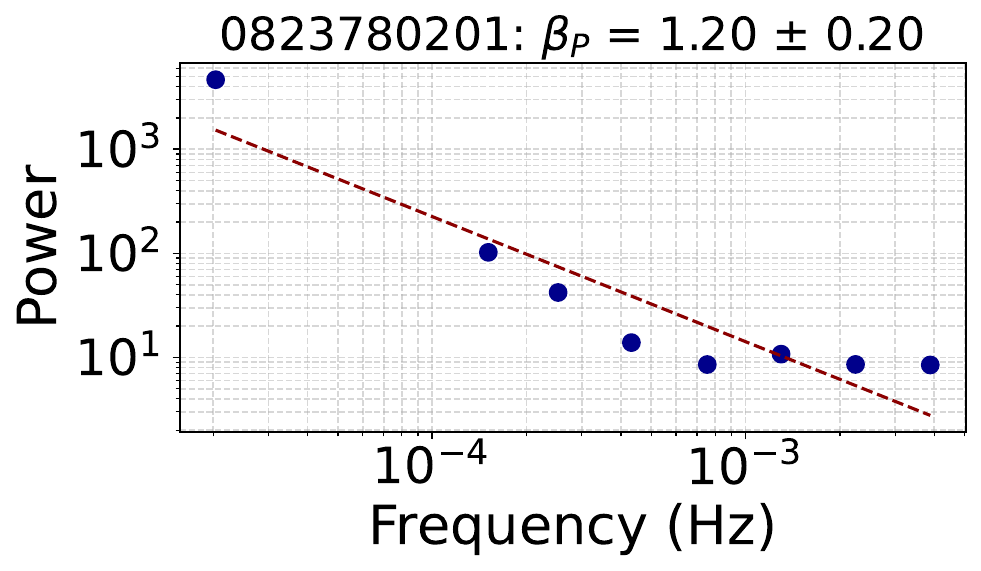}
	\end{minipage}
	\begin{minipage}{.3\textwidth} 
		\centering 
		\includegraphics[width=.99\linewidth]{01PDF/0823780201_PDF.pdf}
	\end{minipage}
	\begin{minipage}{.3\textwidth} 
		\centering 
		\includegraphics[height=.99\linewidth, angle=-90]{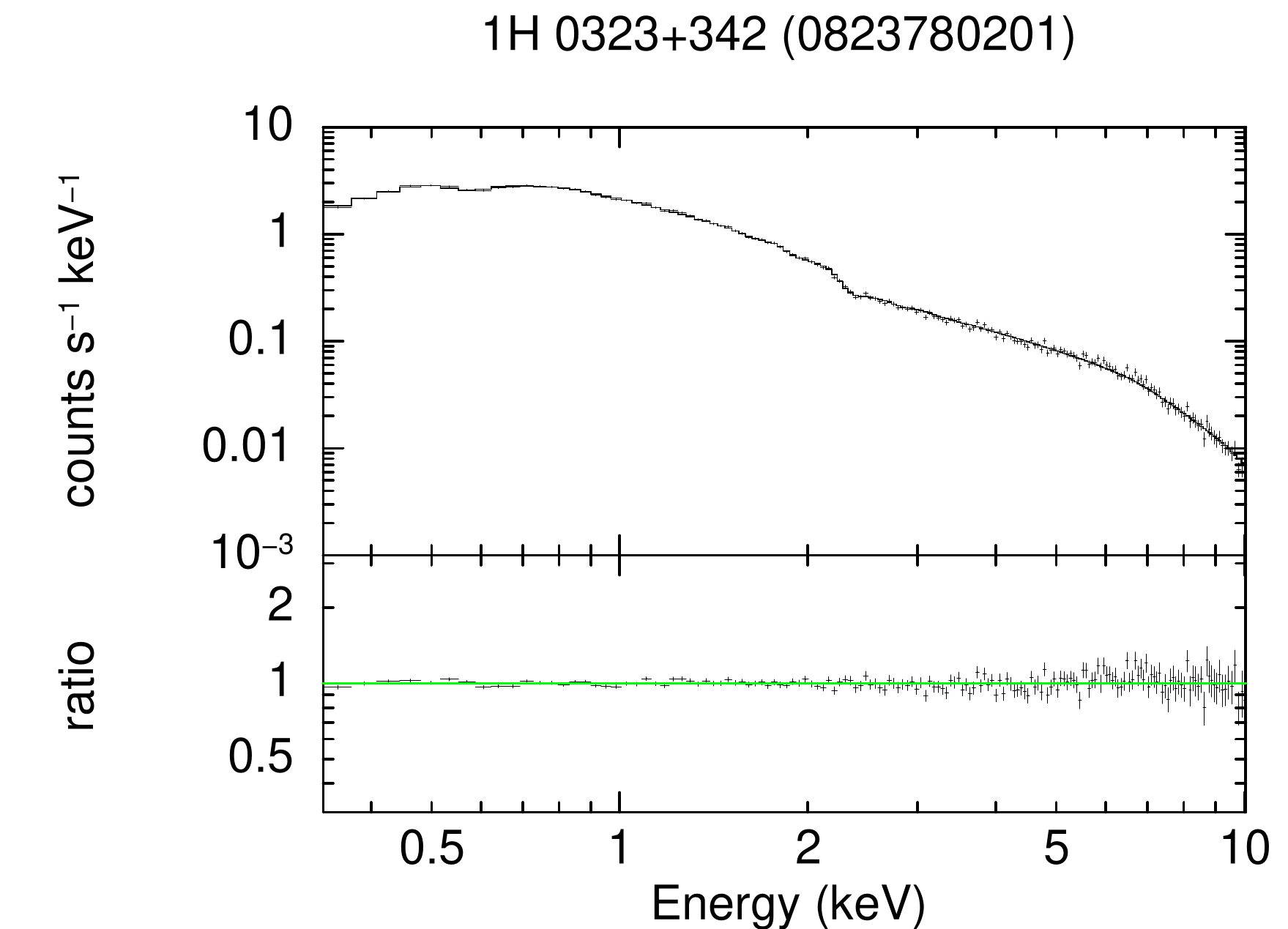}
	\end{minipage}
    \begin{minipage}{.3\textwidth} 
		\centering 
		\includegraphics[width=.99\linewidth]{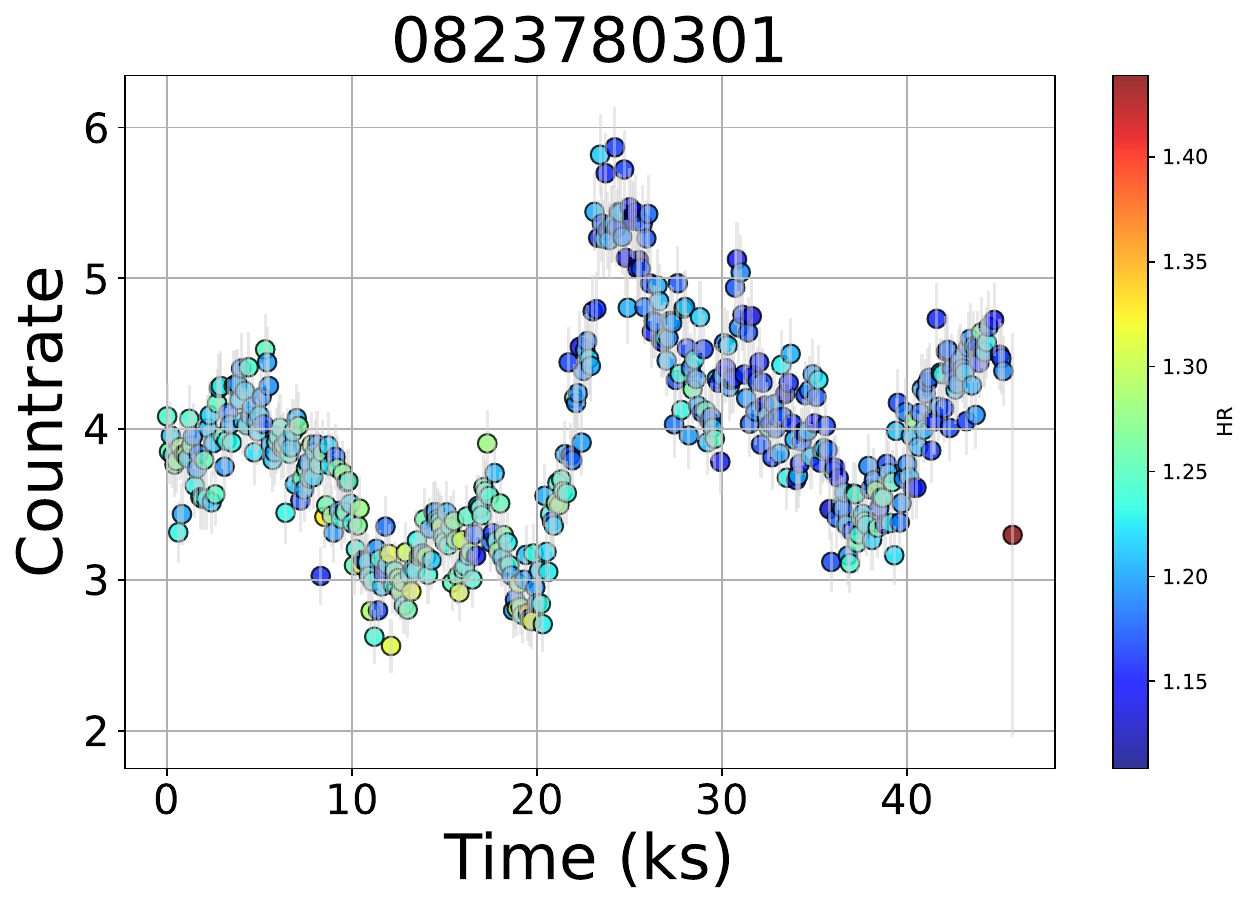}
	\end{minipage}
	\begin{minipage}{.3\textwidth} 
		\centering 
		\includegraphics[width=.99\linewidth]{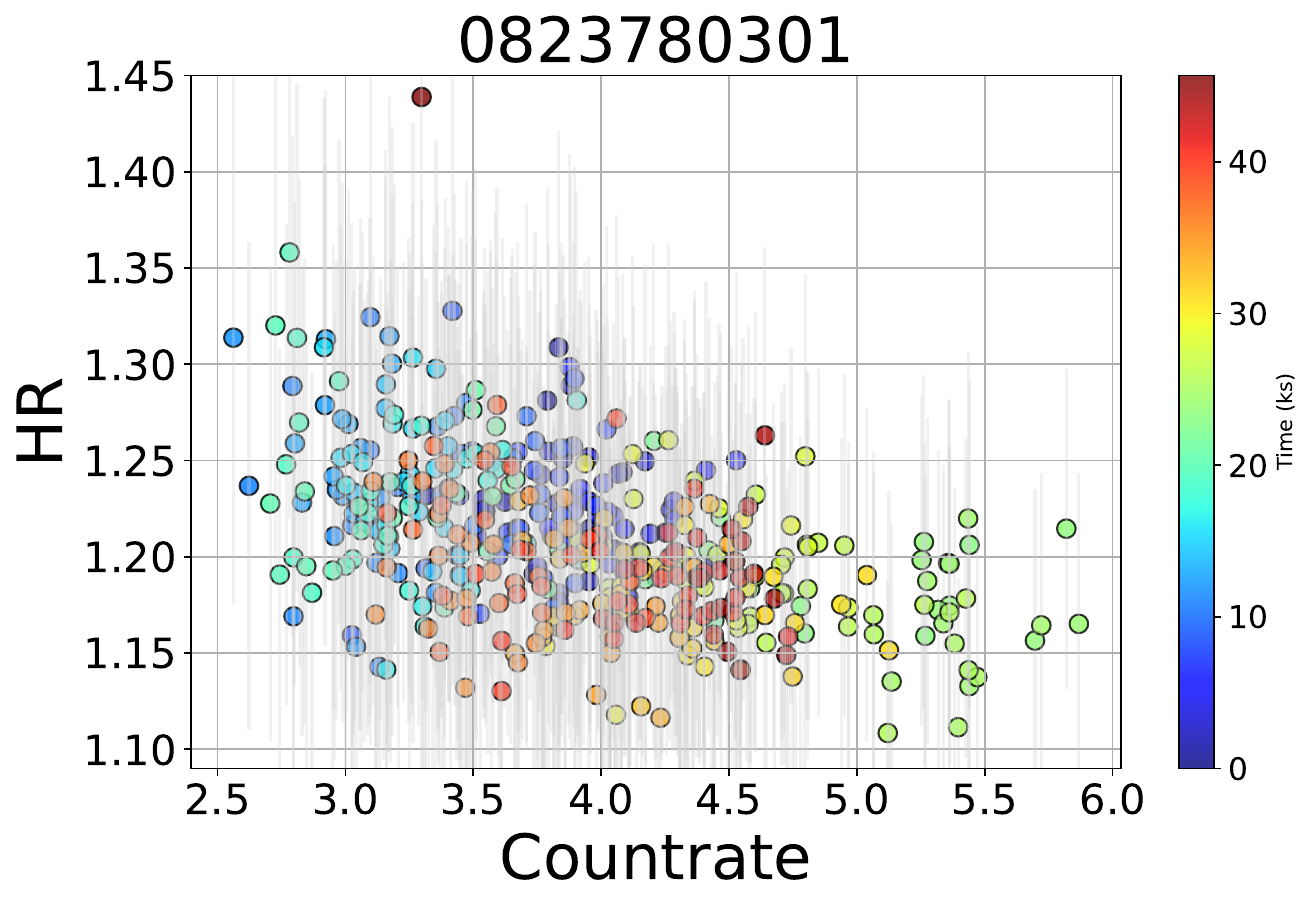}
	\end{minipage}
	\begin{minipage}{.3\textwidth} 
		\centering 
		\includegraphics[width=.99\linewidth, angle=0]{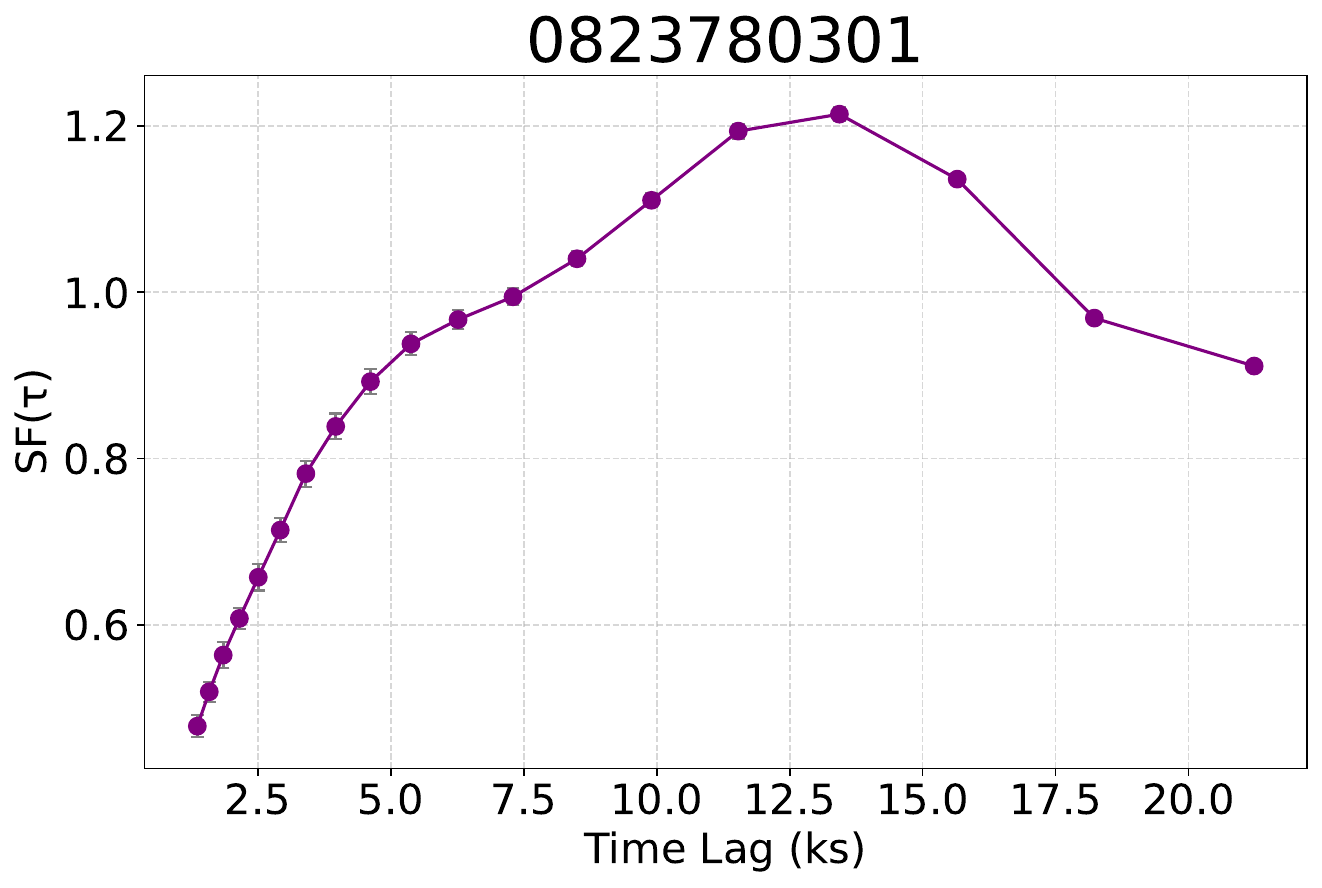}
	\end{minipage}
	\begin{minipage}{.3\textwidth} 
		\centering 
		\includegraphics[width=.99\linewidth]{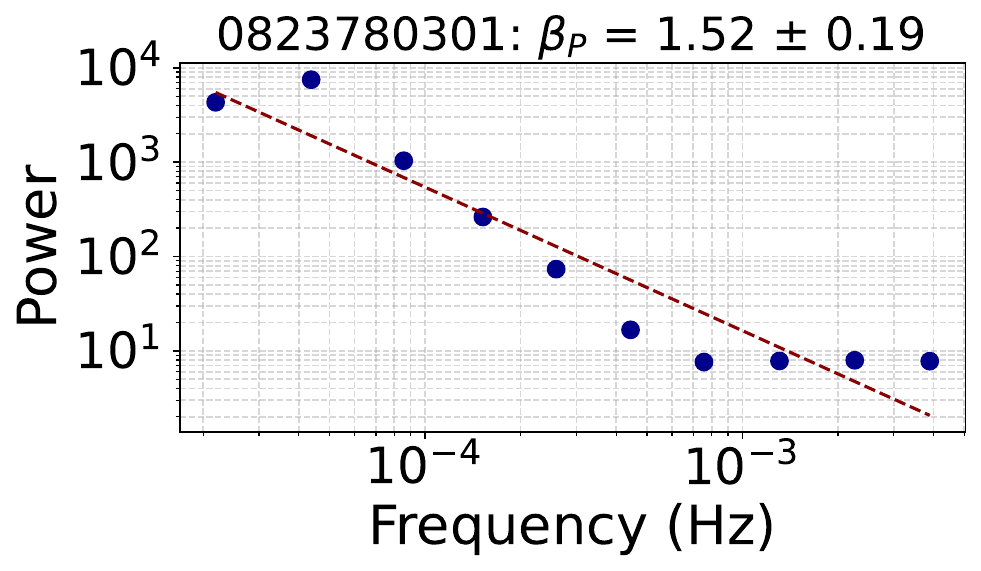}
	\end{minipage}
	\begin{minipage}{.3\textwidth} 
		\centering 
		\includegraphics[width=.99\linewidth]{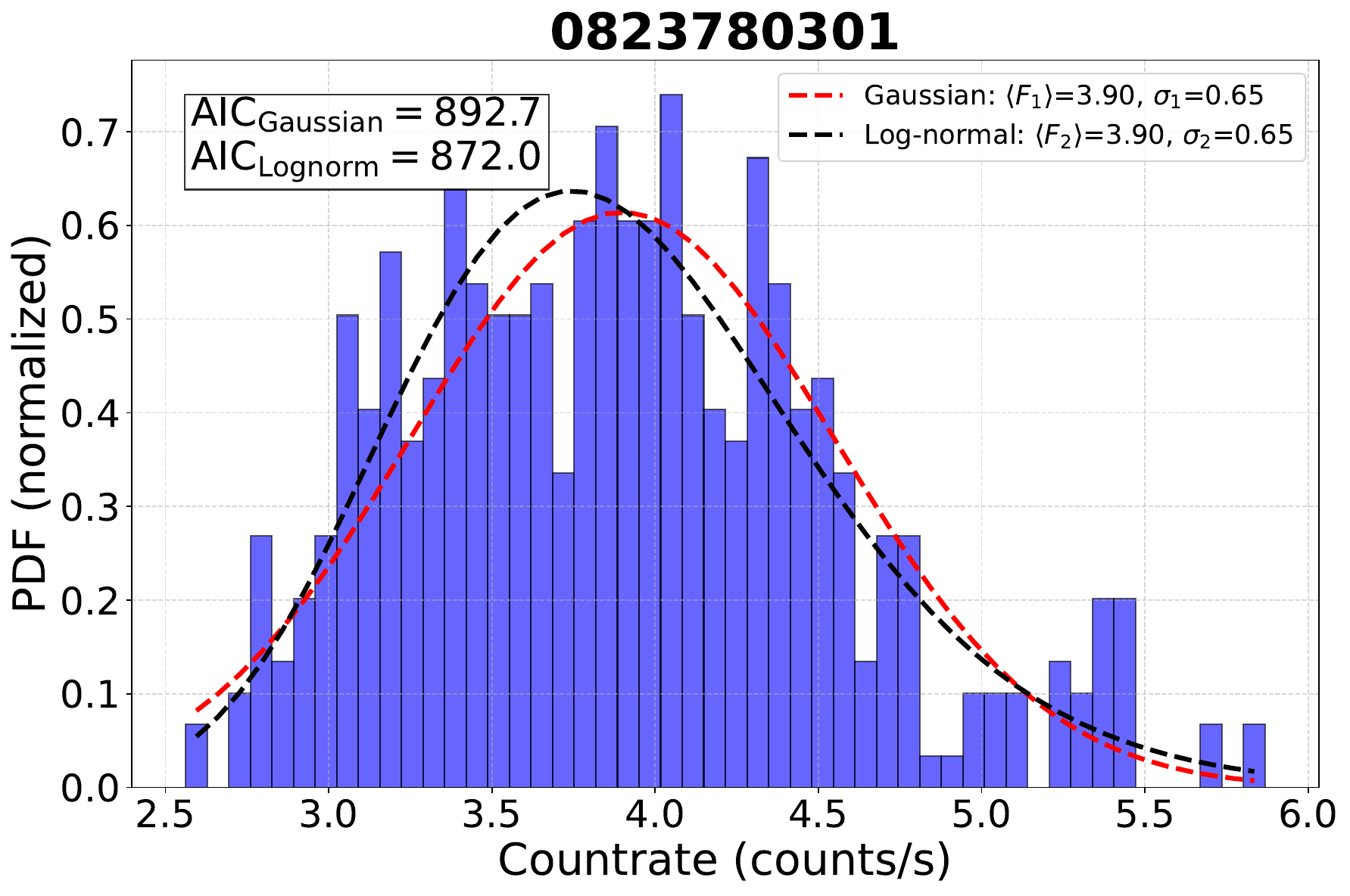}
	\end{minipage}
	\begin{minipage}{.3\textwidth} 
		\centering 
		\includegraphics[height=.99\linewidth, angle=-90]{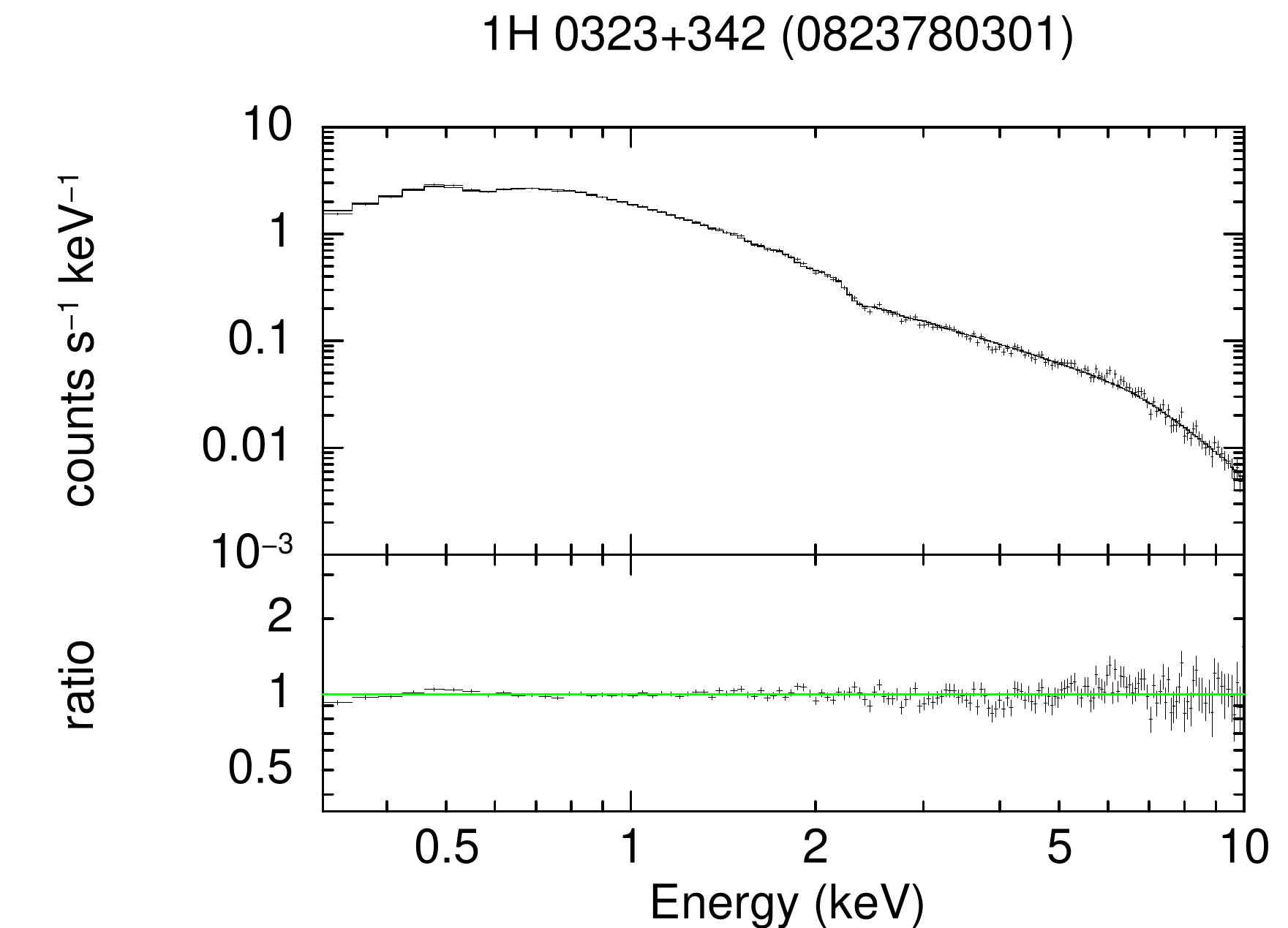}
	\end{minipage}
\end{figure*}

\begin{figure*}
	\centering
	\begin{minipage}{.3\textwidth} 
		\centering 
		\includegraphics[width=.99\linewidth]{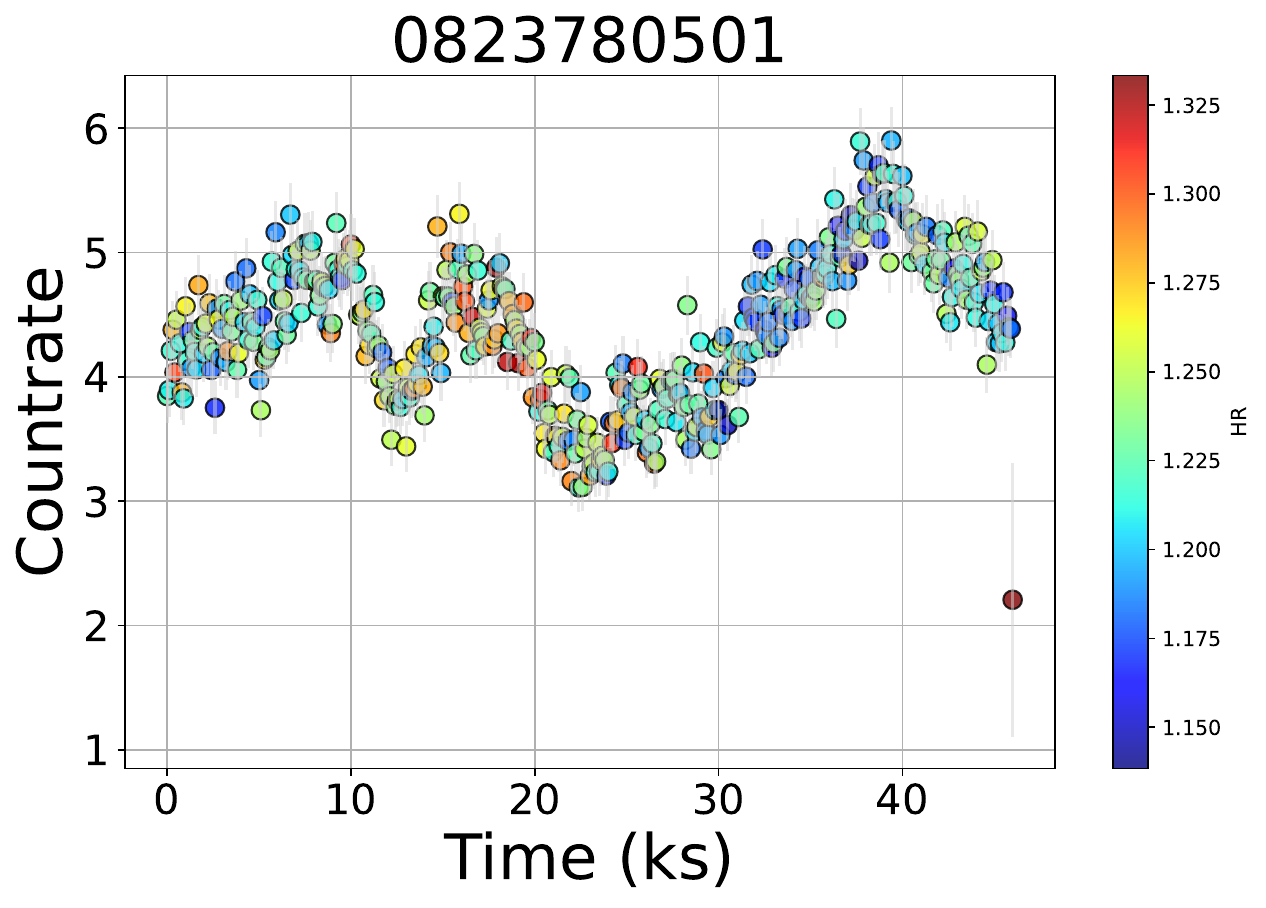}
	\end{minipage}
	\begin{minipage}{.3\textwidth} 
		\centering 
		\includegraphics[width=.99\linewidth]{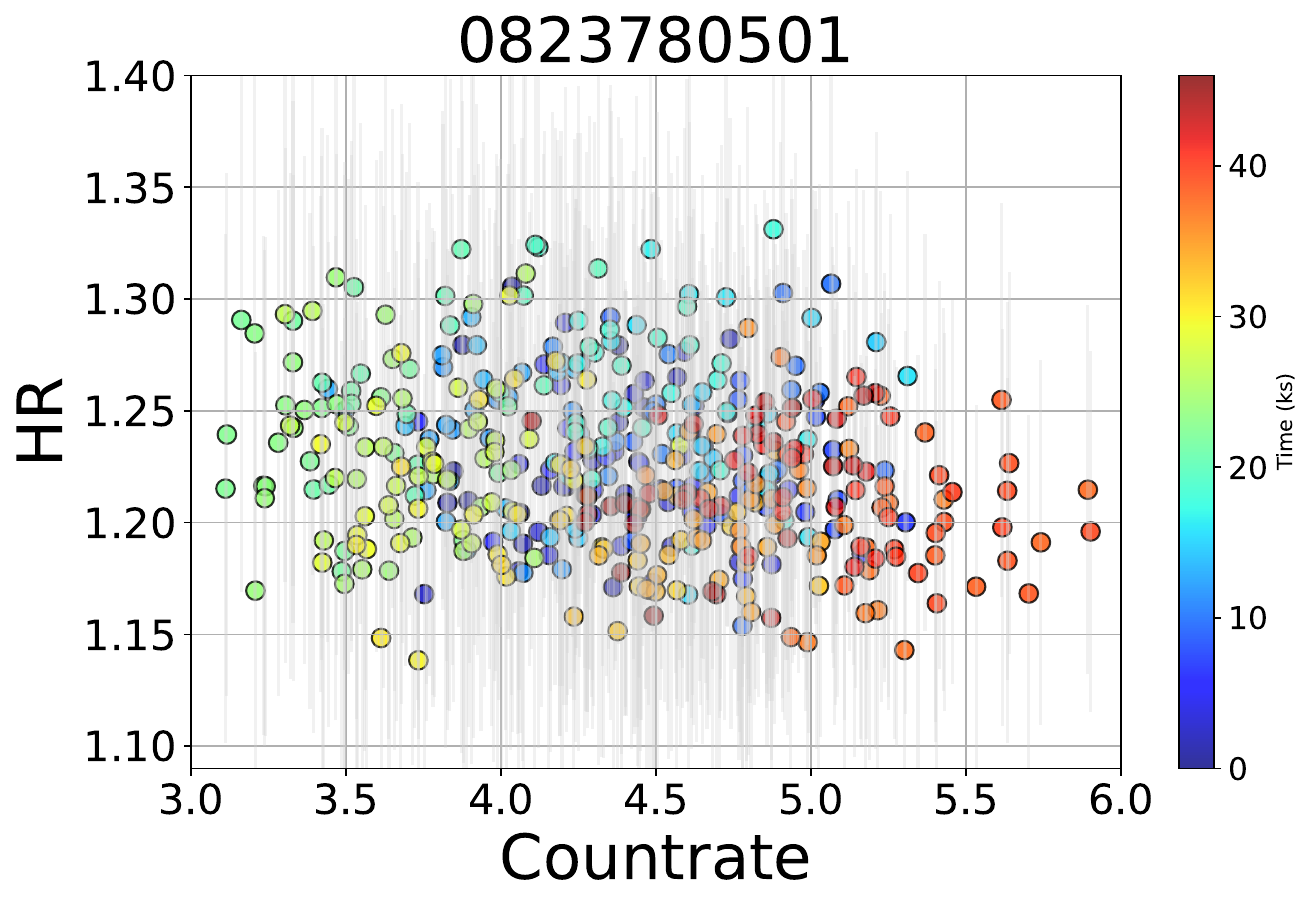}
	\end{minipage}
	\begin{minipage}{.3\textwidth} 
		\centering 
		\includegraphics[width=.99\linewidth, angle=0]{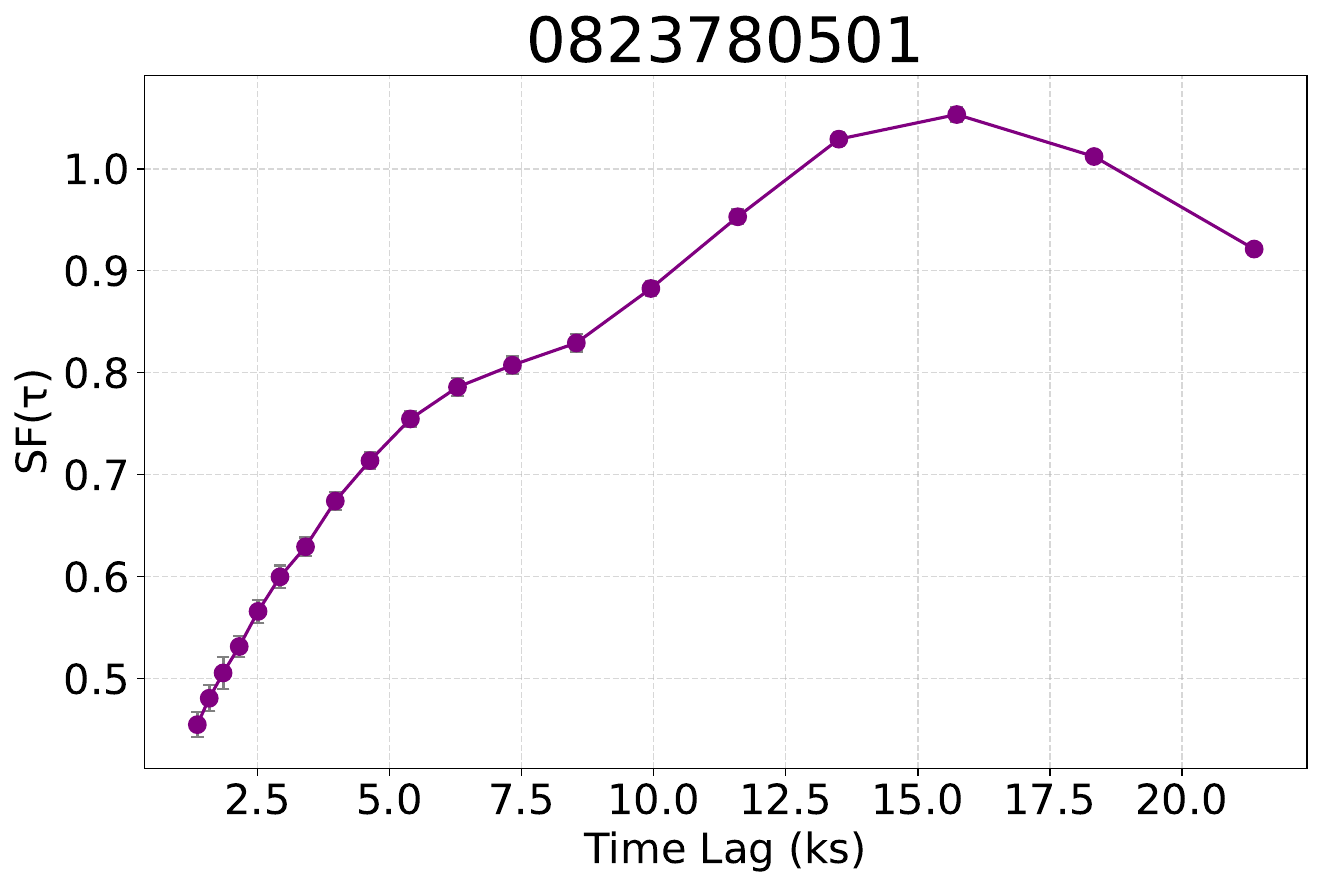}
	\end{minipage}
	\begin{minipage}{.3\textwidth} 
		\centering 
		\includegraphics[width=.99\linewidth]{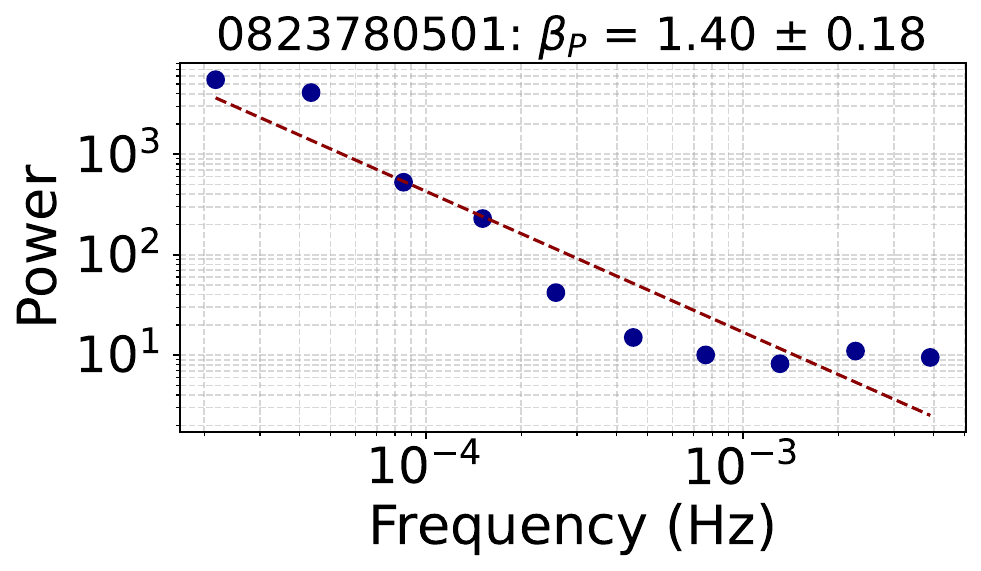}
	\end{minipage}
	\begin{minipage}{.3\textwidth} 
		\centering 
		\includegraphics[width=.99\linewidth]{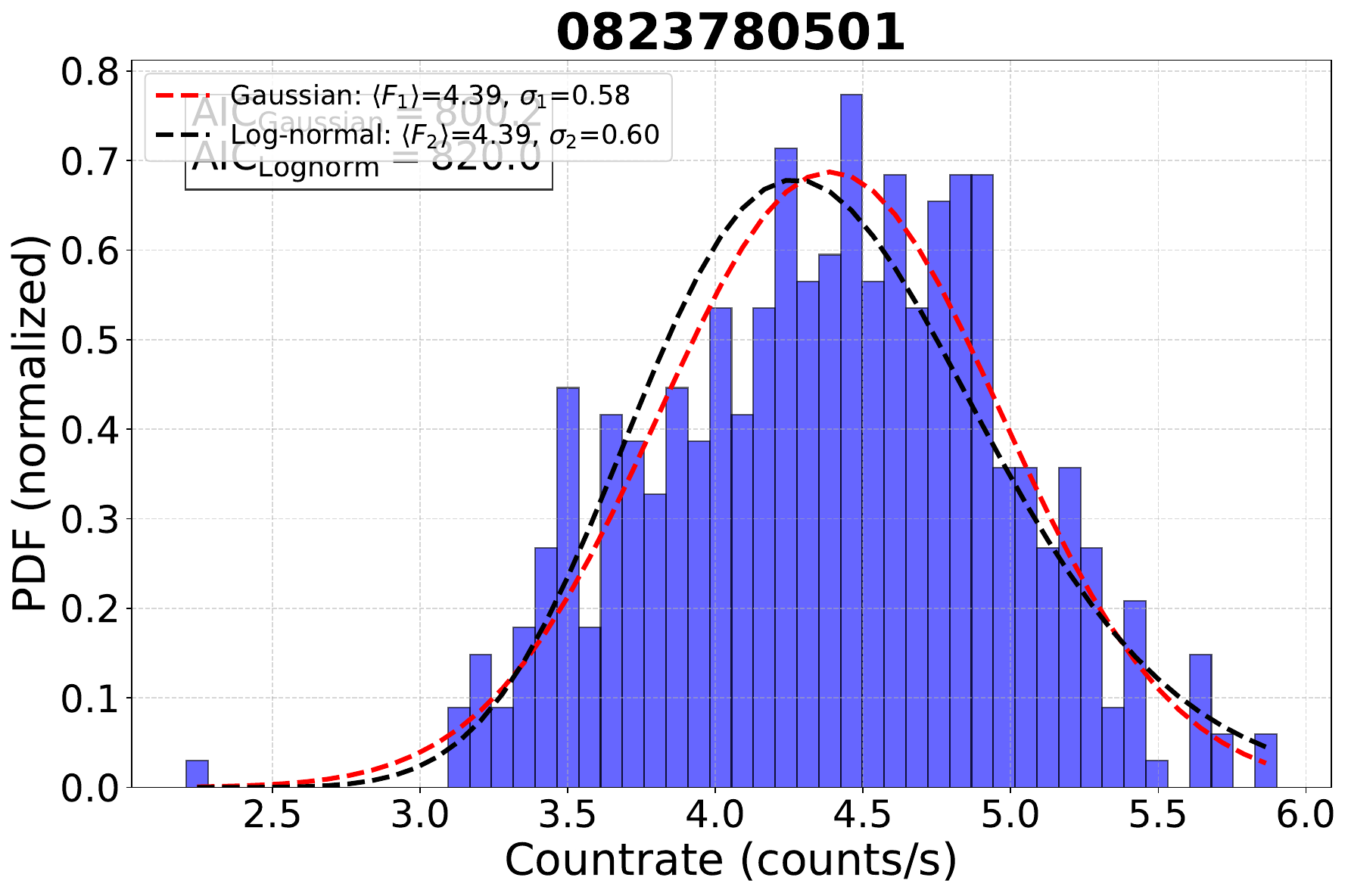}
	\end{minipage}
	\begin{minipage}{.3\textwidth} 
		\centering 
		\includegraphics[height=.99\linewidth, angle=-90]{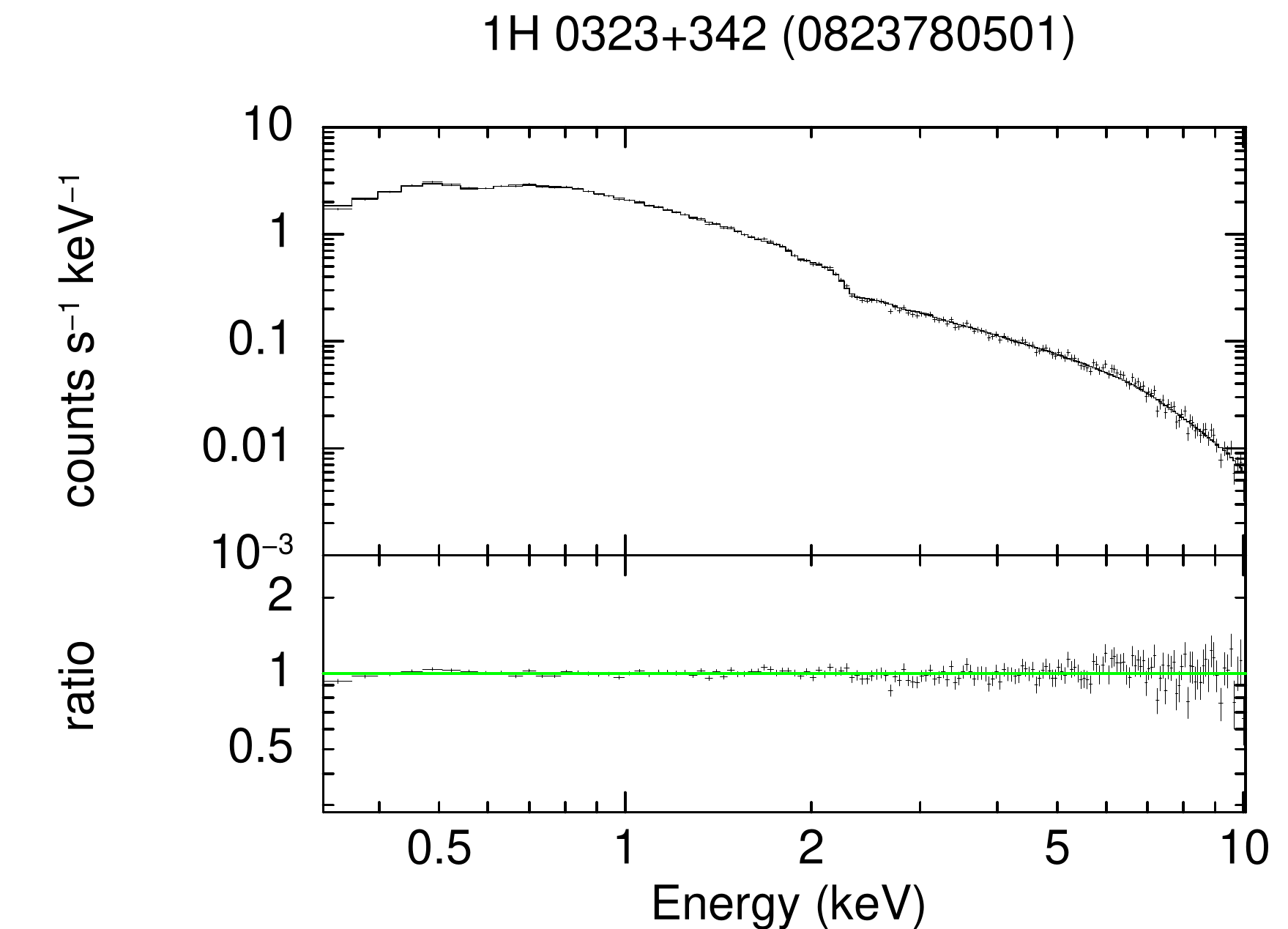}
	\end{minipage}
    	\begin{minipage}{.3\textwidth} 
		\centering 
		\includegraphics[width=.99\linewidth]{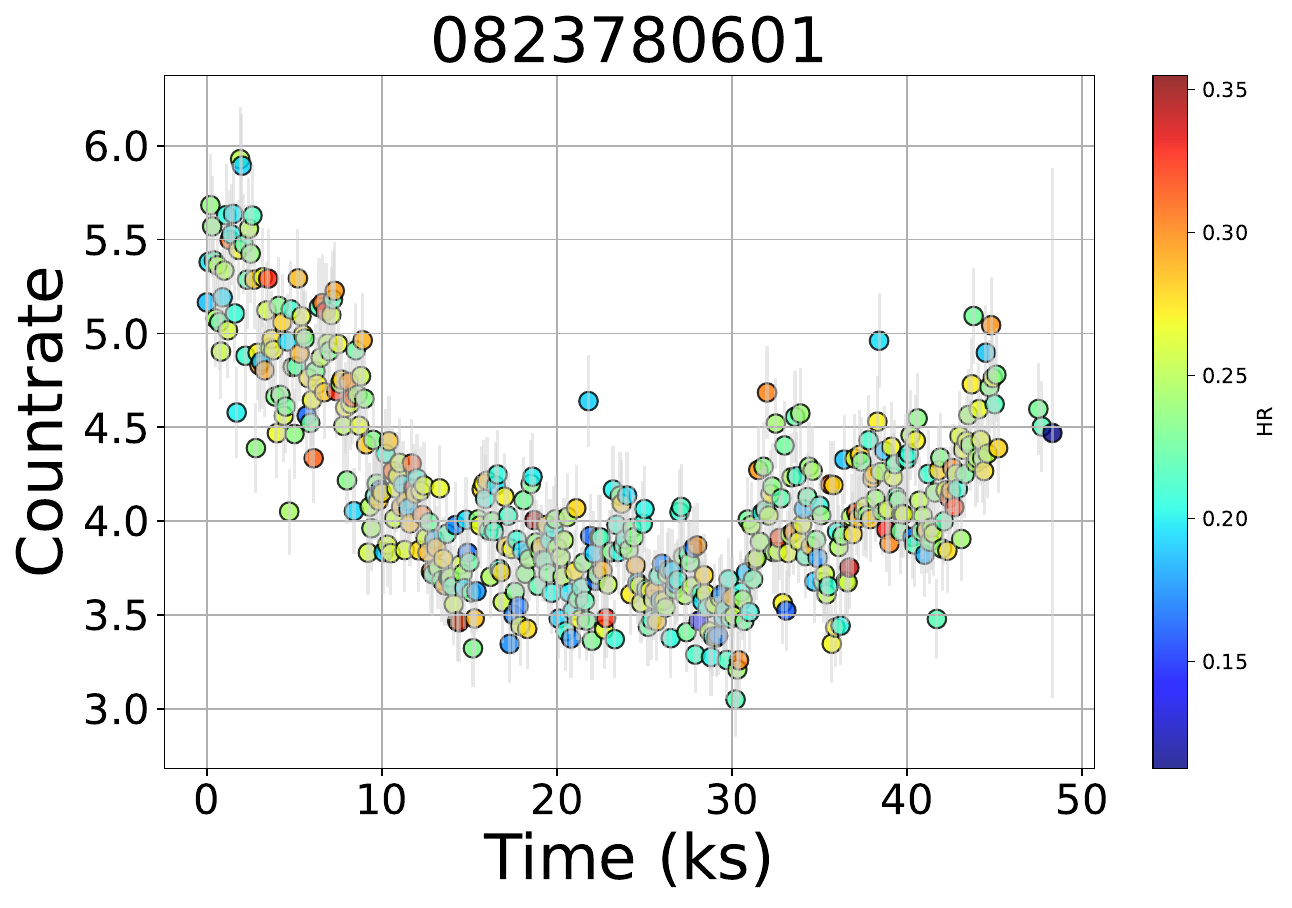}
	\end{minipage}
	\begin{minipage}{.3\textwidth} 
		\centering 
		\includegraphics[width=.99\linewidth]{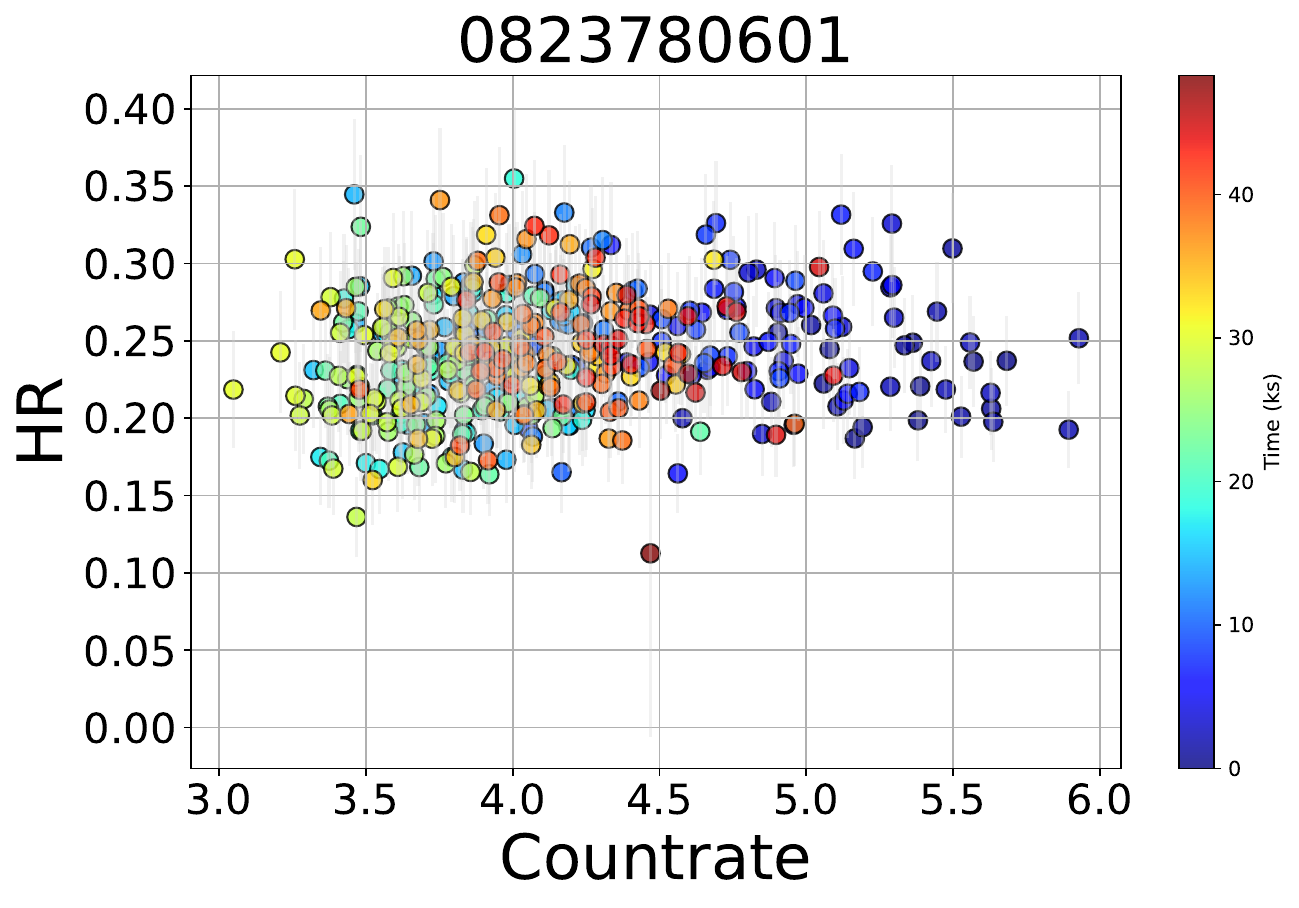}
	\end{minipage}
	\begin{minipage}{.3\textwidth} 
		\centering 
		\includegraphics[width=.99\linewidth, angle=0]{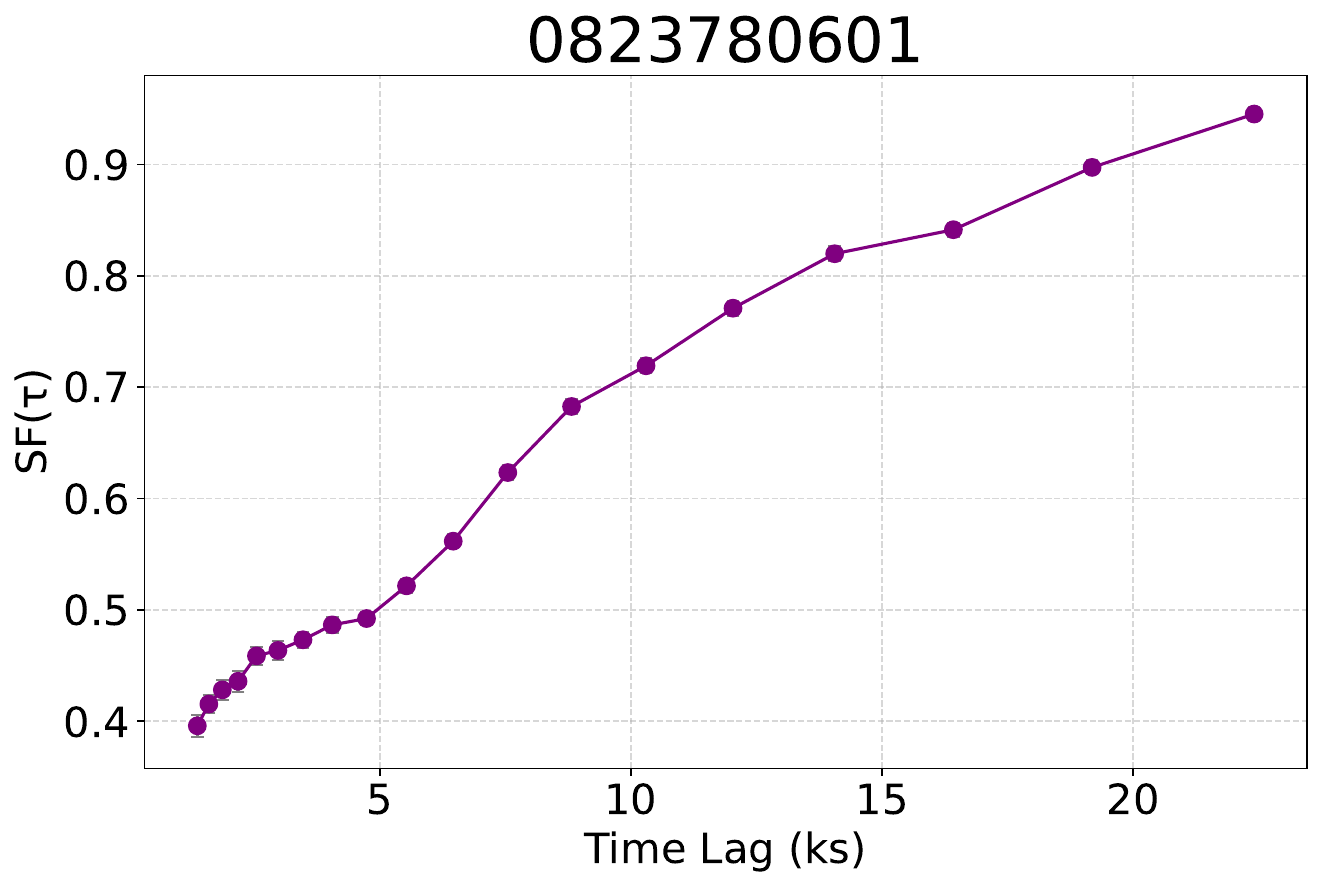}
	\end{minipage}
    \begin{minipage}{.3\textwidth} 
		\centering 
		\includegraphics[width=.99\linewidth]{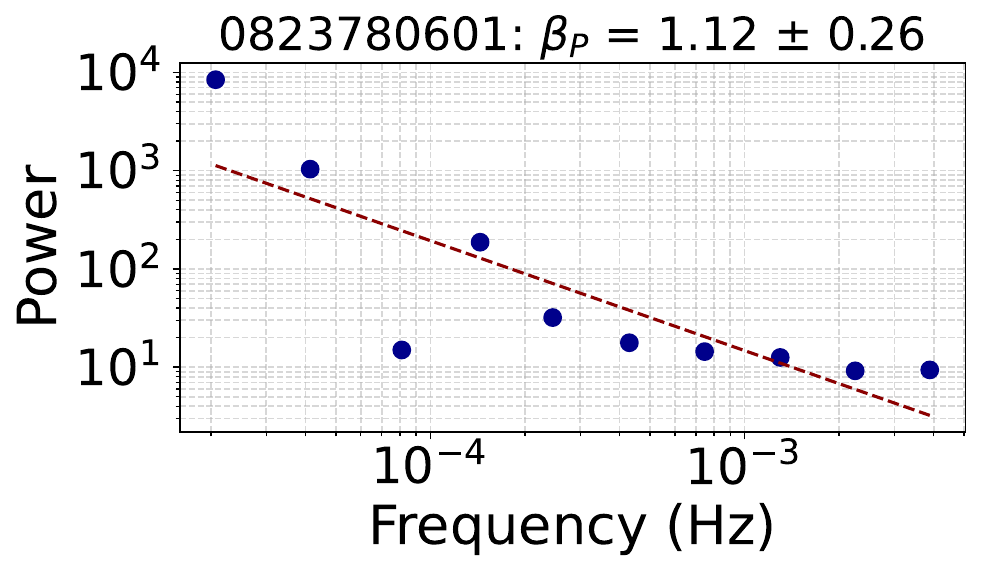}
	\end{minipage}
	\begin{minipage}{.3\textwidth} 
		\centering 
		\includegraphics[width=.99\linewidth]{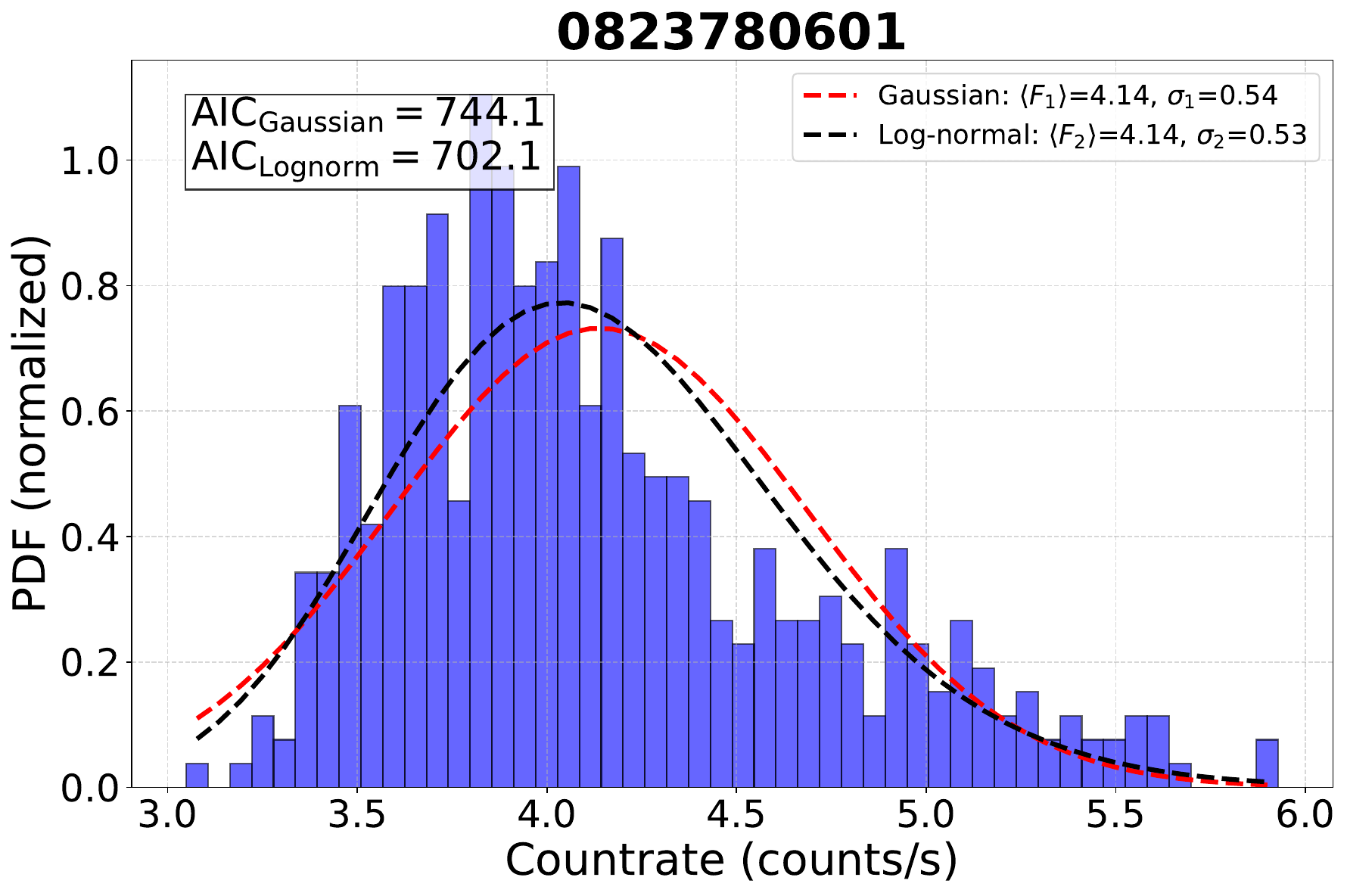}
	\end{minipage}
	\begin{minipage}{.3\textwidth} 
		\centering 
		\includegraphics[height=.99\linewidth, angle=-90]{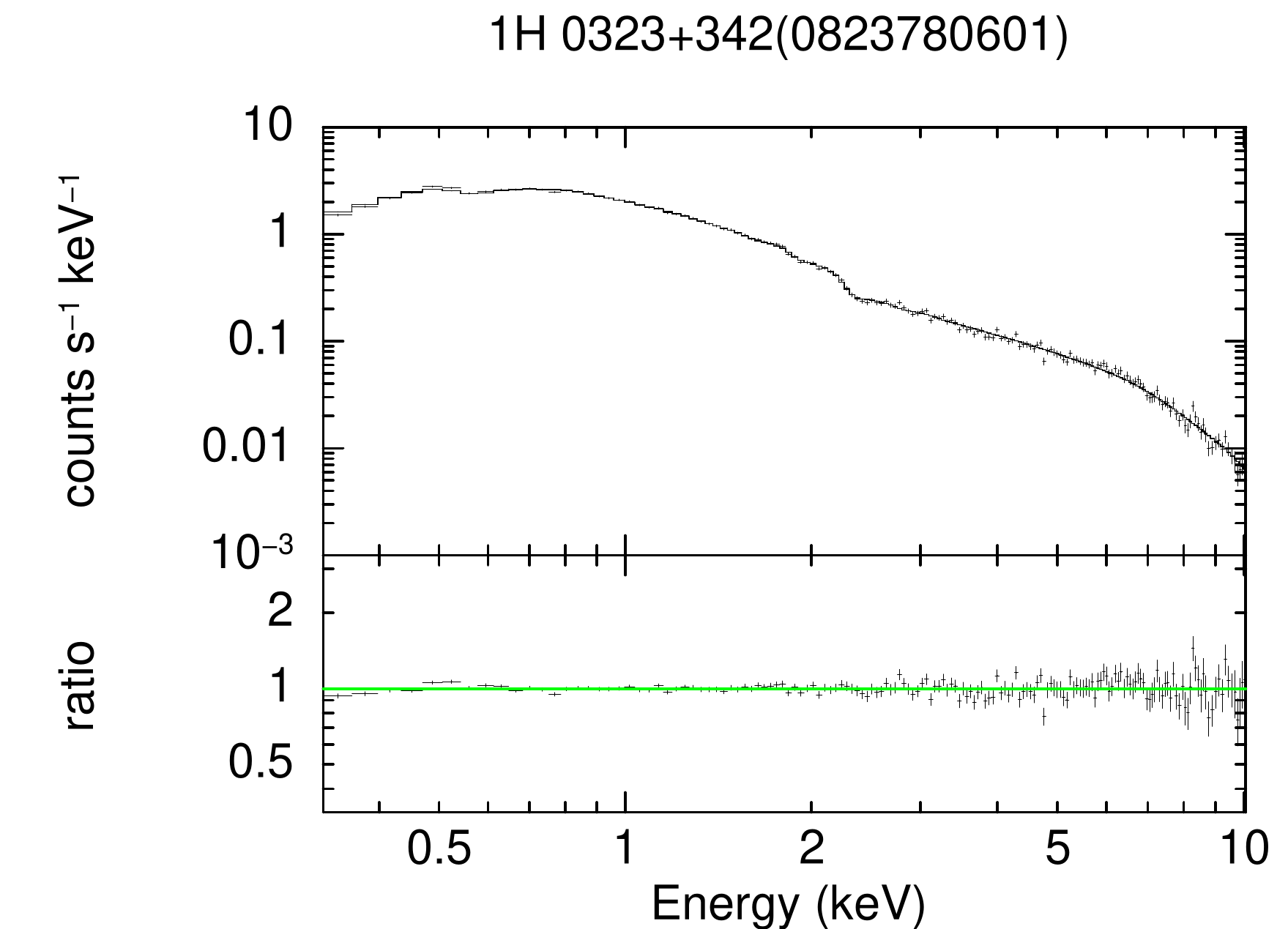}
	\end{minipage}
\begin{minipage}{.3\textwidth} 
		\centering 
		\includegraphics[width=.99\linewidth]{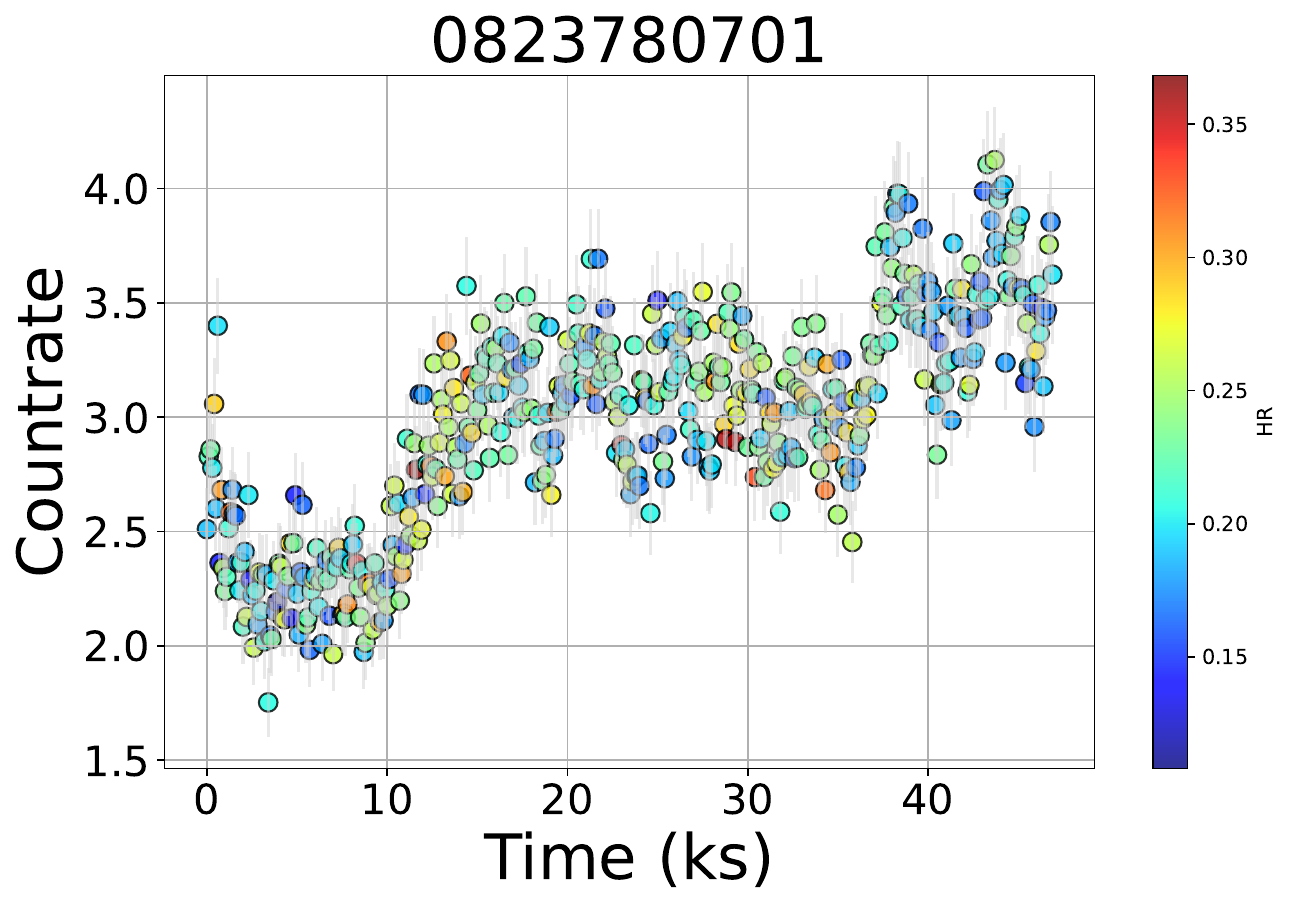}
	\end{minipage}
	\begin{minipage}{.3\textwidth} 
		\centering 
		\includegraphics[width=.99\linewidth]{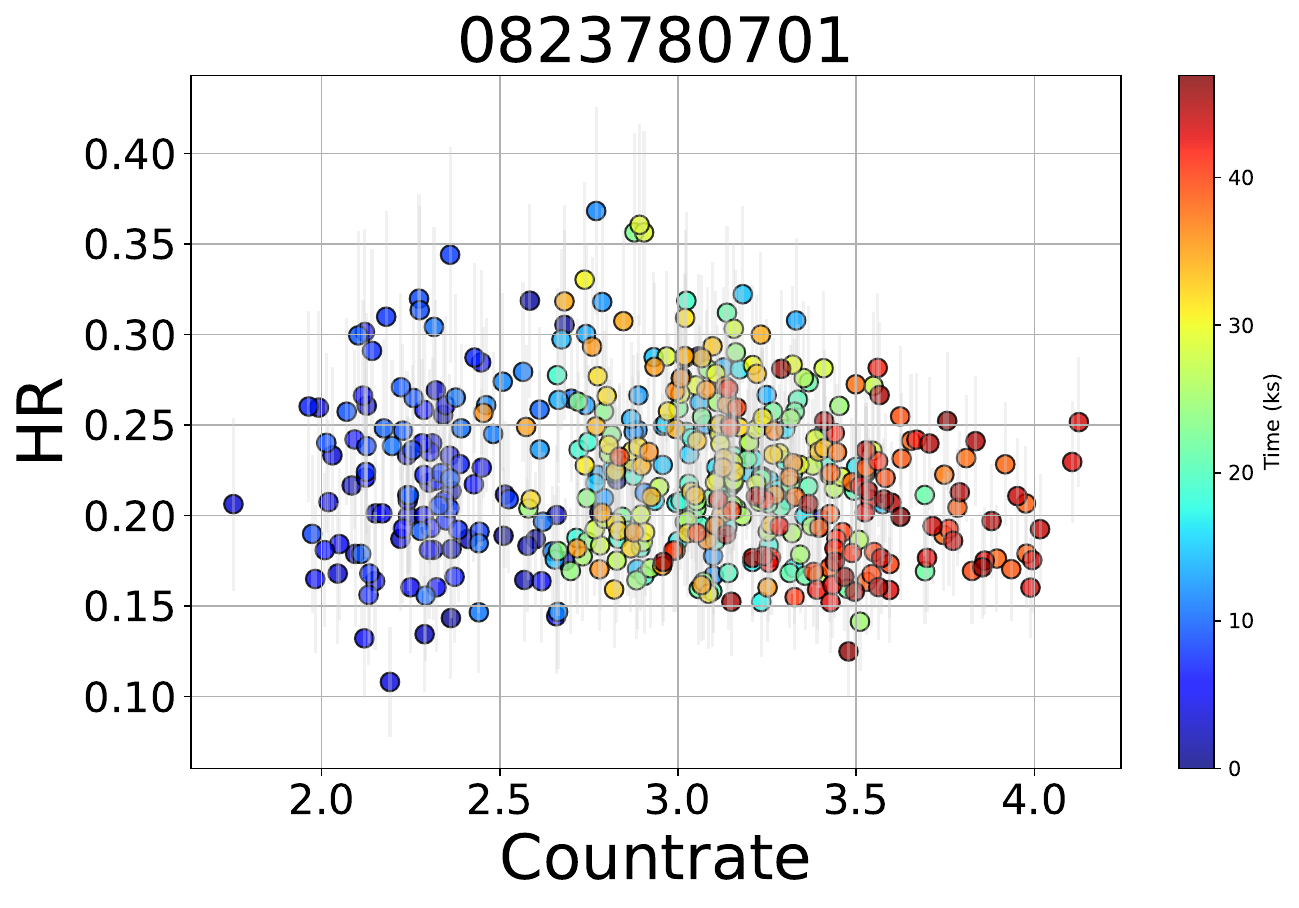}
	\end{minipage}
	\begin{minipage}{.3\textwidth} 
		\centering 
		\includegraphics[width=.99\linewidth, angle=0]{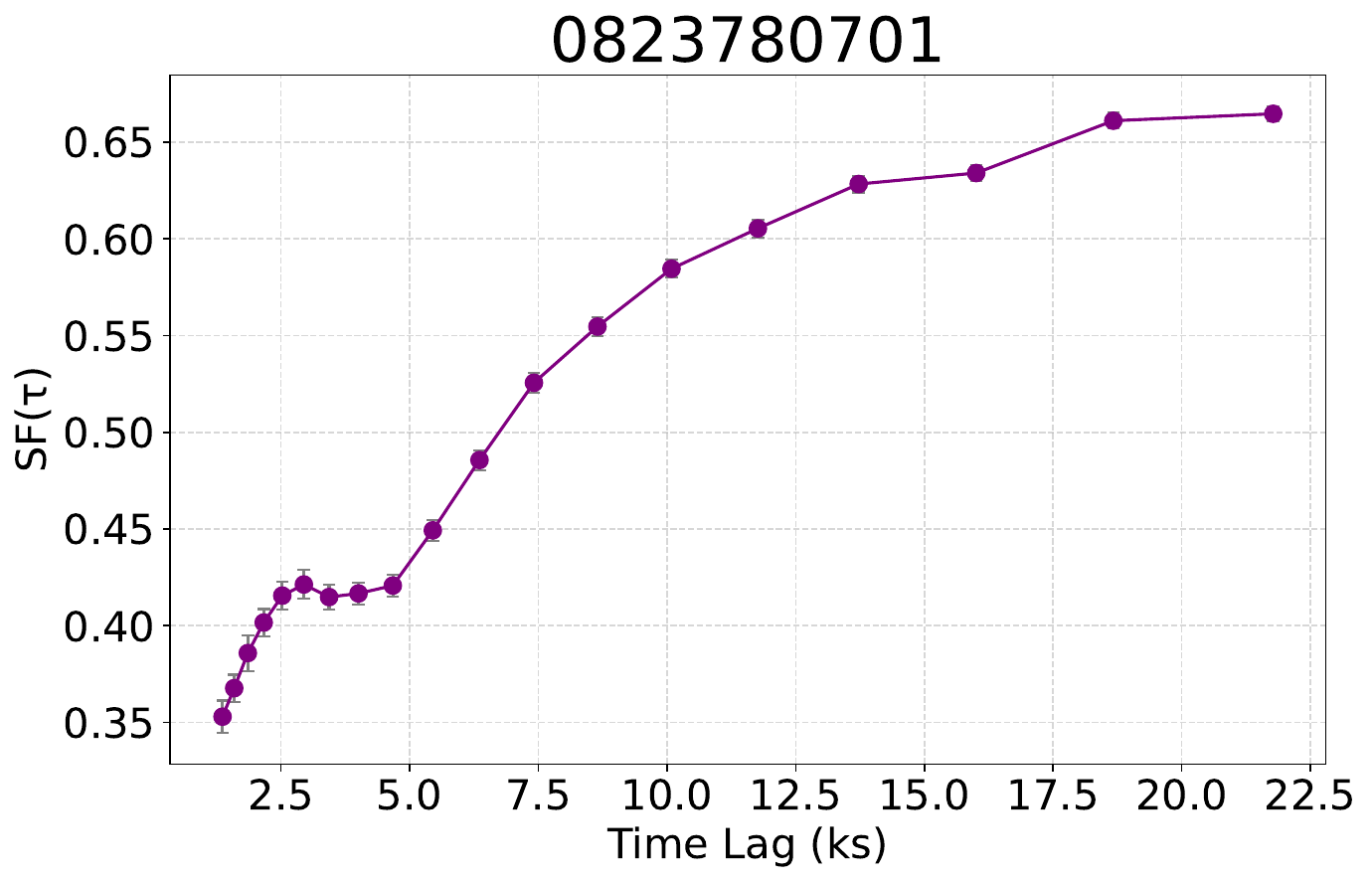}
	\end{minipage}
	\begin{minipage}{.3\textwidth} 
		\centering 
		\includegraphics[width=.99\linewidth]{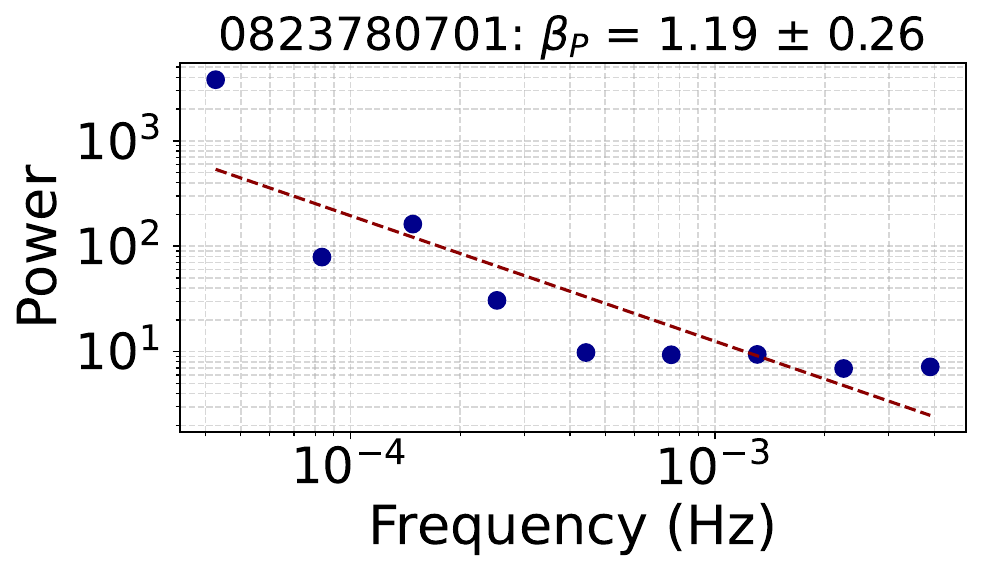}
	\end{minipage}
	\begin{minipage}{.3\textwidth} 
		\centering 
		\includegraphics[width=.99\linewidth]{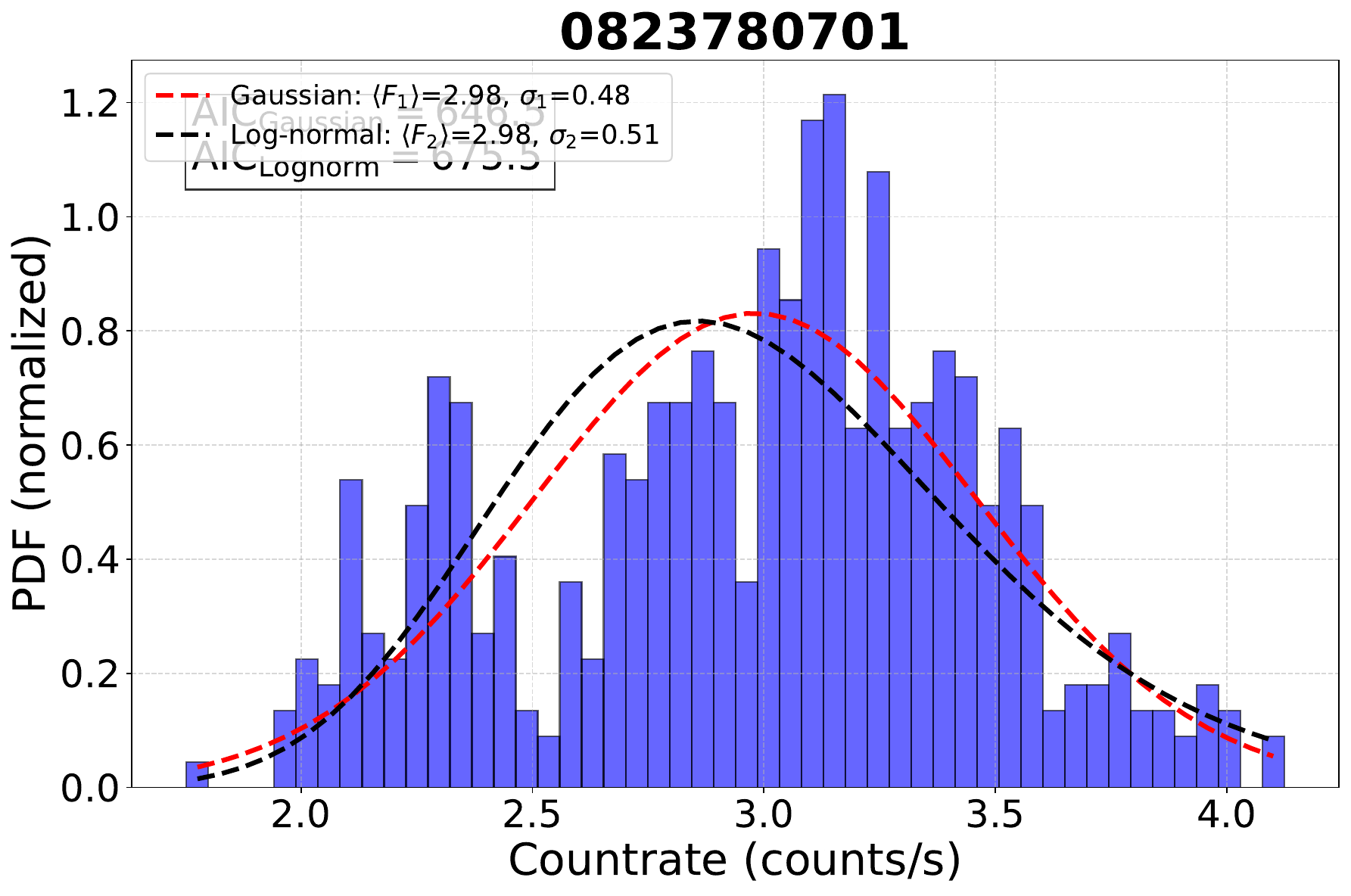}
	\end{minipage}
	\begin{minipage}{.3\textwidth} 
		\centering 
		\includegraphics[height=.99\linewidth, angle=-90]{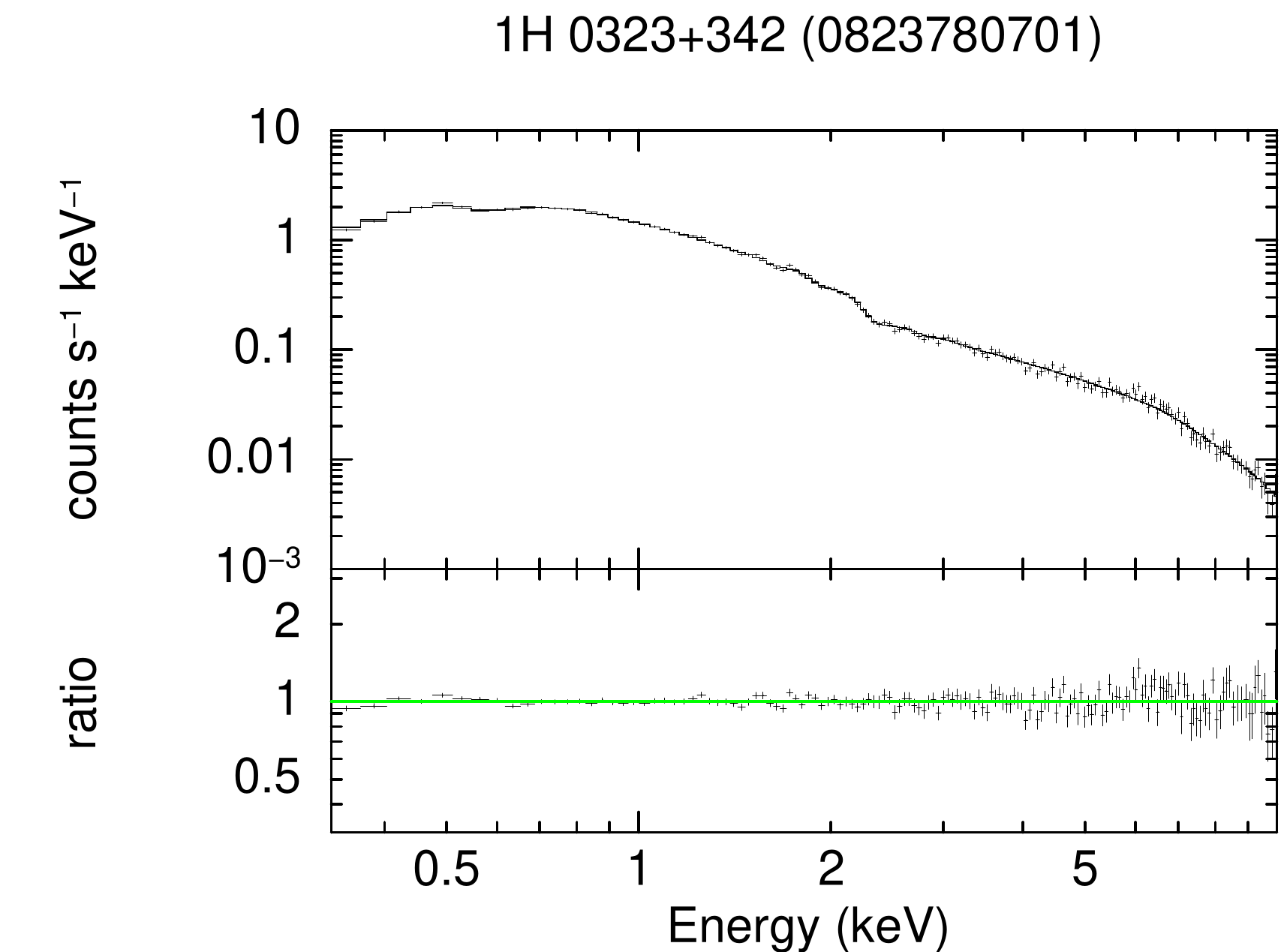}
	\end{minipage}
\end{figure*}

\begin{figure*}\label{app:PKS 2004-447}
	\centering
	\caption{LCs, HR plots, Structure Function, PSD, PDF, and spectral fits derived from observations of PKS 2004-447.}
	\begin{minipage}{.3\textwidth} 
		\centering 
		\includegraphics[width=.99\linewidth]{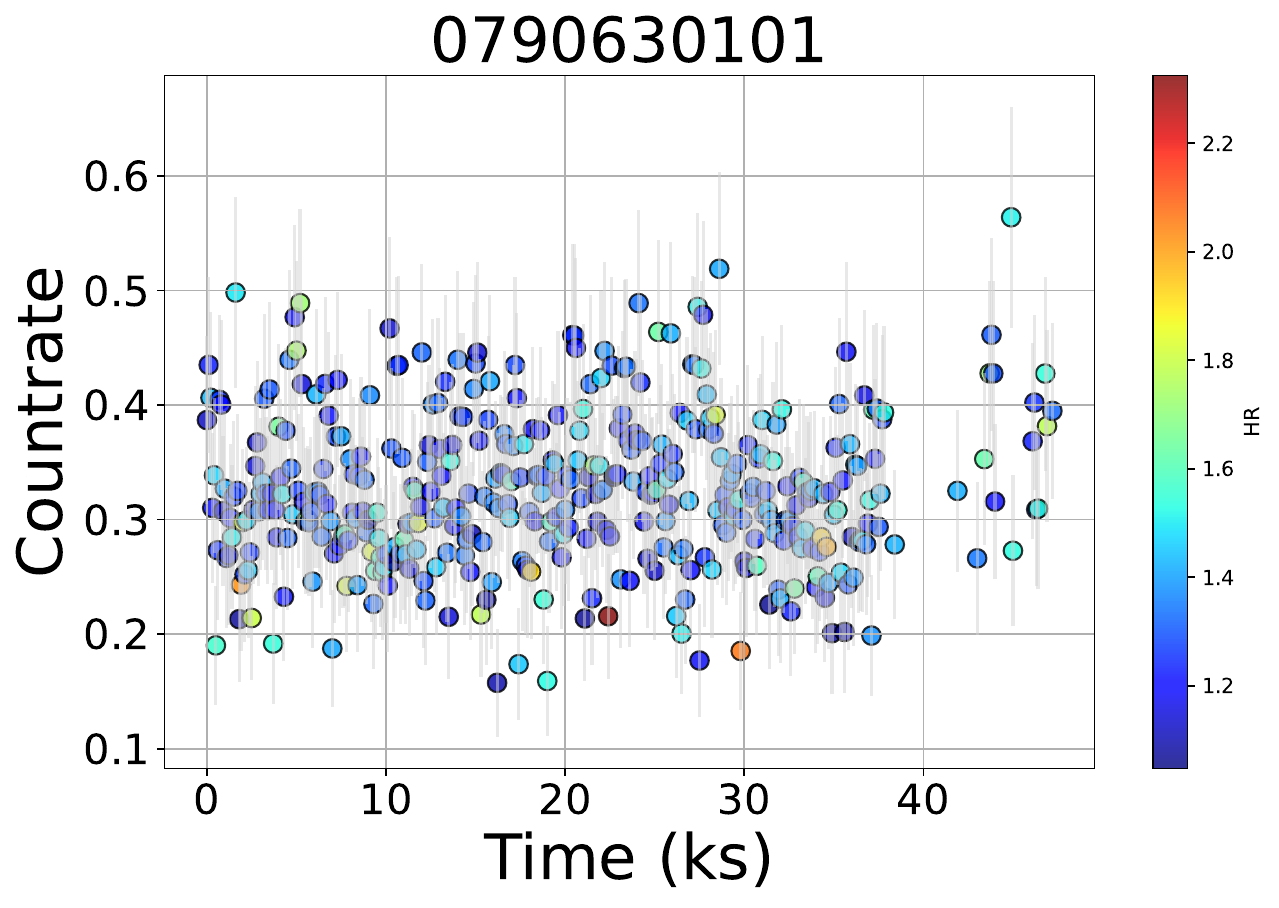}
	\end{minipage}
	\begin{minipage}{.3\textwidth} 
		\centering 
		\includegraphics[width=.99\linewidth]{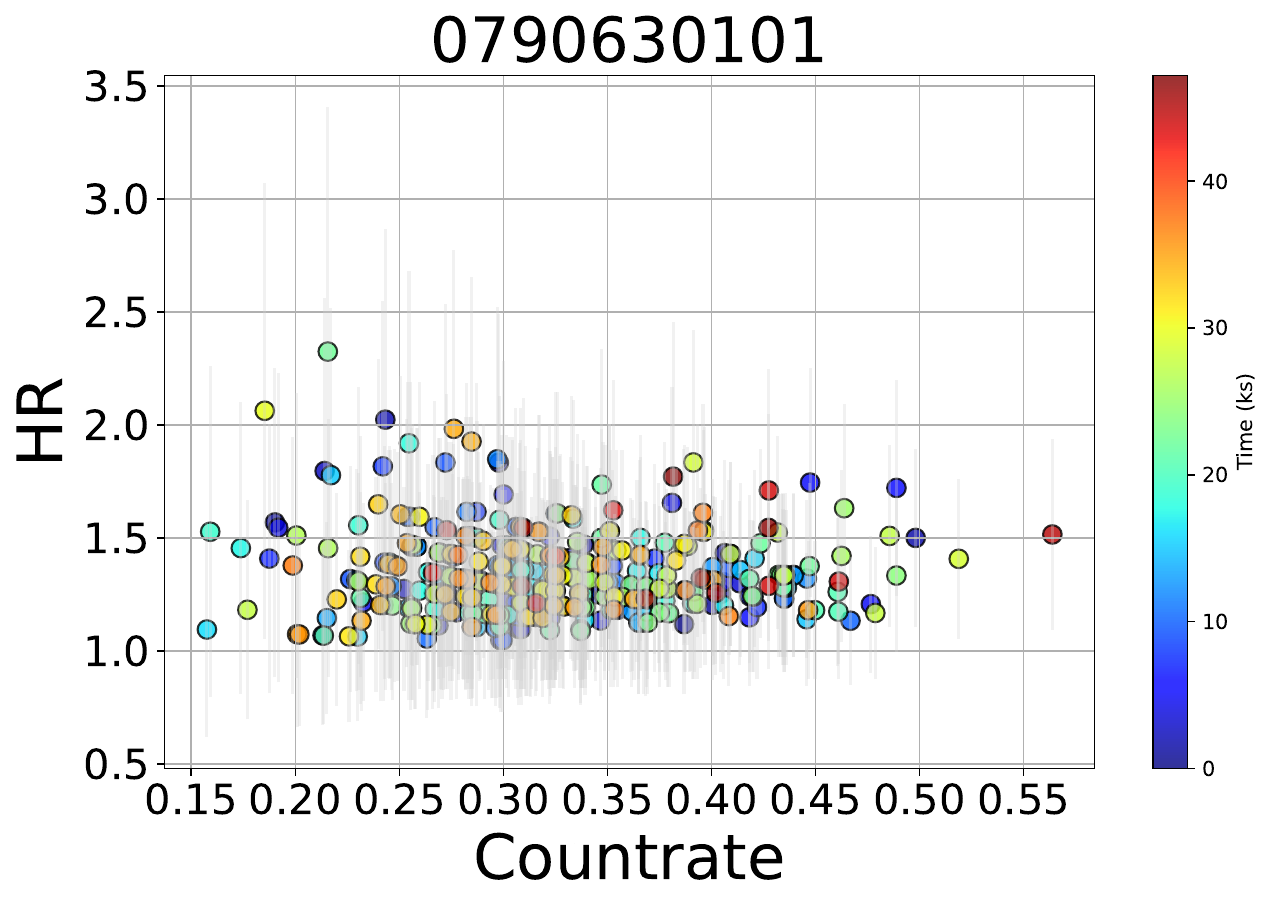}
	\end{minipage}
	\begin{minipage}{.3\textwidth} 
		\centering 
		\includegraphics[width=.99\linewidth, angle=0]{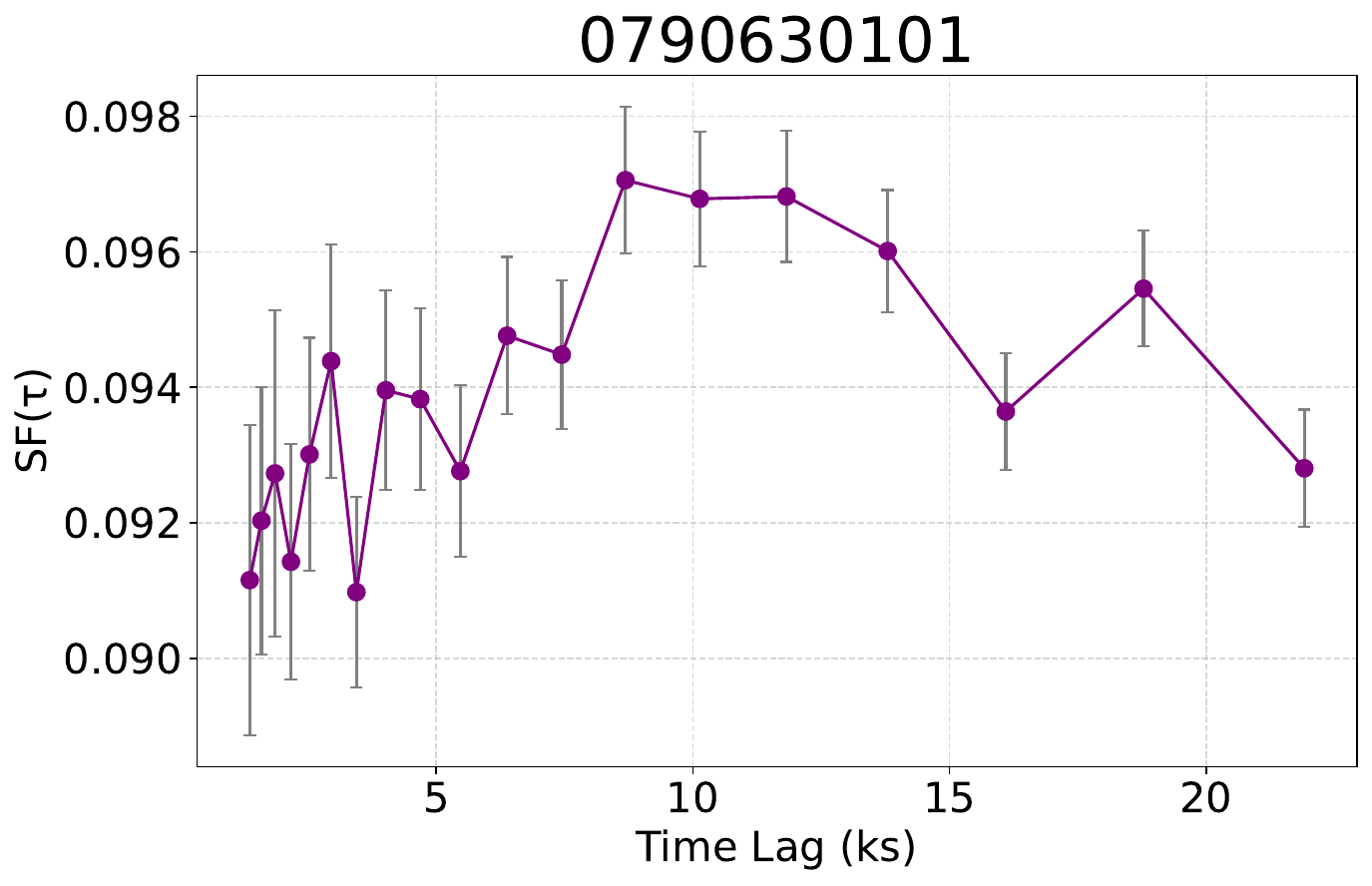}
	\end{minipage}
	\begin{minipage}{.3\textwidth} 
		\centering 
		\includegraphics[width=.99\linewidth]{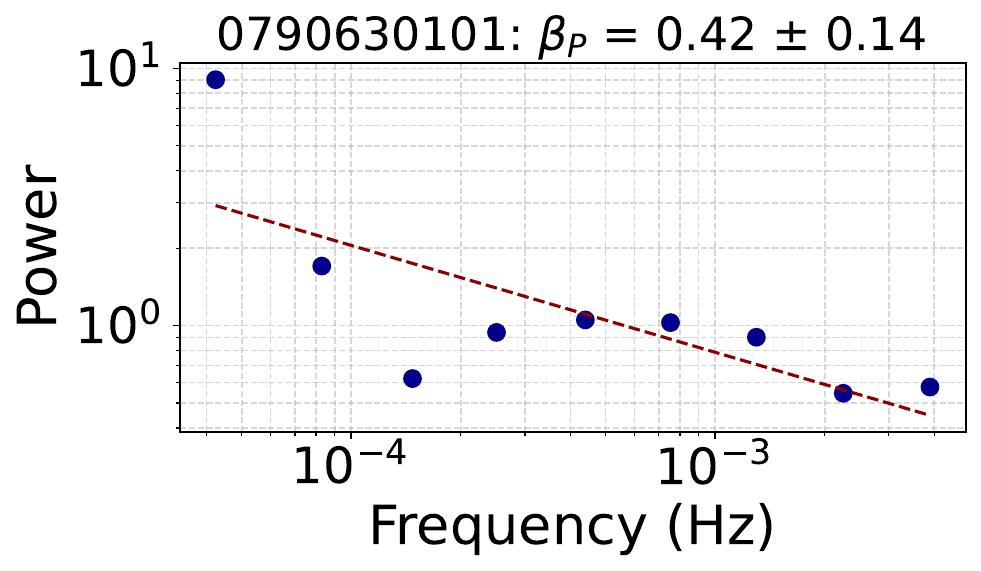}
	\end{minipage}
	\begin{minipage}{.3\textwidth} 
		\centering 
		\includegraphics[width=.99\linewidth]{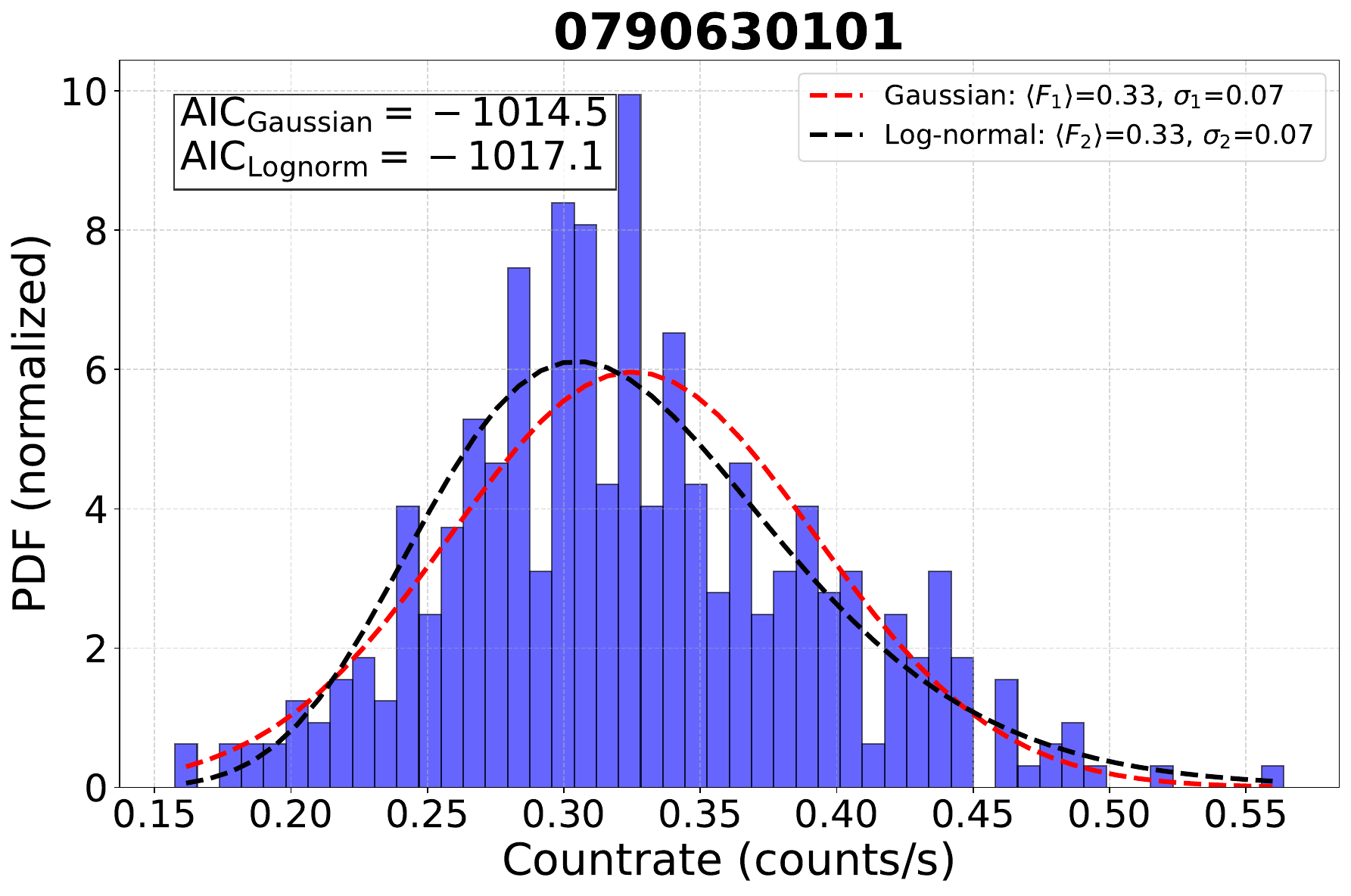}
	\end{minipage}
	\begin{minipage}{.3\textwidth} 
		\centering 
		\includegraphics[height=.99\linewidth, angle=-90]{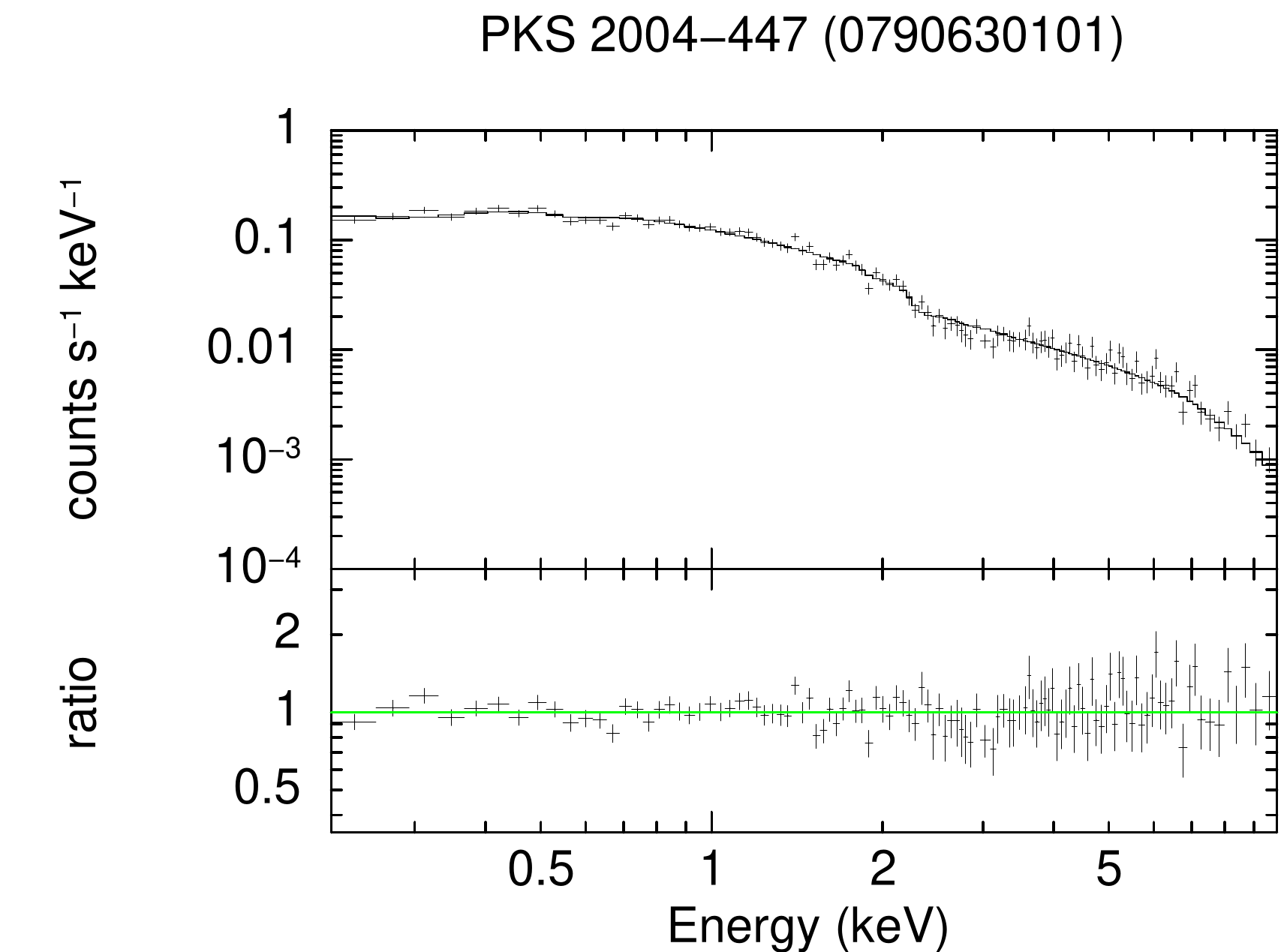}
	\end{minipage}
    	\begin{minipage}{.3\textwidth} 
		\centering 
		\includegraphics[width=.99\linewidth]{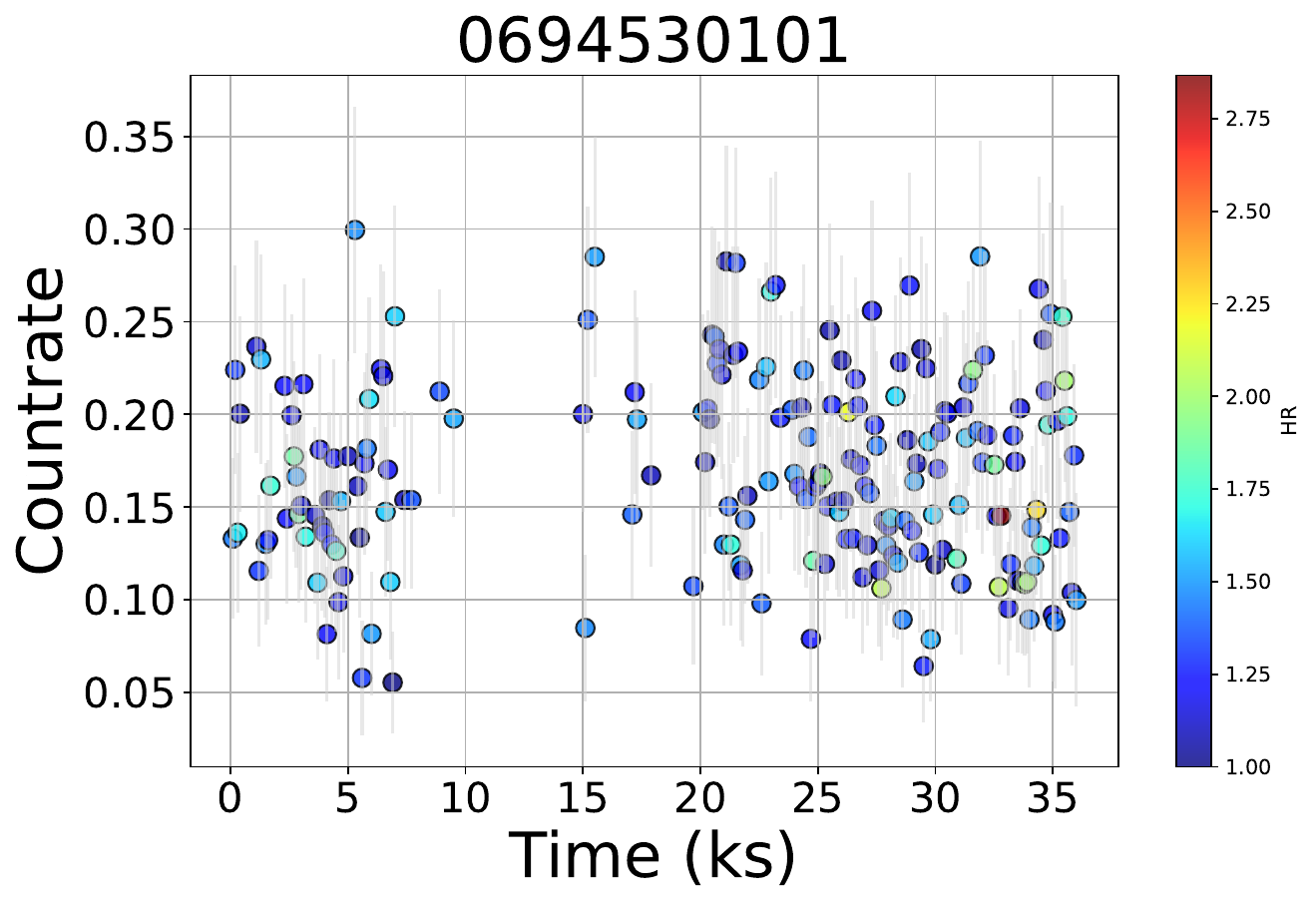}
	\end{minipage}
	\begin{minipage}{.3\textwidth} 
		\centering 
		\includegraphics[width=.99\linewidth]{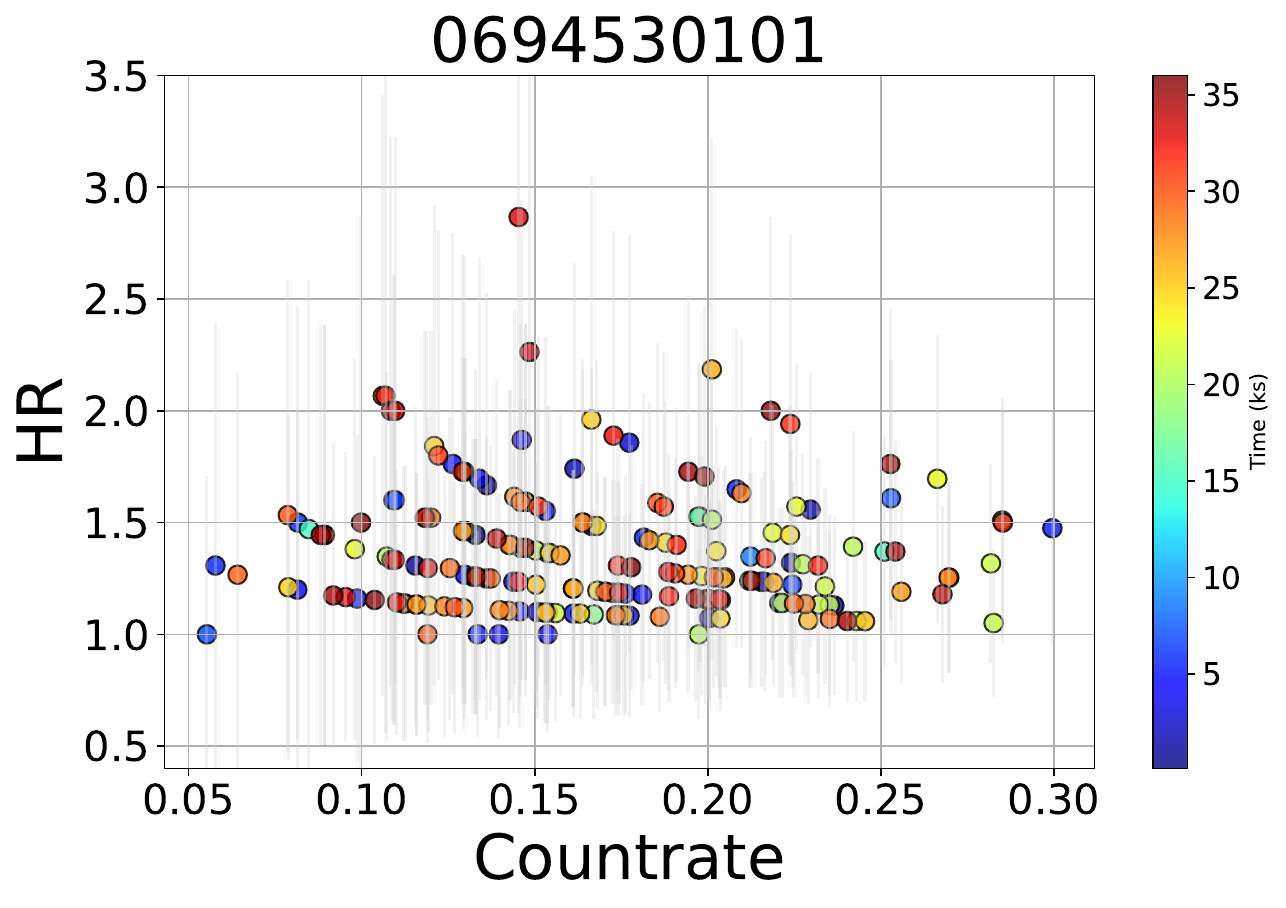}
	\end{minipage}
	\begin{minipage}{.3\textwidth} 
		\centering 
		\includegraphics[width=.99\linewidth, angle=0]{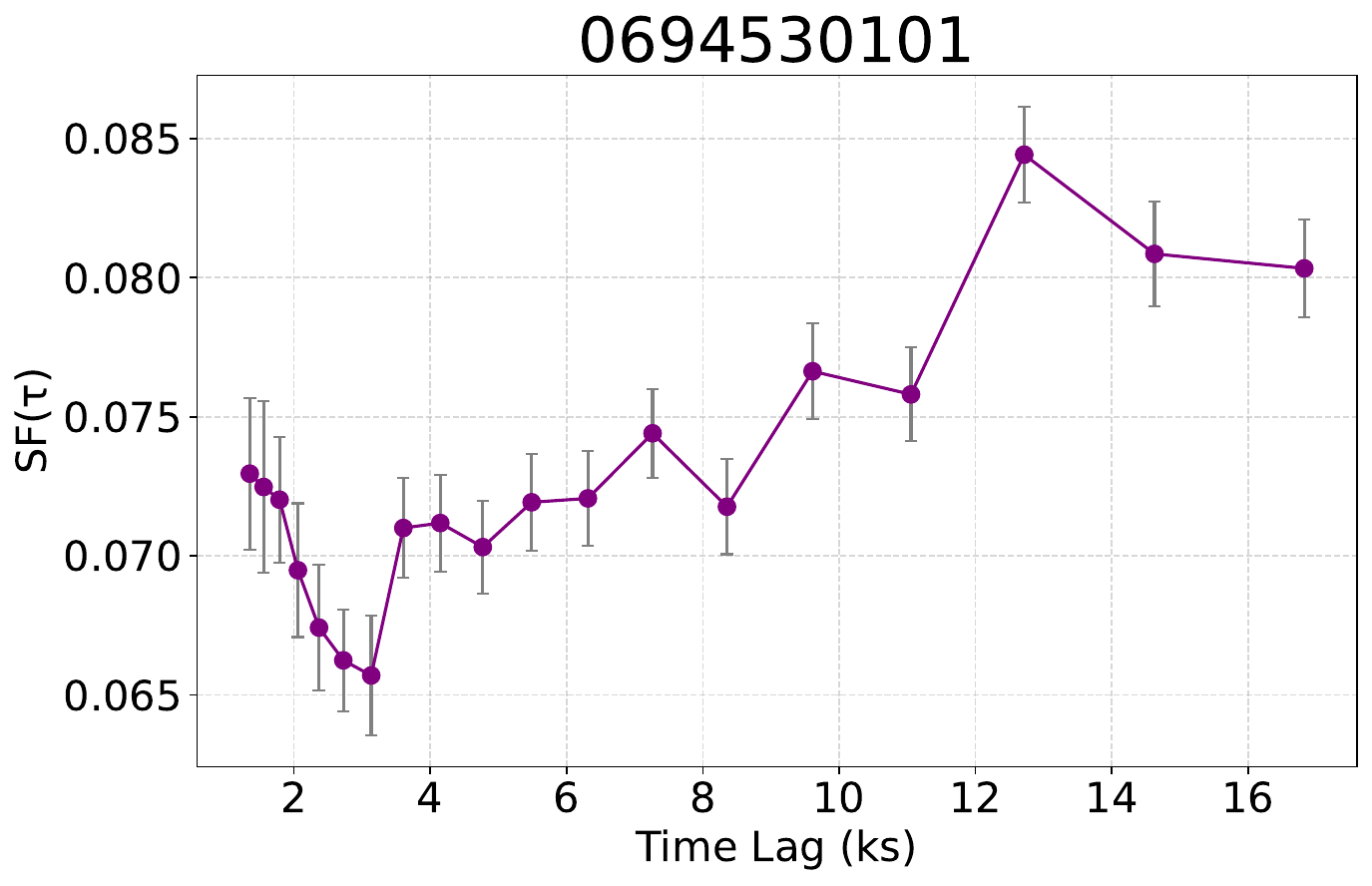}
	\end{minipage}
    \begin{minipage}{.3\textwidth} 
		\centering 
		\includegraphics[width=.99\linewidth]{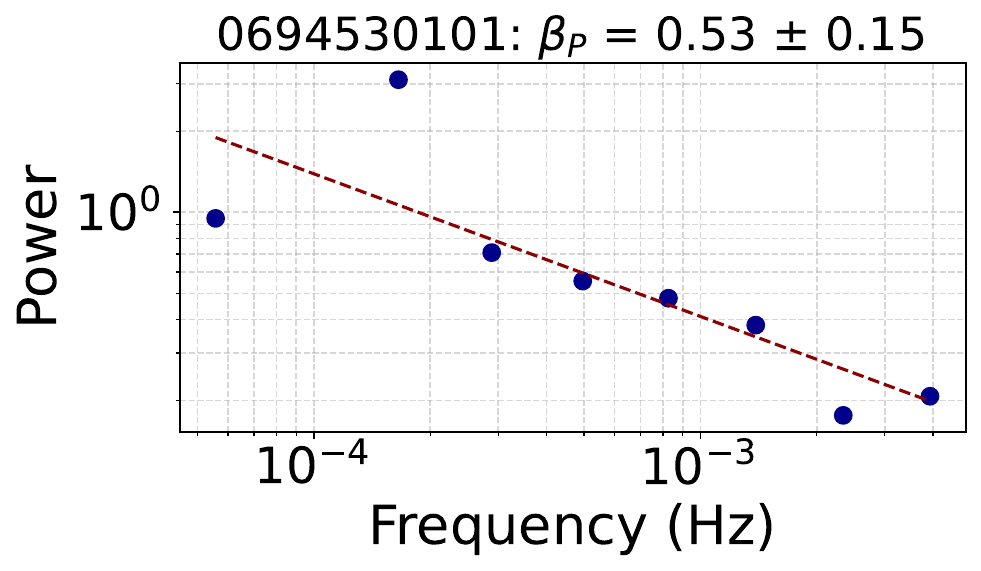}
	\end{minipage}
	\begin{minipage}{.3\textwidth} 
		\centering 
		\includegraphics[width=.99\linewidth]{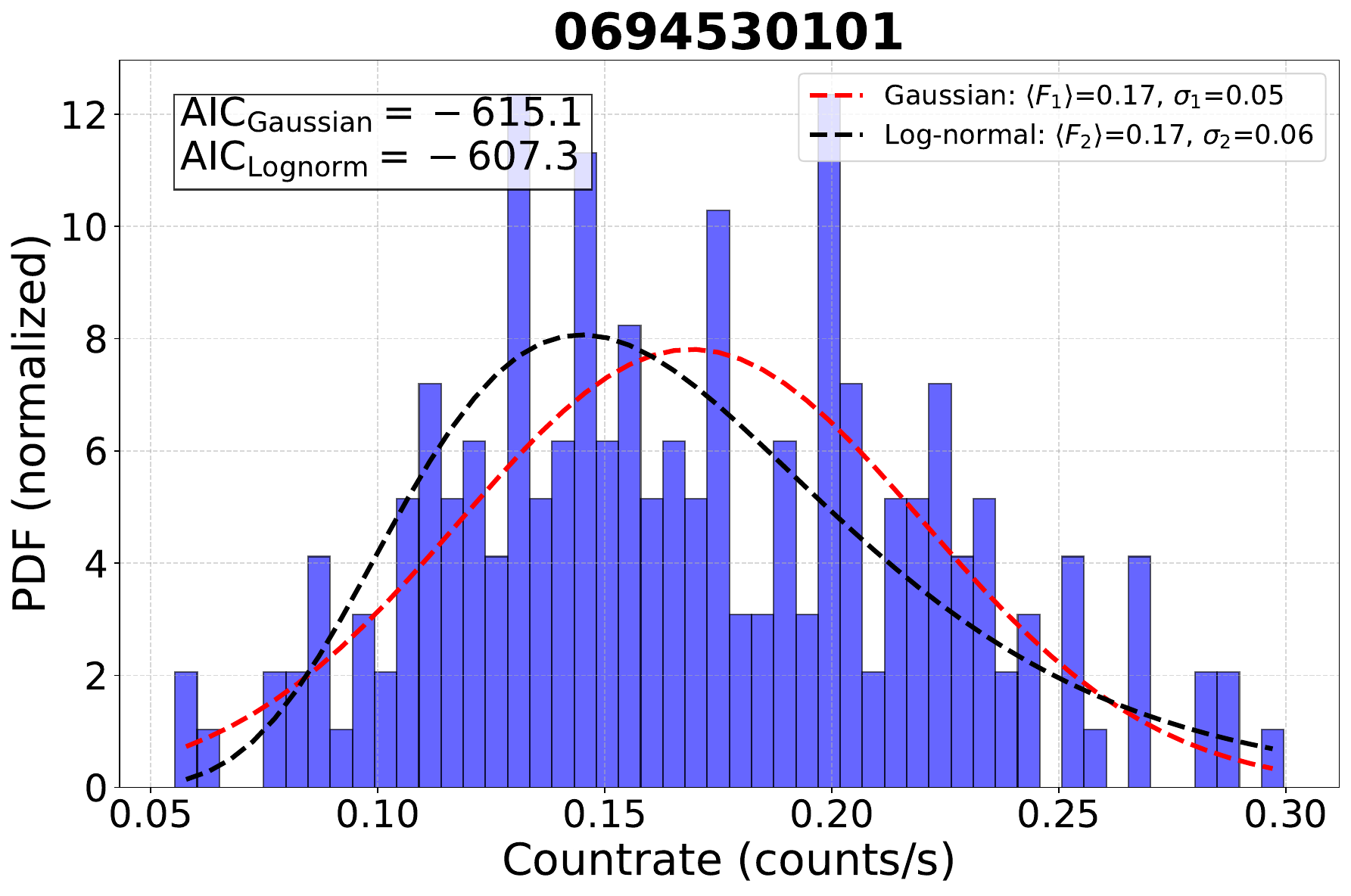}
	\end{minipage}
	\begin{minipage}{.3\textwidth} 
		\centering 
		\includegraphics[height=.99\linewidth, angle=-90]{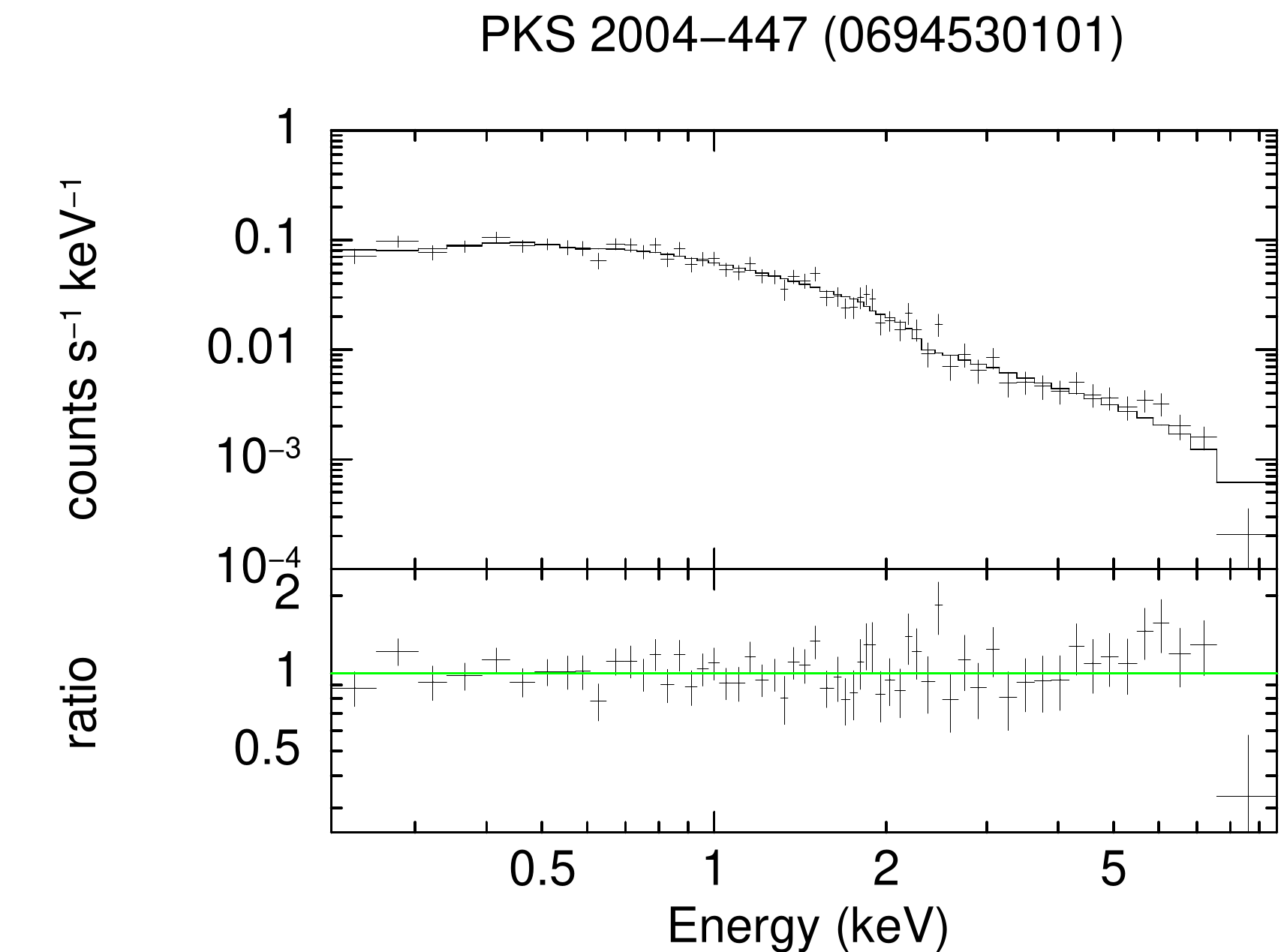}
	\end{minipage}

\begin{minipage}{.3\textwidth} 
		\centering 
		\includegraphics[width=.99\linewidth]{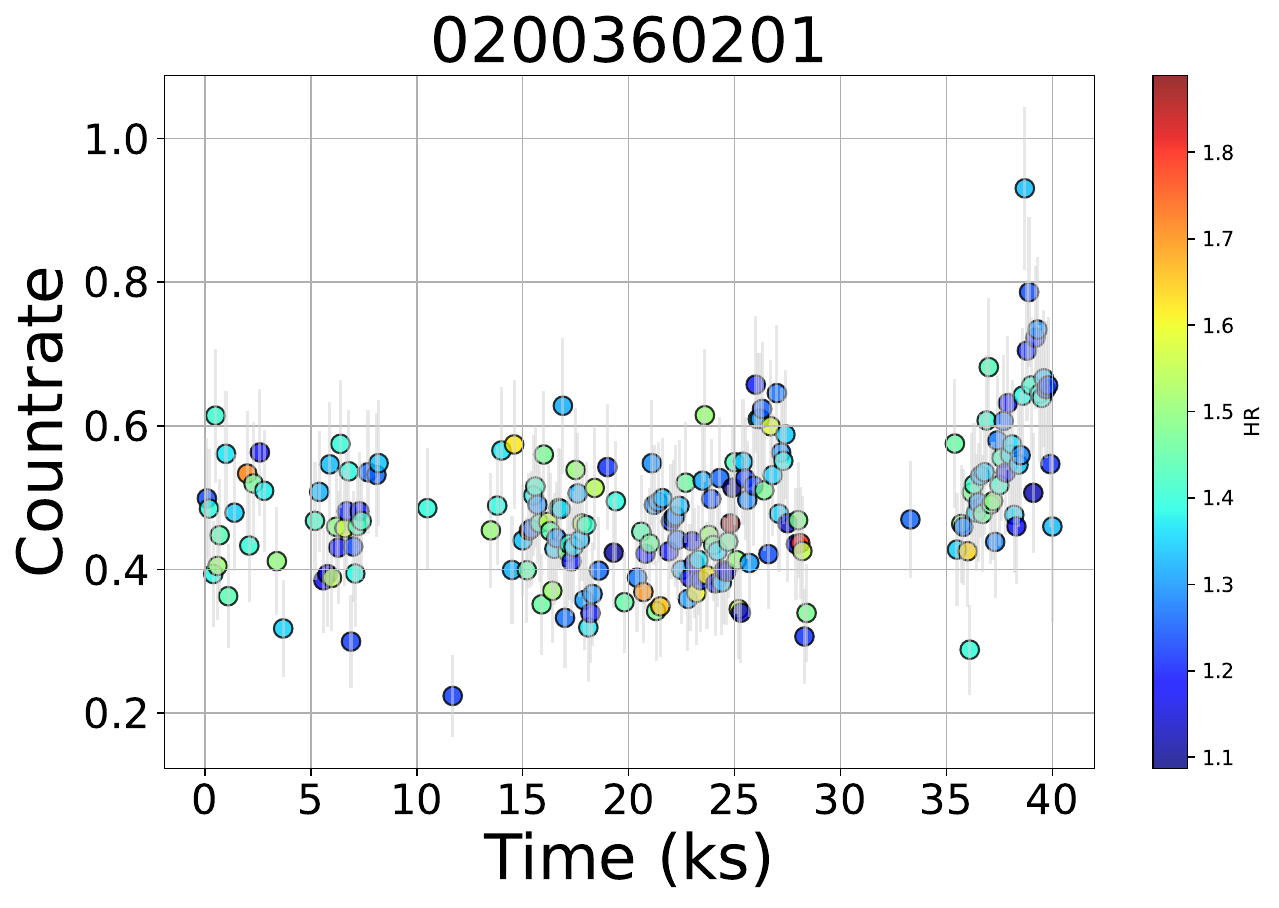}
	\end{minipage}
	\begin{minipage}{.3\textwidth} 
		\centering 
		\includegraphics[width=.99\linewidth]{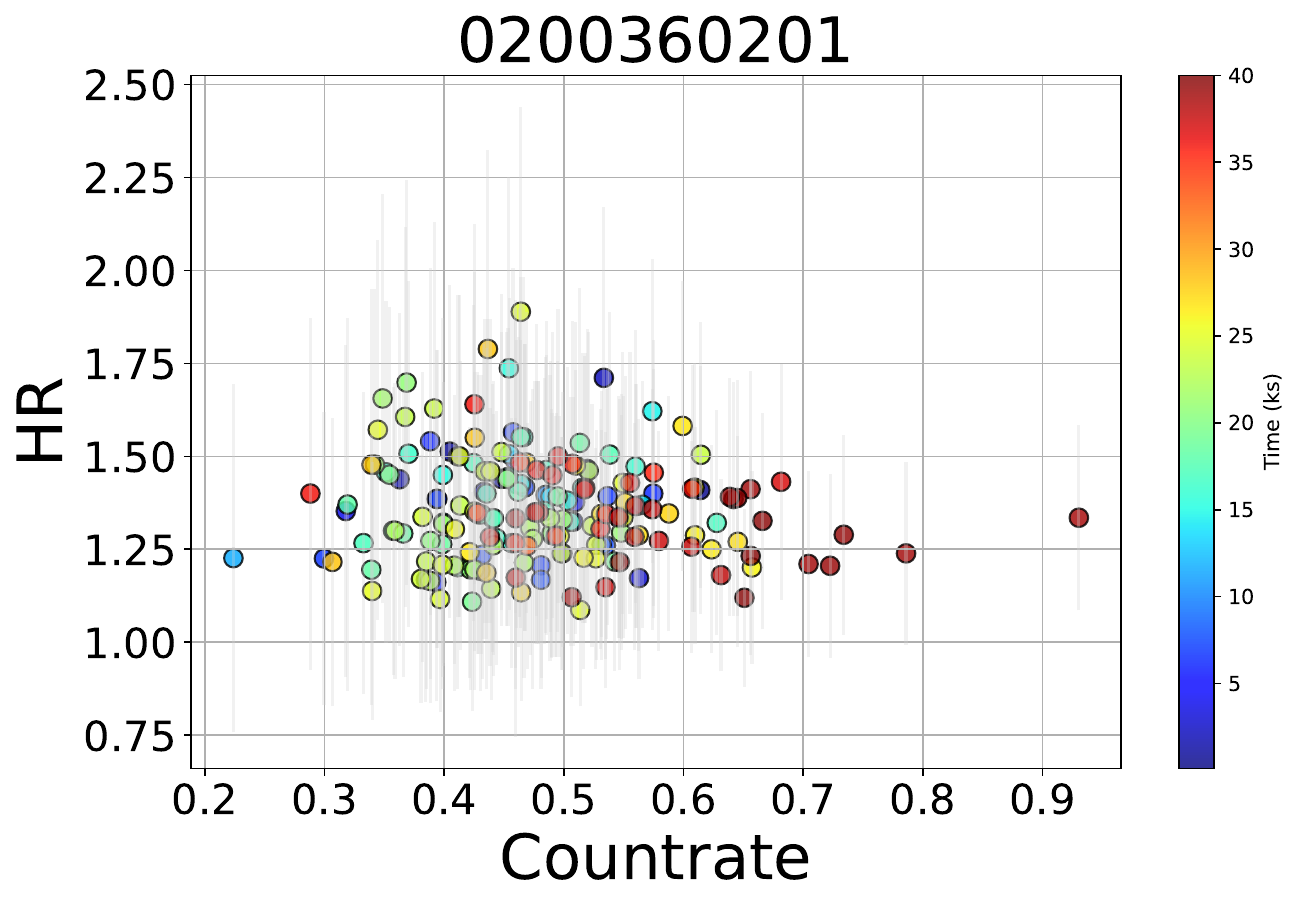}
	\end{minipage}
	\begin{minipage}{.3\textwidth} 
		\centering 
		\includegraphics[width=.99\linewidth, angle=0]{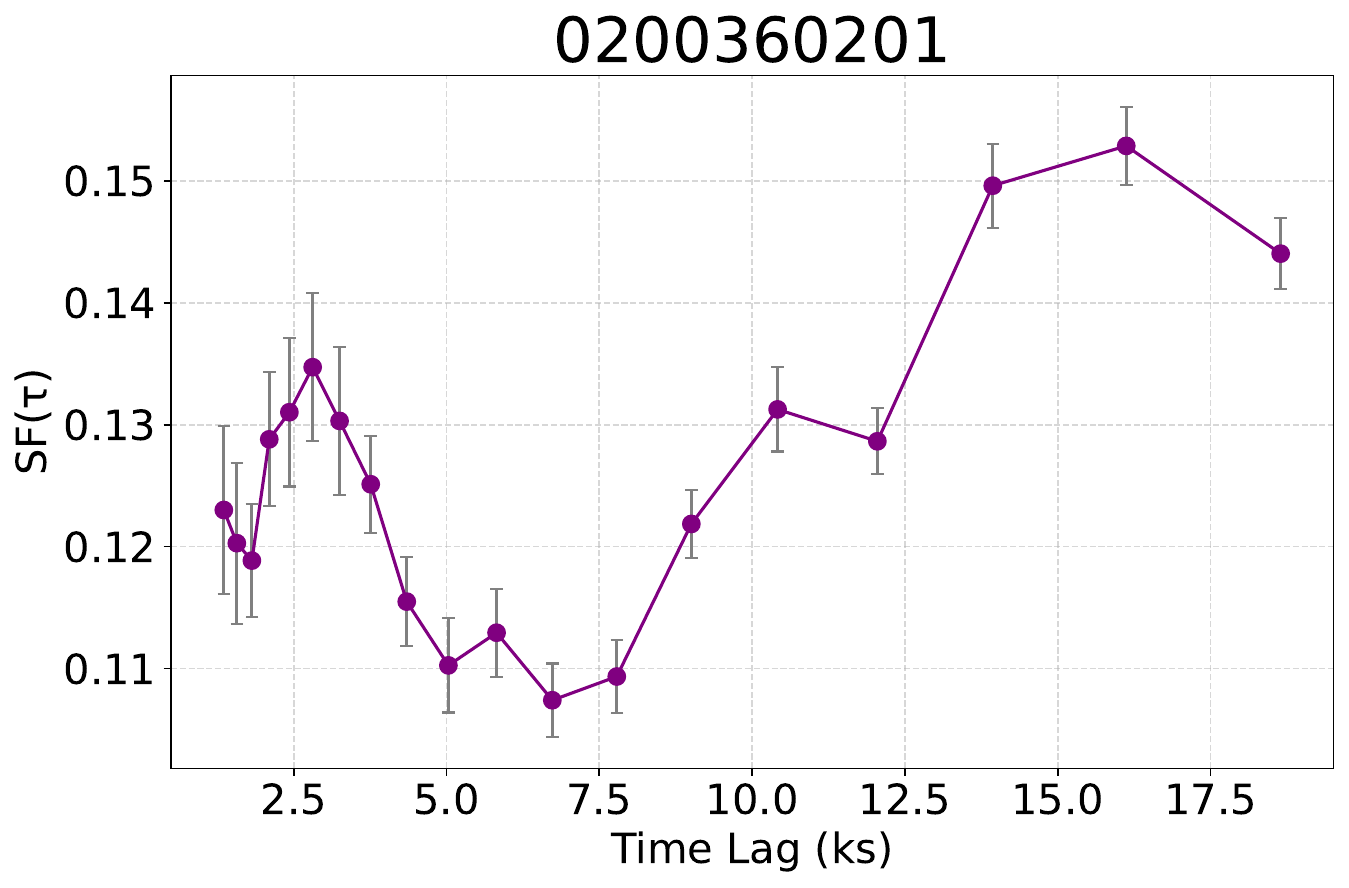}
	\end{minipage}
	\begin{minipage}{.3\textwidth} 
		\centering 
		\includegraphics[width=.99\linewidth]{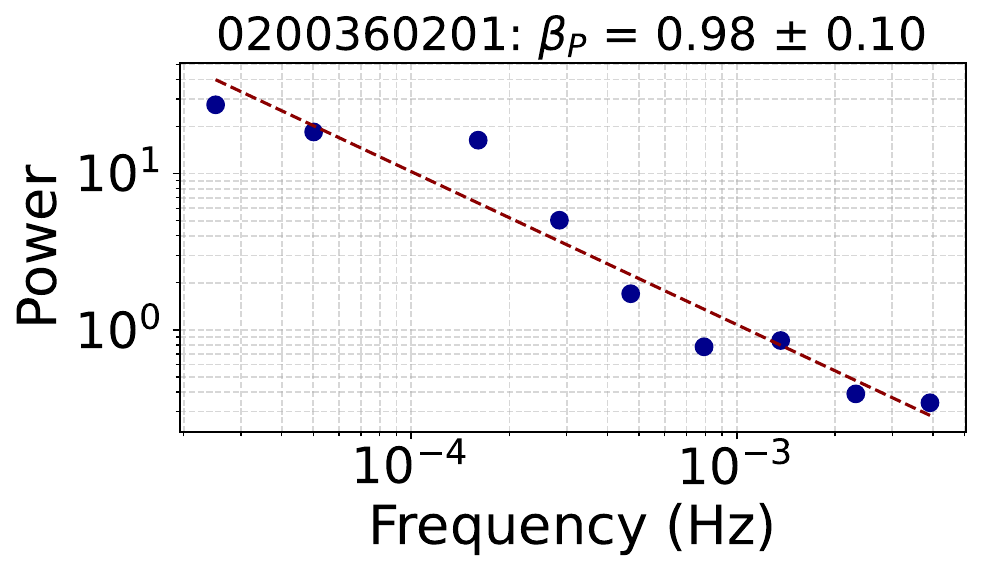}
	\end{minipage}
	\begin{minipage}{.3\textwidth} 
		\centering 
		\includegraphics[width=.99\linewidth]{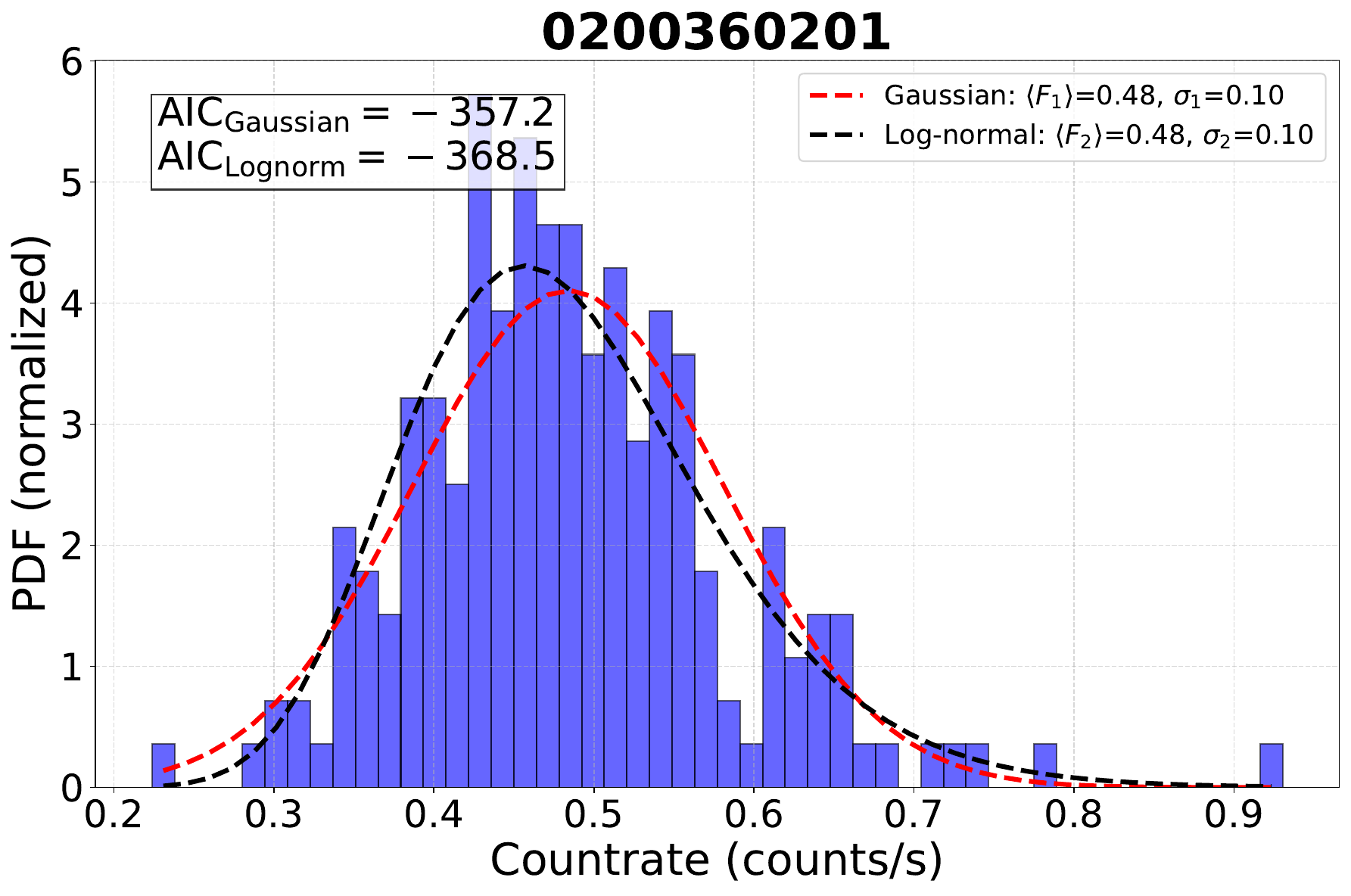}
	\end{minipage}
	\begin{minipage}{.3\textwidth} 
		\centering 
		\includegraphics[height=.99\linewidth, angle=-90]{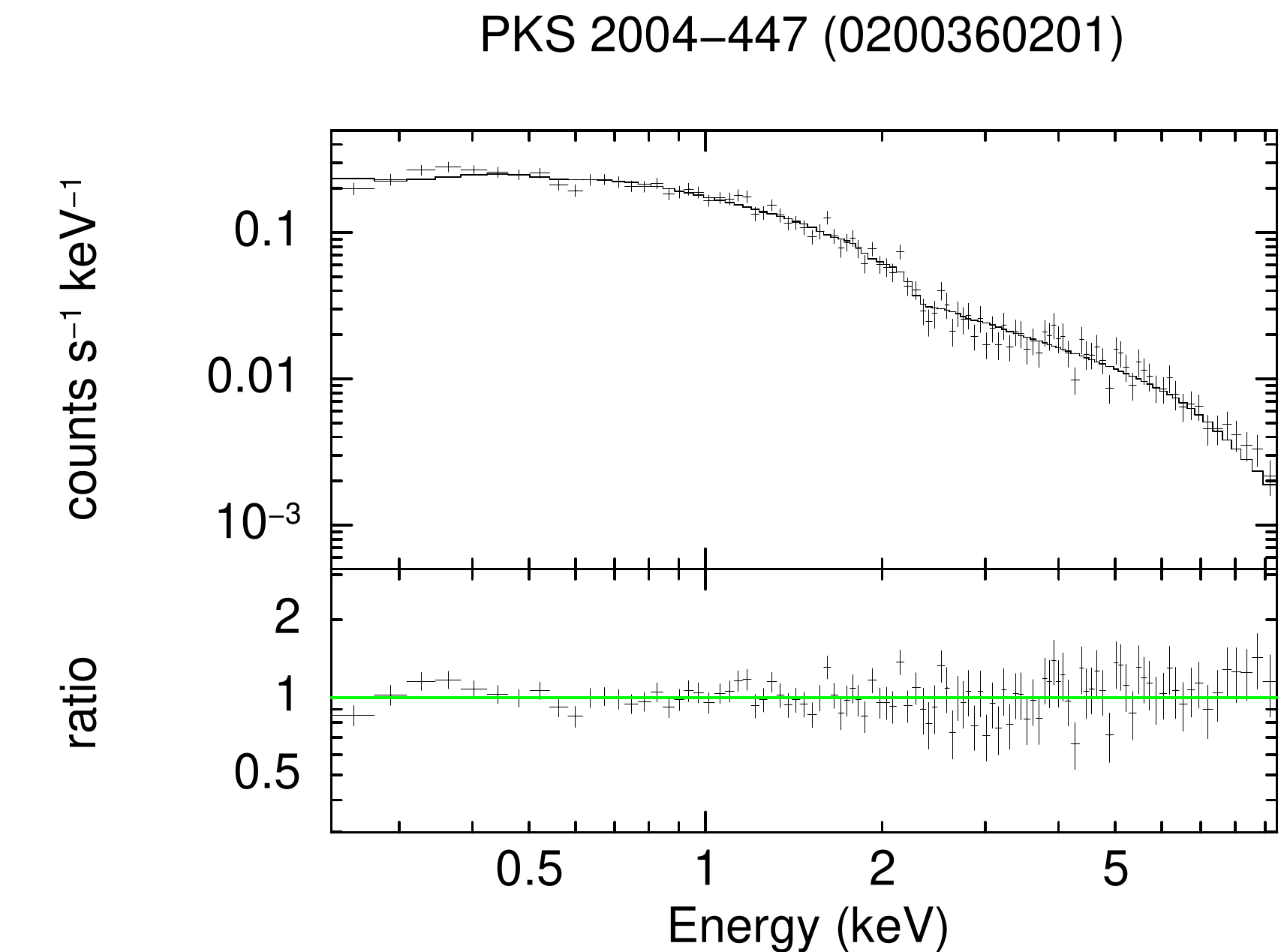}
	\end{minipage}
\end{figure*}

\begin{figure*}
	\centering
	
	\begin{minipage}{.3\textwidth} 
		\centering 
		\includegraphics[width=.99\linewidth]{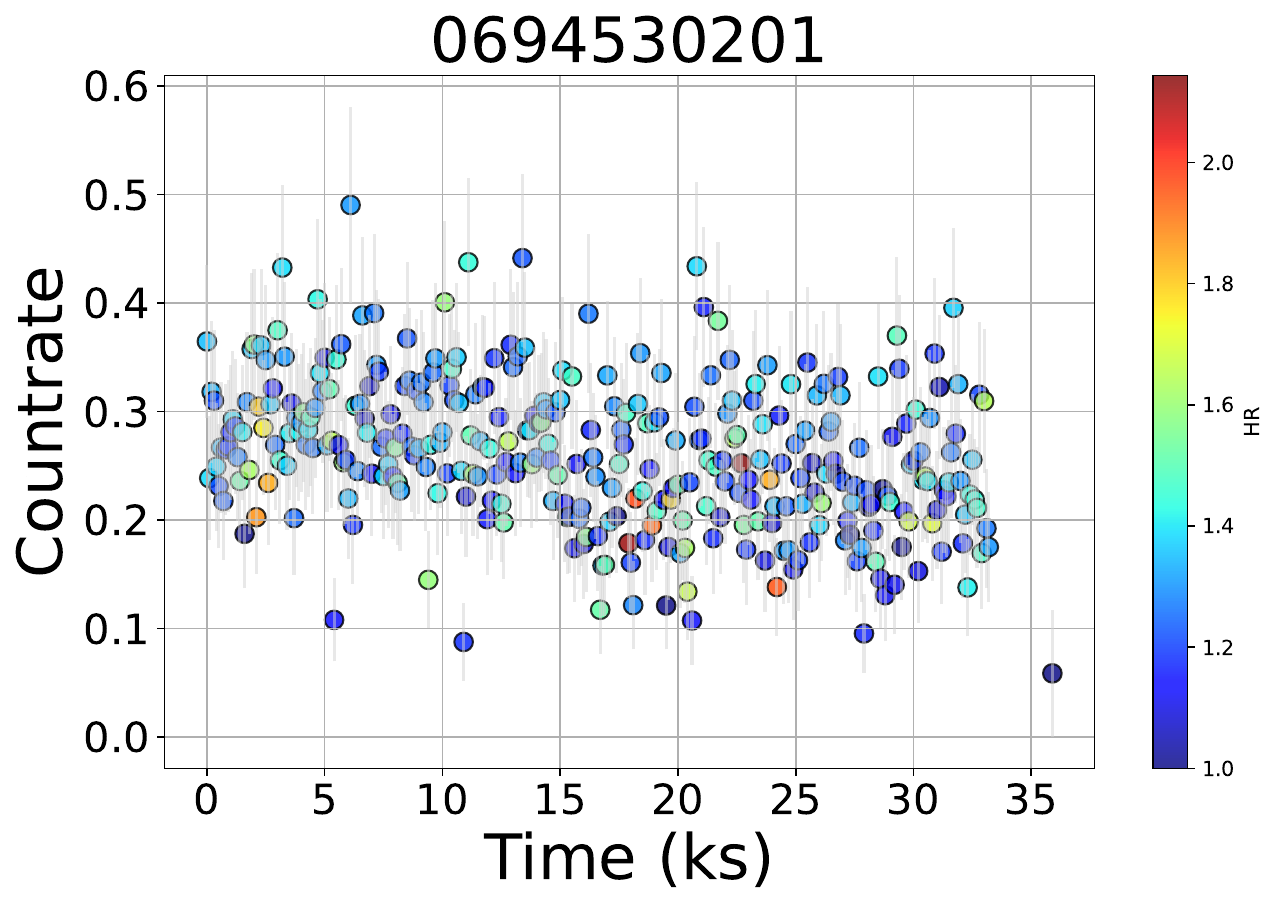}
	\end{minipage}
	\begin{minipage}{.3\textwidth} 
		\centering 
		\includegraphics[width=.99\linewidth]{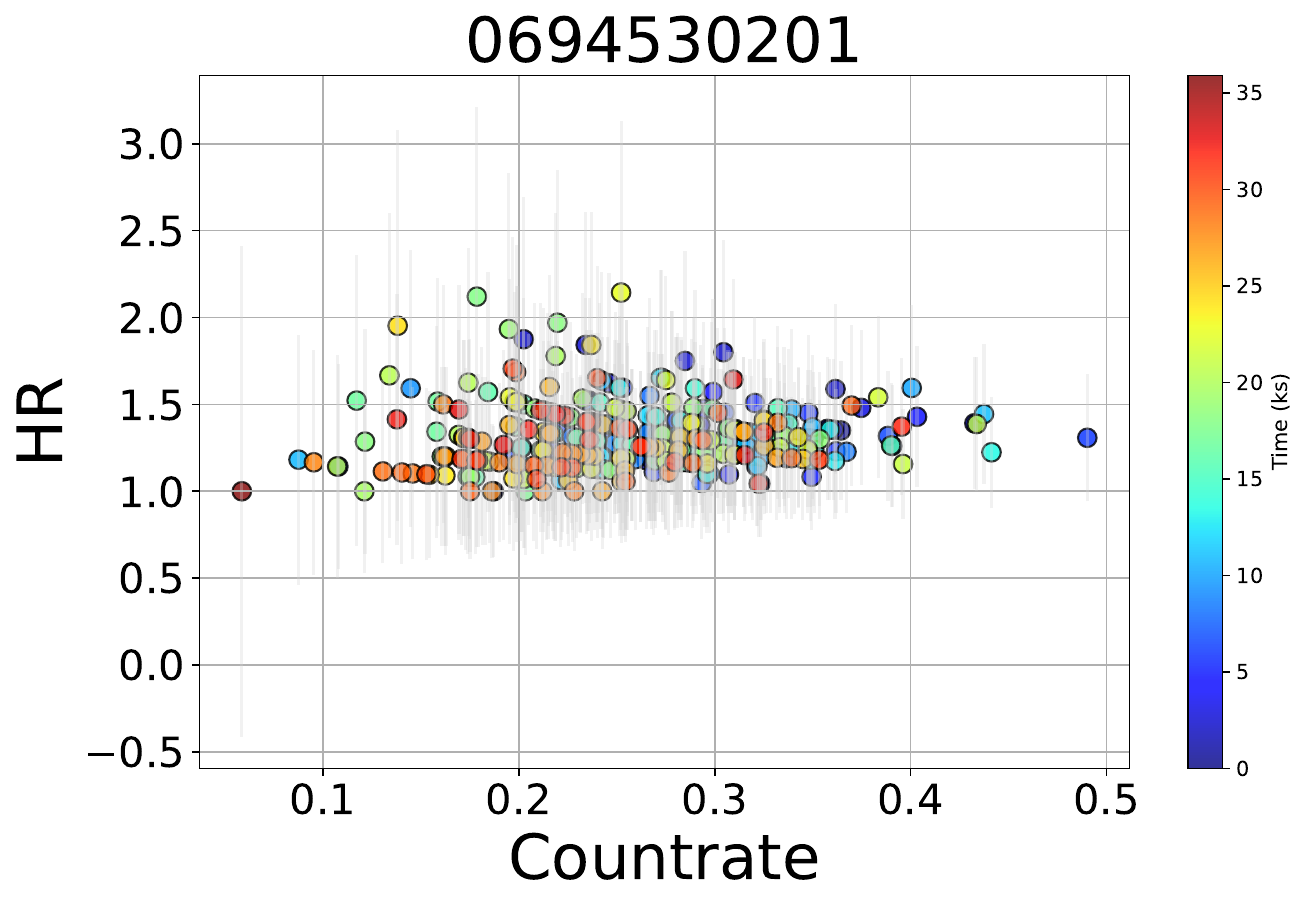}
	\end{minipage}
	\begin{minipage}{.3\textwidth} 
		\centering 
		\includegraphics[width=.99\linewidth, angle=0]{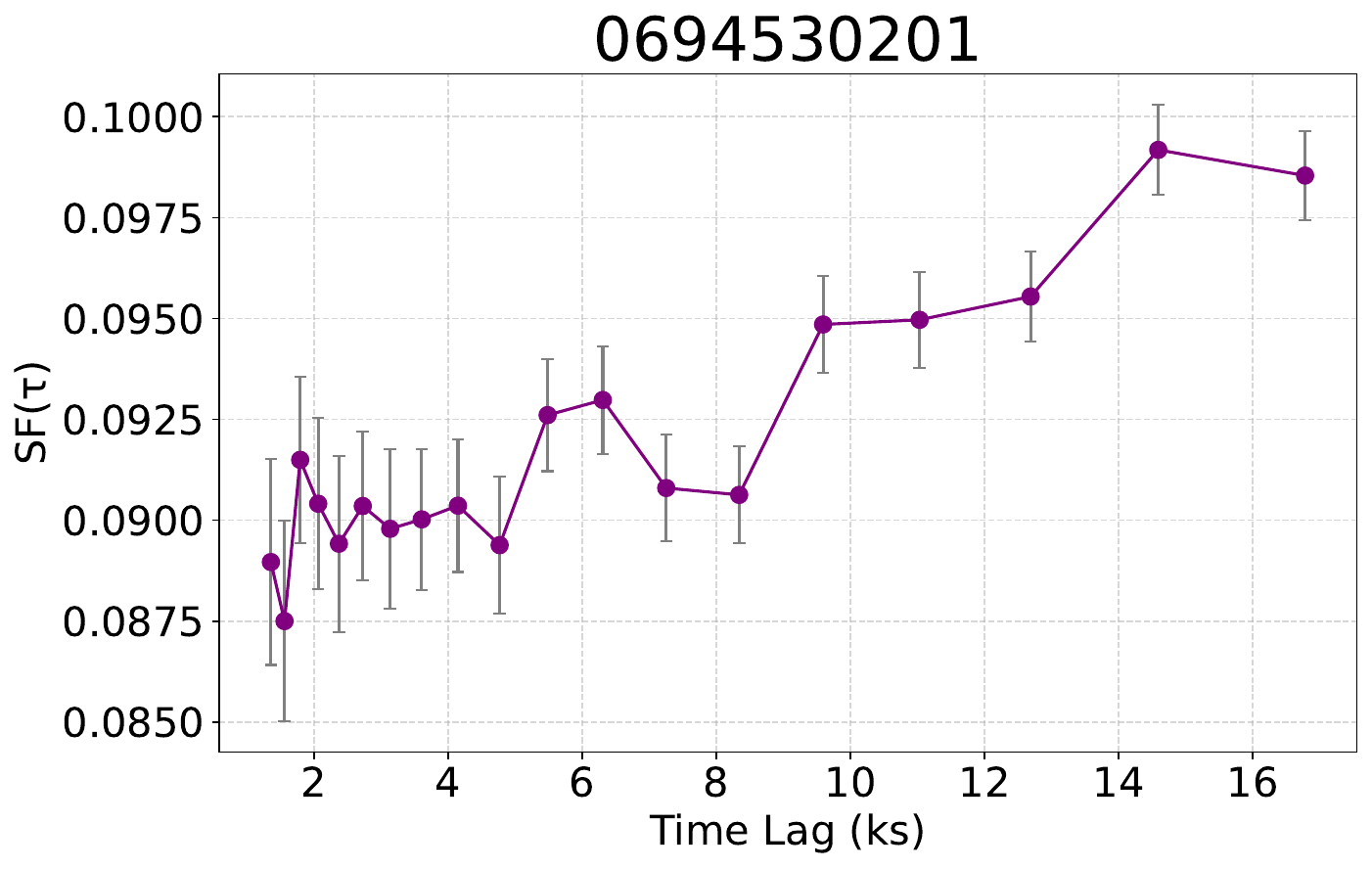}
	\end{minipage}
	\begin{minipage}{.3\textwidth} 
		\centering 
		\includegraphics[width=.99\linewidth]{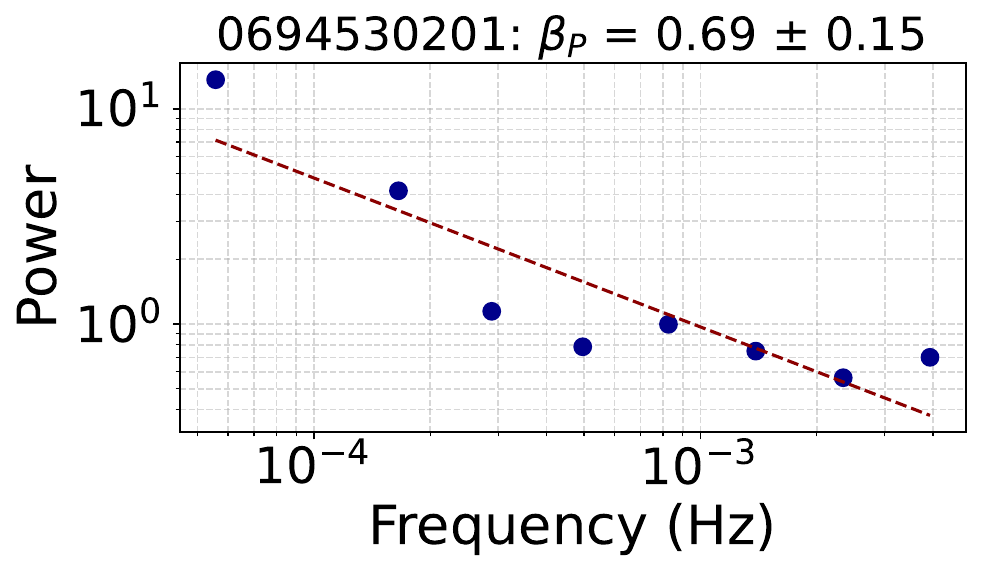}
	\end{minipage}
	\begin{minipage}{.3\textwidth} 
		\centering 
		\includegraphics[width=.99\linewidth]{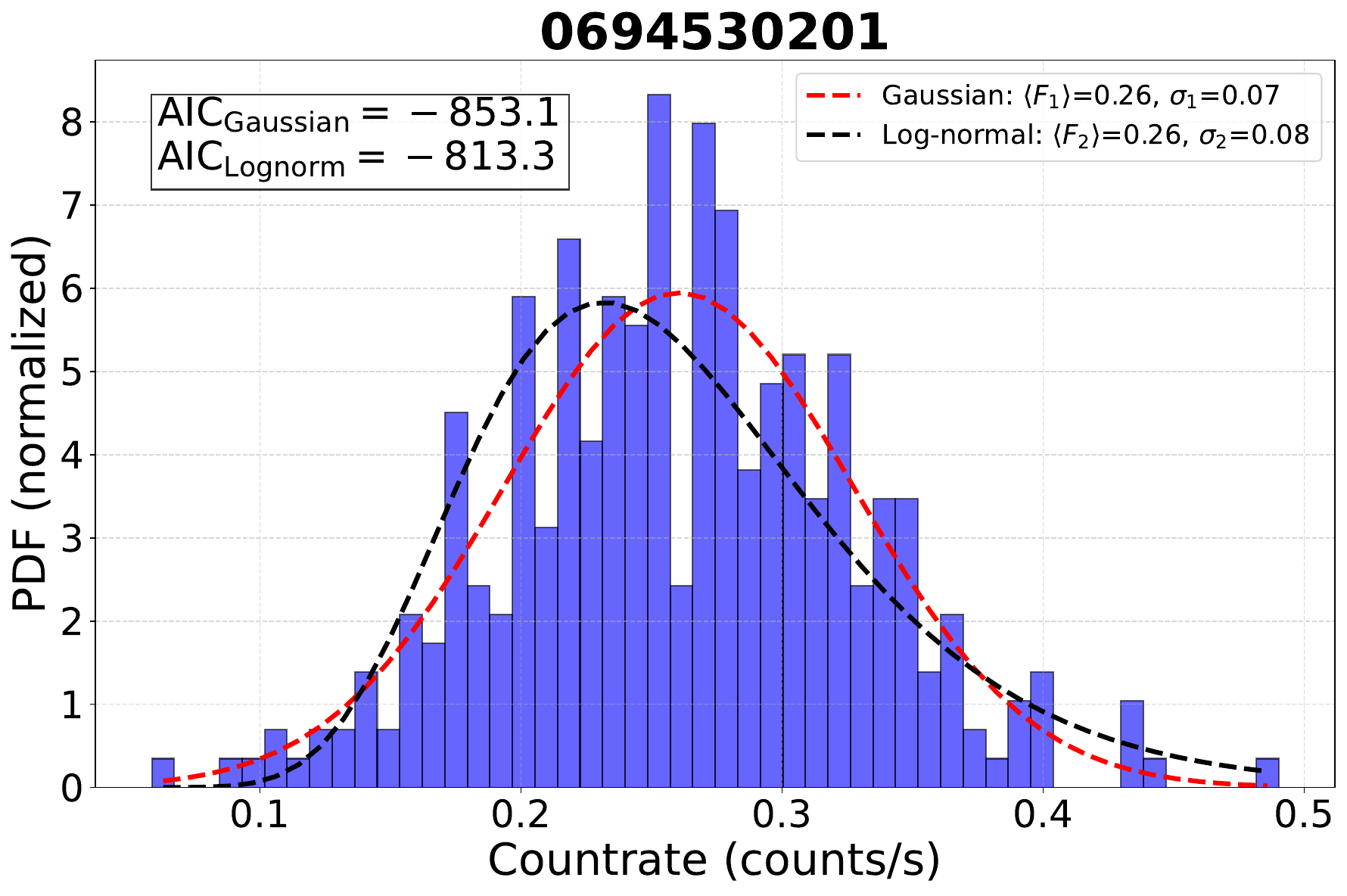}
	\end{minipage}
	\begin{minipage}{.3\textwidth} 
		\centering 
		\includegraphics[height=.99\linewidth, angle=-90]{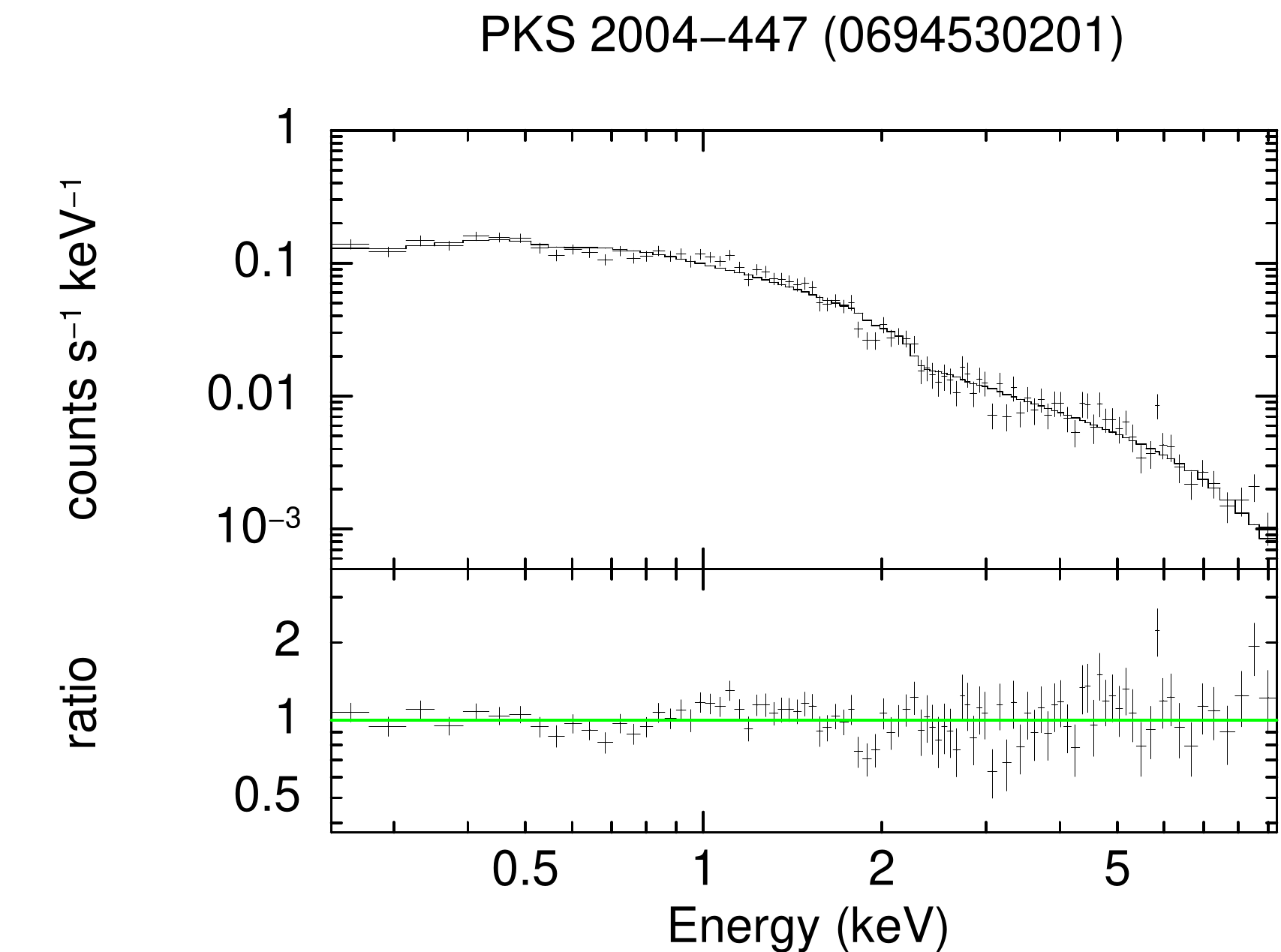}
	\end{minipage}
    	\begin{minipage}{.3\textwidth} 
		\centering 
		\includegraphics[width=.99\linewidth]{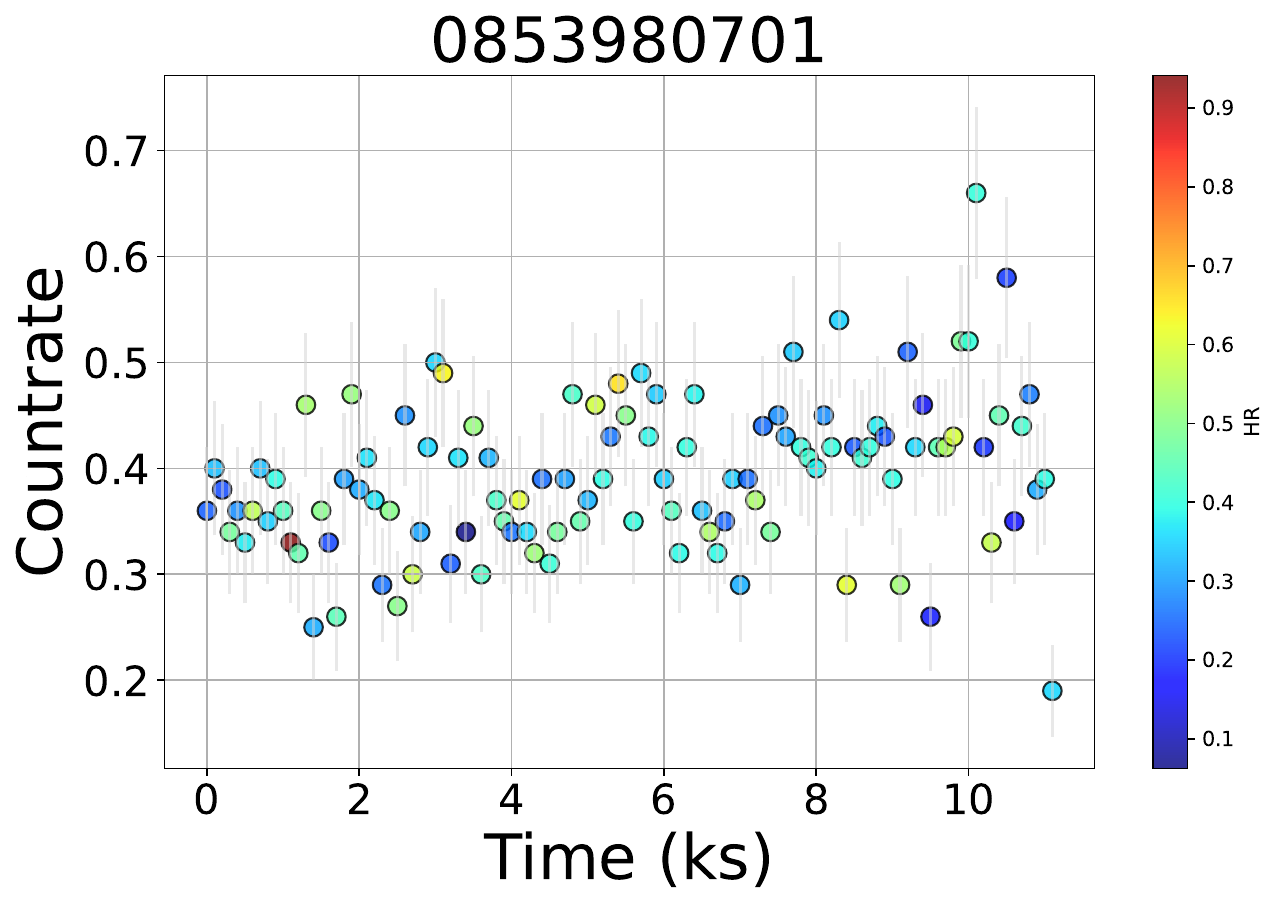}
	\end{minipage}
	\begin{minipage}{.3\textwidth} 
		\centering 
		\includegraphics[width=.99\linewidth]{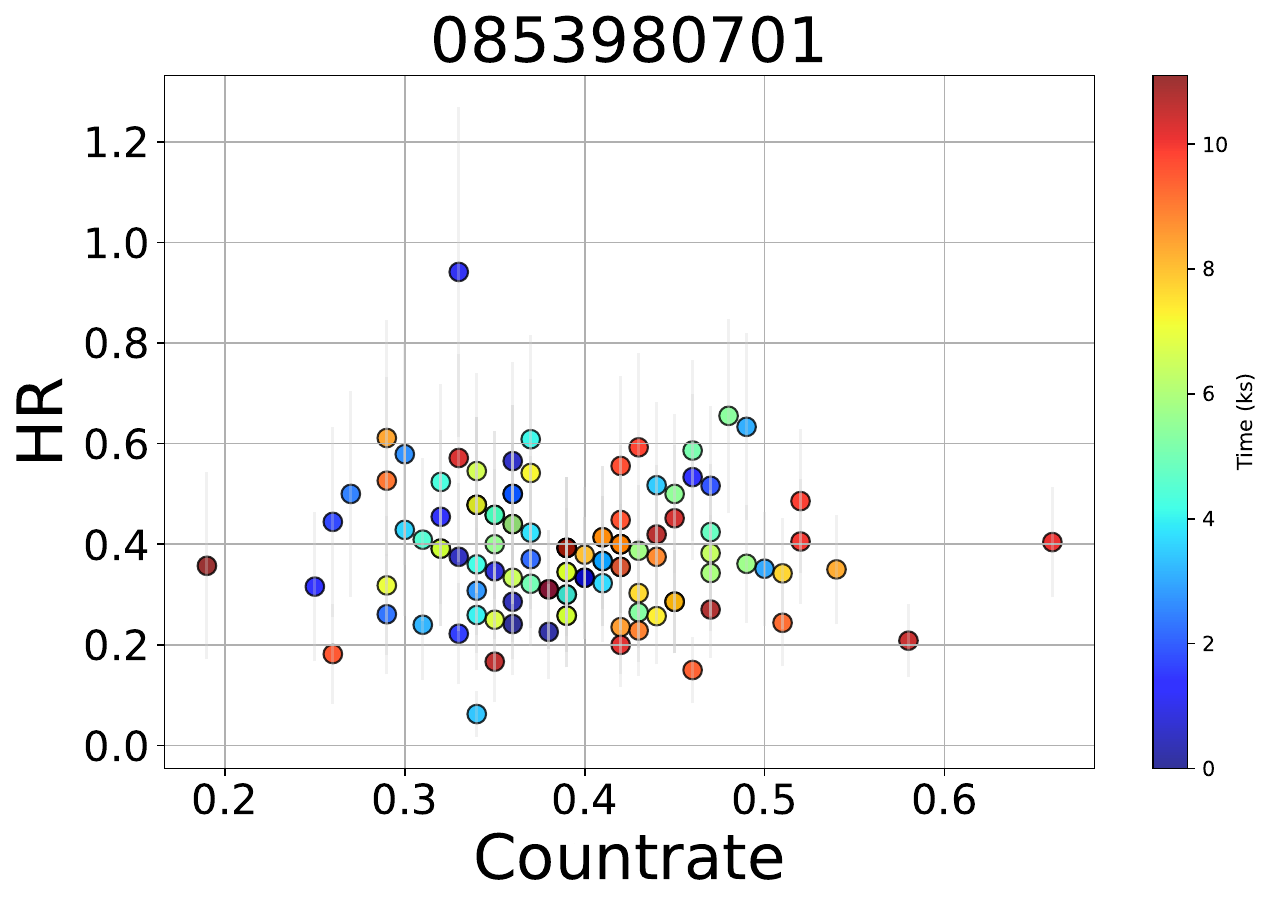}
	\end{minipage}
	\begin{minipage}{.3\textwidth} 
		\centering 
		\includegraphics[width=.99\linewidth, angle=0]{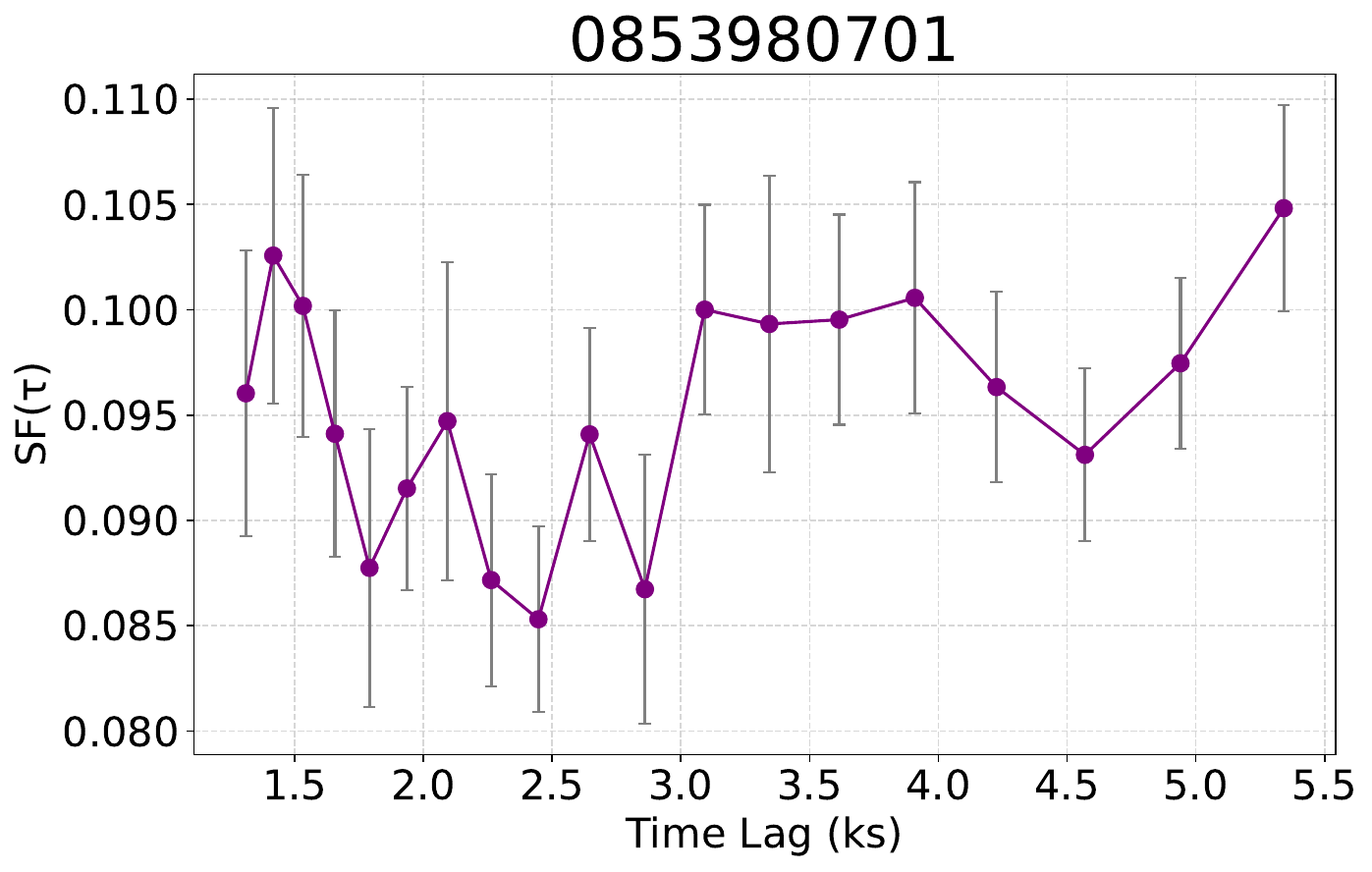}
	\end{minipage}
    \begin{minipage}{.3\textwidth} 
		\centering 
		\includegraphics[width=.99\linewidth]{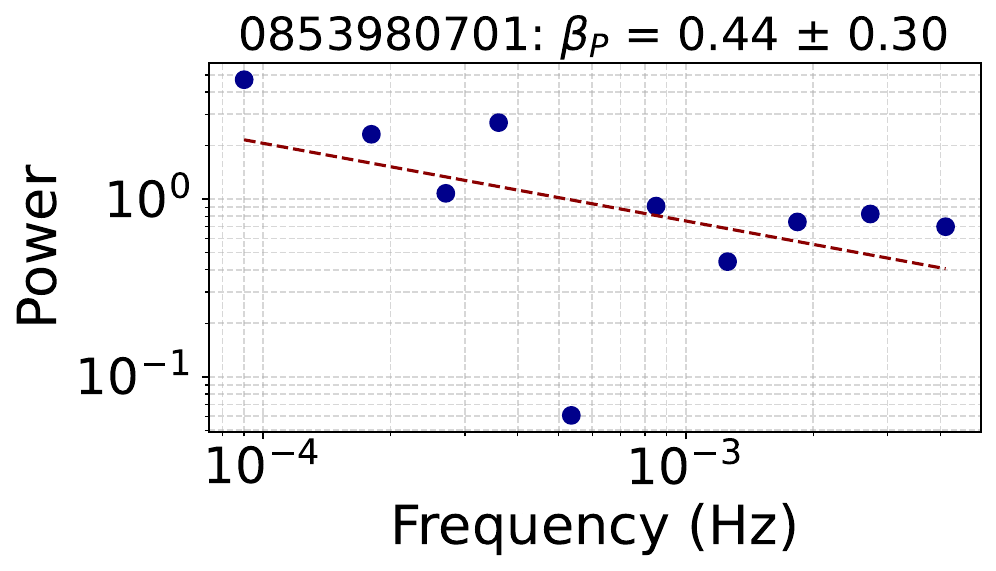}
	\end{minipage}
	\begin{minipage}{.3\textwidth} 
		\centering 
		\includegraphics[width=.99\linewidth]{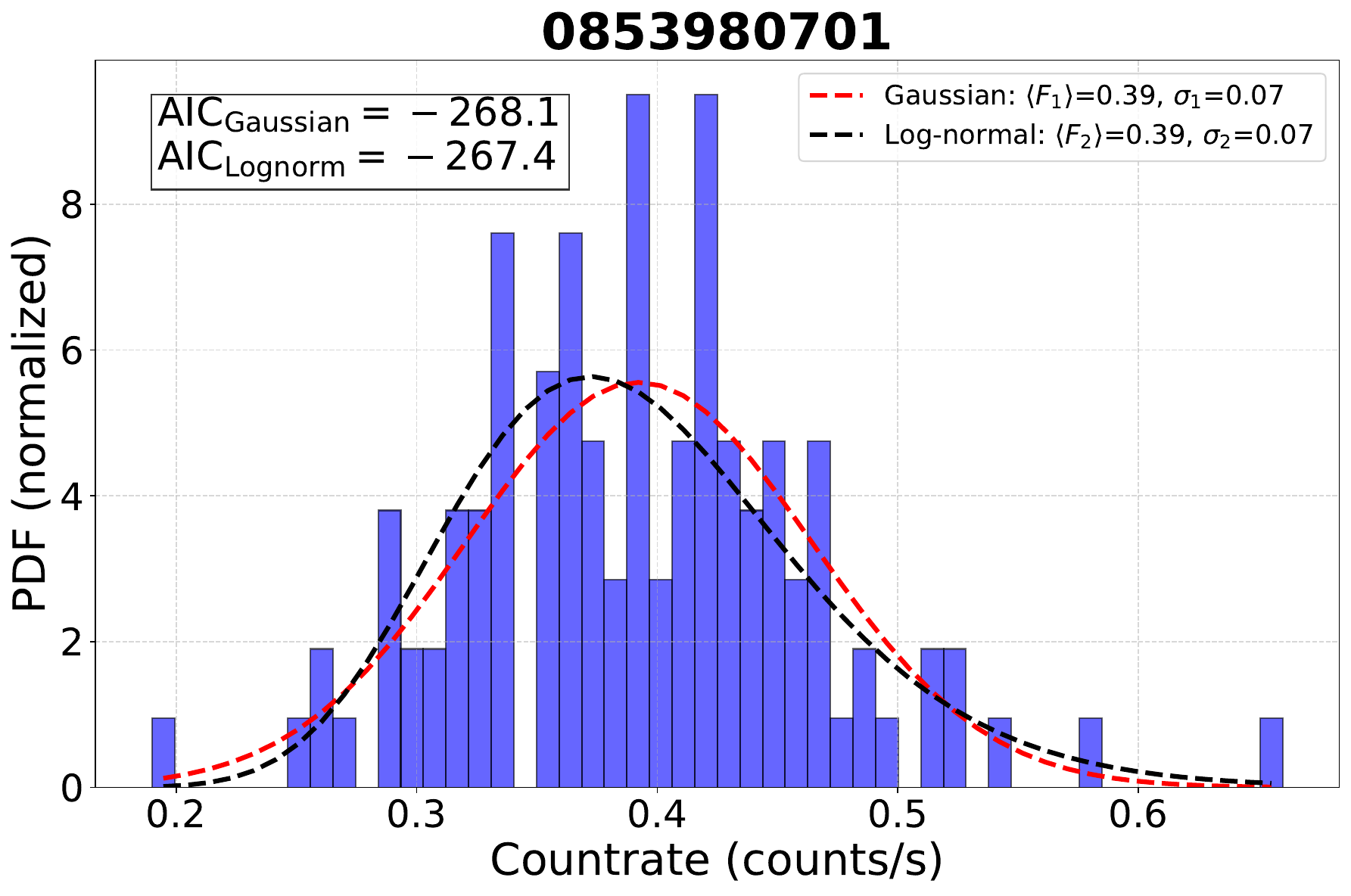}
	\end{minipage}
	\begin{minipage}{.3\textwidth} 
		\centering 
		\includegraphics[height=.99\linewidth, angle=-90]{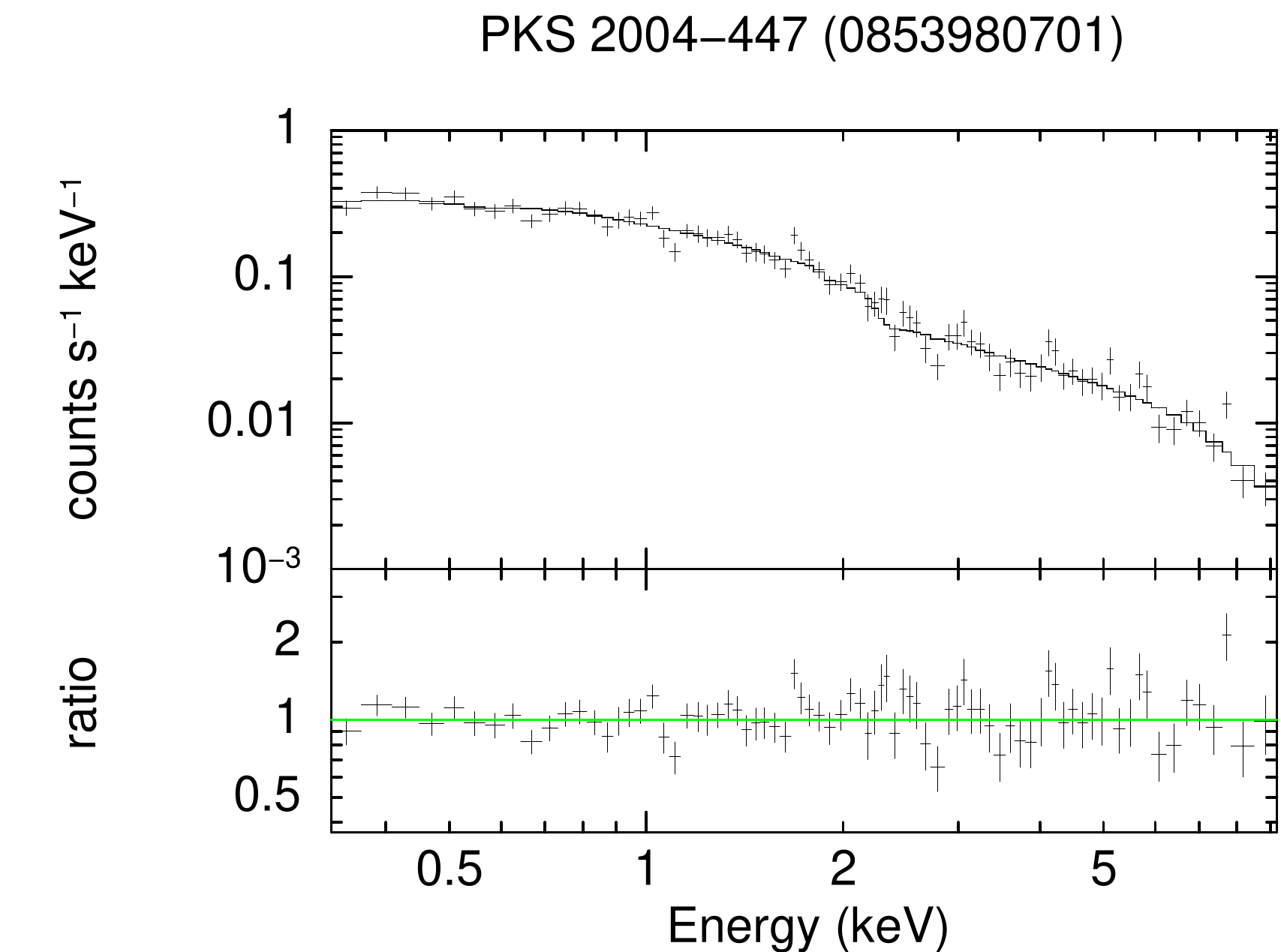}
	\end{minipage}
\end{figure*}

\begin{figure*}\label{app:PMN J0948+0022}
	\centering
	\caption{LCs, HR plots, Structure Function, PSD, PDF, and spectral fits derived from observations of PMN J0948+0022.}
	\begin{minipage}{.3\textwidth} 
		\centering 
		\includegraphics[width=.99\linewidth]{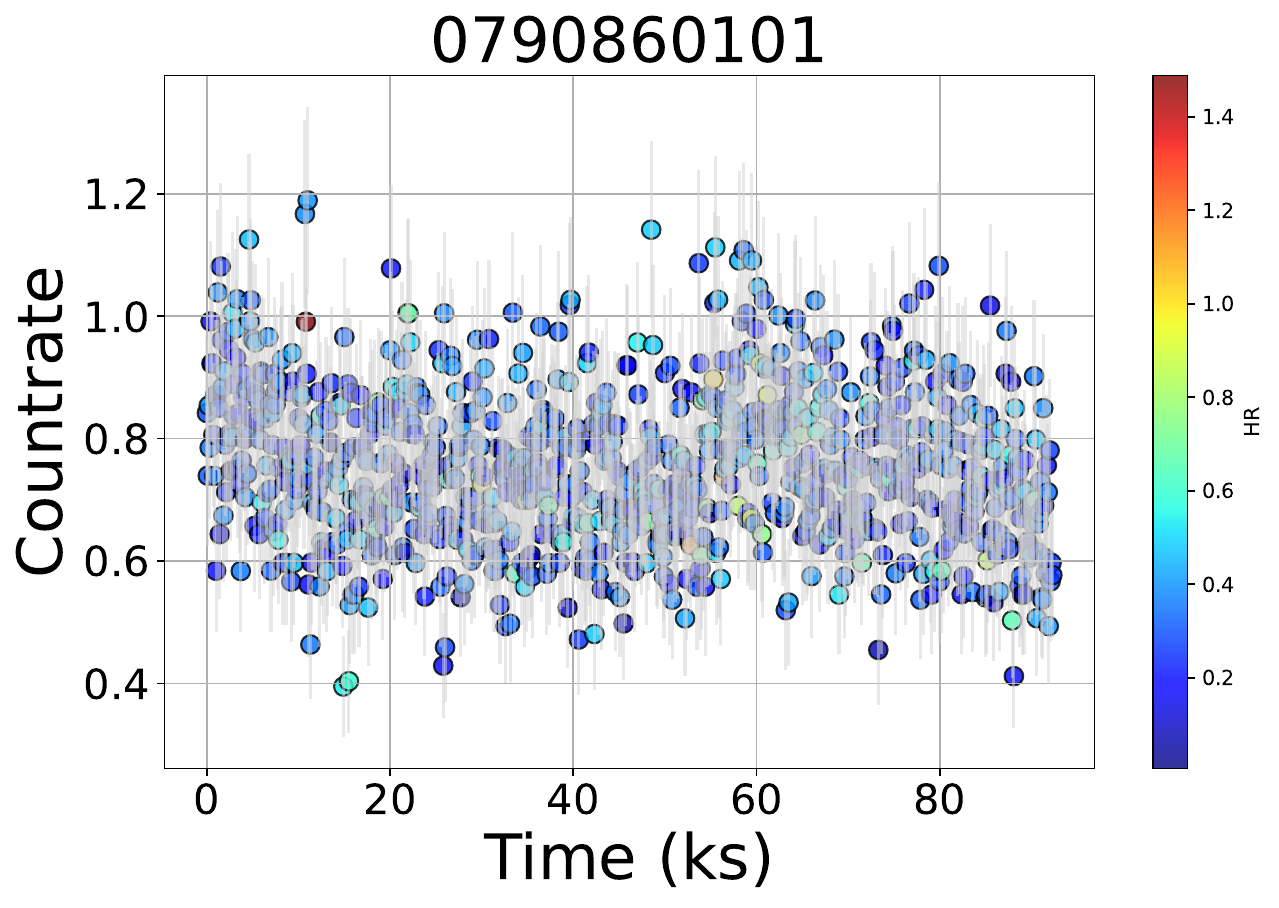}
	\end{minipage}
	\begin{minipage}{.3\textwidth} 
		\centering 
		\includegraphics[width=.99\linewidth]{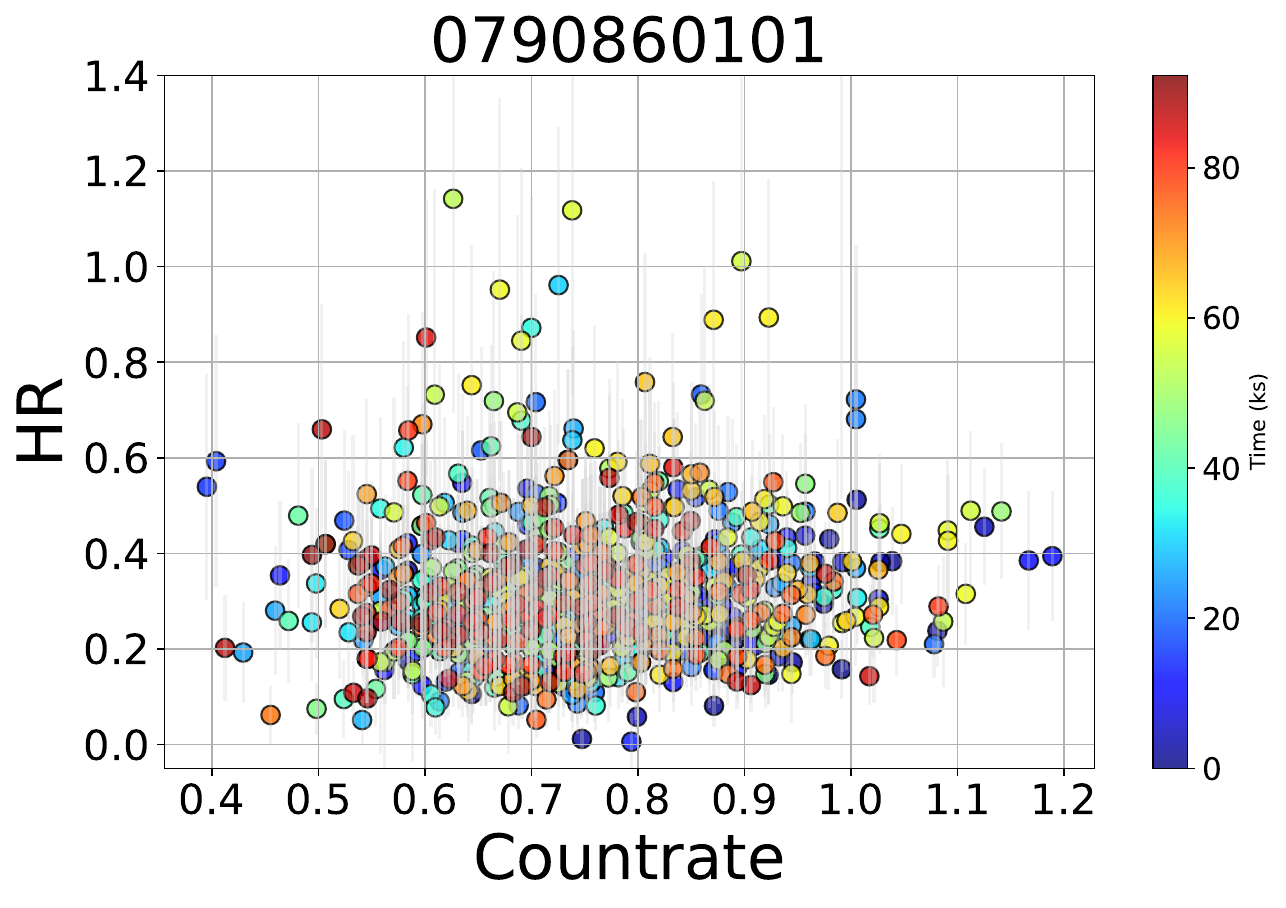}
	\end{minipage}
	\begin{minipage}{.3\textwidth} 
		\centering 
		\includegraphics[width=.99\linewidth, angle=0]{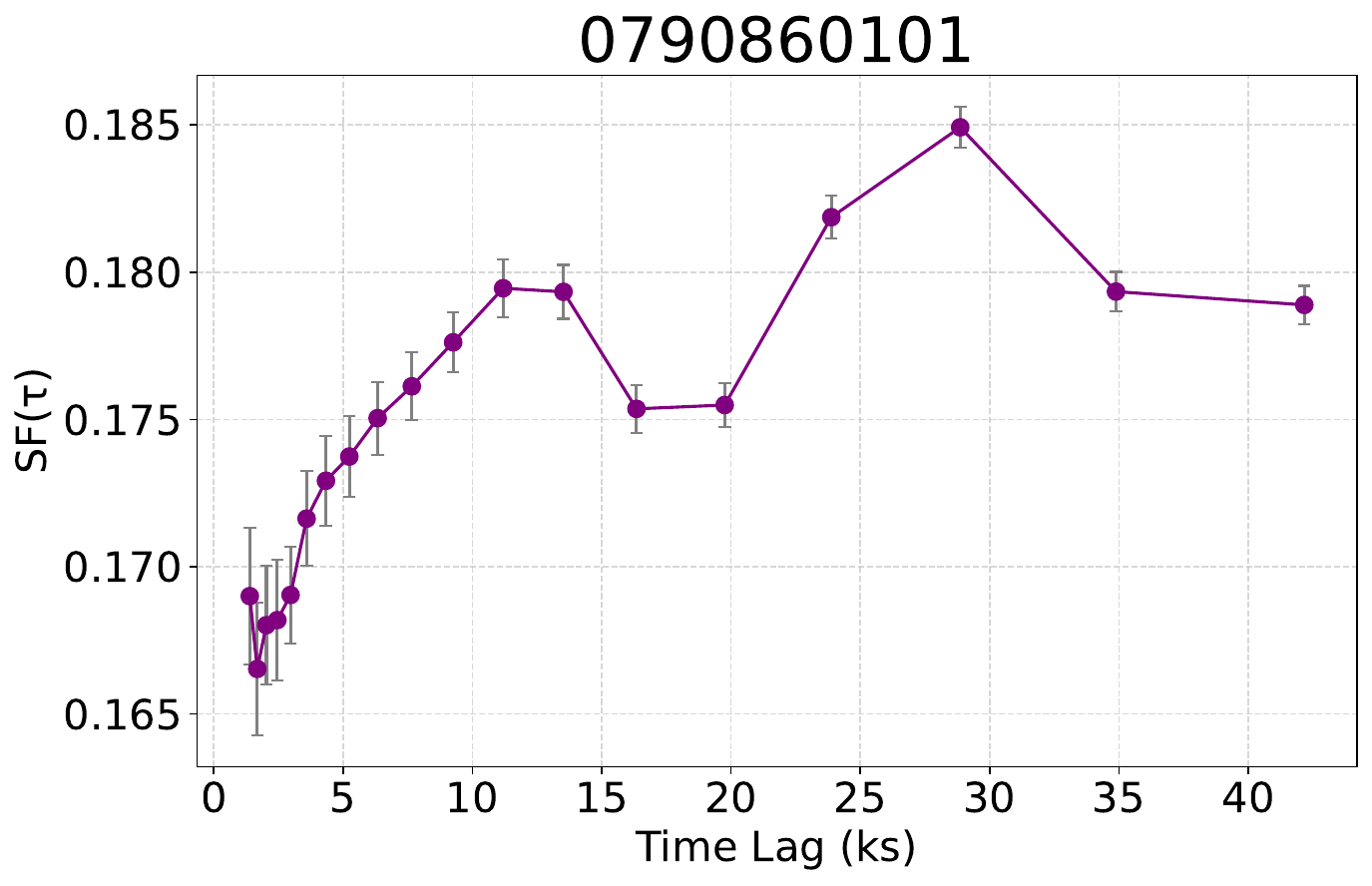}
	\end{minipage}
	\begin{minipage}{.3\textwidth} 
		\centering 
		\includegraphics[width=.99\linewidth]{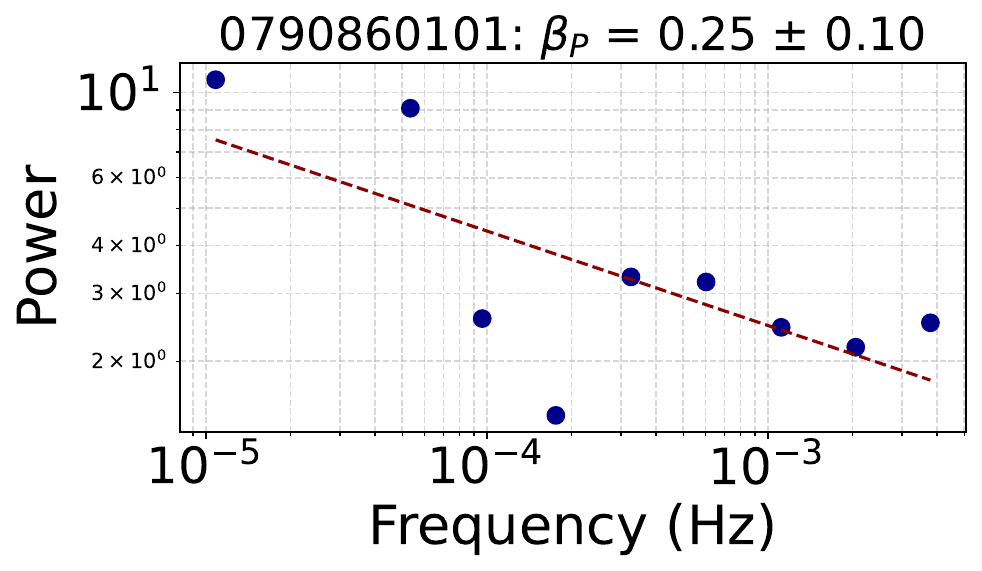}
	\end{minipage}
	\begin{minipage}{.3\textwidth} 
		\centering 
		\includegraphics[width=.99\linewidth]{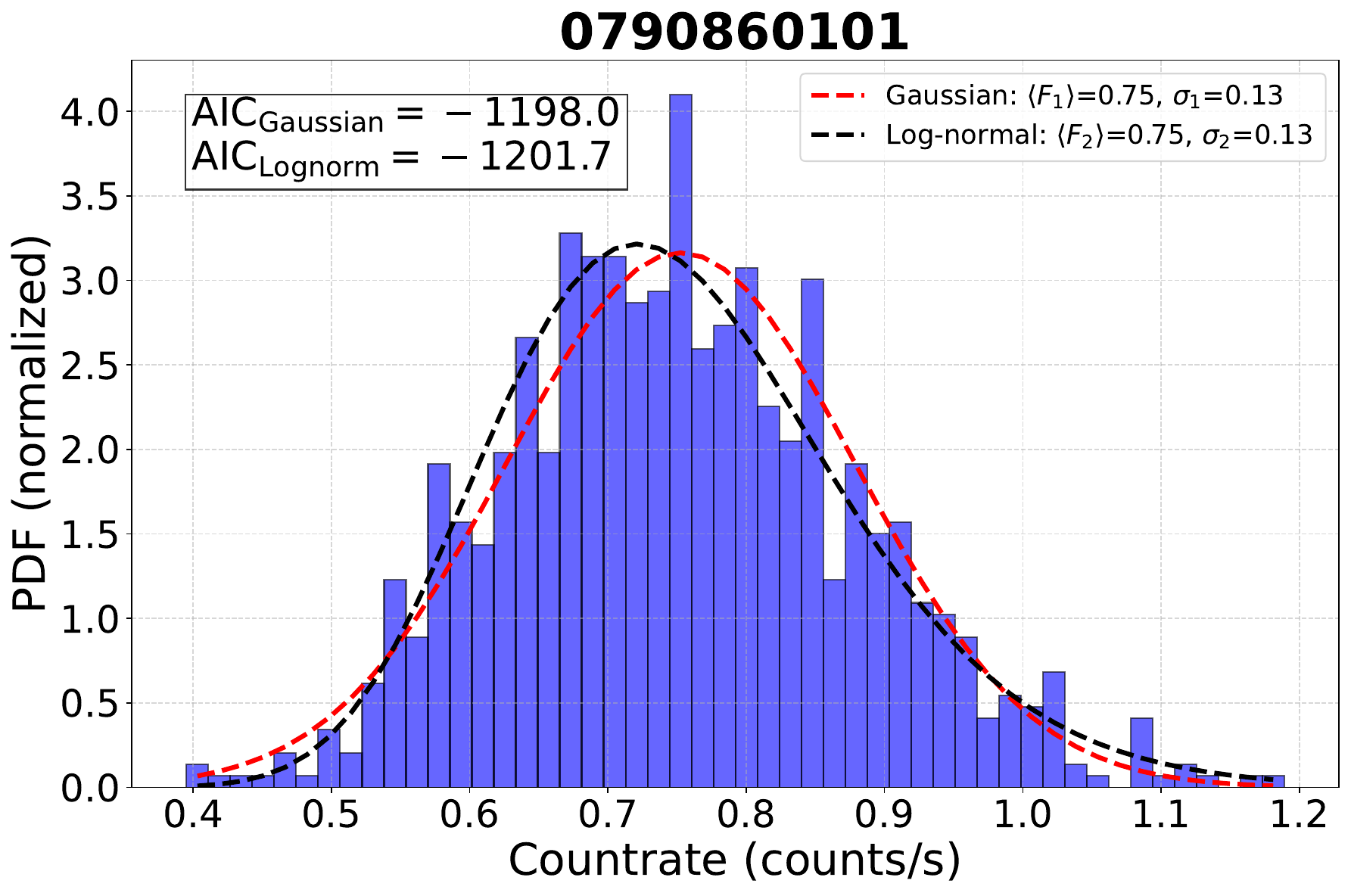}
	\end{minipage}
	\begin{minipage}{.3\textwidth} 
		\centering 
		\includegraphics[height=.99\linewidth, angle=-90]{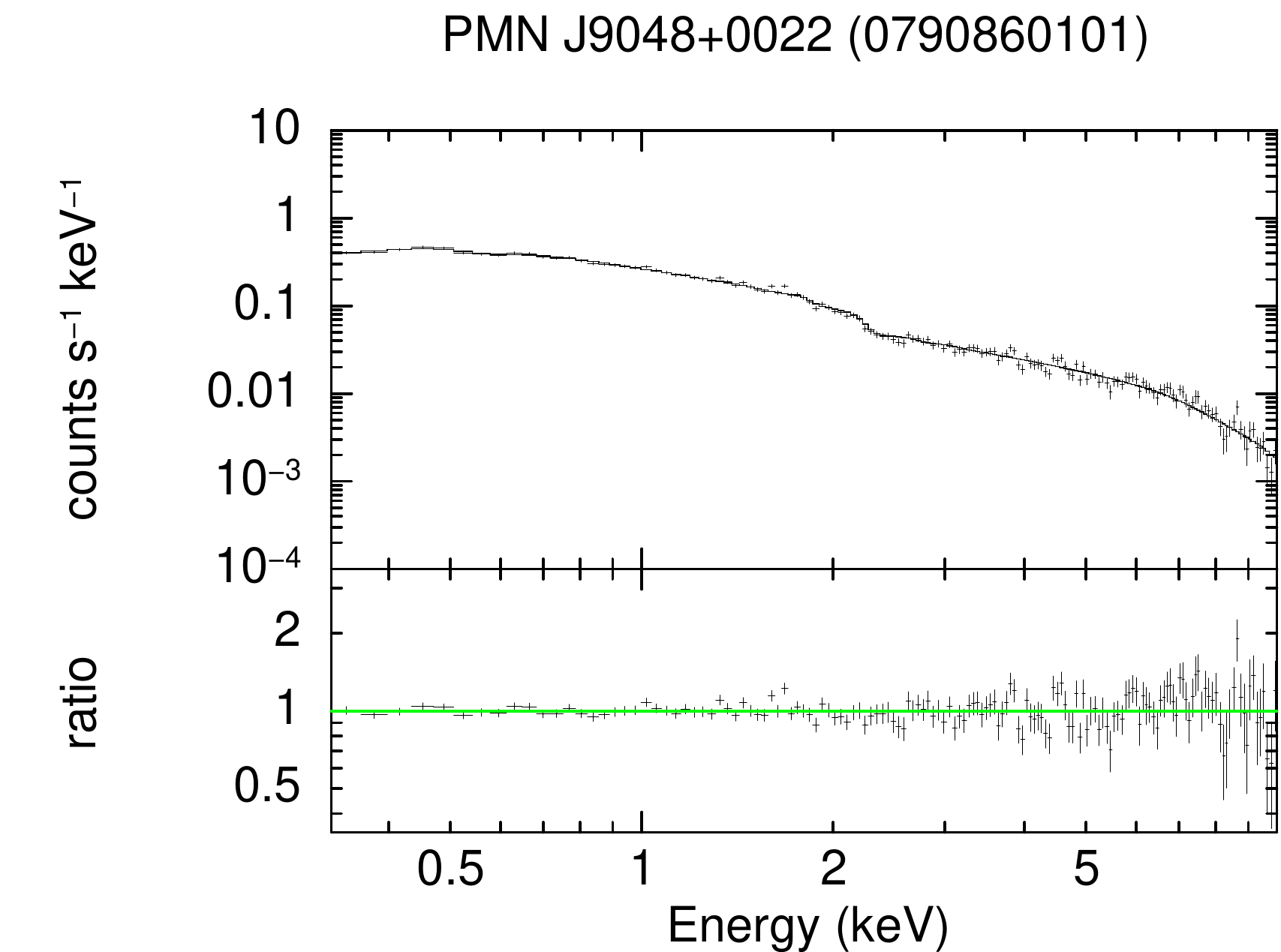}
	\end{minipage}
	\begin{minipage}{.3\textwidth} 
		\centering 
		\includegraphics[width=.99\linewidth]{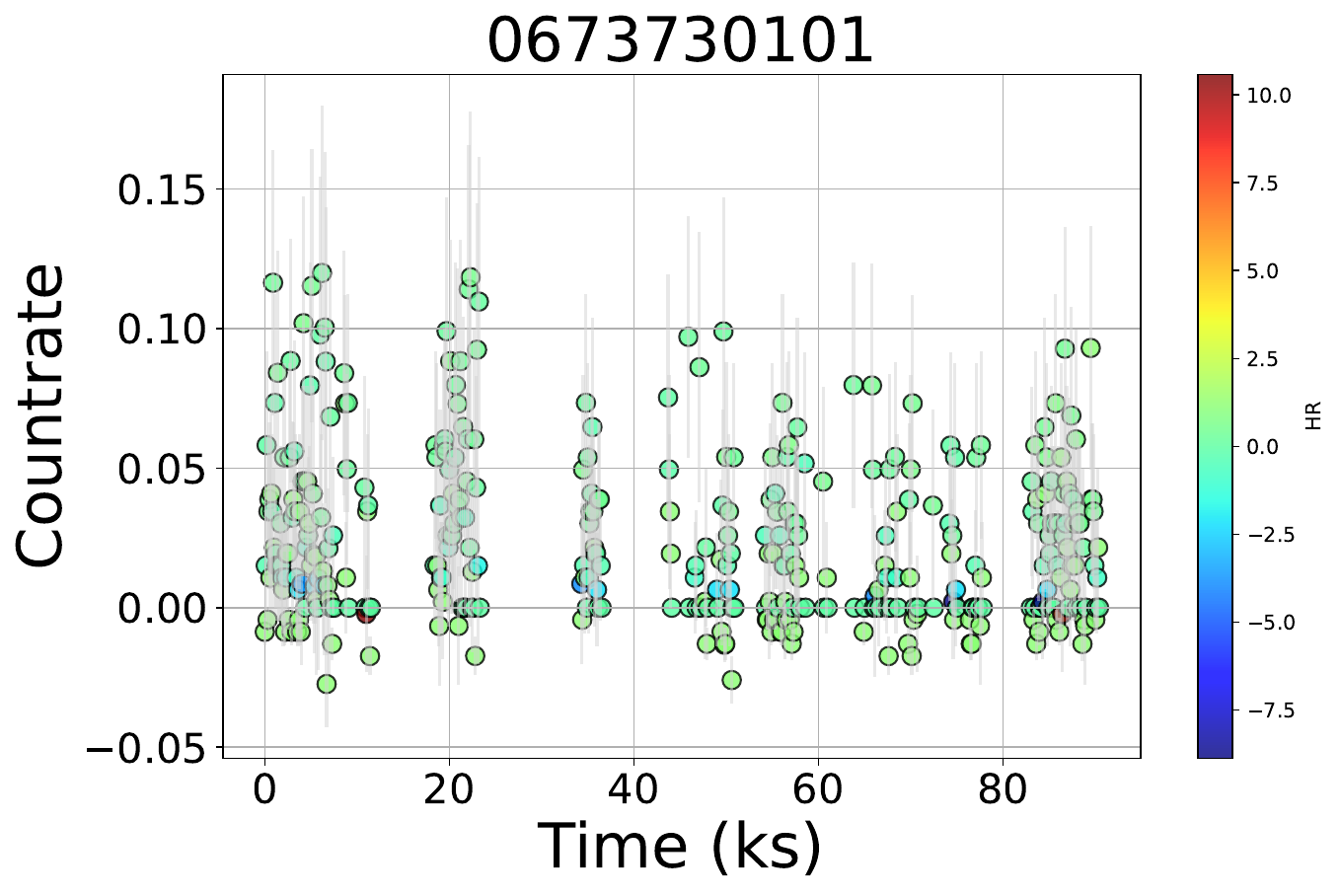}
	\end{minipage}
	\begin{minipage}{.3\textwidth} 
		\centering 
		\includegraphics[width=.99\linewidth]{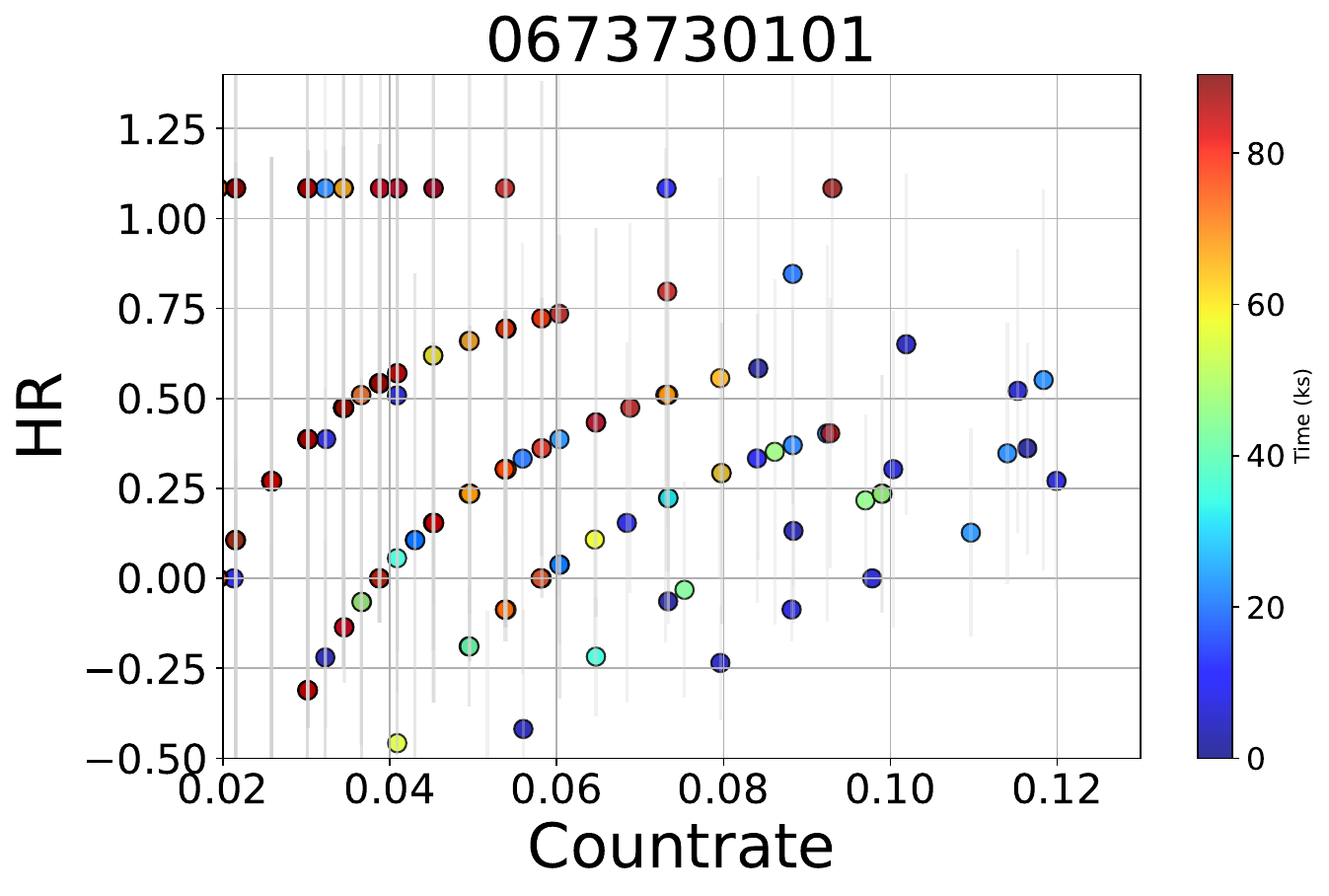}
	\end{minipage}
	\begin{minipage}{.3\textwidth} 
		\centering 
		\includegraphics[width=.99\linewidth, angle=0]{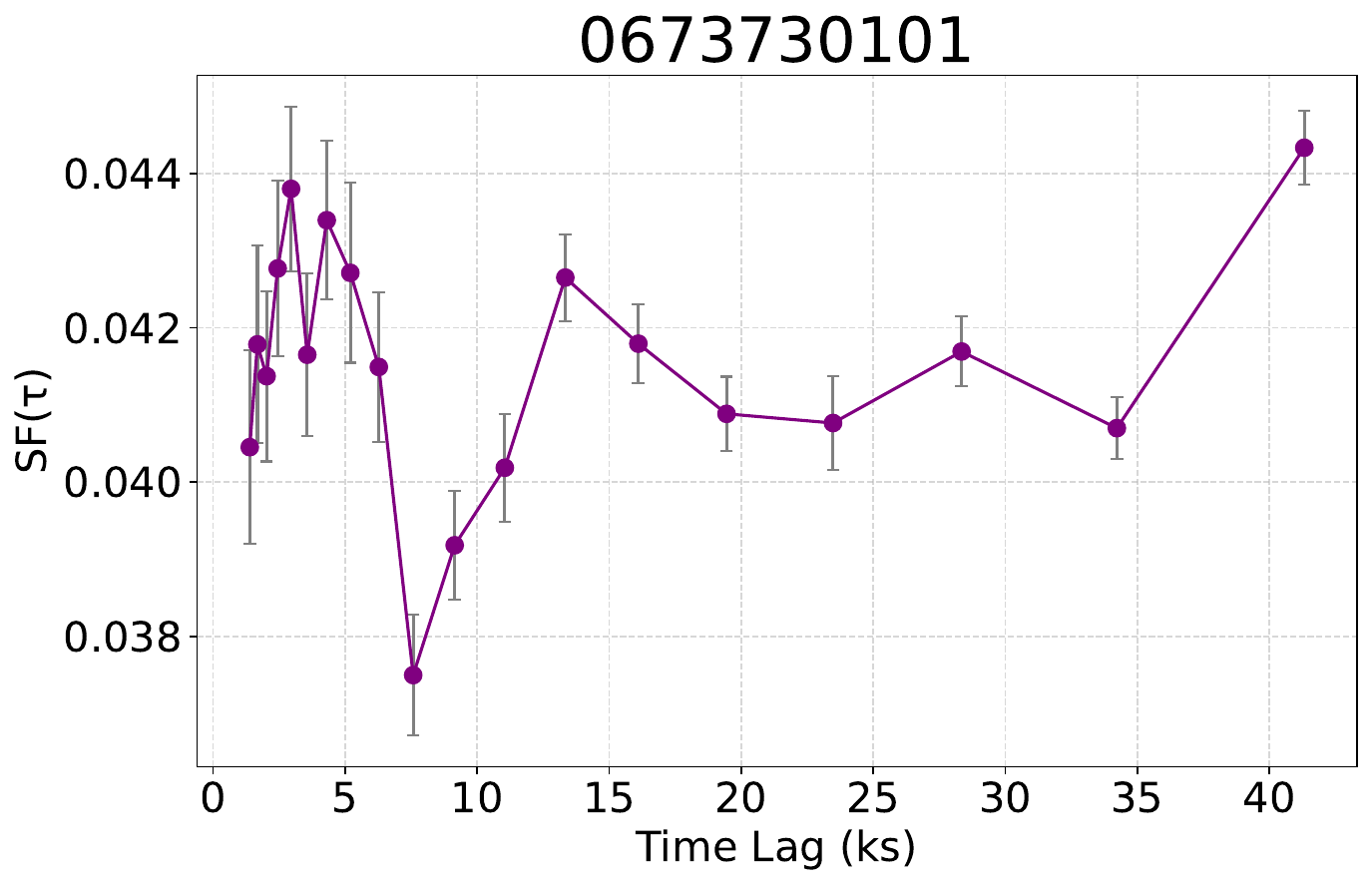}
	\end{minipage}
	\begin{minipage}{.3\textwidth} 
		\centering 
		\includegraphics[width=.99\linewidth]{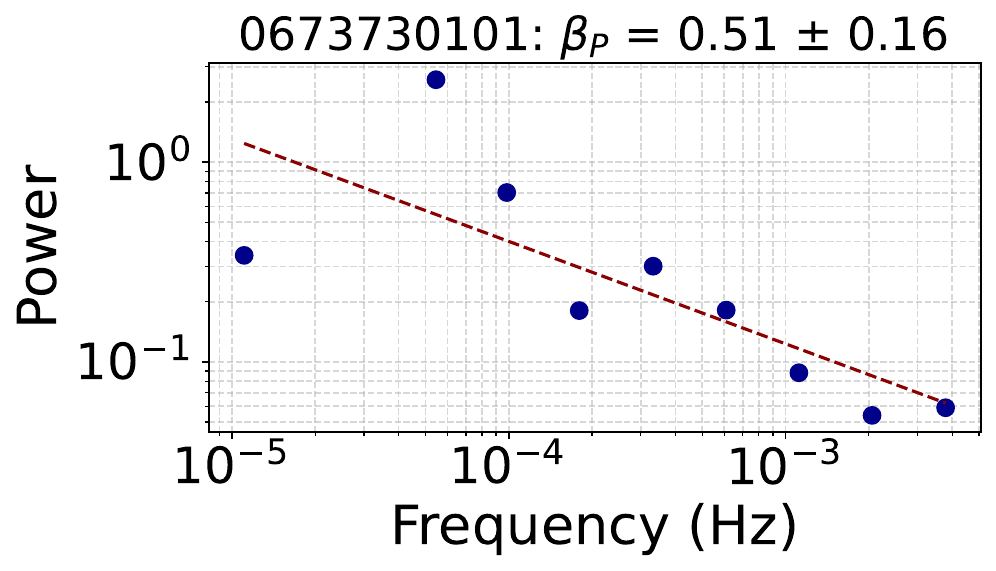}
	\end{minipage}
	\begin{minipage}{.3\textwidth} 
		\centering 
		\includegraphics[width=.99\linewidth]{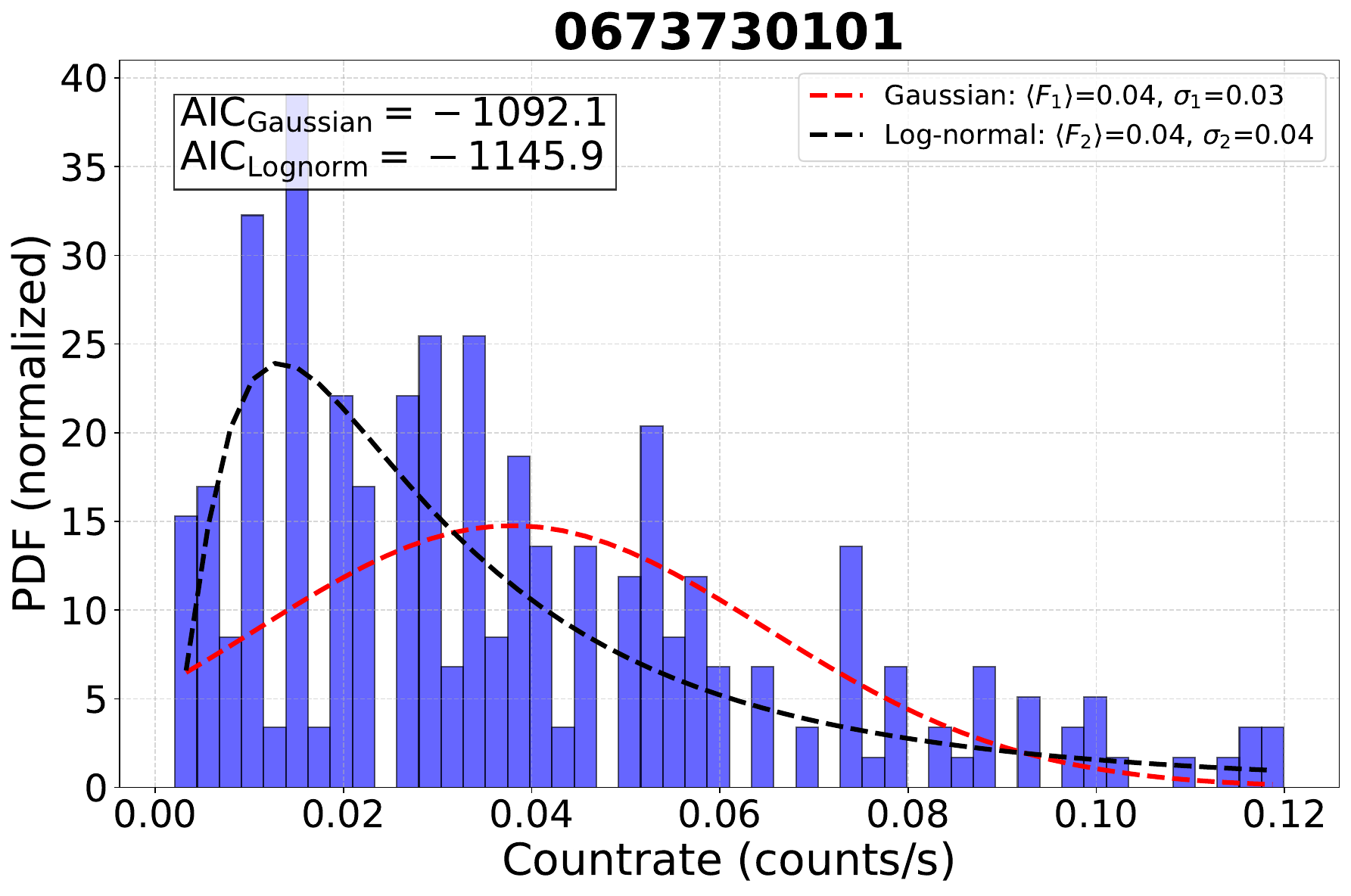}
	\end{minipage}
	\begin{minipage}{.3\textwidth} 
		\centering 
		\includegraphics[height=.99\linewidth, angle=-90]{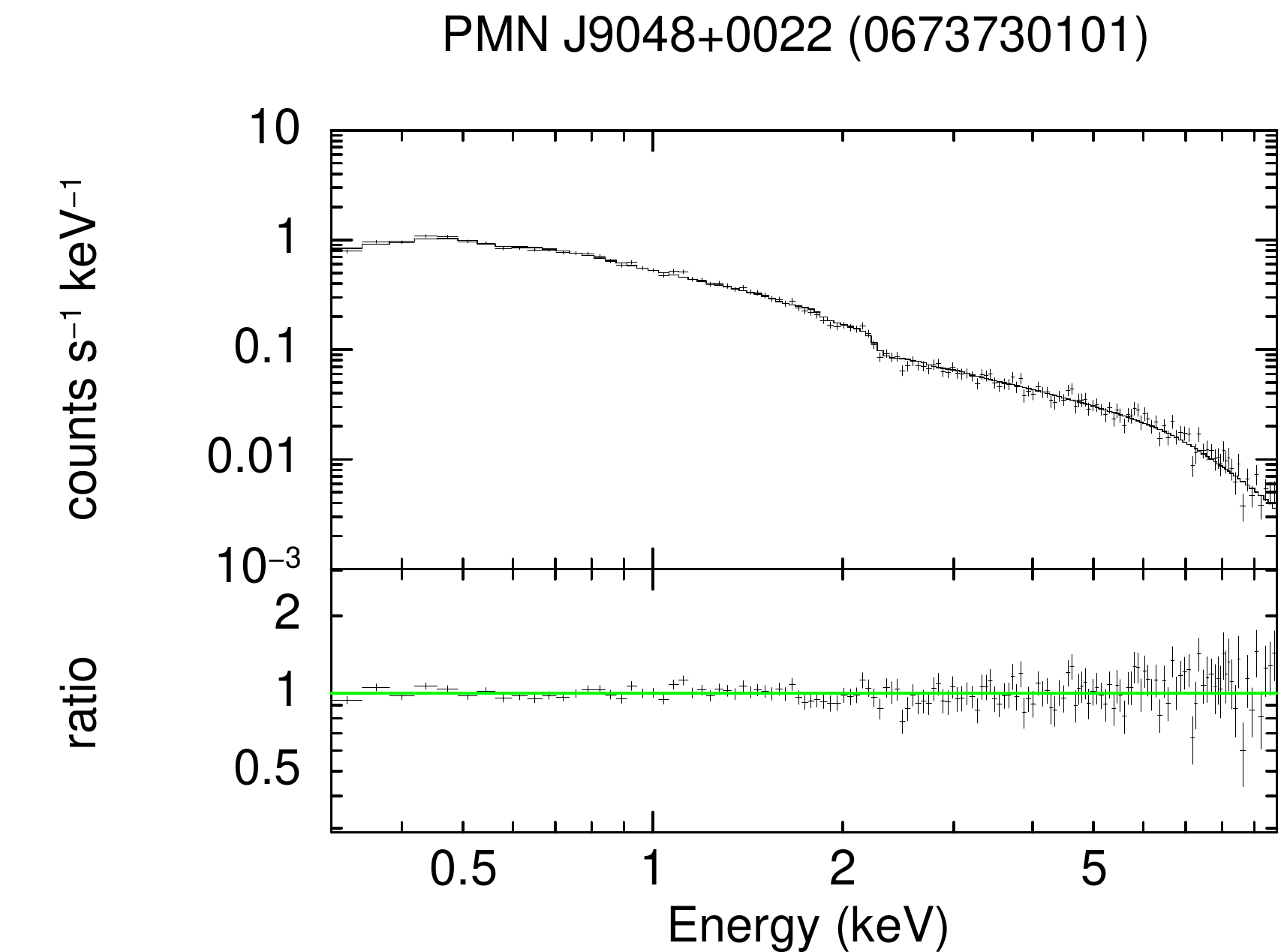}
	\end{minipage}
    	\begin{minipage}{.3\textwidth} 
		\centering 
		\includegraphics[width=.99\linewidth]{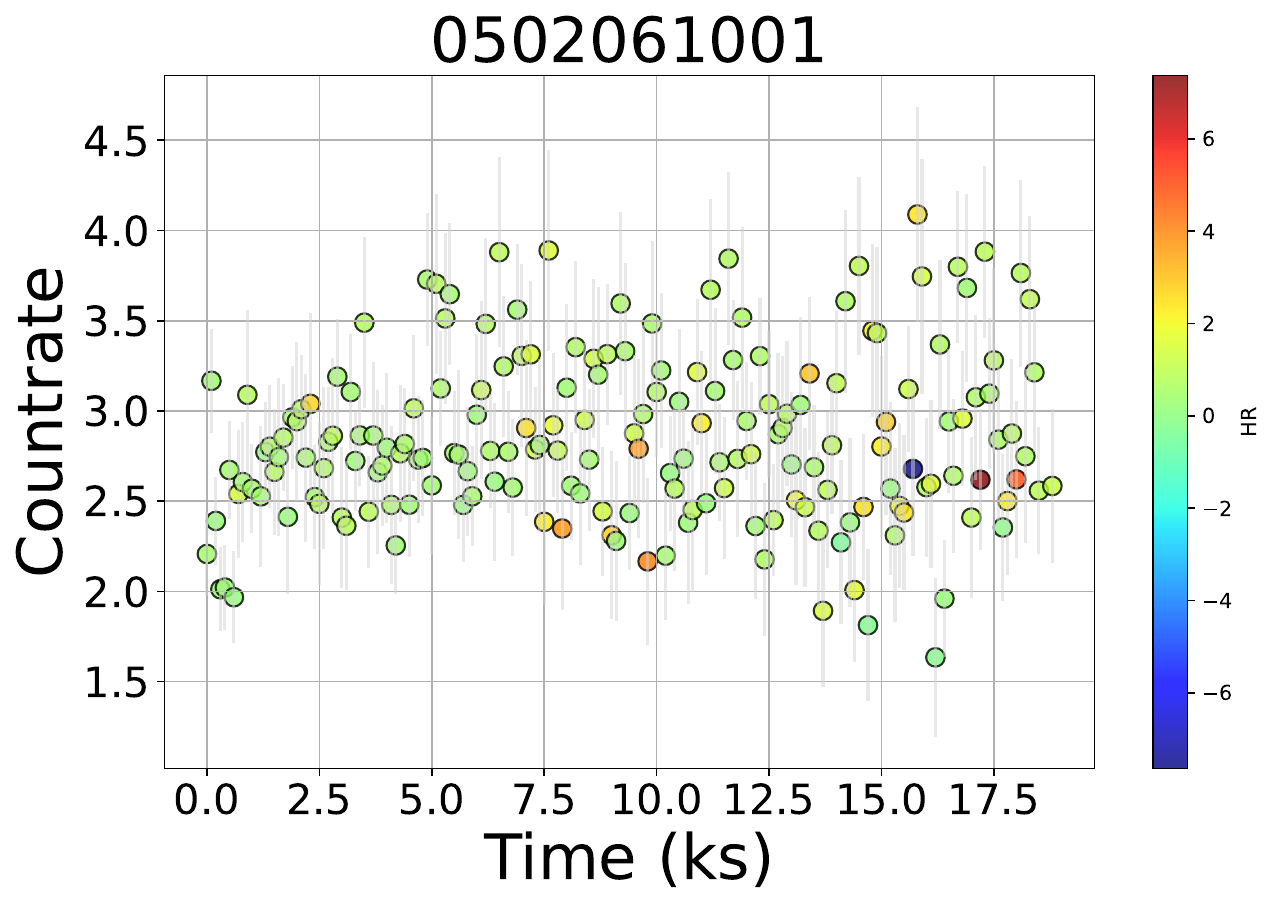}
	\end{minipage}
	\begin{minipage}{.3\textwidth} 
		\centering 
		\includegraphics[width=.99\linewidth]{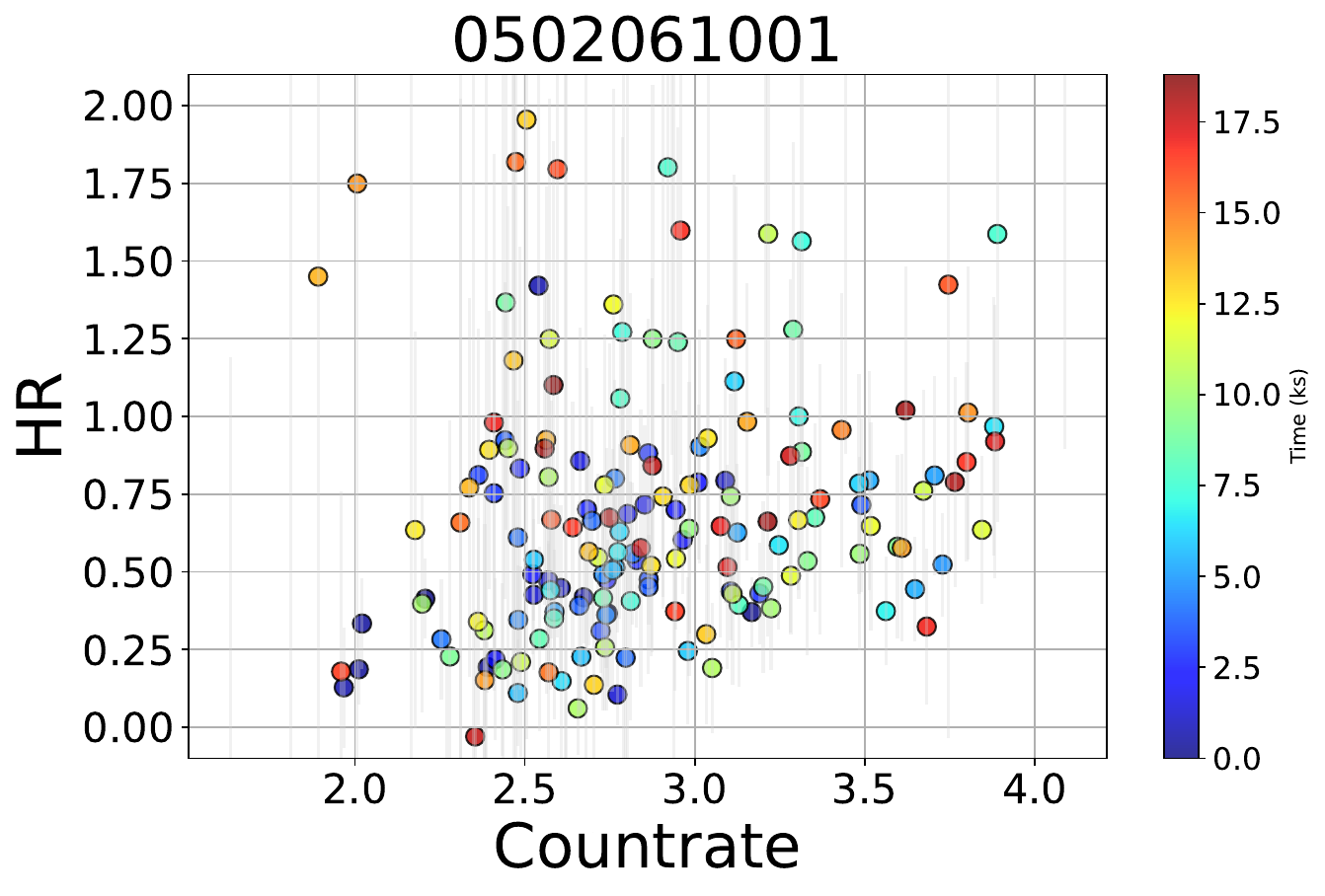}
	\end{minipage}
	\begin{minipage}{.3\textwidth} 
		\centering 
		\includegraphics[width=.99\linewidth, angle=0]{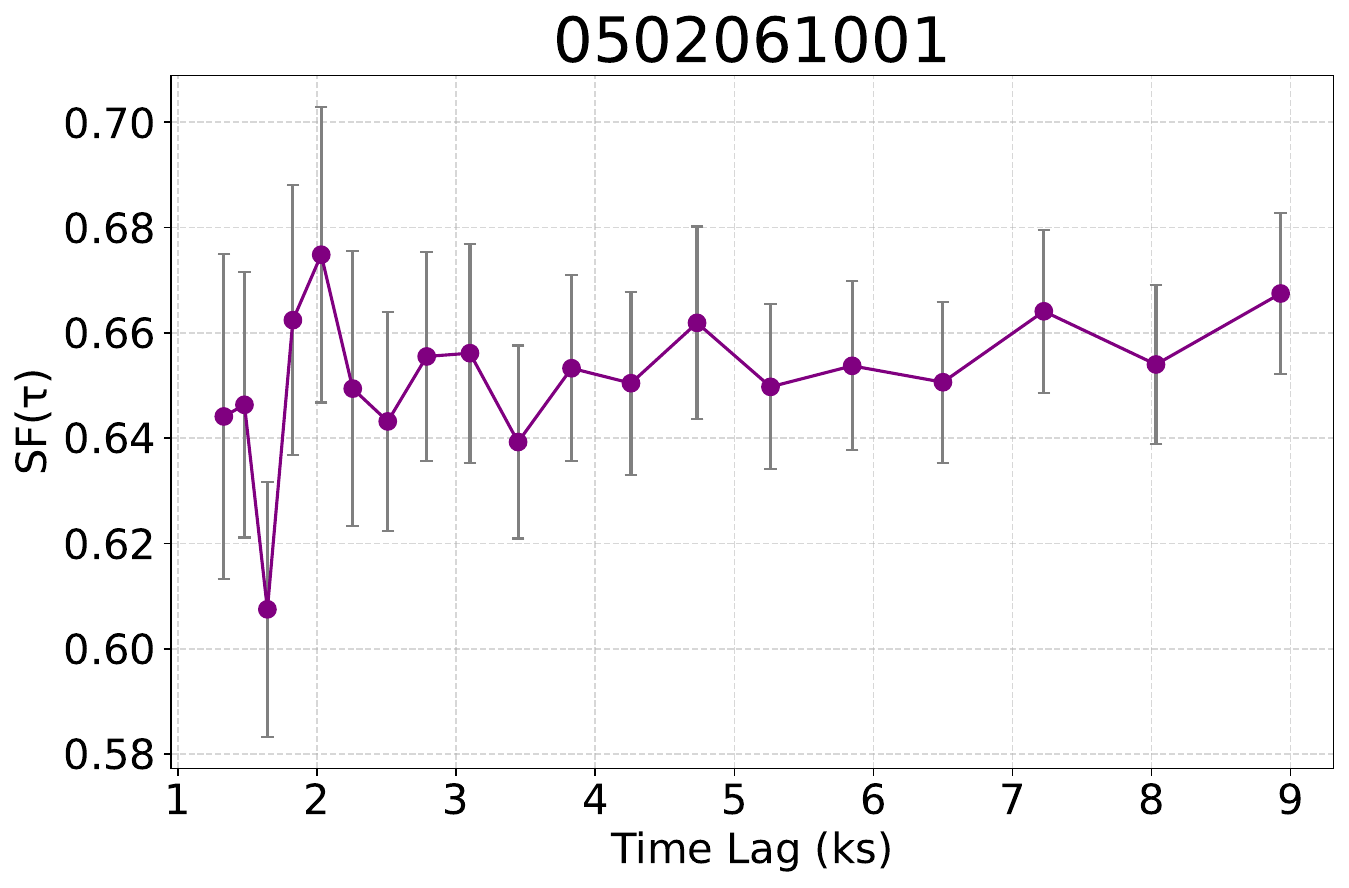}
	\end{minipage}
    \begin{minipage}{.3\textwidth} 
		\centering 
		\includegraphics[width=.99\linewidth]{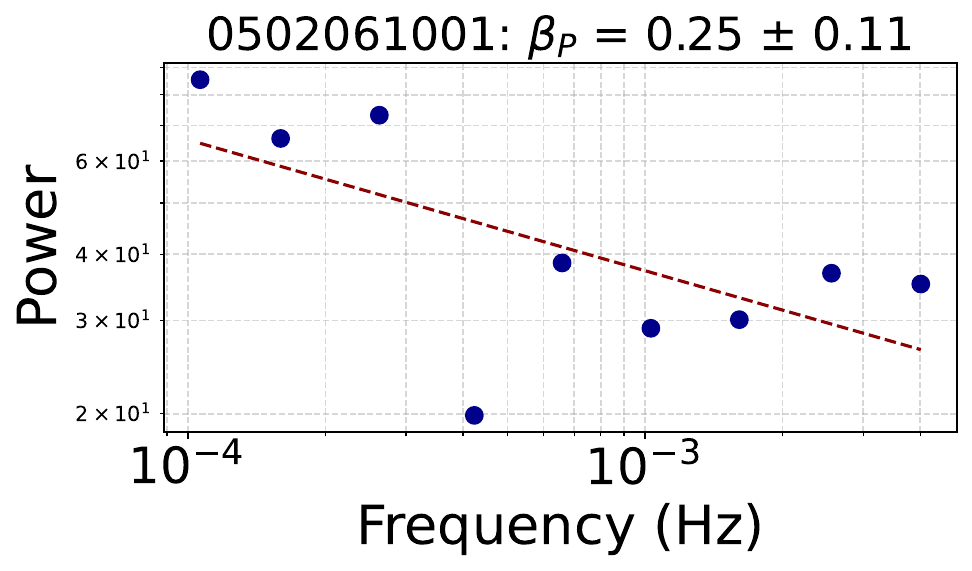}
	\end{minipage}
	\begin{minipage}{.3\textwidth} 
		\centering 
		\includegraphics[width=.99\linewidth]{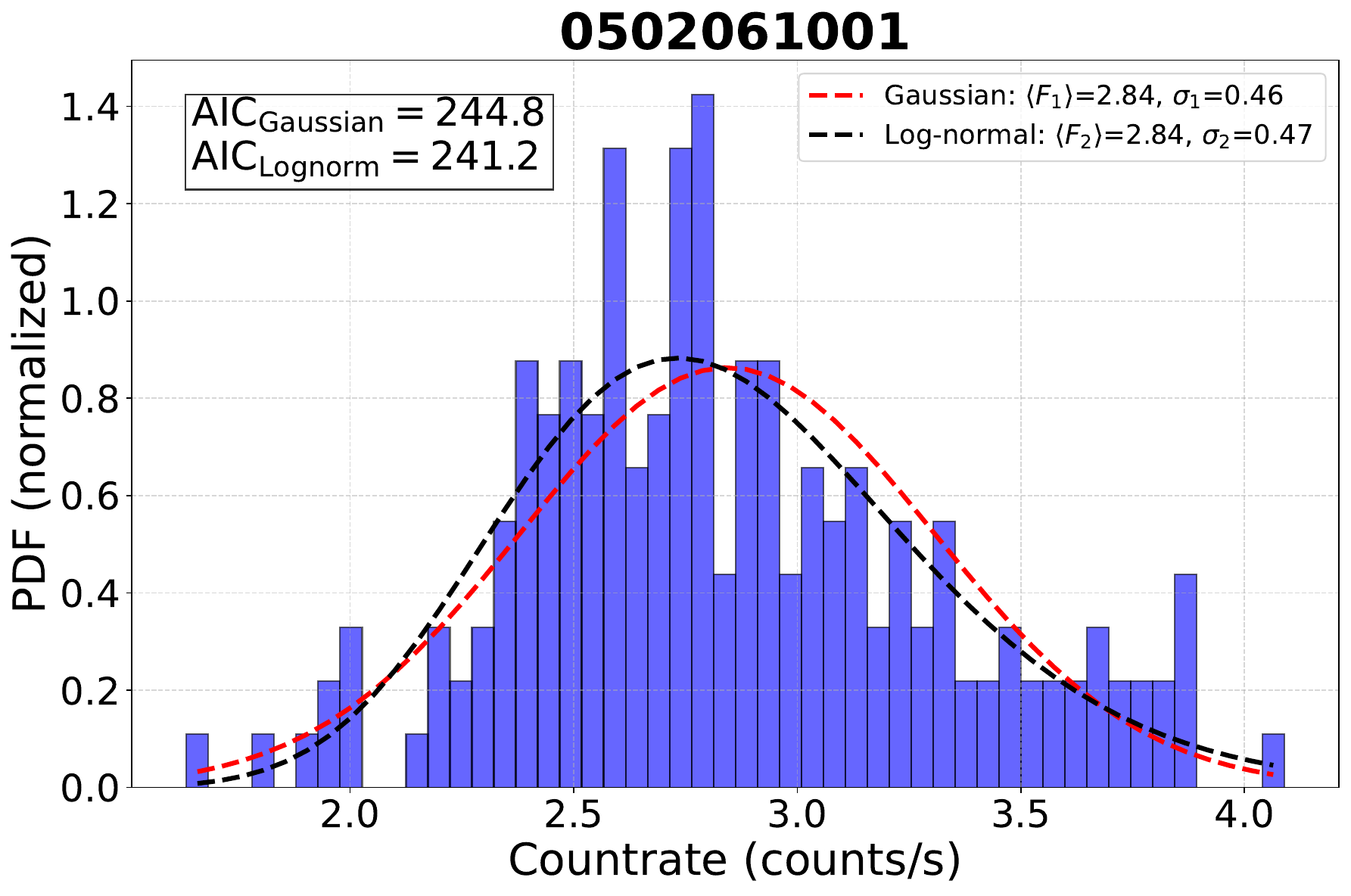}
	\end{minipage}
	\begin{minipage}{.3\textwidth} 
		\centering 
		\includegraphics[height=.99\linewidth, angle=-90]{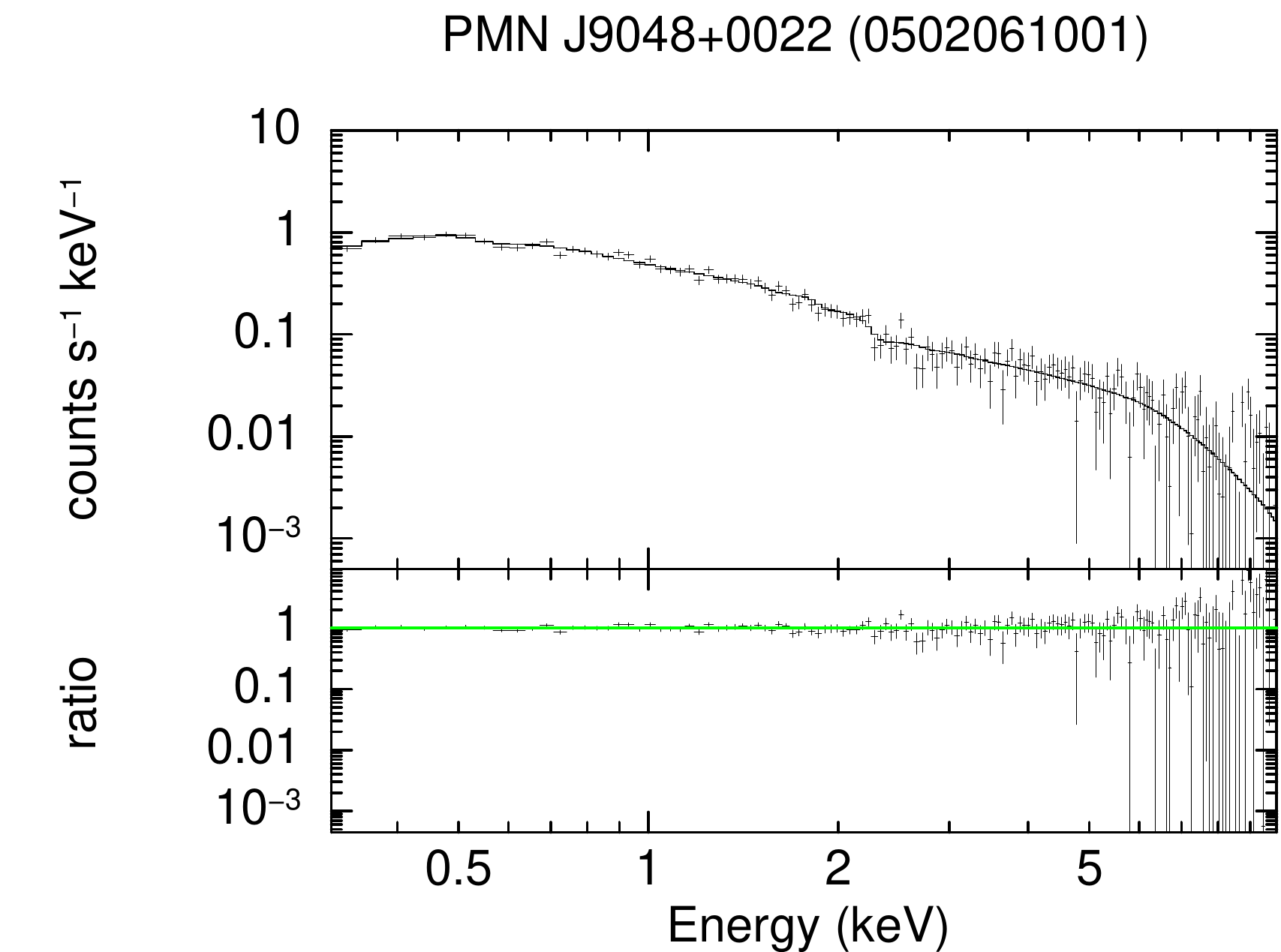}
	\end{minipage}
\end{figure*}

\clearpage
\begin{figure*}\label{app:J1641+3454}
	\centering
	\caption{LCs, HR plots, Structure Function, PSD, PDF, and spectral fits derived from observations of J1641+3454.}
	\begin{minipage}{.3\textwidth} 
		\centering 
		\includegraphics[width=.99\linewidth]{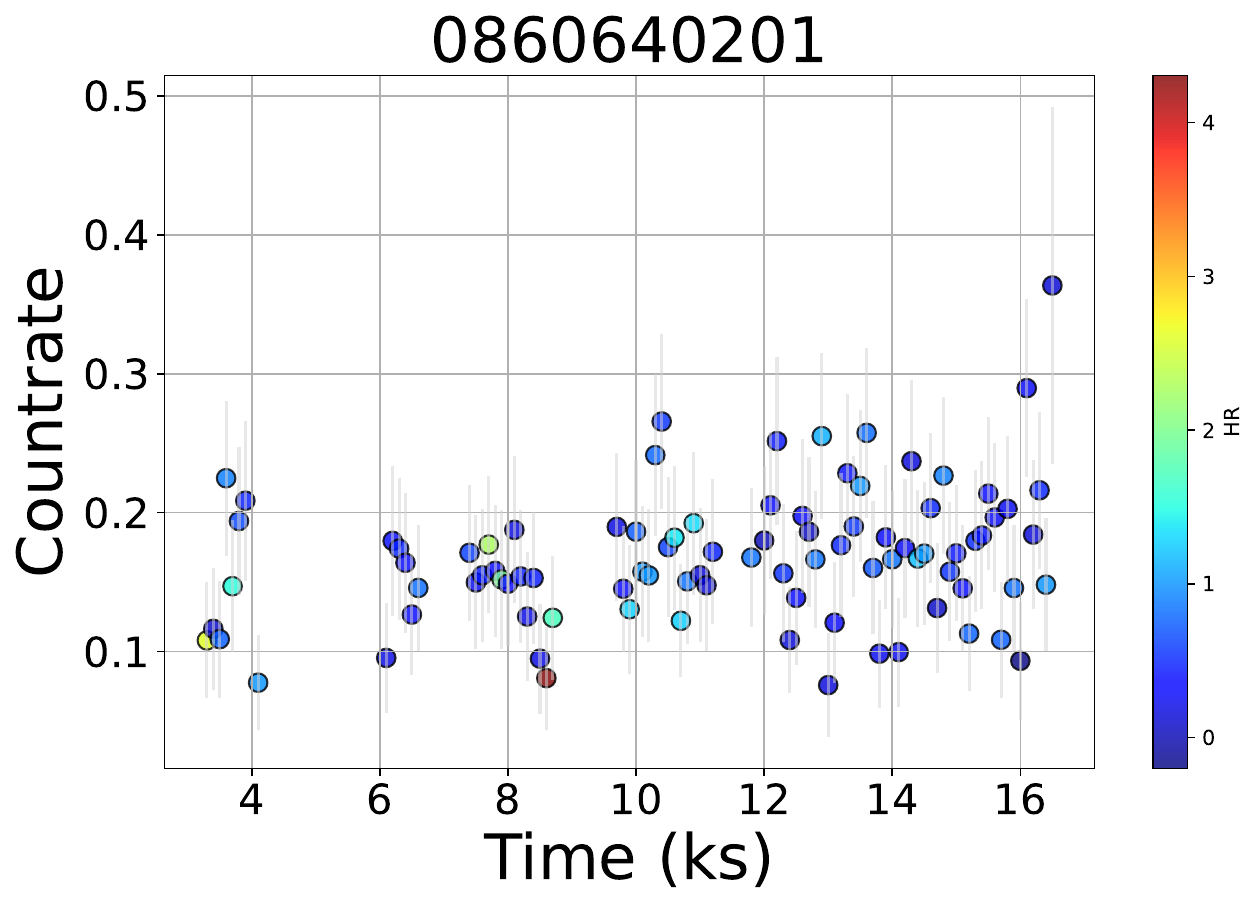}
	\end{minipage}
	\begin{minipage}{.3\textwidth} 
		\centering 
		\includegraphics[width=.99\linewidth]{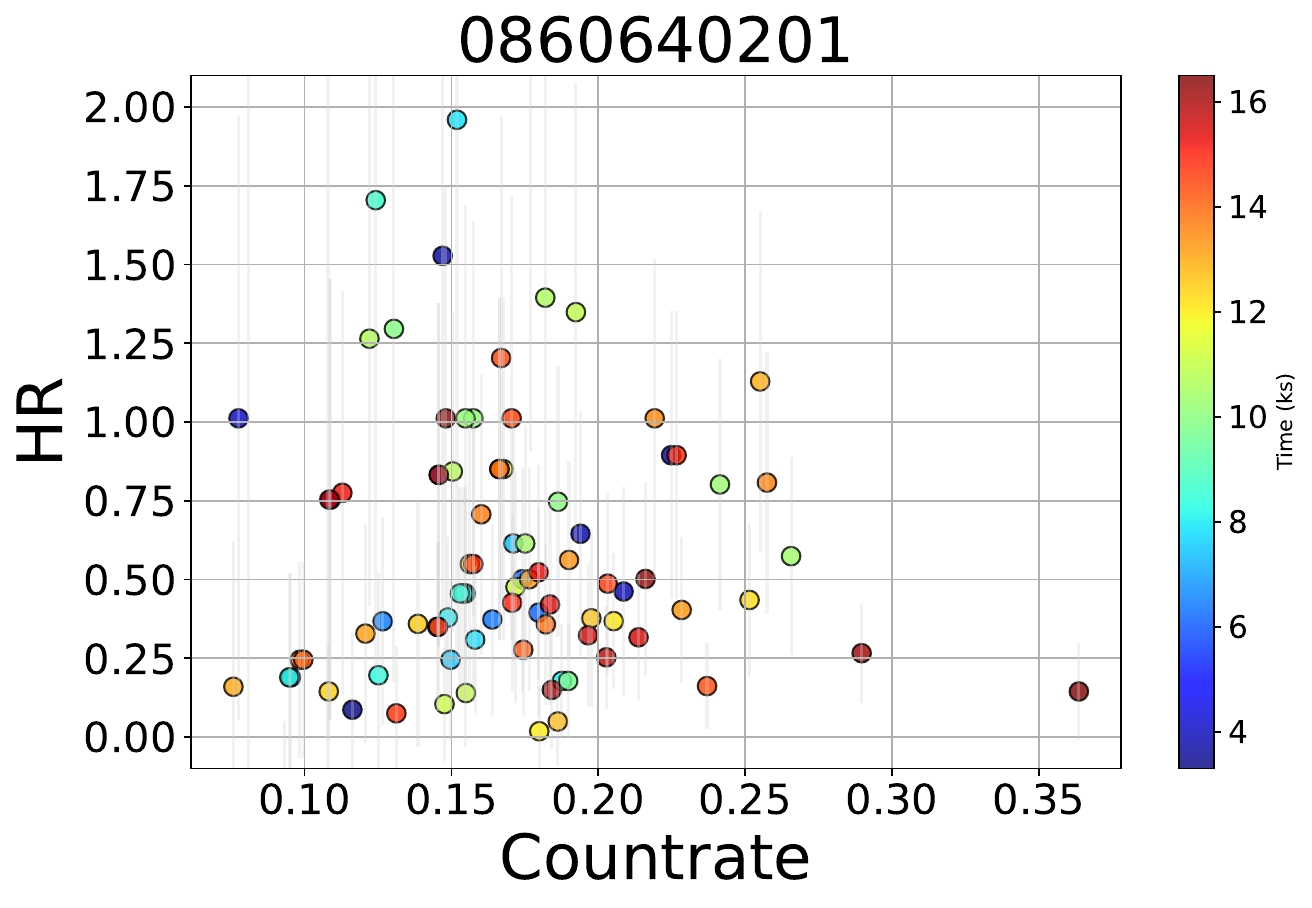}
	\end{minipage}
	\begin{minipage}{.3\textwidth} 
		\centering 
		\includegraphics[width=.99\linewidth, angle=0]{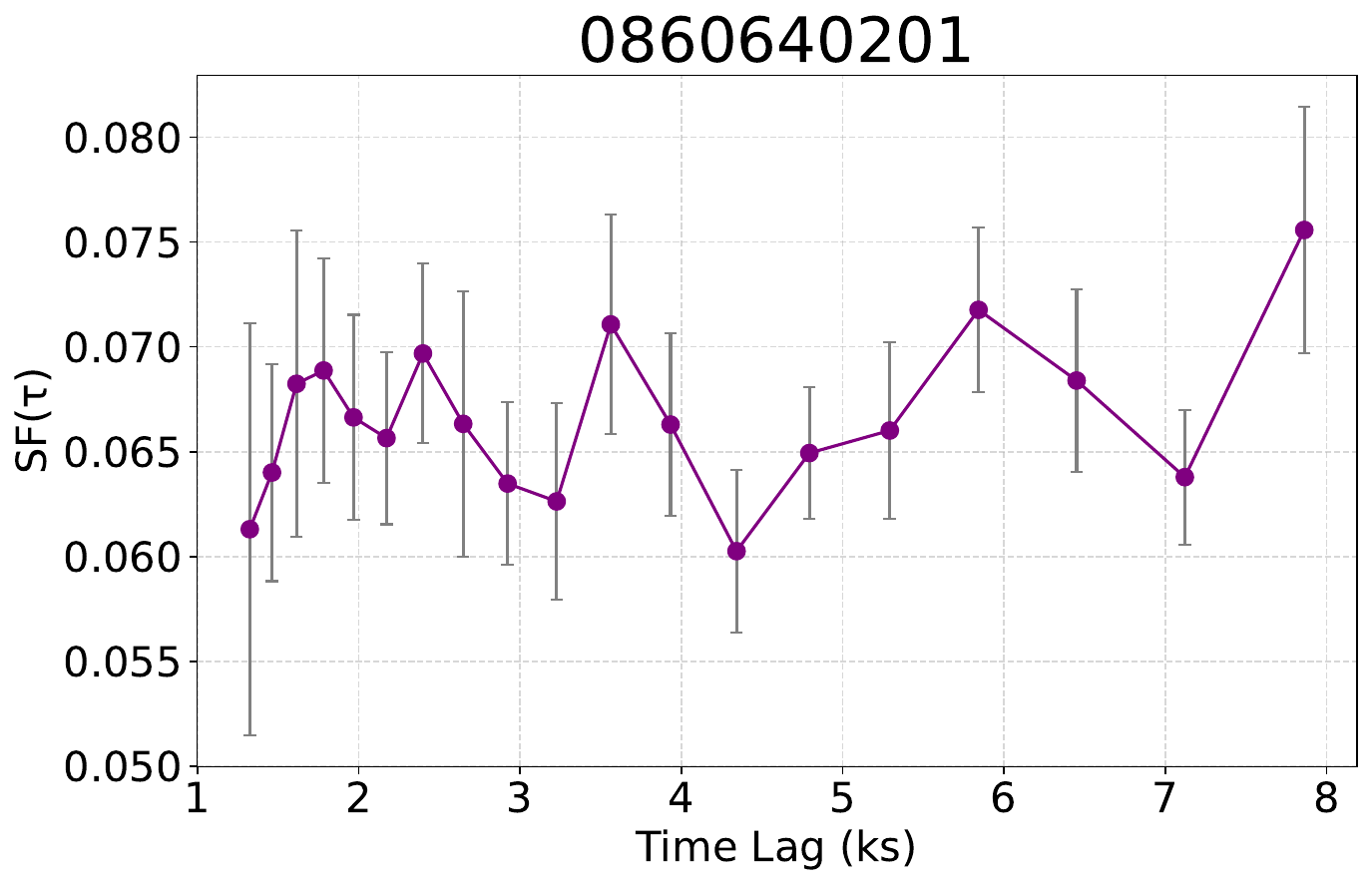}
	\end{minipage}
	\begin{minipage}{.3\textwidth} 
		\centering 
		\includegraphics[width=.99\linewidth]{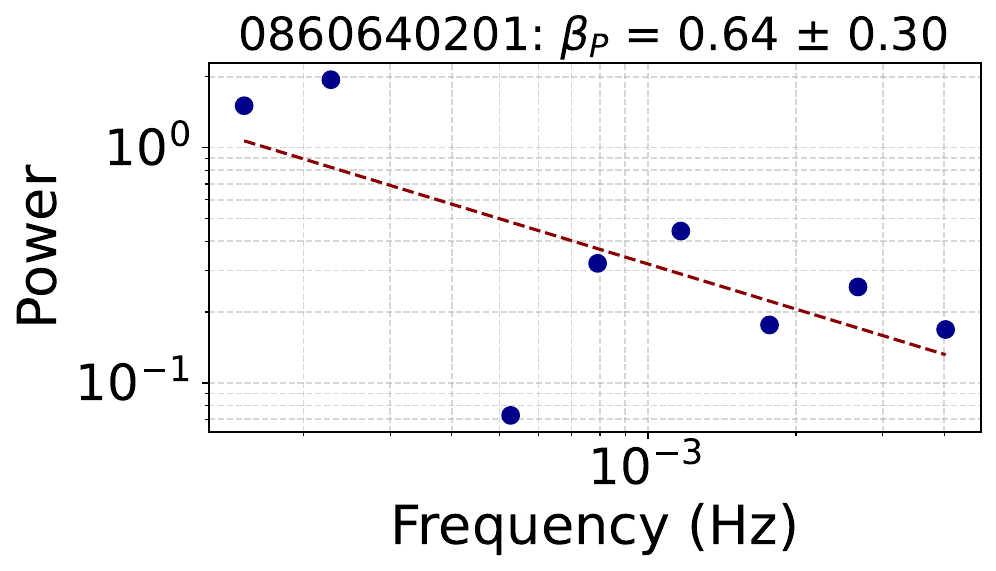}
	\end{minipage}
	\begin{minipage}{.3\textwidth} 
		\centering 
		\includegraphics[width=.99\linewidth]{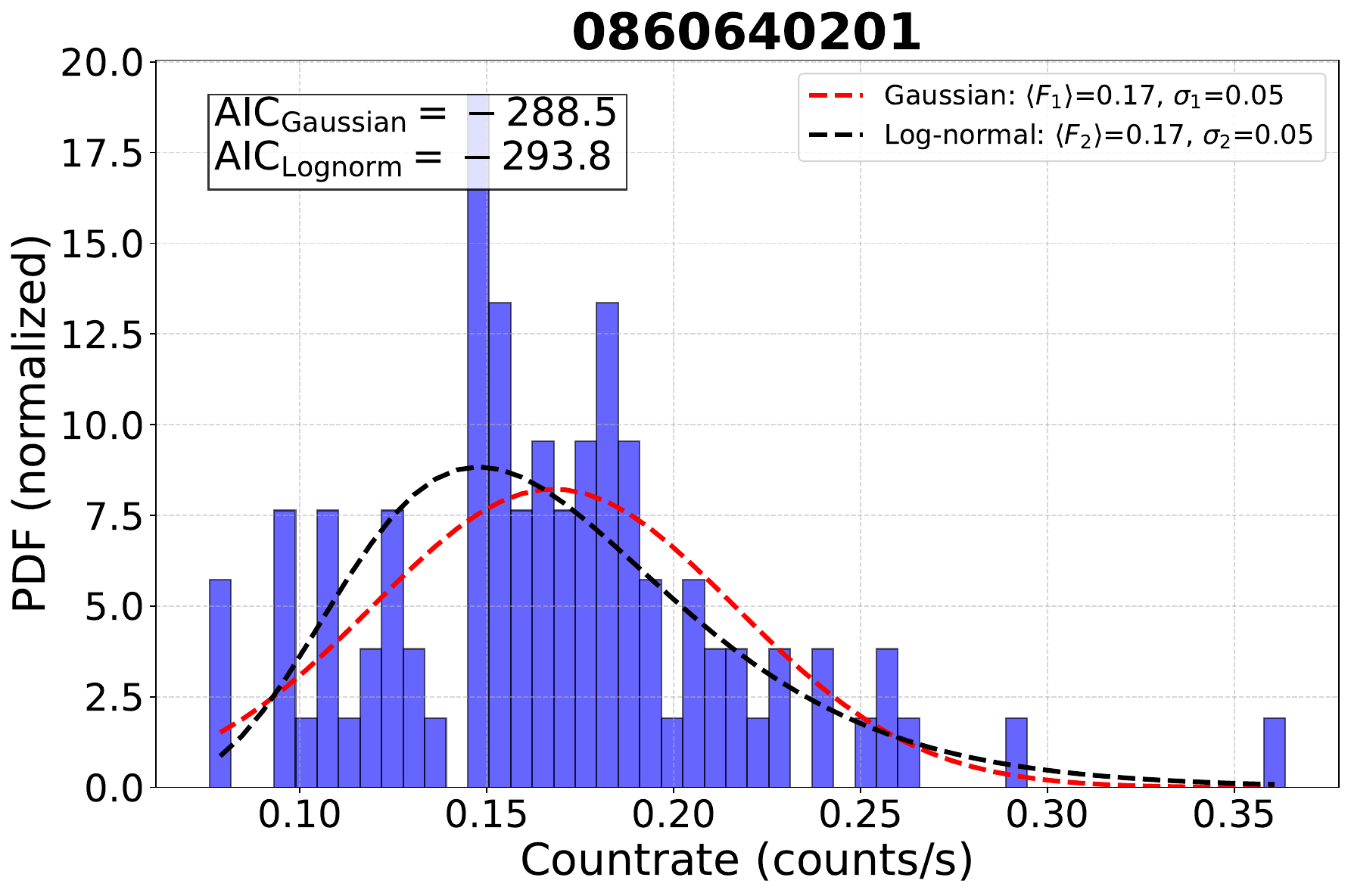}
	\end{minipage}
	\begin{minipage}{.3\textwidth} 
		\centering 
		\includegraphics[height=.99\linewidth, angle=-90]{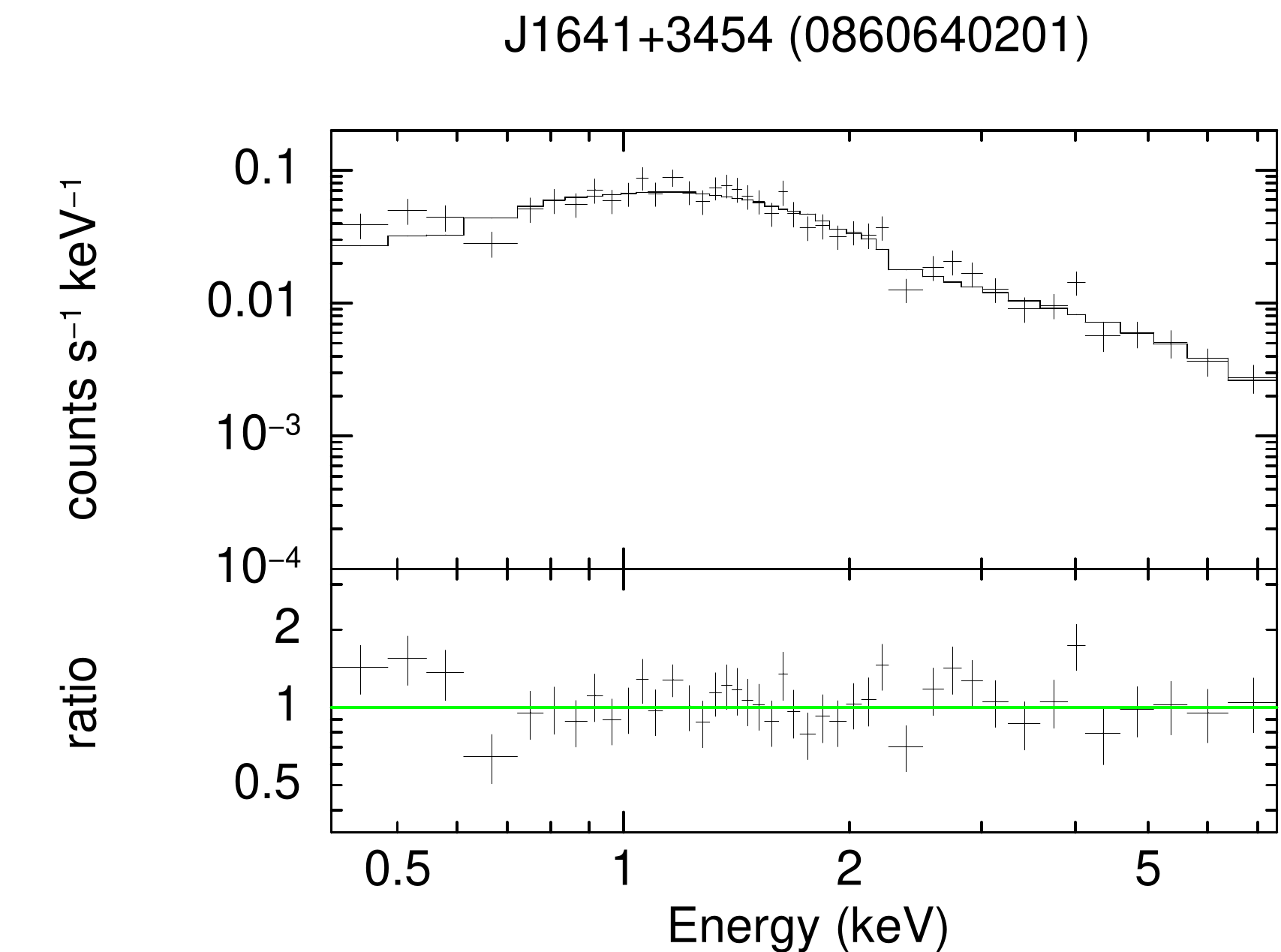}
	\end{minipage}
	\begin{minipage}{.3\textwidth} 
		\centering 
		\includegraphics[width=.99\linewidth]{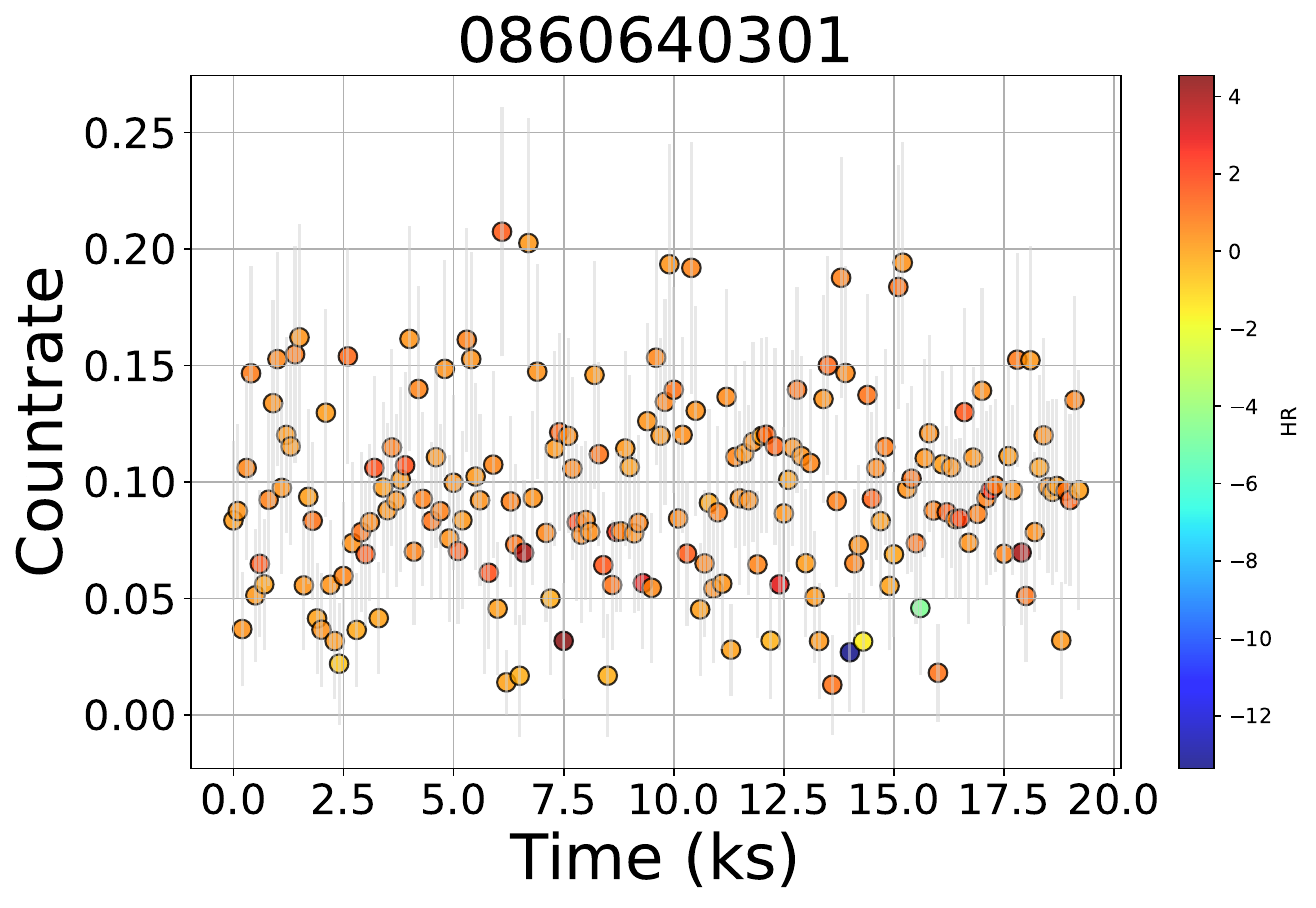}
	\end{minipage}
	\begin{minipage}{.3\textwidth} 
		\centering 
		\includegraphics[width=.99\linewidth]{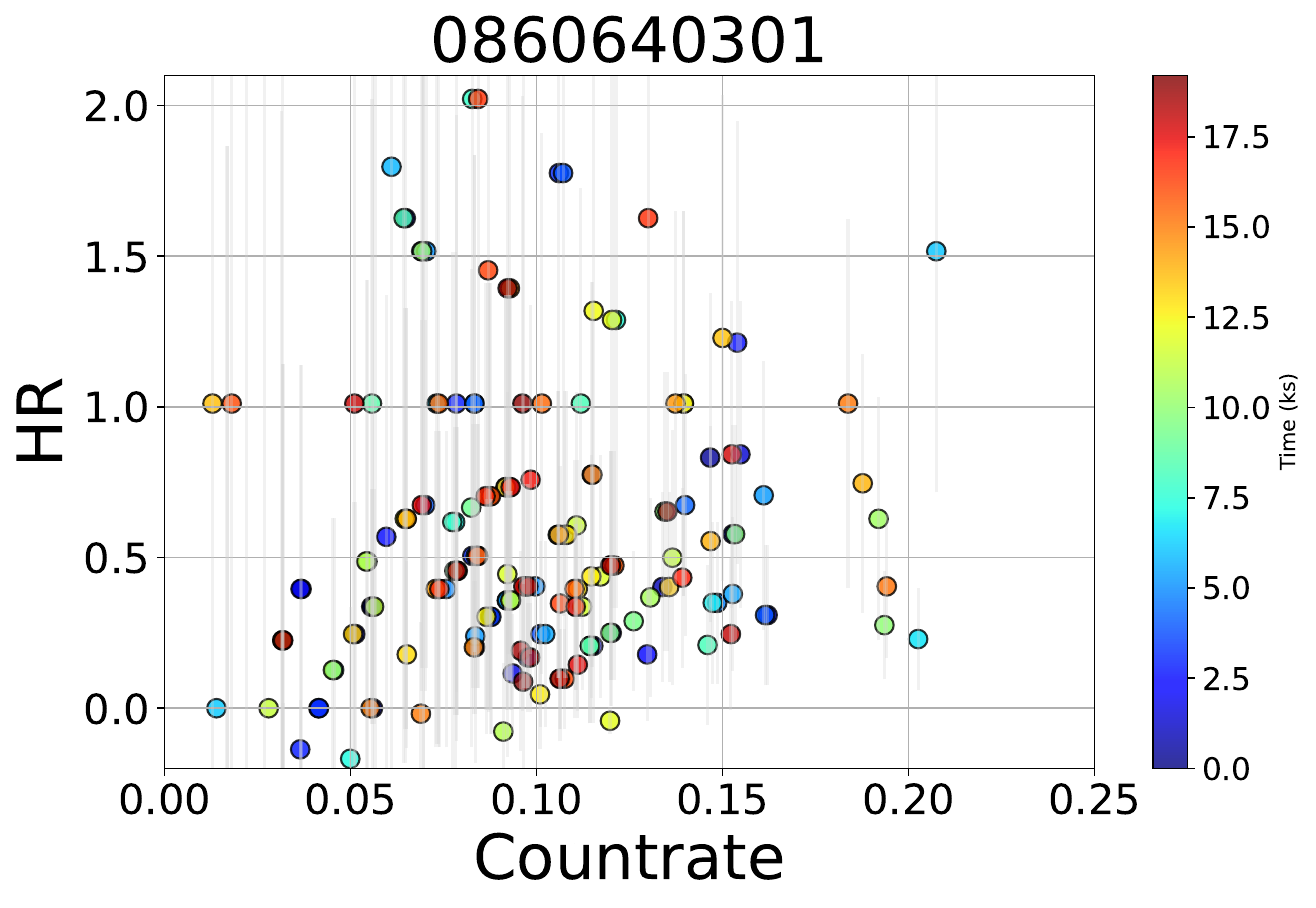}
	\end{minipage}
	\begin{minipage}{.3\textwidth} 
		\centering 
		\includegraphics[width=.99\linewidth, angle=0]{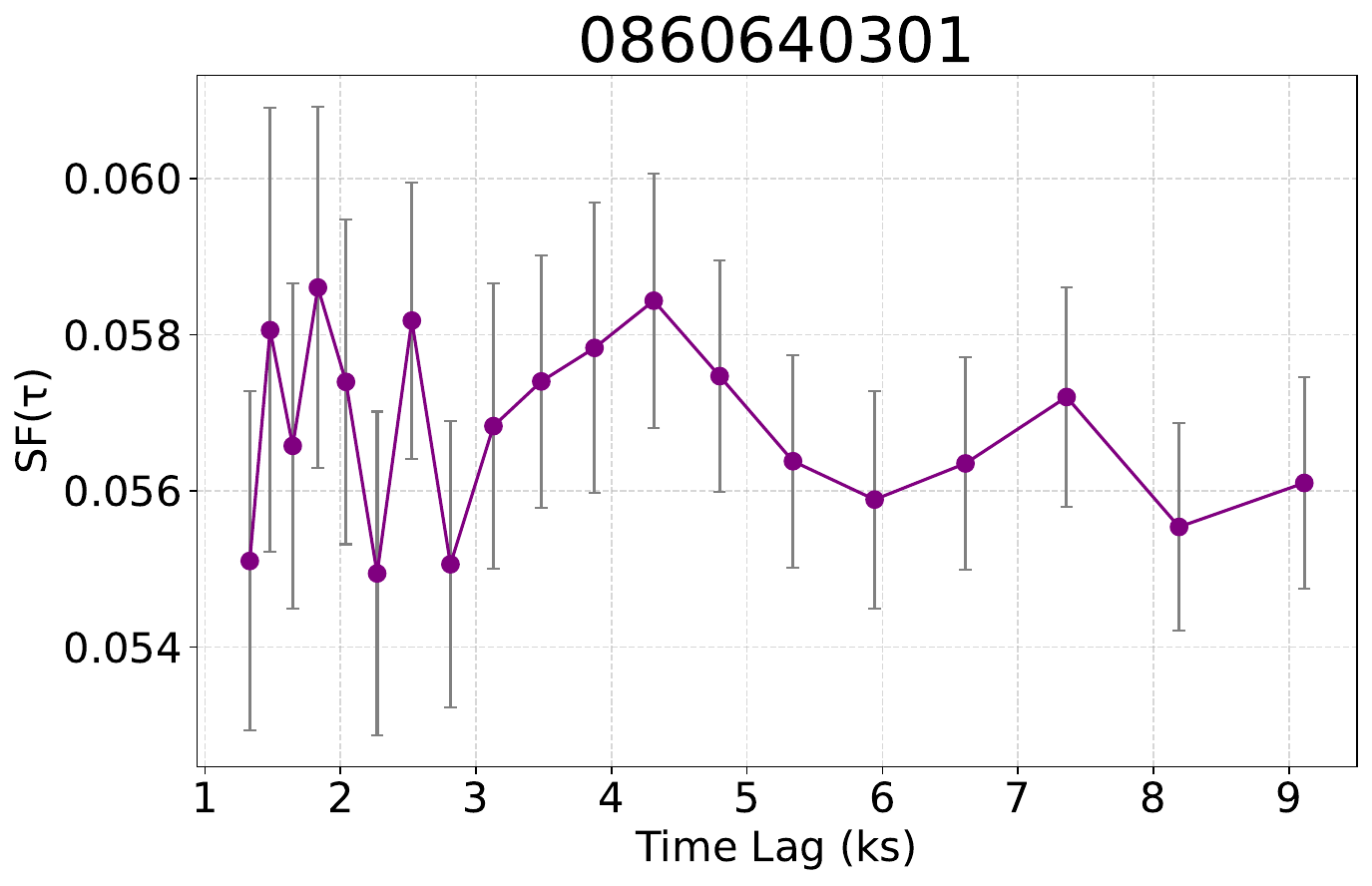}
	\end{minipage}
	\begin{minipage}{.3\textwidth} 
		\centering 
		\includegraphics[width=.99\linewidth]{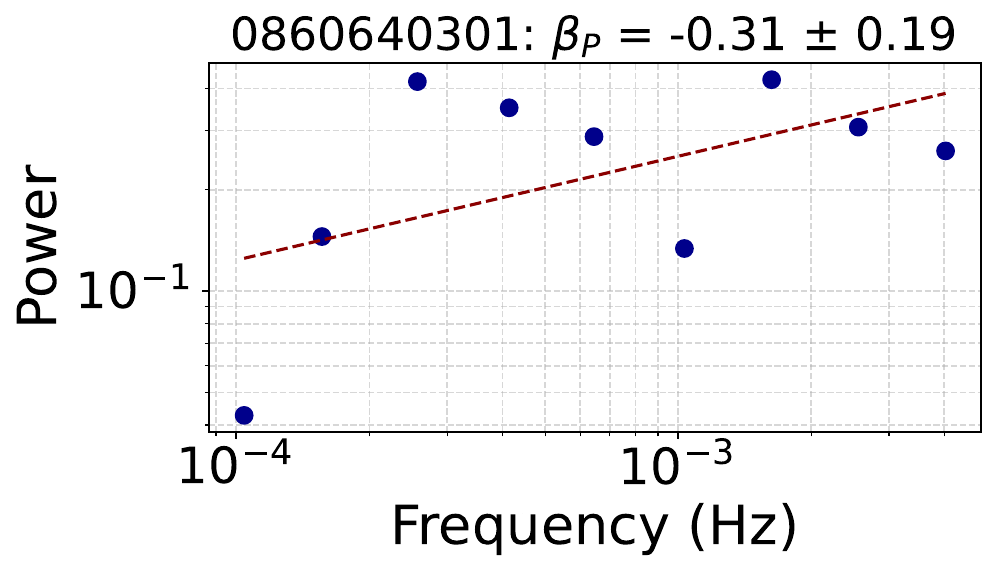}
	\end{minipage}
	\begin{minipage}{.3\textwidth} 
		\centering 
		\includegraphics[width=.99\linewidth]{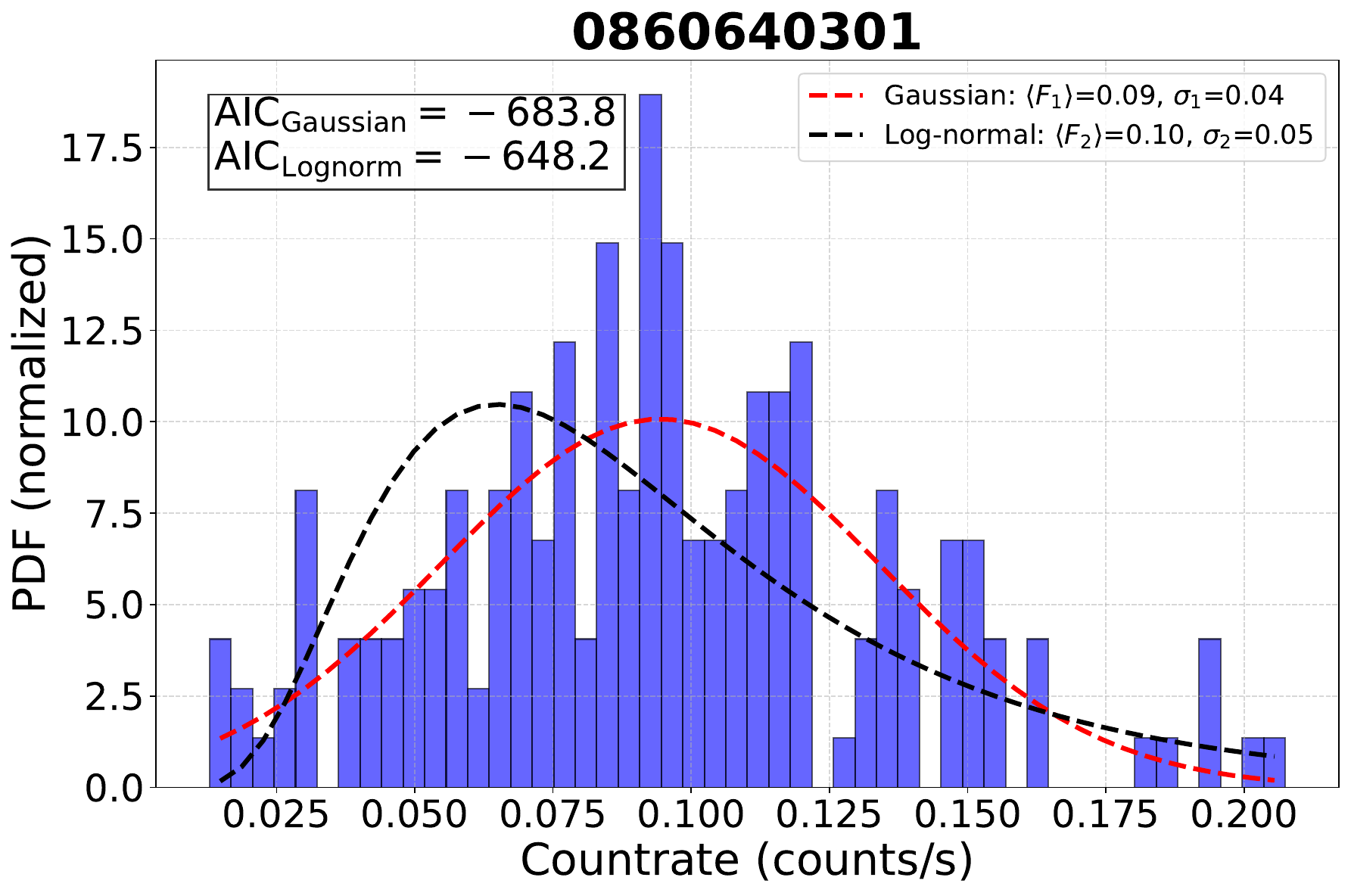}
	\end{minipage}
	\begin{minipage}{.3\textwidth} 
		\centering 
		\includegraphics[height=.99\linewidth, angle=-90]{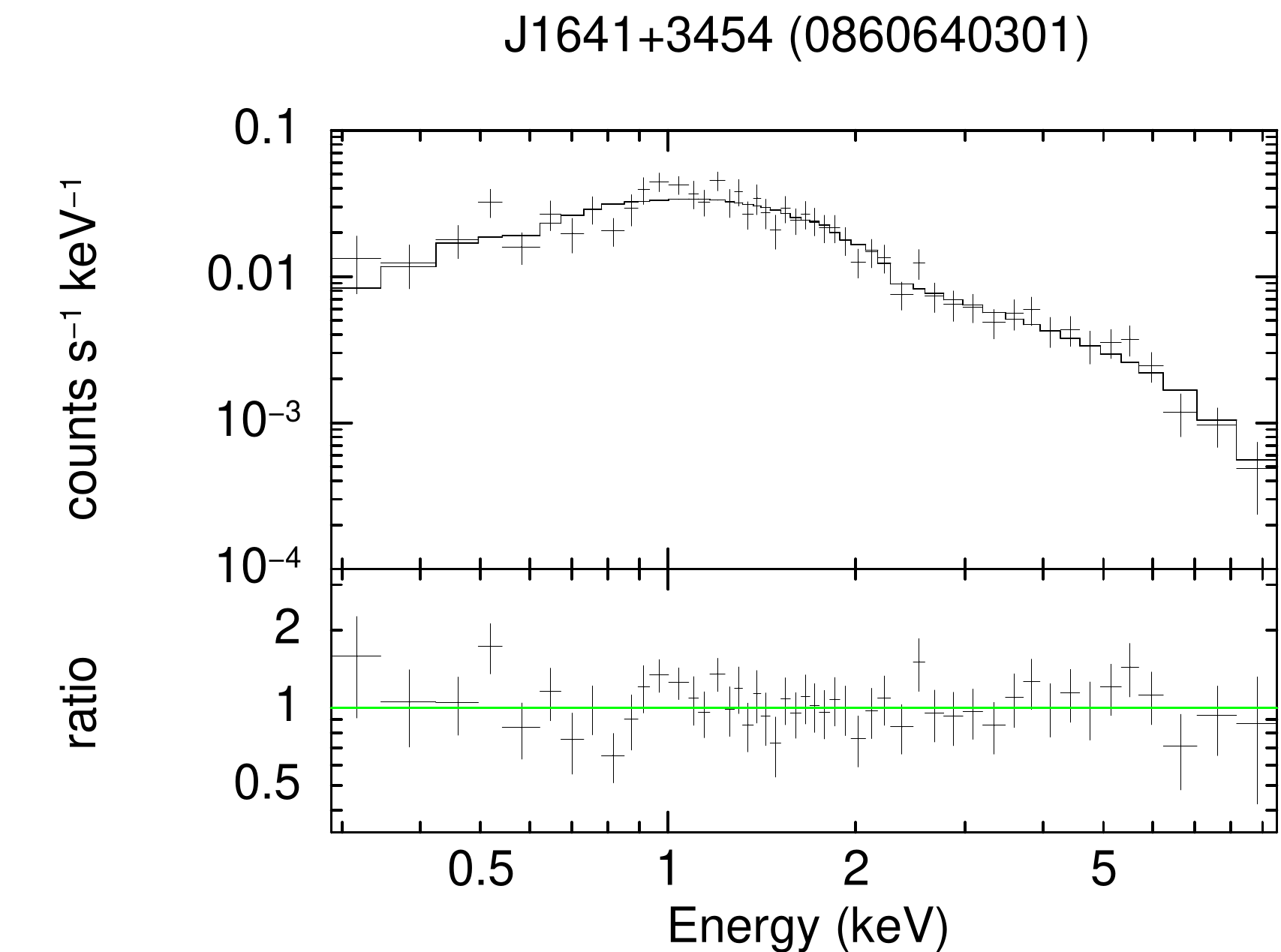}
	\end{minipage}
\end{figure*}

\begin{figure*}\label{app: 3C 286}
	\centering
	\caption{LCs, HR plots, Structure Function, PSD, PDF, and spectral fits derived from observations of 3C 286.}
    	\begin{minipage}{.3\textwidth} 
		\centering 
		\includegraphics[width=.99\linewidth]{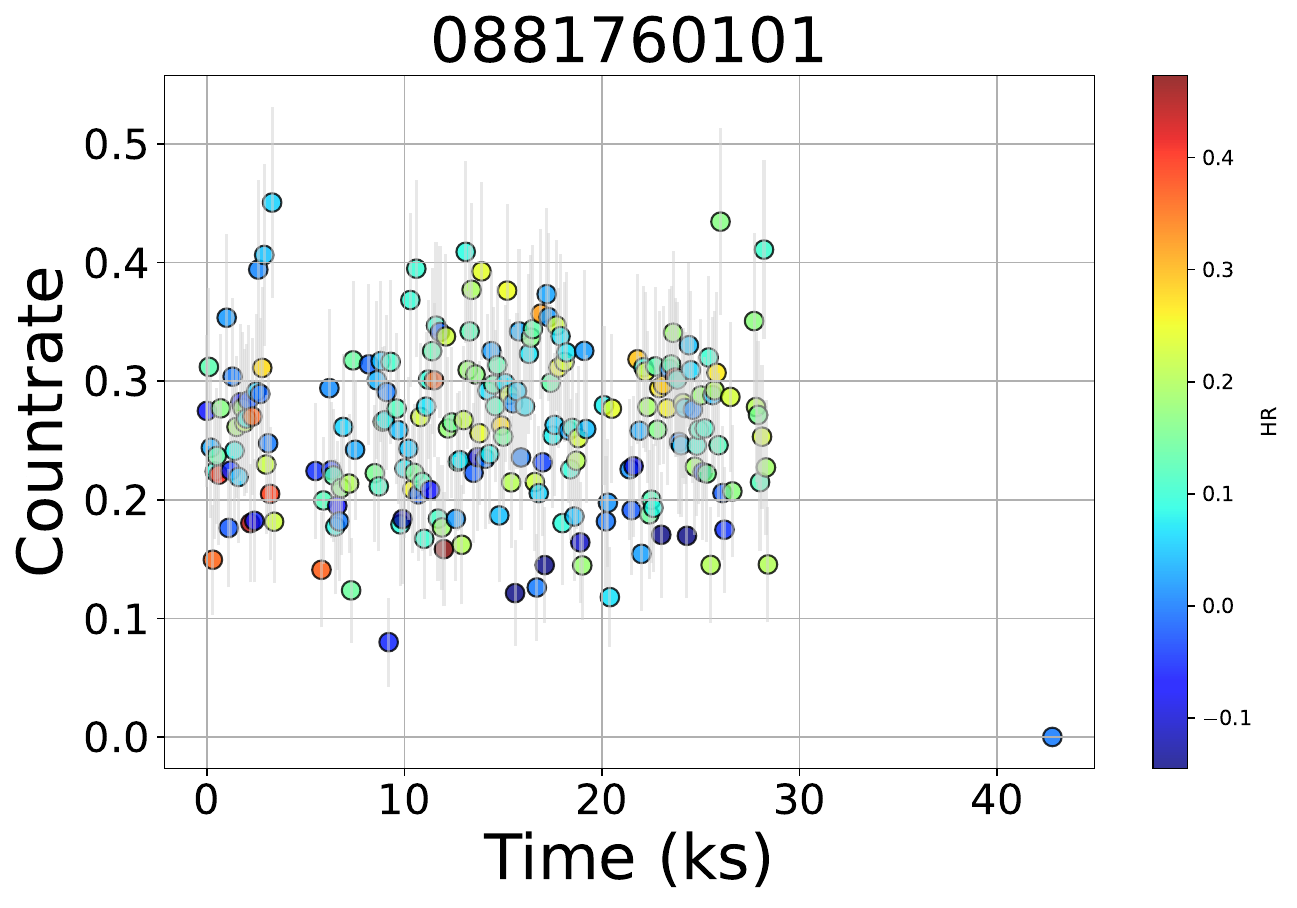}
	\end{minipage}
	\begin{minipage}{.3\textwidth} 
		\centering 
		\includegraphics[width=.99\linewidth]{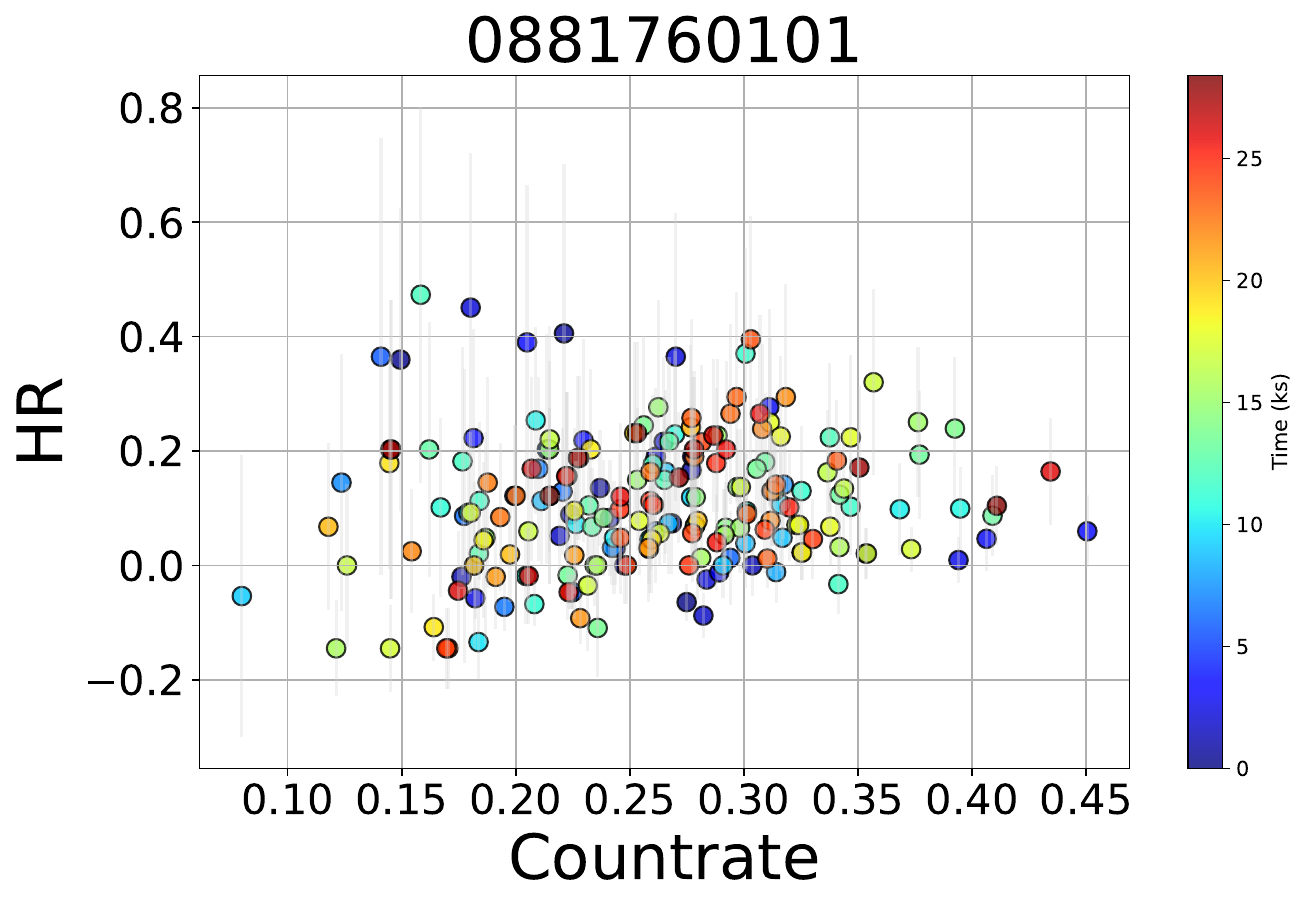}
	\end{minipage}
	\begin{minipage}{.3\textwidth} 
		\centering 
		\includegraphics[width=.99\linewidth, angle=0]{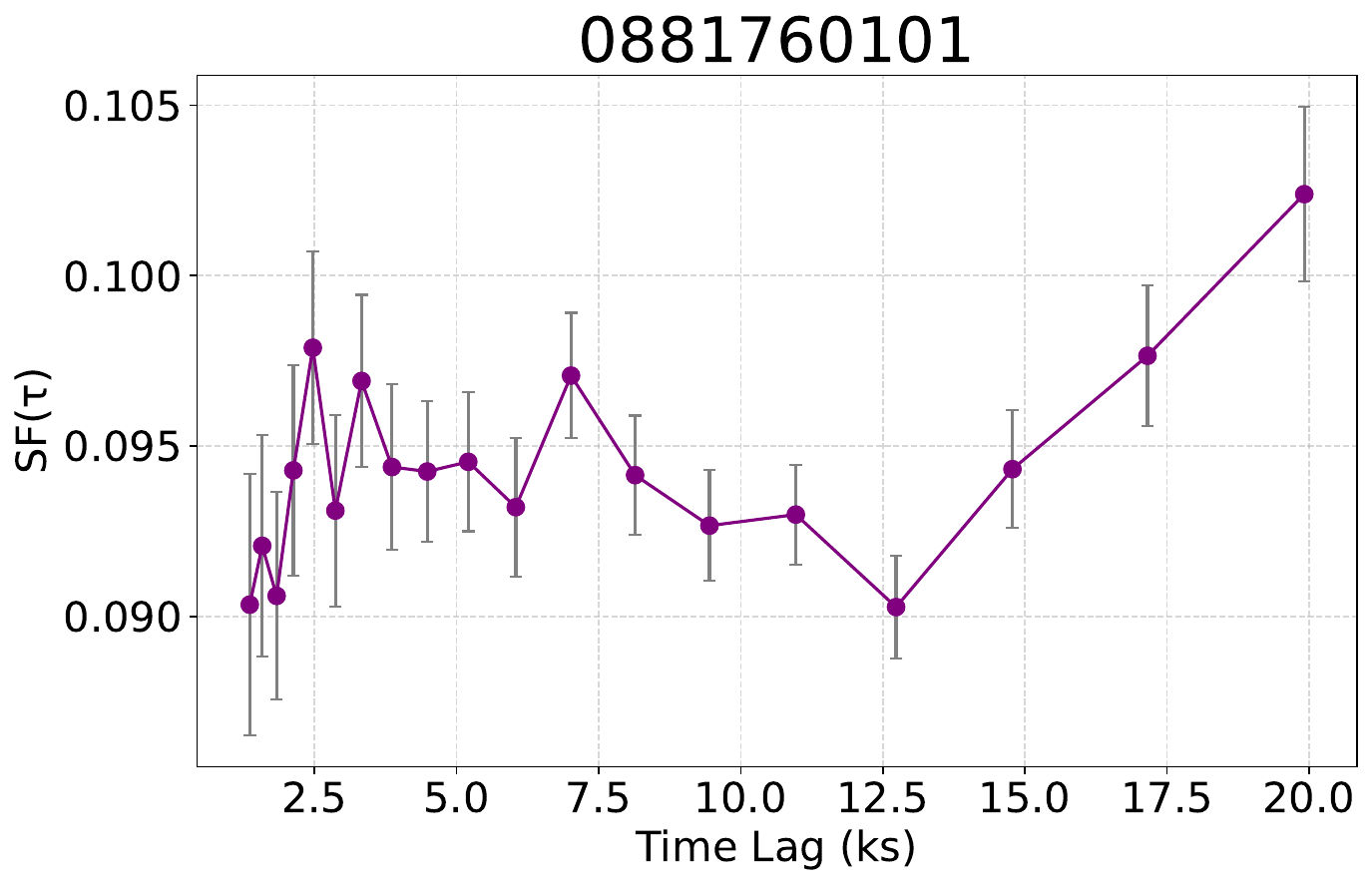}
	\end{minipage}
    \begin{minipage}{.3\textwidth} 
		\centering 
		\includegraphics[width=.99\linewidth]{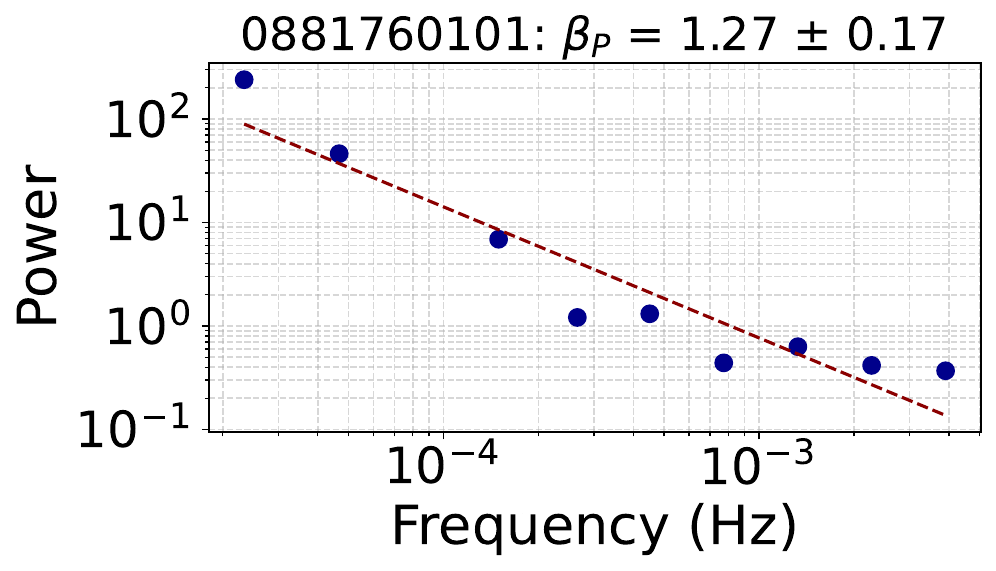}
	\end{minipage}
	\begin{minipage}{.3\textwidth} 
		\centering 
		\includegraphics[width=.99\linewidth]{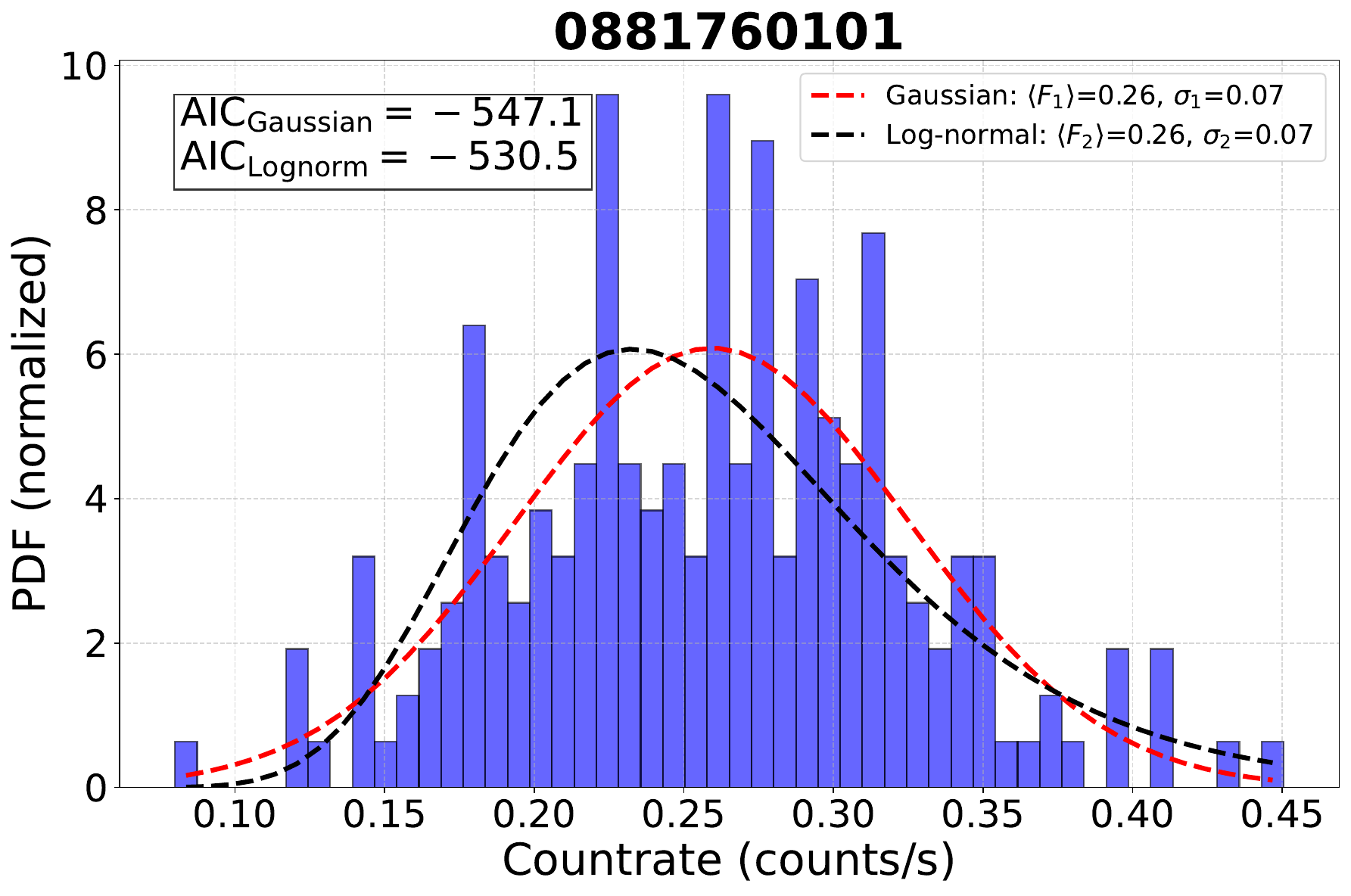}
	\end{minipage}
	\begin{minipage}{.3\textwidth} 
		\centering 
		\includegraphics[height=.99\linewidth, angle=-90]{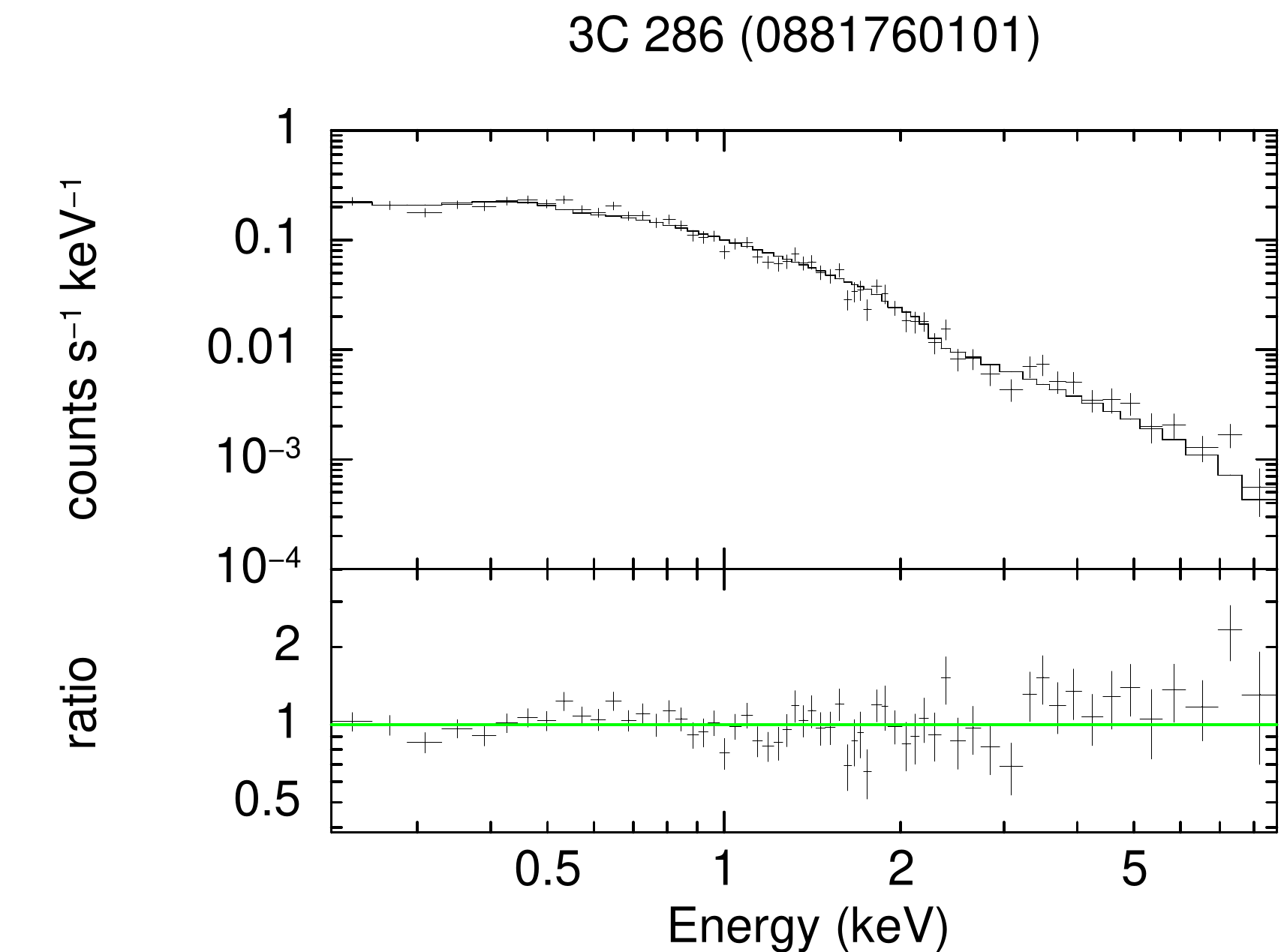}
	\end{minipage}
\end{figure*}

\begin{figure*}\label{app:TXS 1419+391}
	\centering
	\caption{LCs, HR plots, Structure Function, PSD, PDF, and spectral fits derived from observation of TXS 1419+391.}
	\label{fig:LC_RBS}
	\begin{minipage}{.3\textwidth} 
		\centering 
		\includegraphics[width=.99\linewidth]{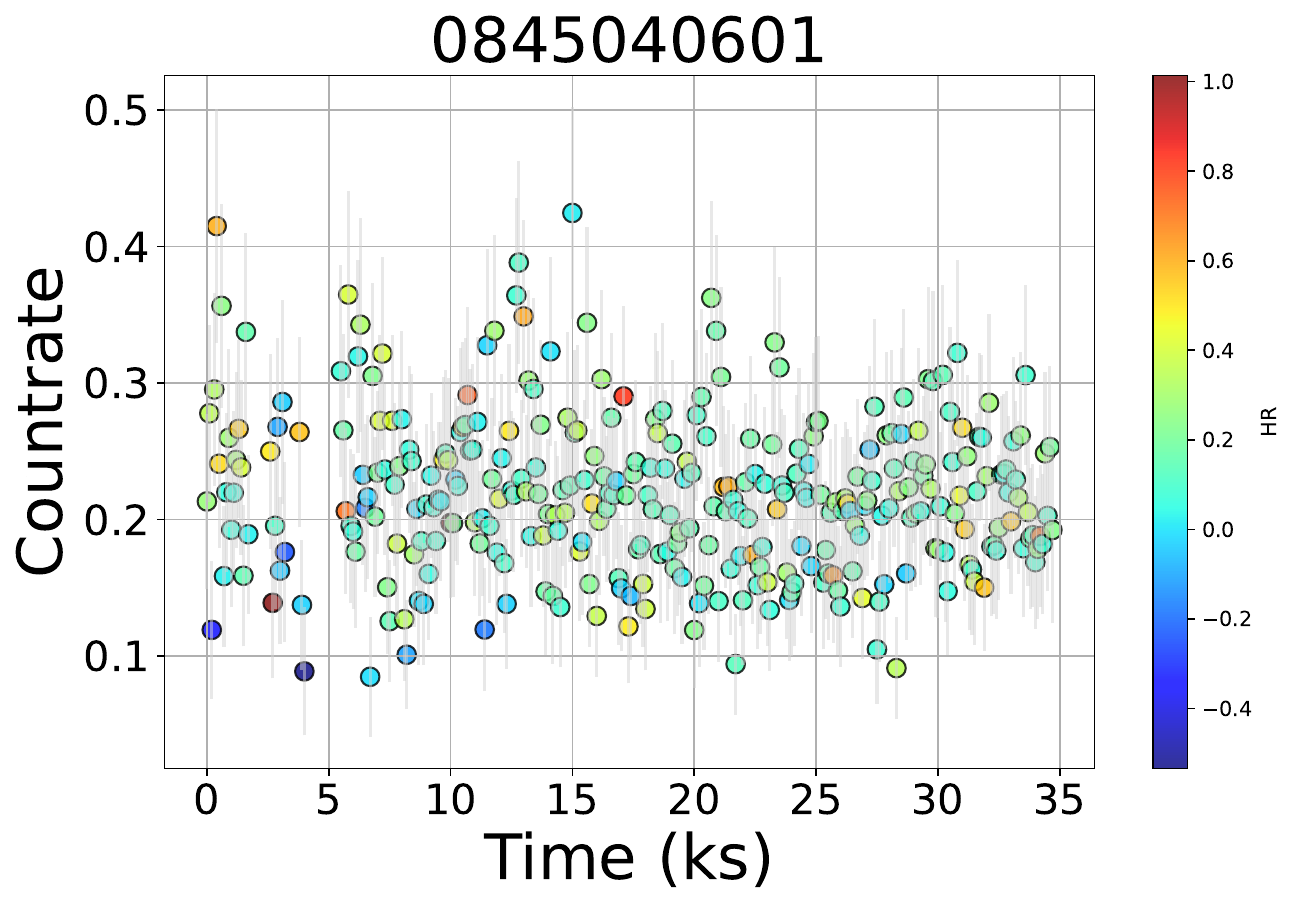}
	\end{minipage}
	\begin{minipage}{.3\textwidth} 
		\centering 
		\includegraphics[width=.99\linewidth]{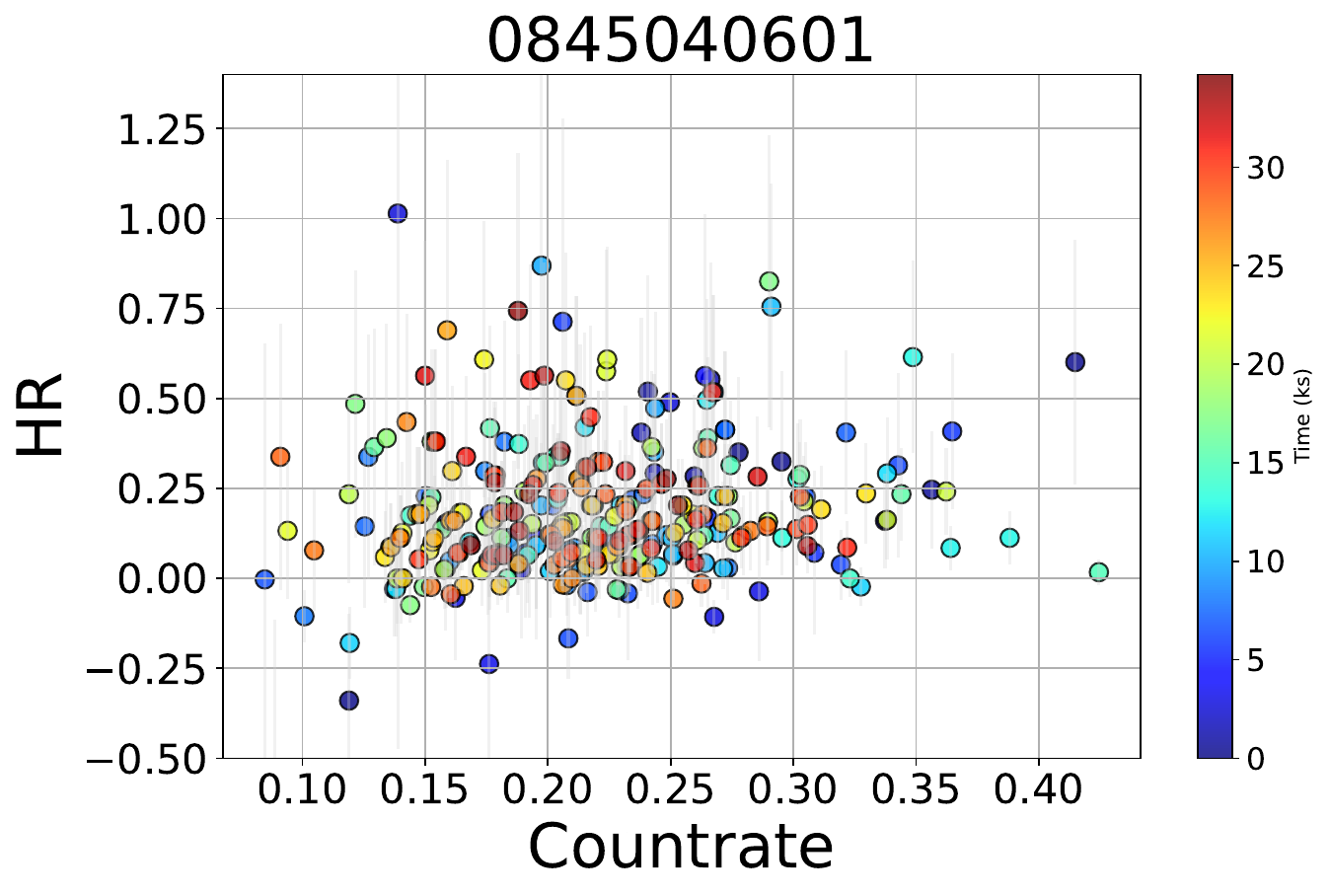}
	\end{minipage}
	\begin{minipage}{.3\textwidth} 
		\centering 
		\includegraphics[width=.99\linewidth, angle=0]{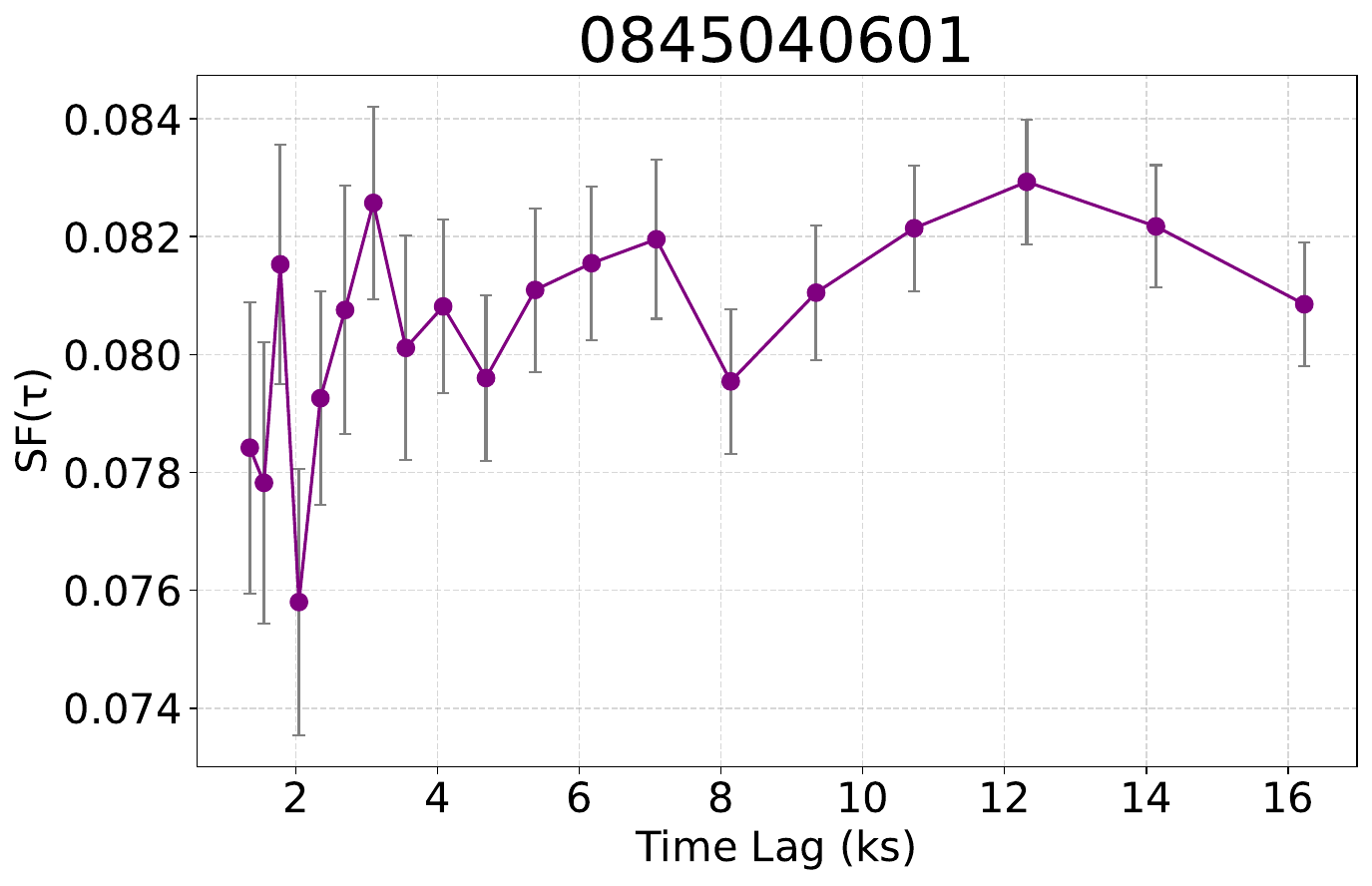}
	\end{minipage}
    	\begin{minipage}{.3\textwidth} 
		\centering 
		\includegraphics[width=.99\linewidth]{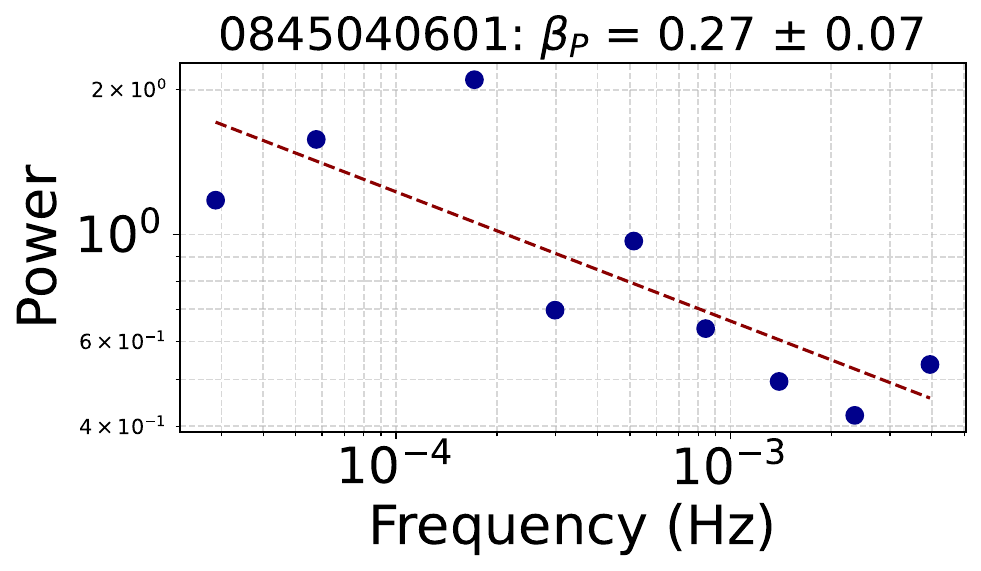}
	\end{minipage}
	\begin{minipage}{.3\textwidth} 
		\centering 
		\includegraphics[width=.99\linewidth]{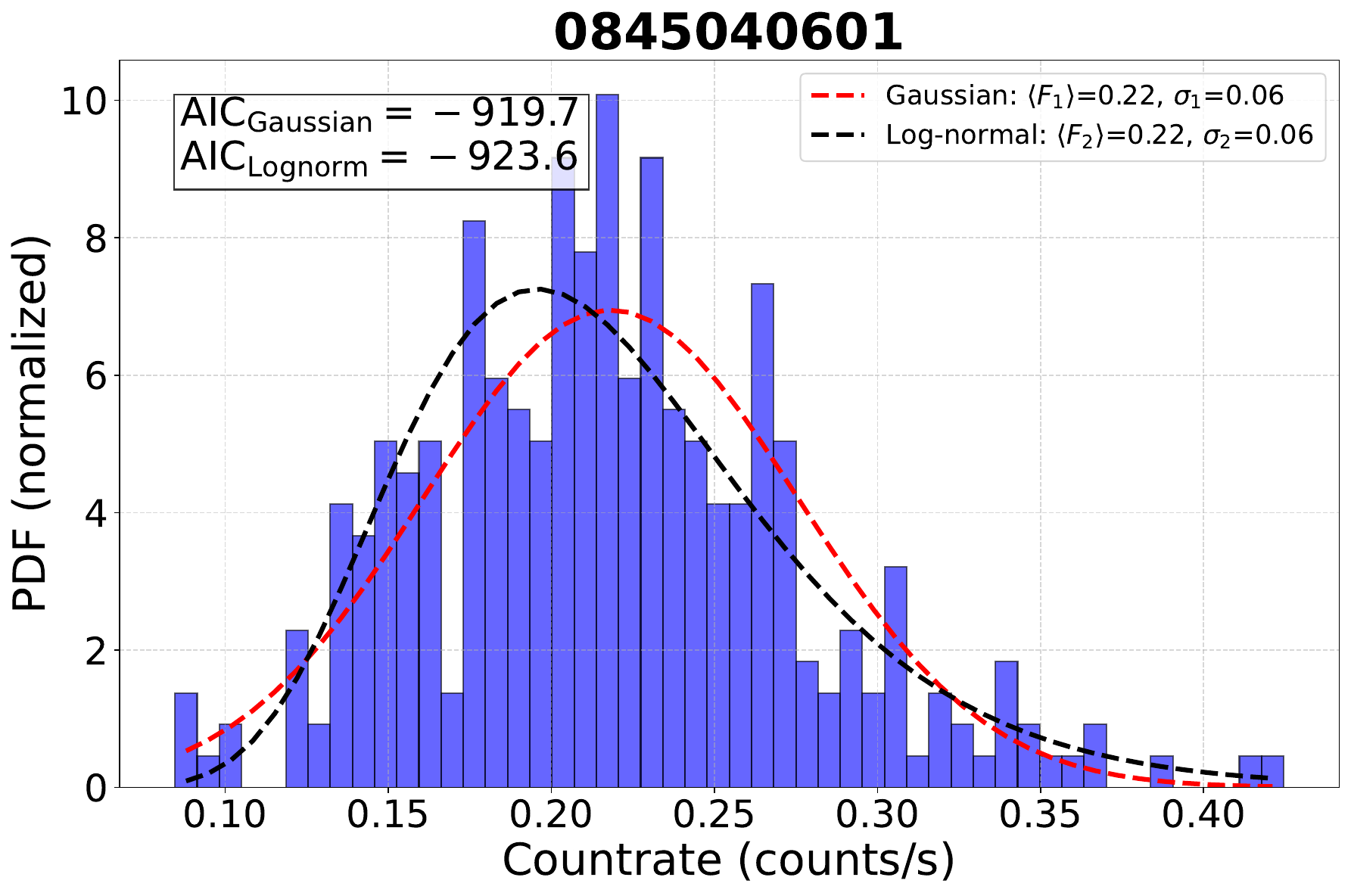}
	\end{minipage}
	\begin{minipage}{.3\textwidth} 
		\centering 
		\includegraphics[height=.99\linewidth, angle=-90]{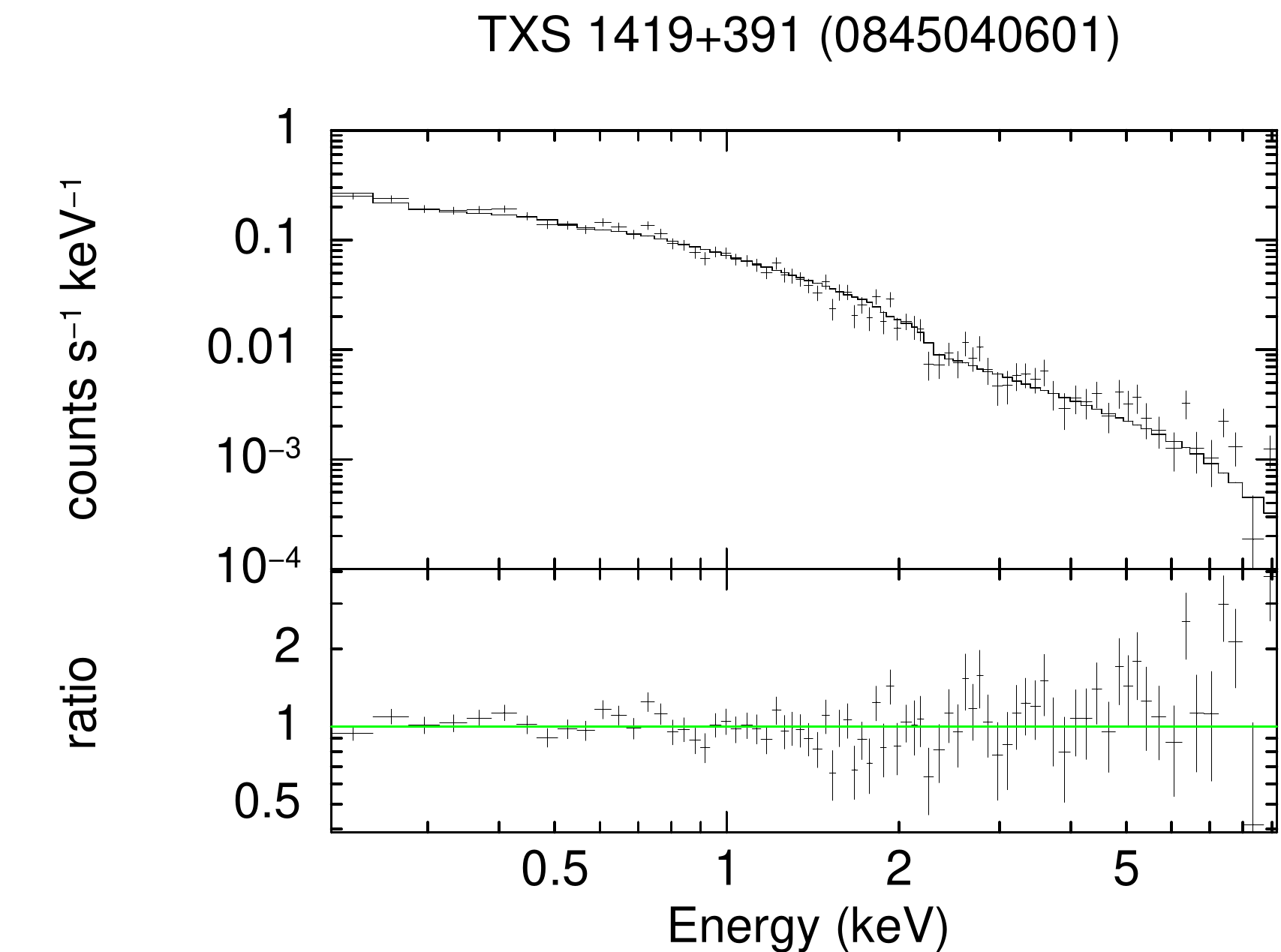}
	\end{minipage}
\end{figure*}

\begin{figure*}\label{app:J0946+1017}
	\centering
	\caption{LCs, HR plots, Structure Function, PSD, PDF, and spectral fits derived from observation of J0946+1017}
	\begin{minipage}{.3\textwidth} 
		\centering 
		\includegraphics[width=.99\linewidth]{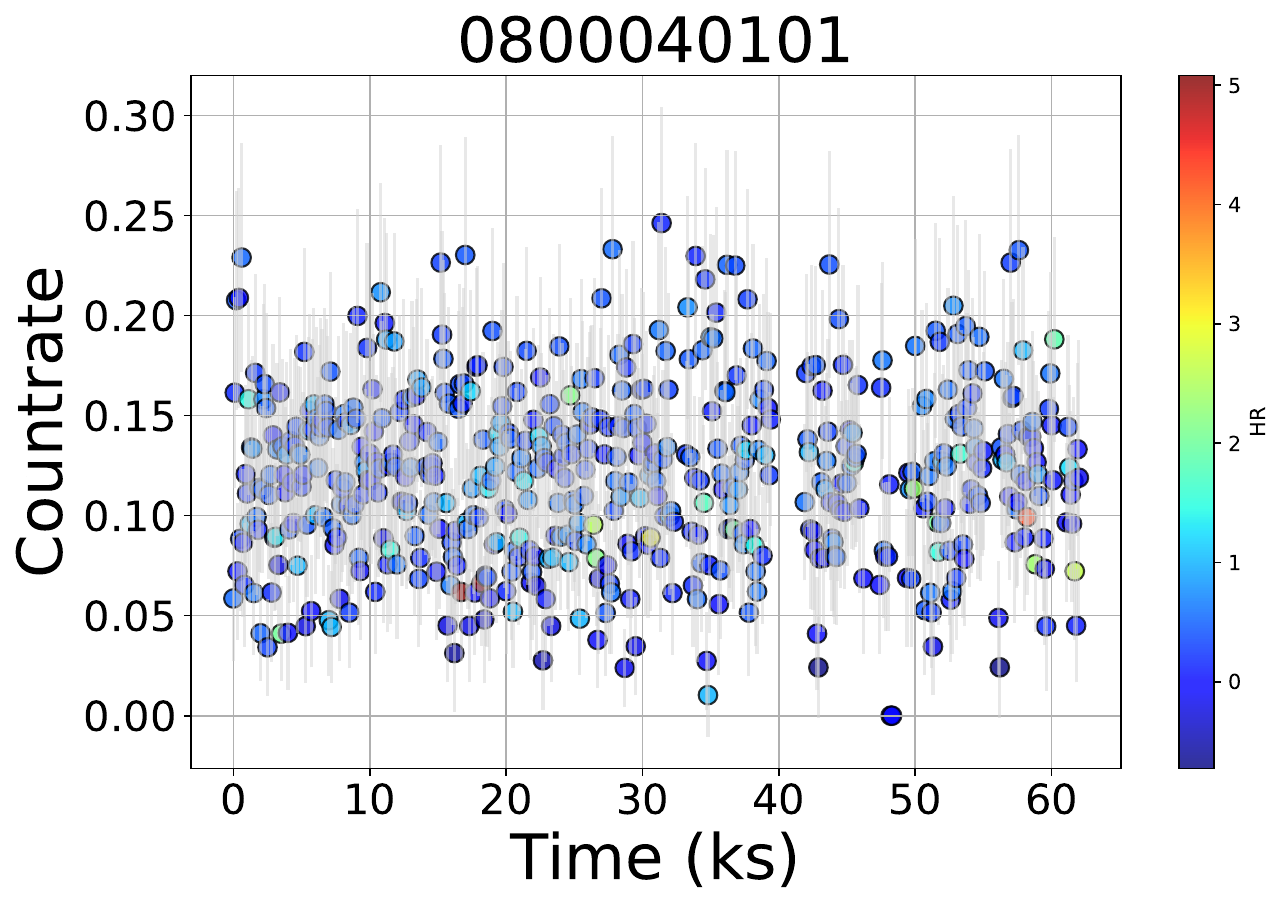}
	\end{minipage}
	\begin{minipage}{.3\textwidth} 
		\centering 
		\includegraphics[width=.99\linewidth]{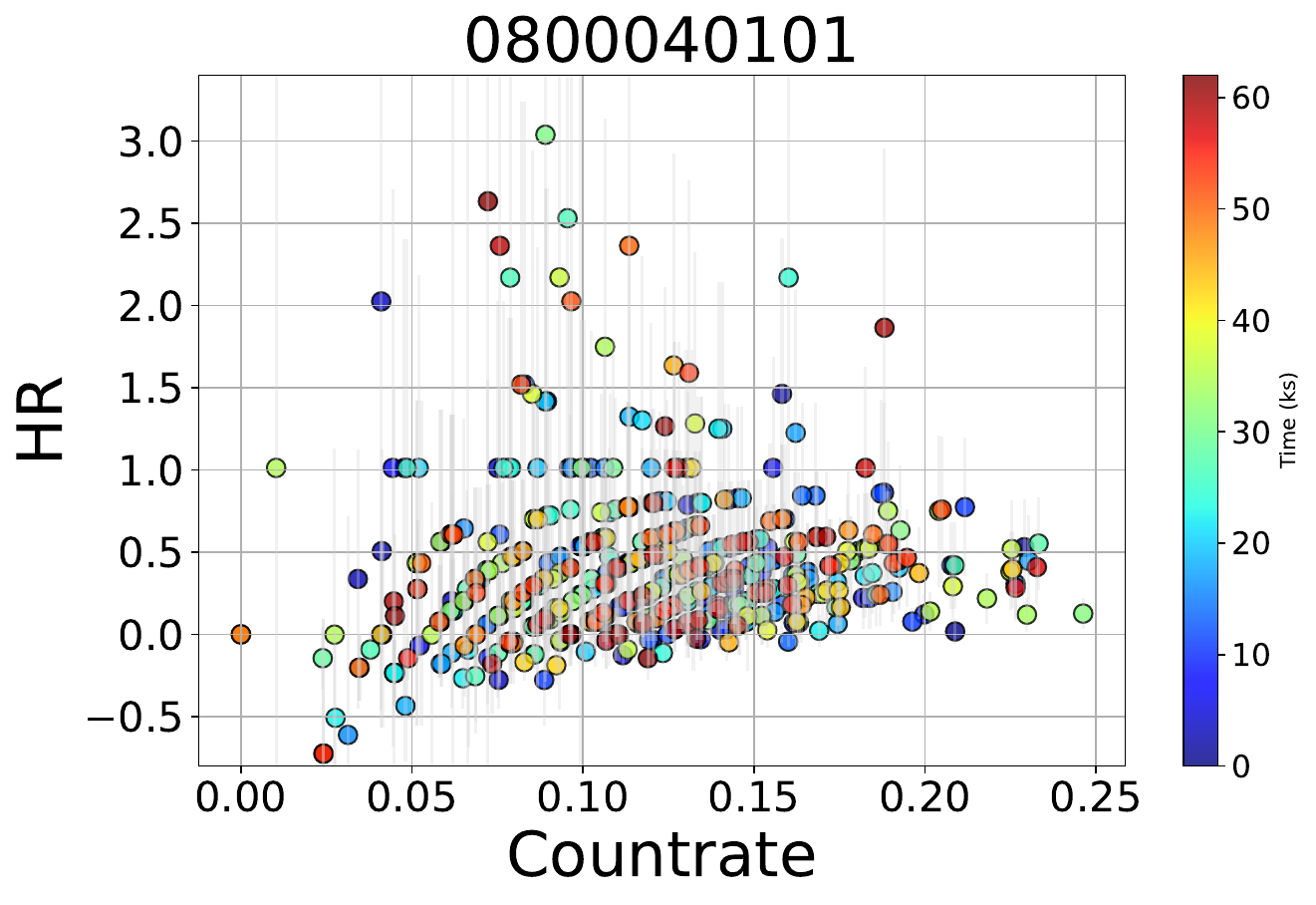}
	\end{minipage}
	\begin{minipage}{.3\textwidth} 
		\centering 
		\includegraphics[width=.99\linewidth, angle=0]{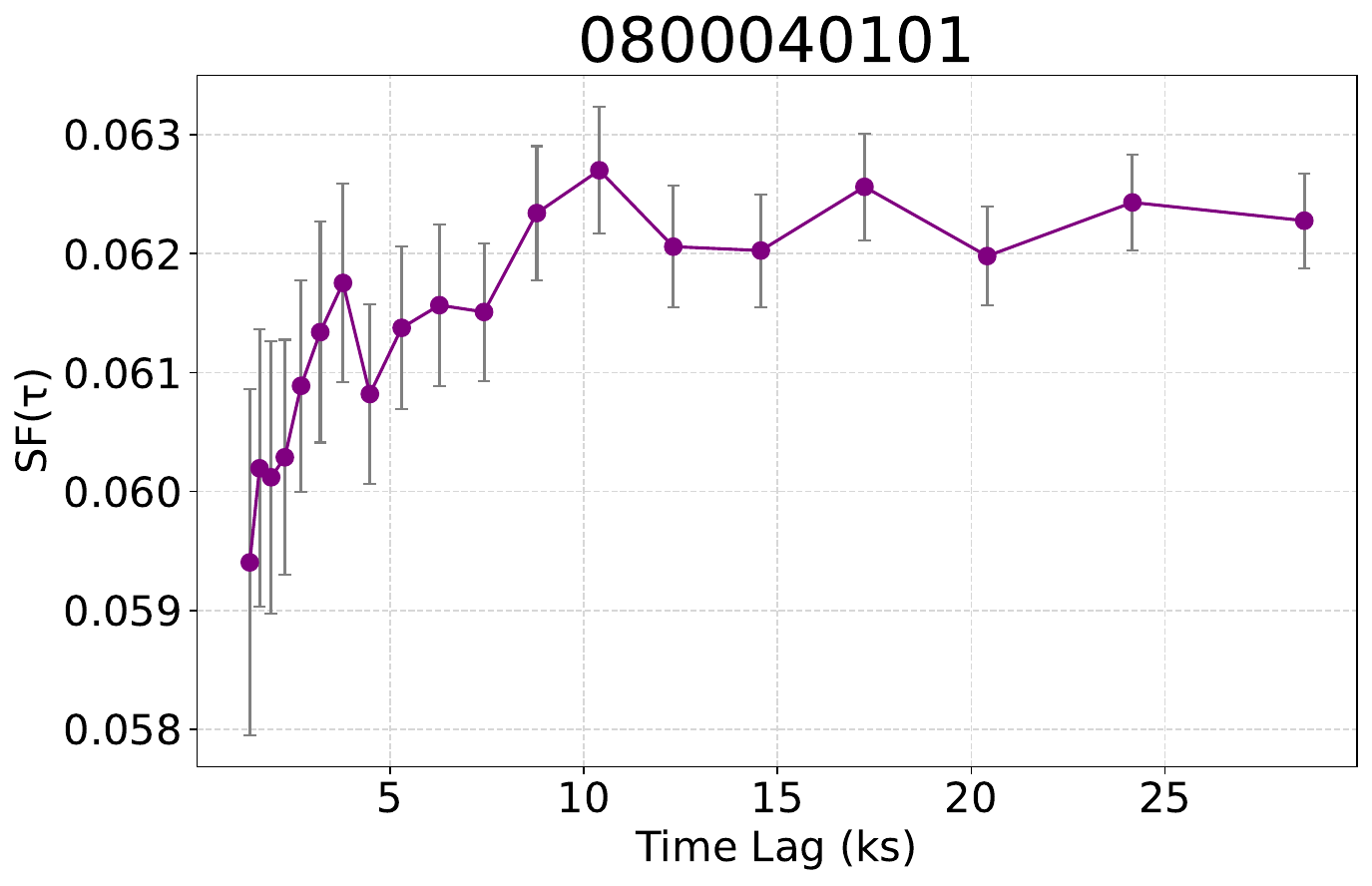}
	\end{minipage}
	\begin{minipage}{.3\textwidth} 
		\centering 
		\includegraphics[width=.99\linewidth]{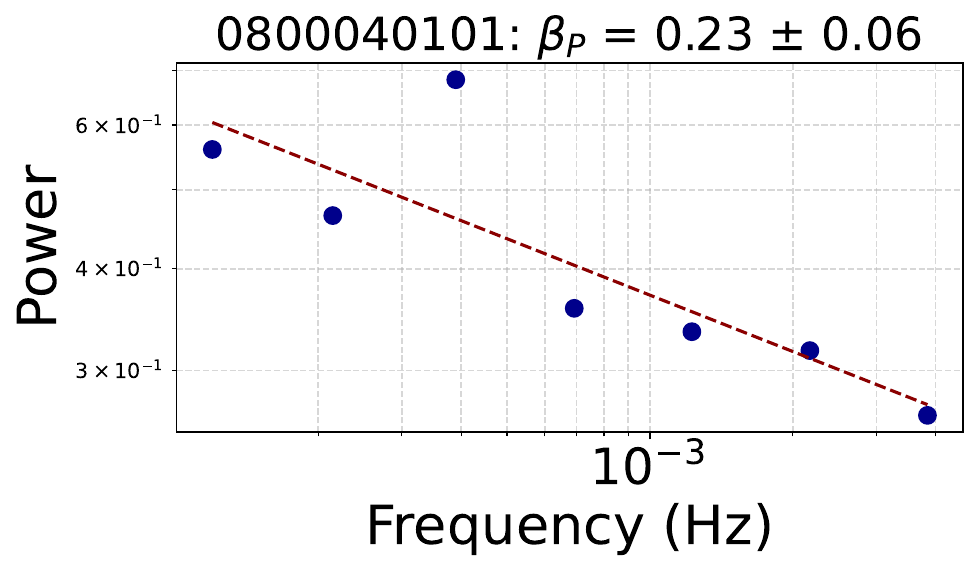}
	\end{minipage}
	\begin{minipage}{.3\textwidth} 
		\centering 
		\includegraphics[width=.99\linewidth]{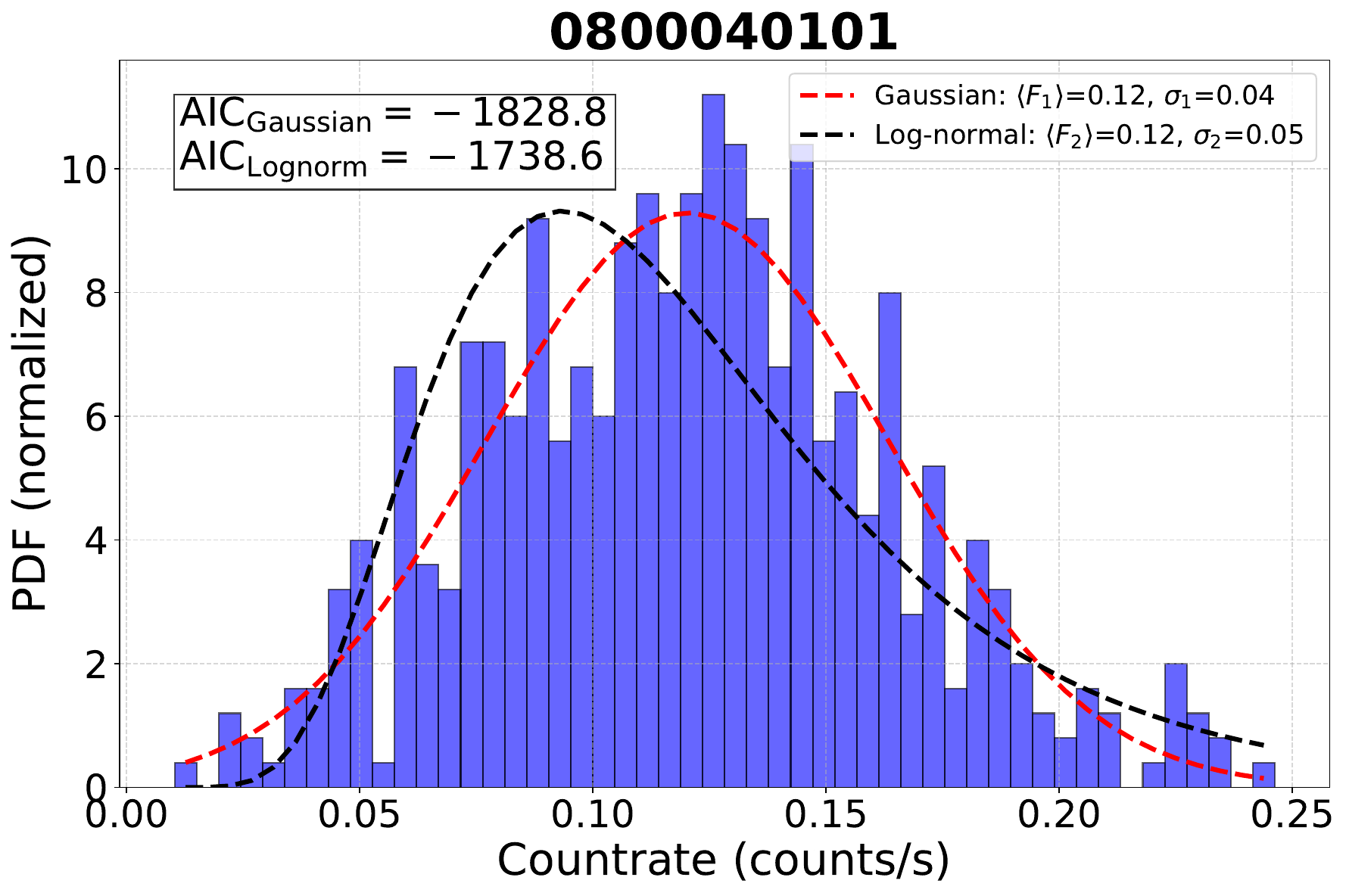}
	\end{minipage}
	\begin{minipage}{.3\textwidth} 
		\centering 
		\includegraphics[height=.99\linewidth, angle=-90]{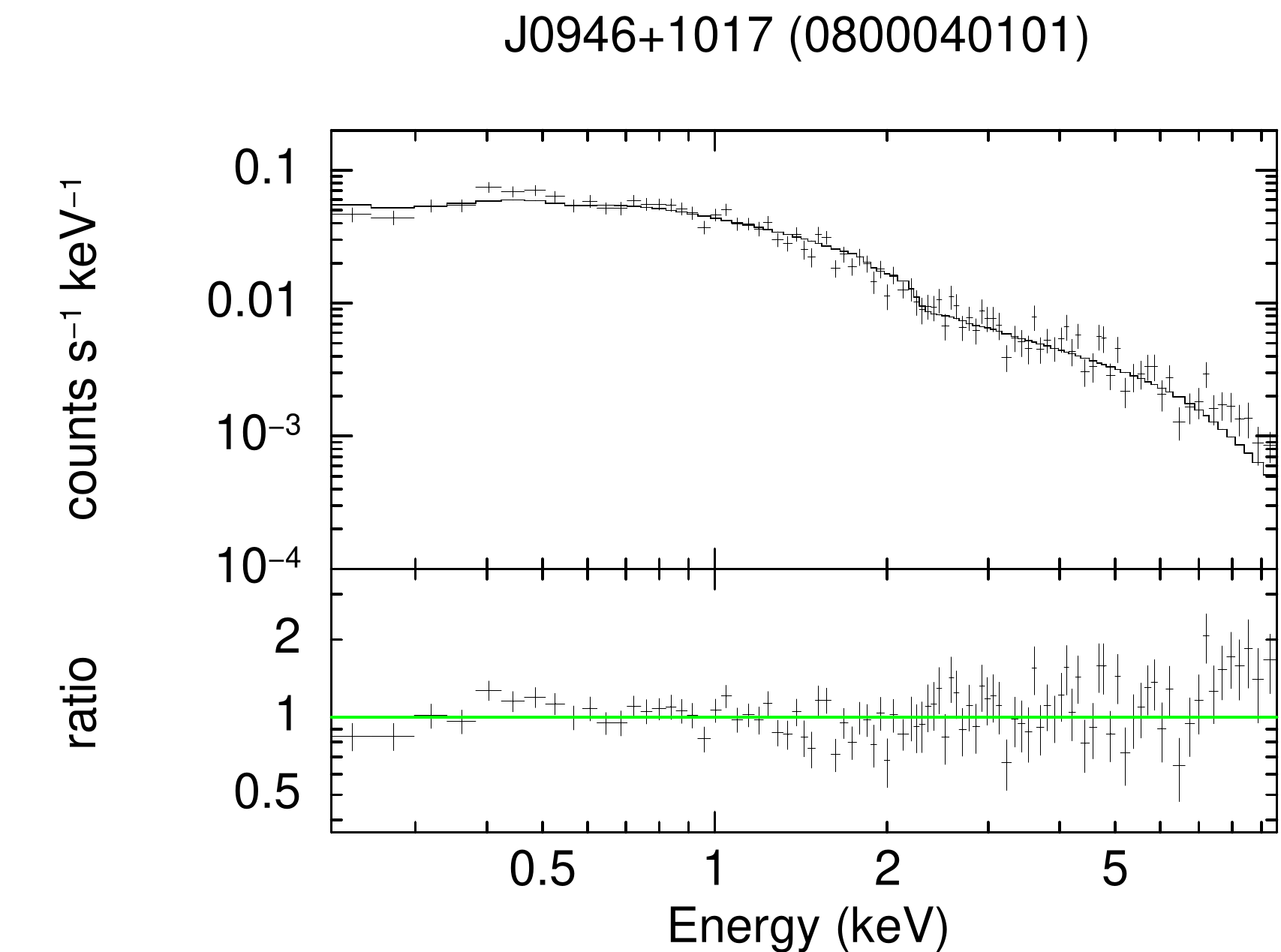}
	\end{minipage}
\end{figure*}

\begin{figure*}\label{app:PKS 1502+036}
	\centering
	\caption{LCs, HR plots, Structure Function, PSD, PDF, and spectral fits derived from observation of PKS 1502+036.}
    	\begin{minipage}{.3\textwidth} 
		\centering 
		\includegraphics[width=.99\linewidth]{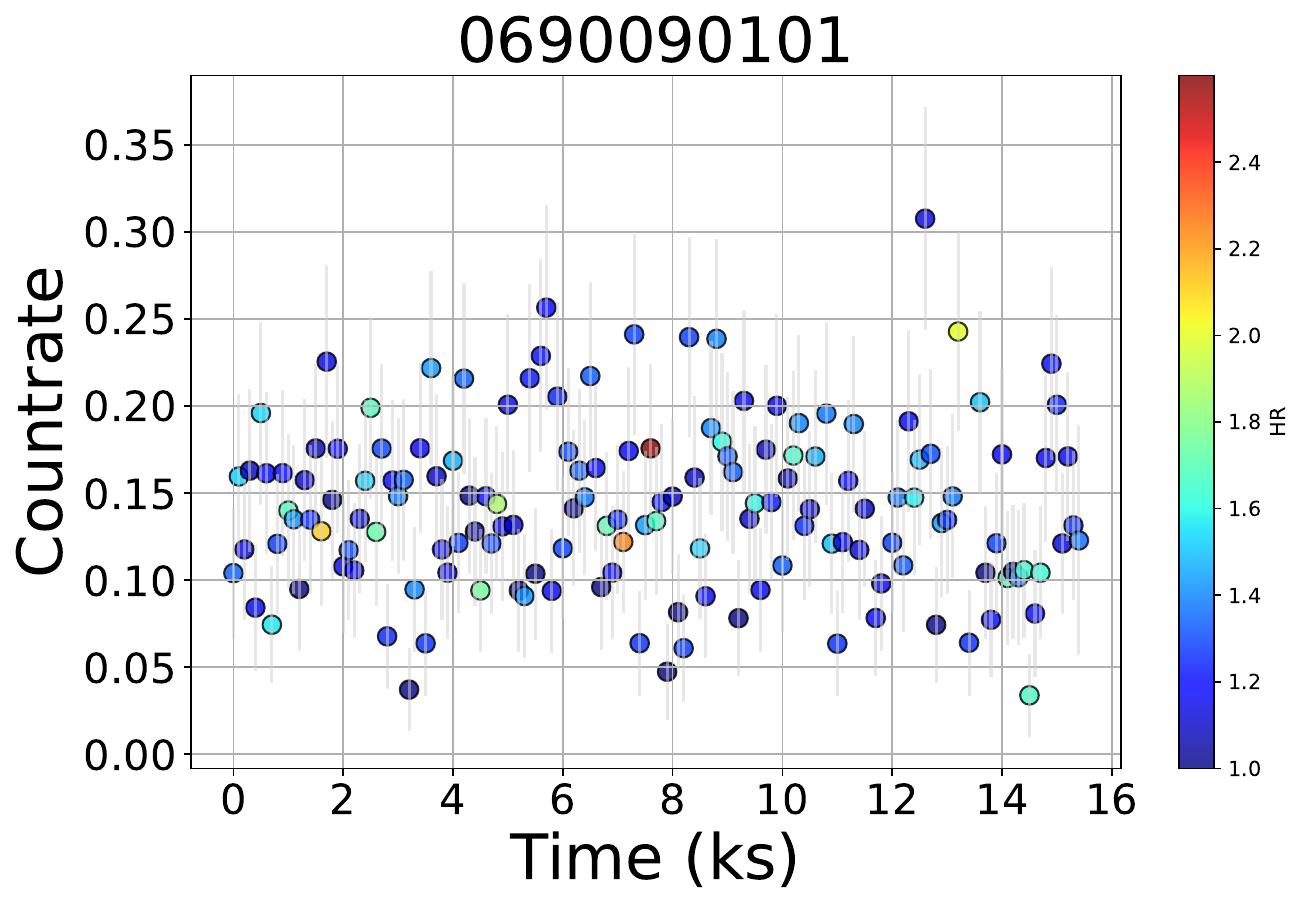}
	\end{minipage}
	\begin{minipage}{.3\textwidth} 
		\centering 
		\includegraphics[width=.99\linewidth]{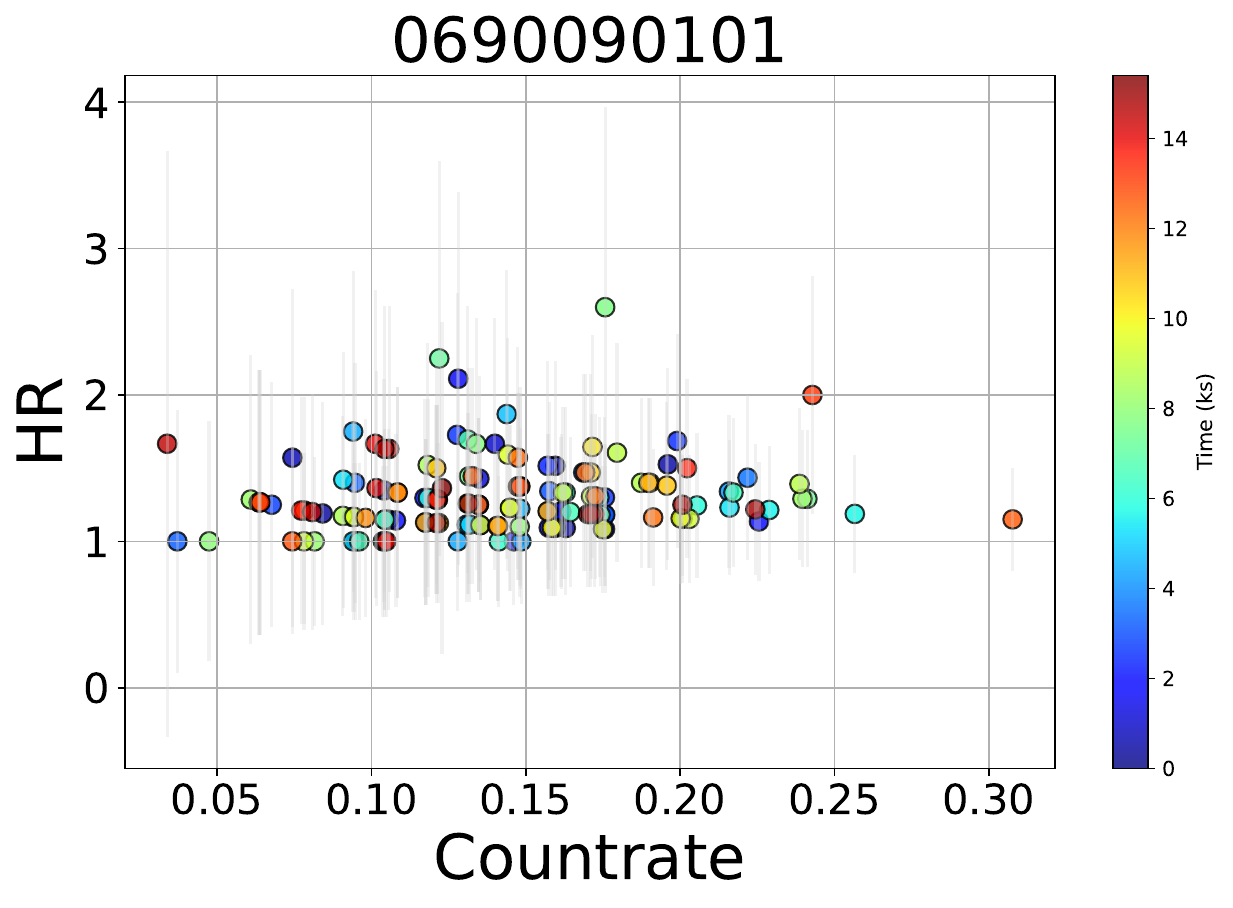}
	\end{minipage}
	\begin{minipage}{.3\textwidth} 
		\centering 
		\includegraphics[width=.99\linewidth, angle=0]{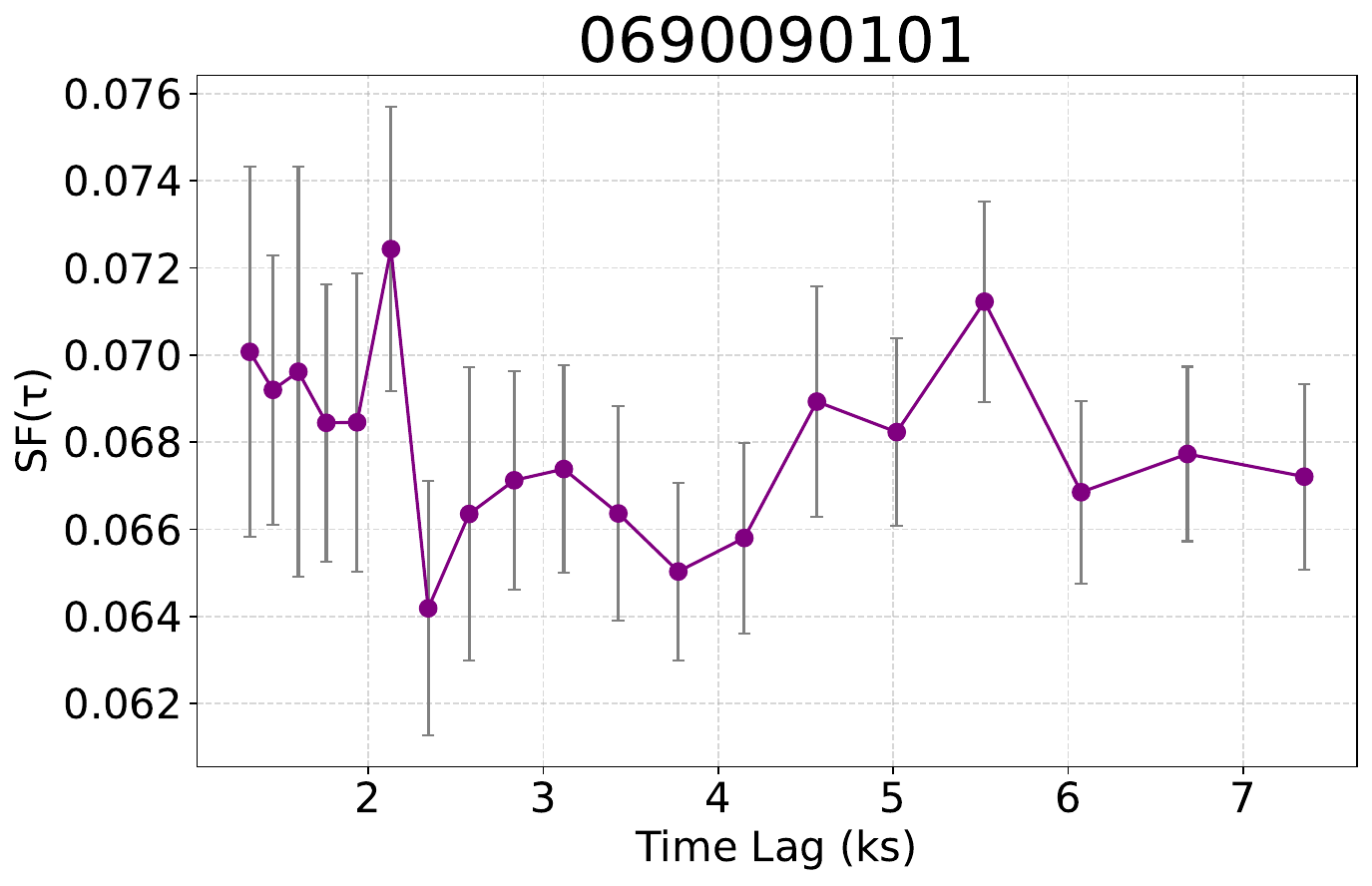}
	\end{minipage}
    \begin{minipage}{.3\textwidth} 
		\centering 
		\includegraphics[width=.99\linewidth]{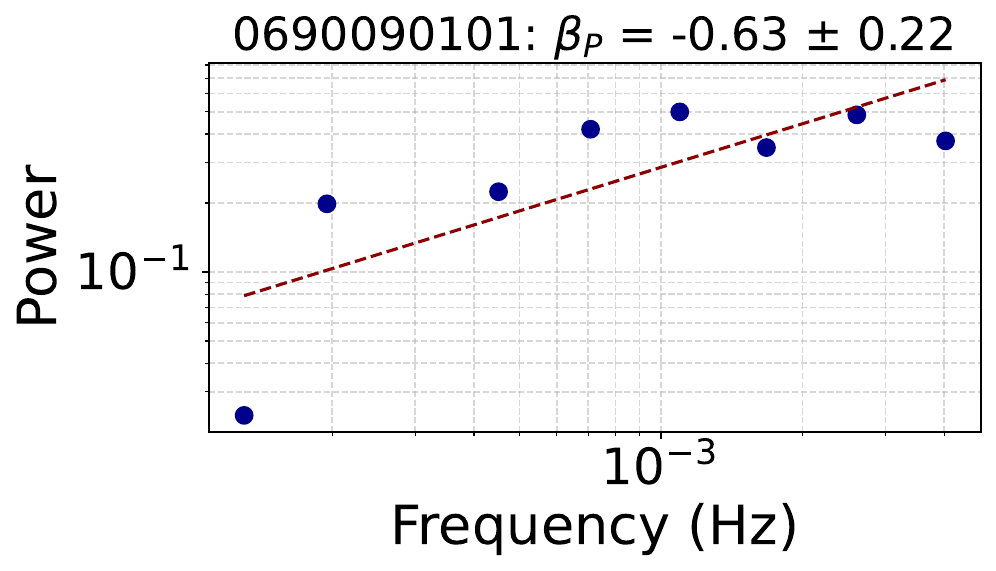}
	\end{minipage}
	\begin{minipage}{.3\textwidth} 
		\centering 
		\includegraphics[width=.99\linewidth]{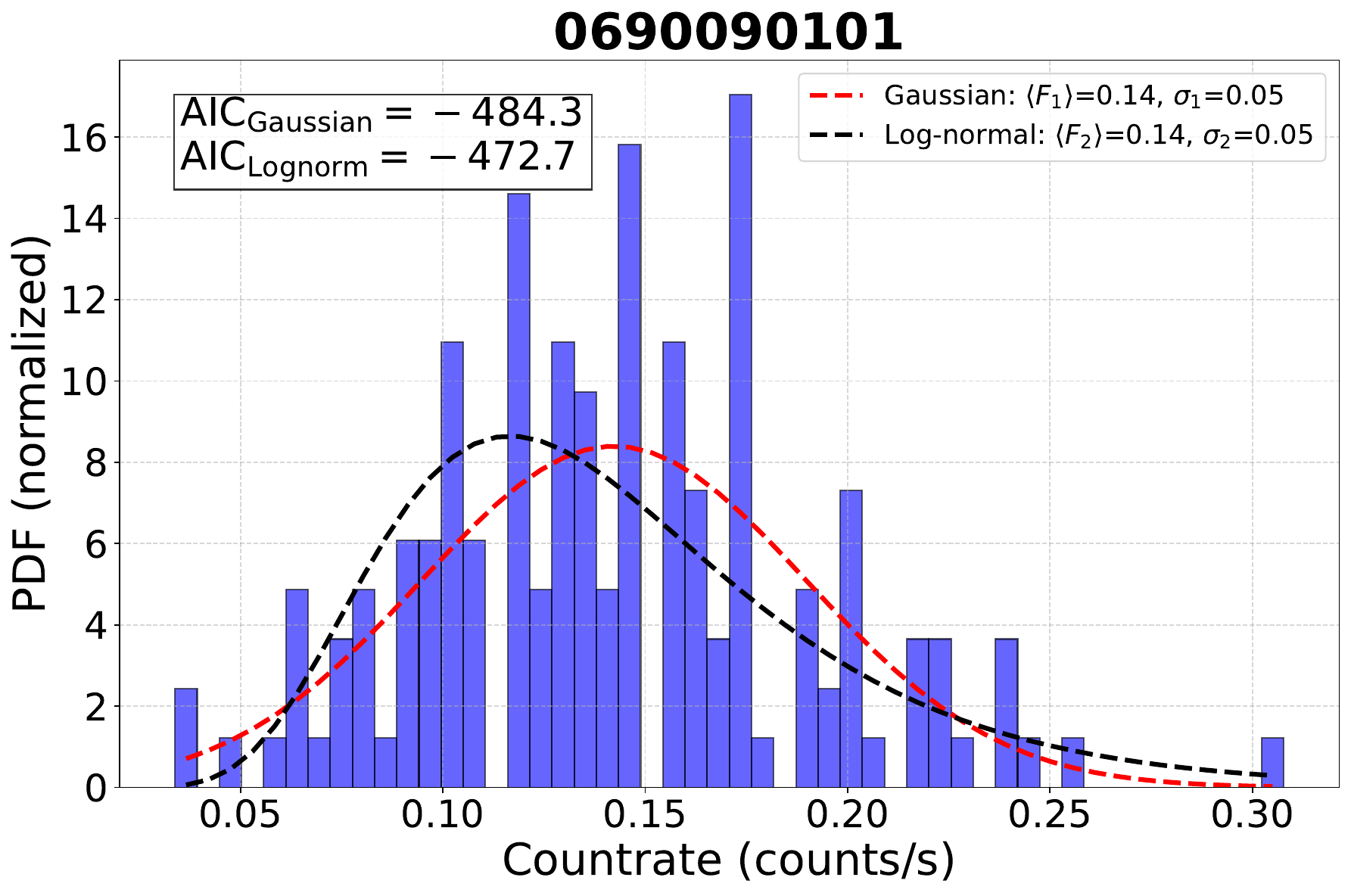}
	\end{minipage}
	\begin{minipage}{.3\textwidth} 
		\centering 
		\includegraphics[height=.99\linewidth, angle=-90]{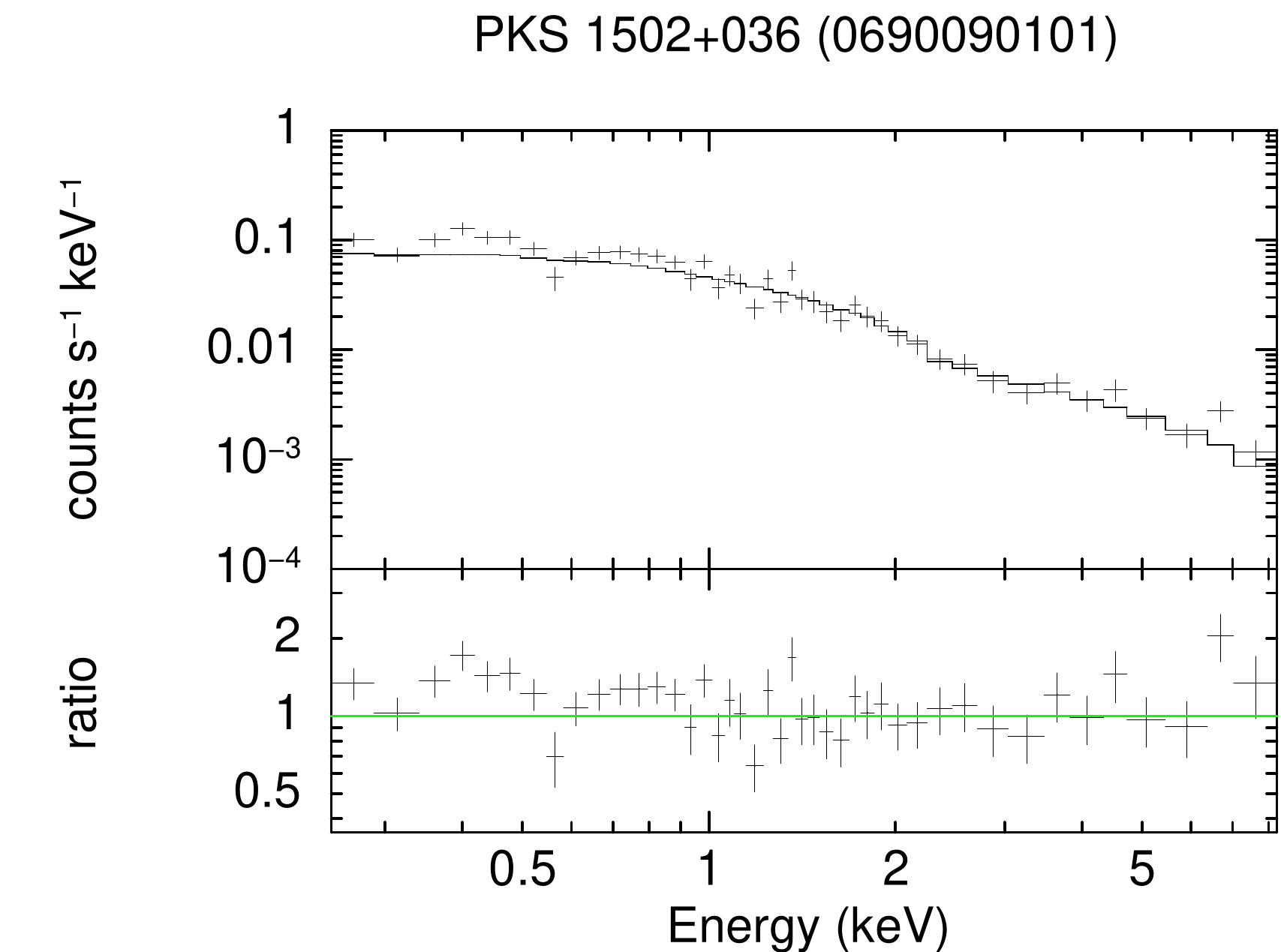}
	\end{minipage}
\end{figure*}

\begin{figure*}\label{app:J1644+263}
	\centering
	\caption{LCs, HR plots, Structure Function, PSD, PDF, and spectral fits derived from observation of J1644+263.}
	\begin{minipage}{.3\textwidth} 
		\centering 
		\includegraphics[width=.99\linewidth]{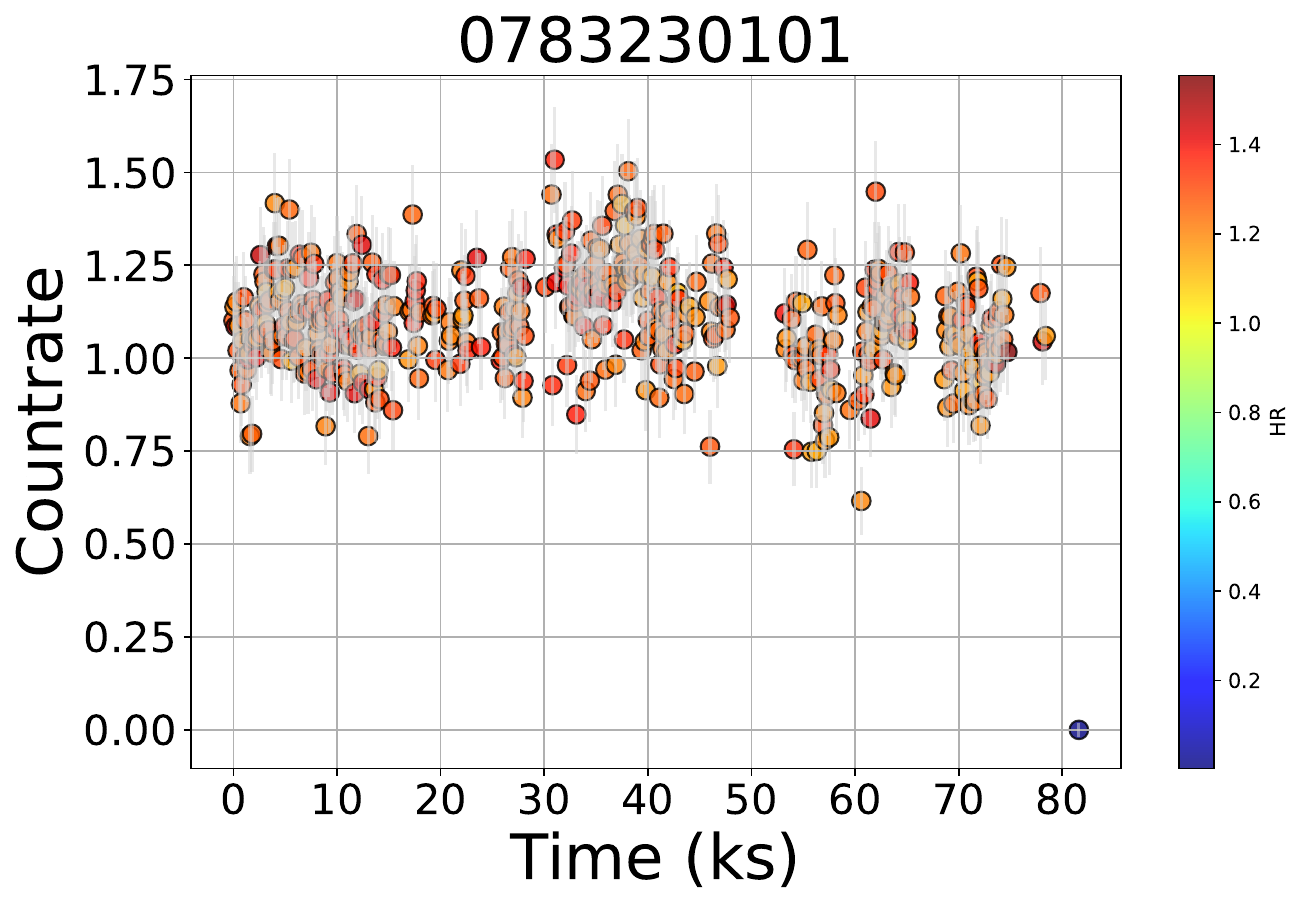}
	\end{minipage}
	\begin{minipage}{.3\textwidth} 
		\centering 
		\includegraphics[width=.99\linewidth]{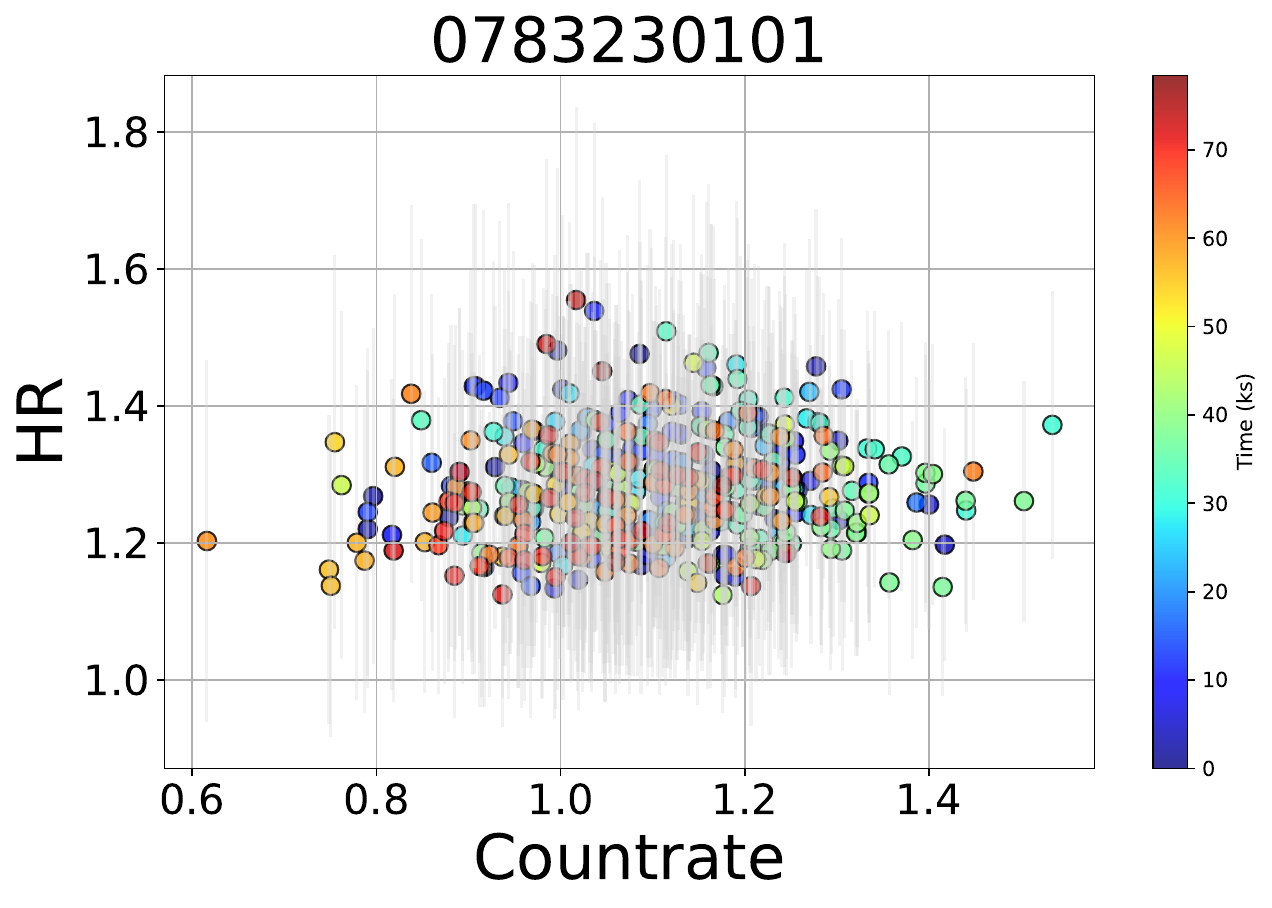}
	\end{minipage}
	\begin{minipage}{.3\textwidth} 
		\centering 
		\includegraphics[width=.99\linewidth, angle=0]{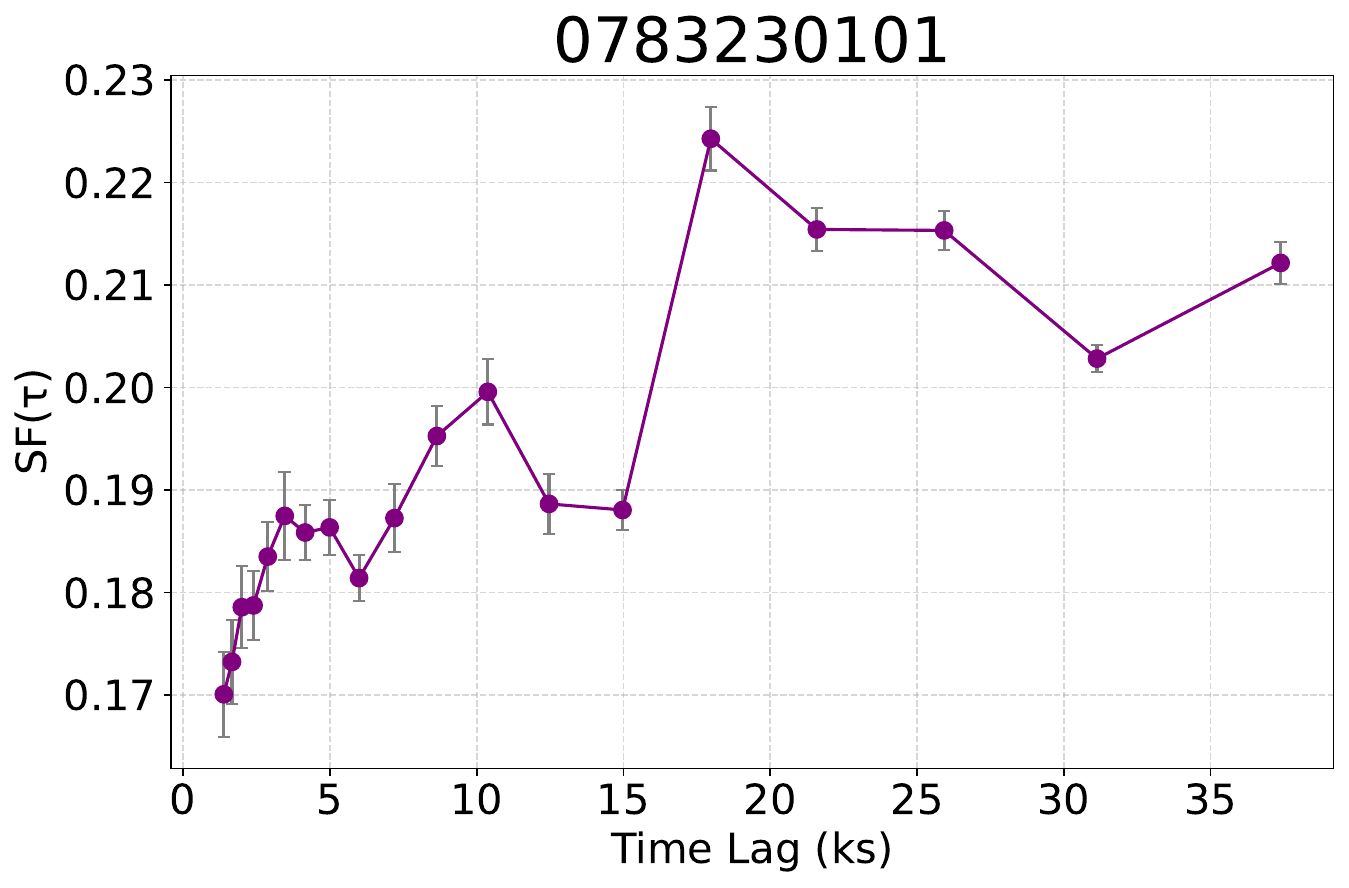}
	\end{minipage}
	\begin{minipage}{.3\textwidth} 
		\centering 
		\includegraphics[width=.99\linewidth]{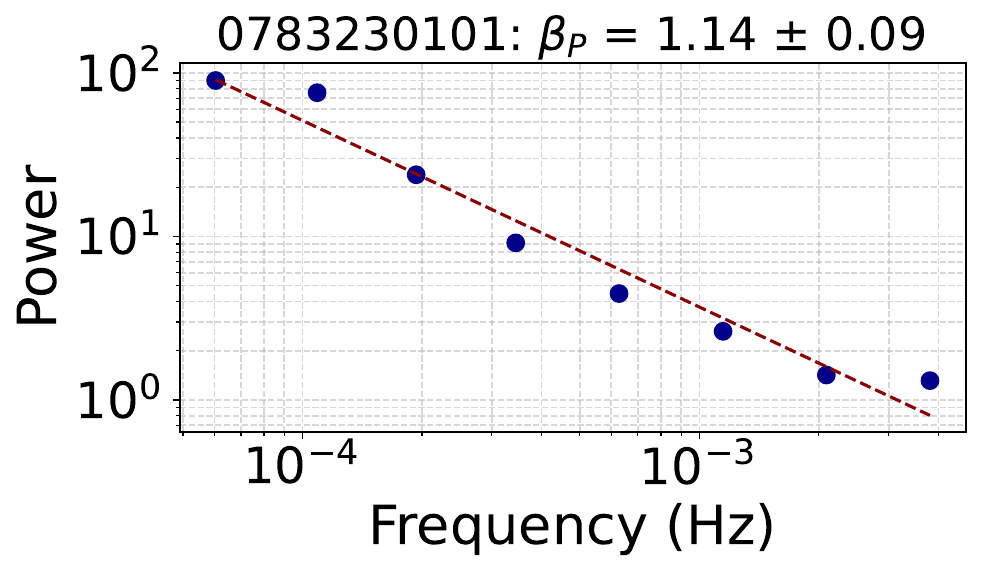}
	\end{minipage}
	\begin{minipage}{.3\textwidth} 
		\centering 
		\includegraphics[width=.99\linewidth]{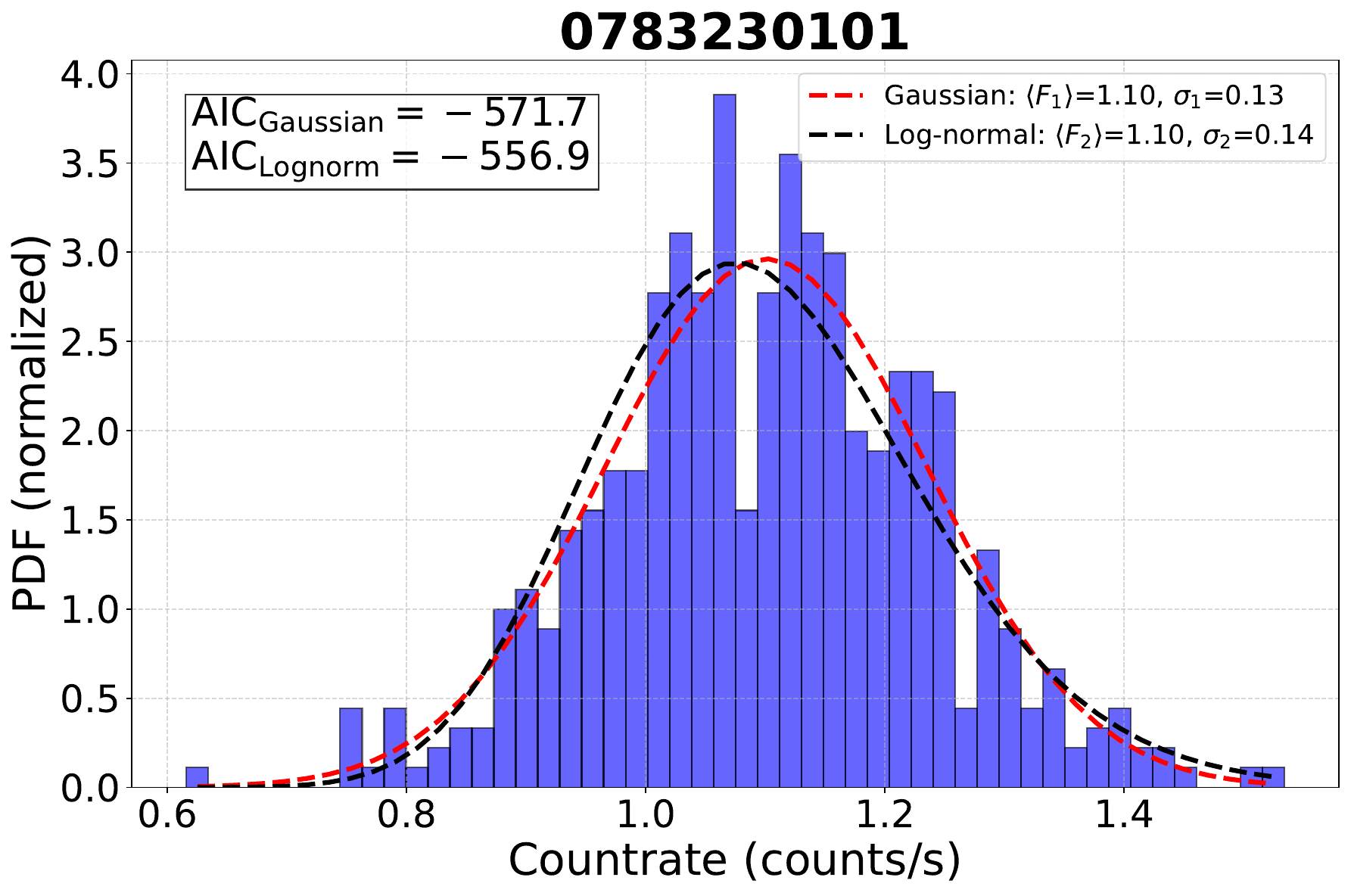}
	\end{minipage}
	\begin{minipage}{.3\textwidth} 
		\centering 
		\includegraphics[height=.99\linewidth, angle=-90]{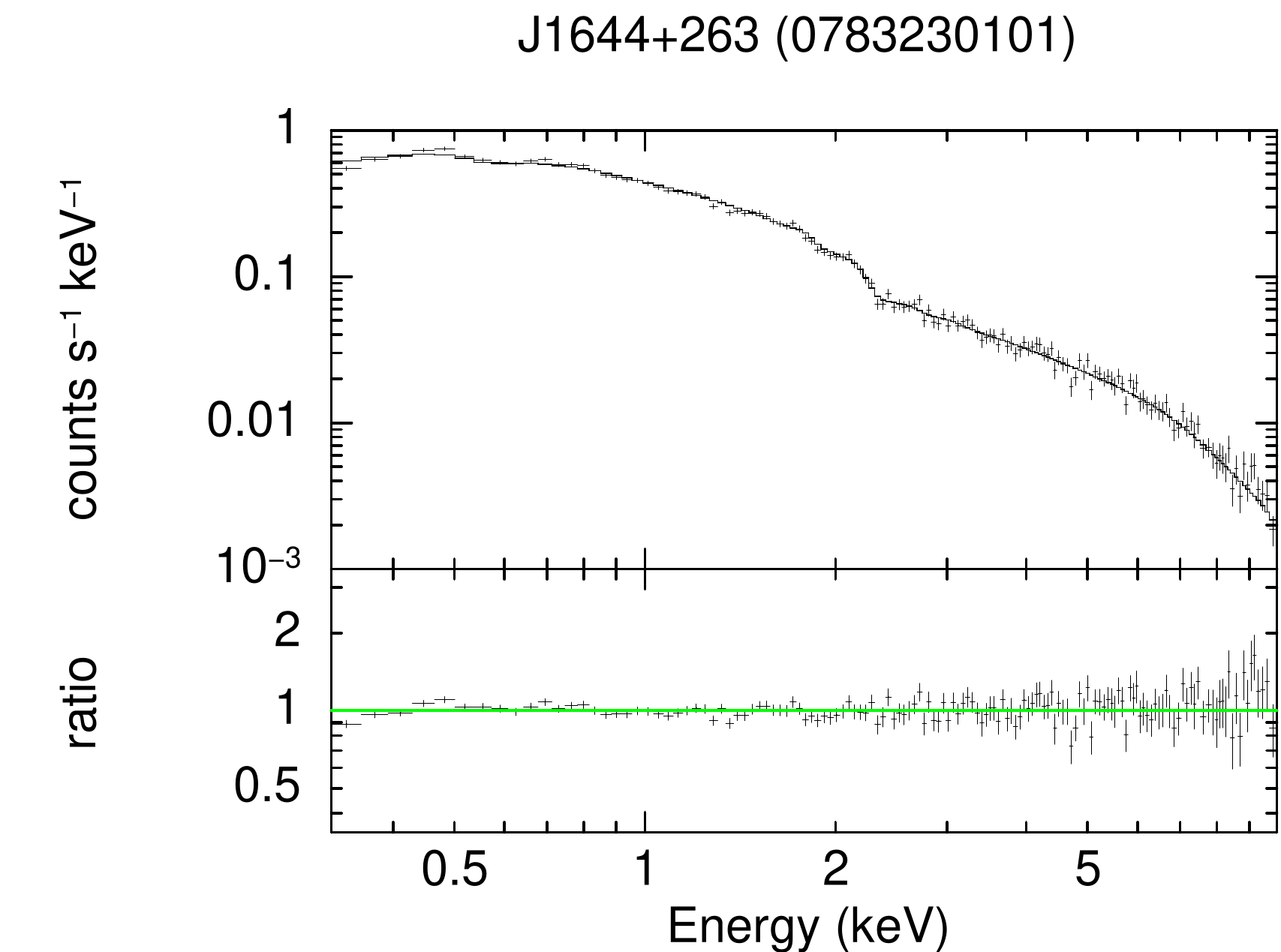}
	\end{minipage}
\end{figure*}

\begin{figure*}\label{app:TXS 2116-077}
	\centering
	\caption{LCs, HR plots, Structure Function, PSD, PDF, and spectral fits derived from observation of TXS 2116-077.}
	\begin{minipage}{.3\textwidth} 
		\centering 
		\includegraphics[width=.99\linewidth]{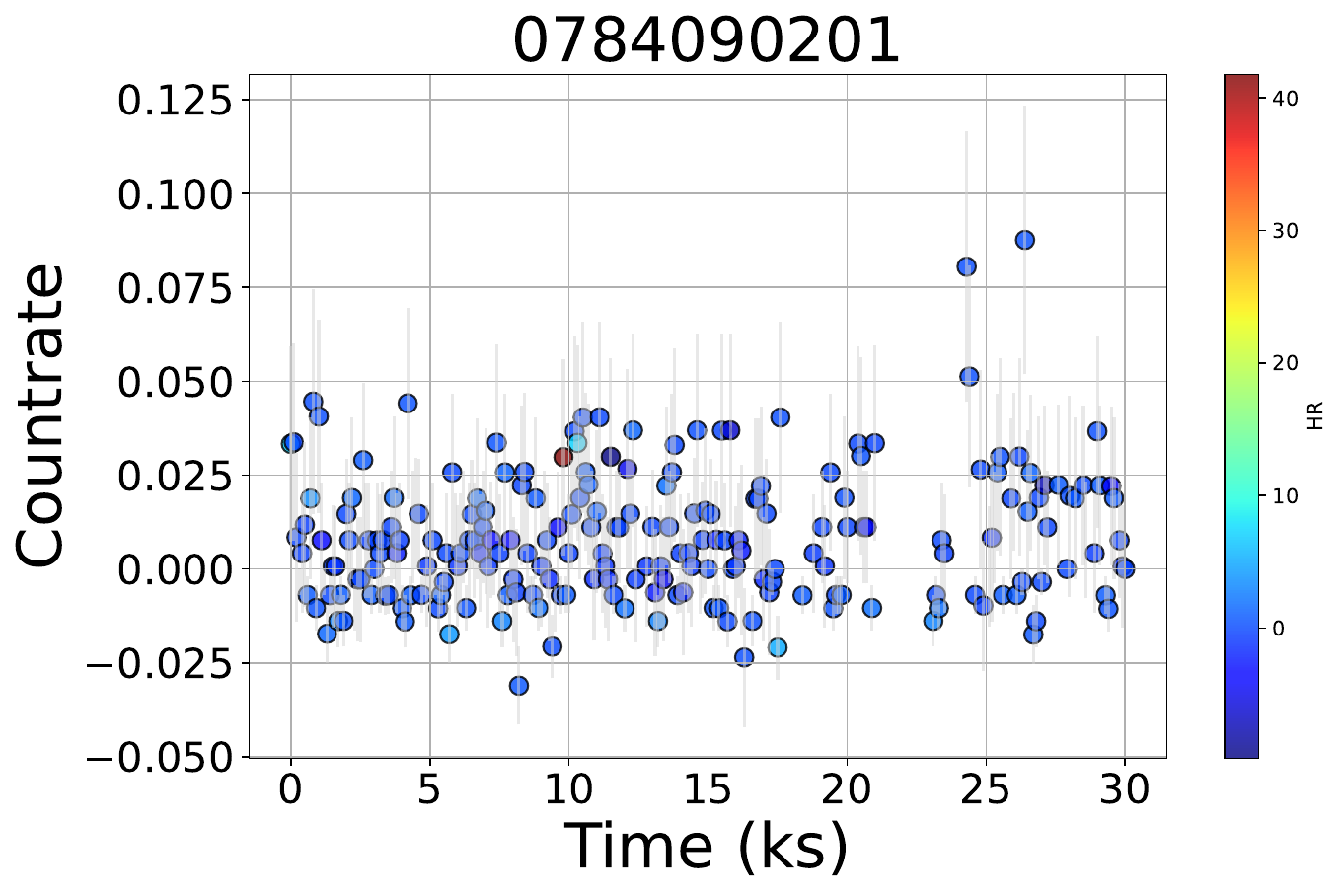}
	\end{minipage}
	\begin{minipage}{.3\textwidth} 
		\centering 
		\includegraphics[width=.99\linewidth]{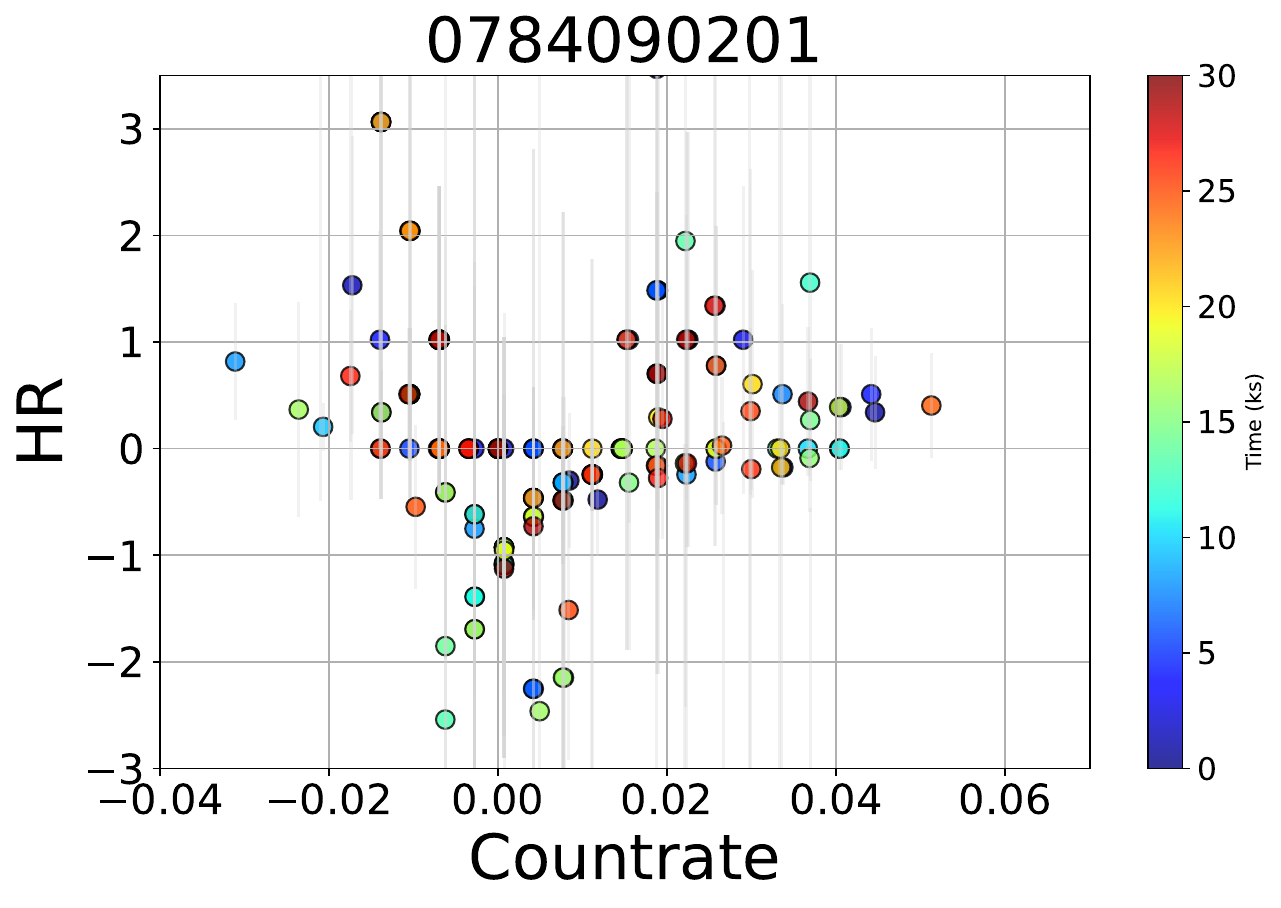}
	\end{minipage}
	\begin{minipage}{.3\textwidth} 
		\centering 
		\includegraphics[width=.99\linewidth, angle=0]{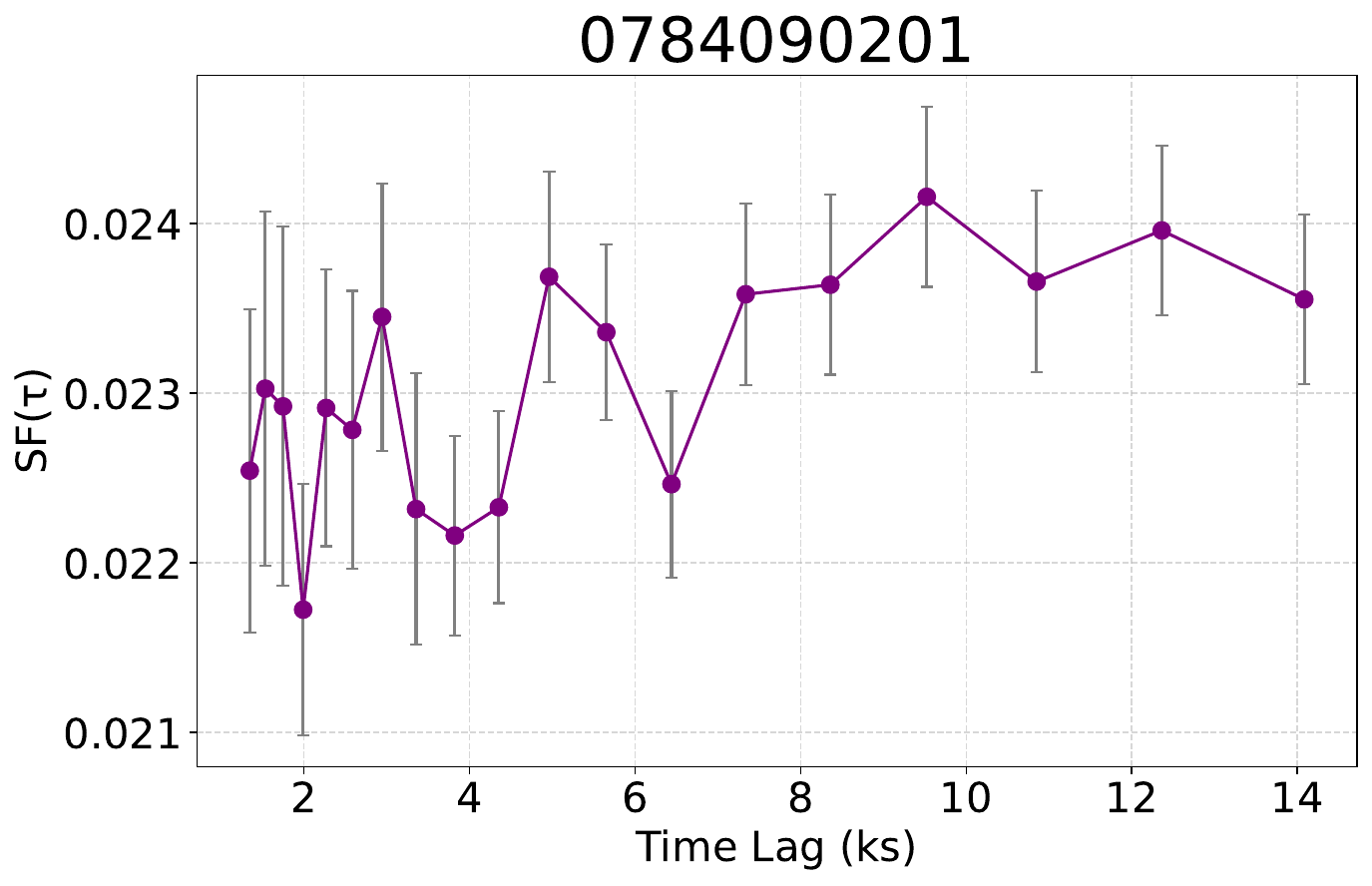}
	\end{minipage}
	\begin{minipage}{.3\textwidth} 
		\centering 
		\includegraphics[width=.99\linewidth]{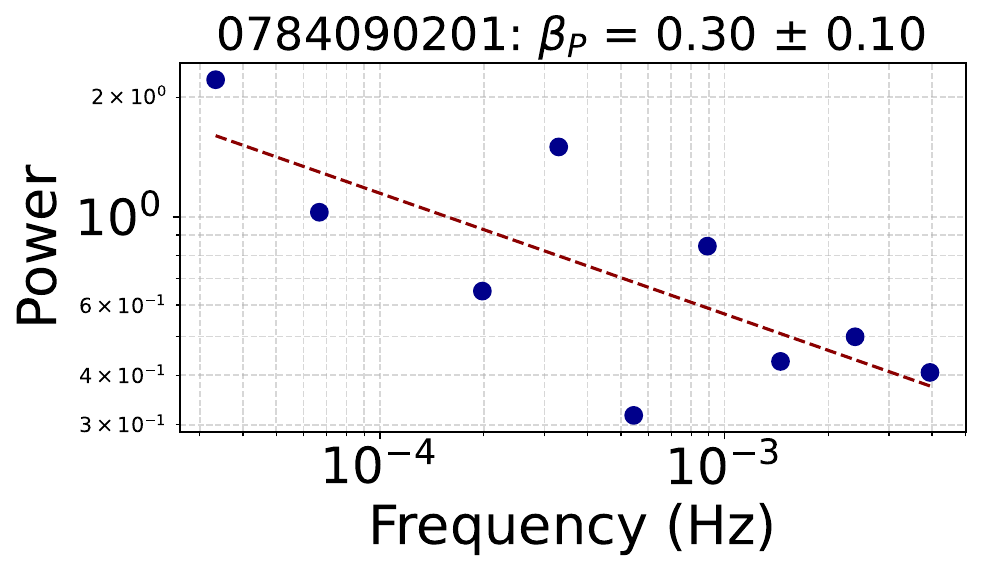}
	\end{minipage}
	\begin{minipage}{.3\textwidth} 
		\centering 
		\includegraphics[width=.99\linewidth]{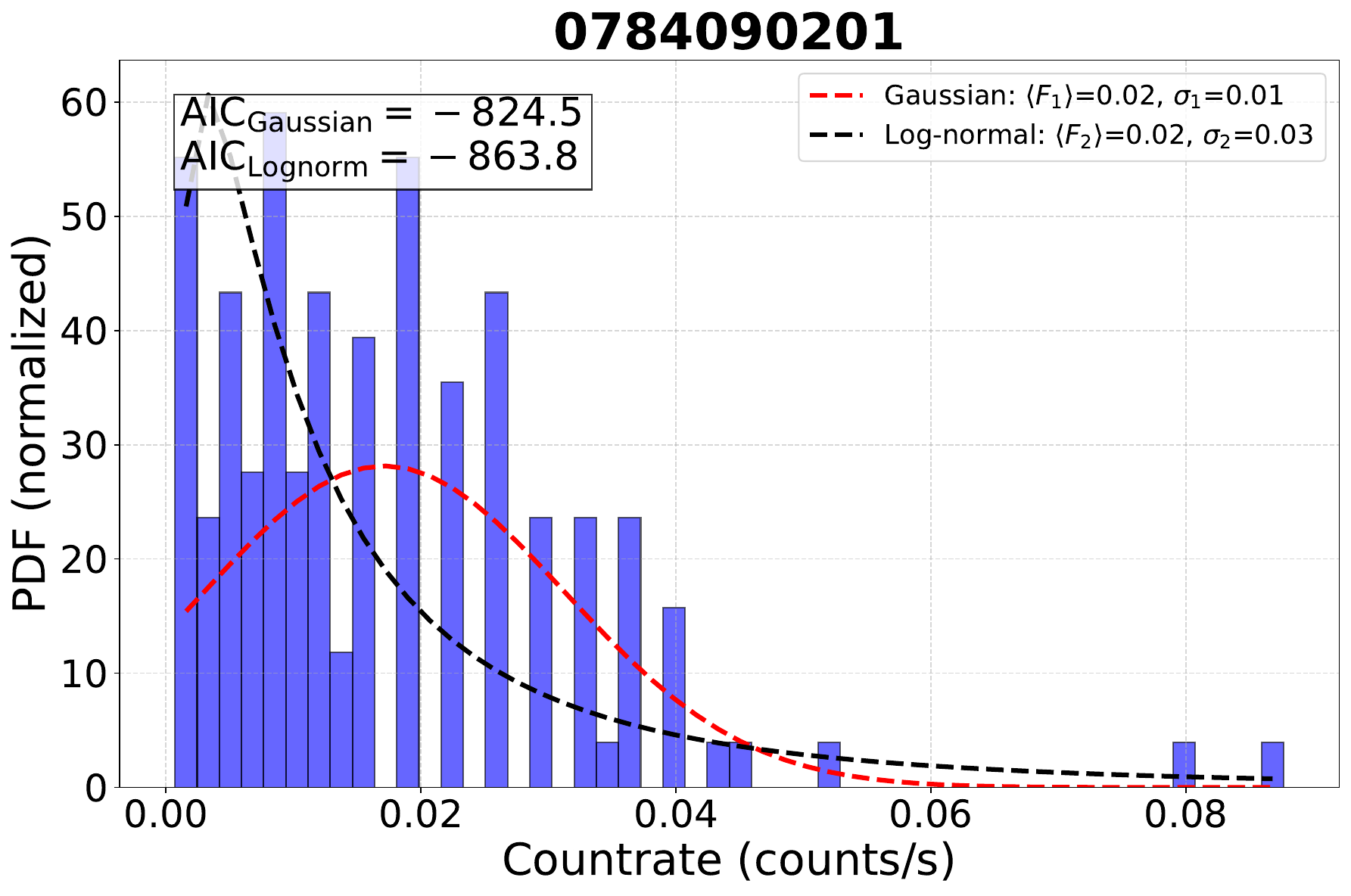}
	\end{minipage}
	\begin{minipage}{.3\textwidth} 
		\centering 
		\includegraphics[height=.99\linewidth, angle=-90]{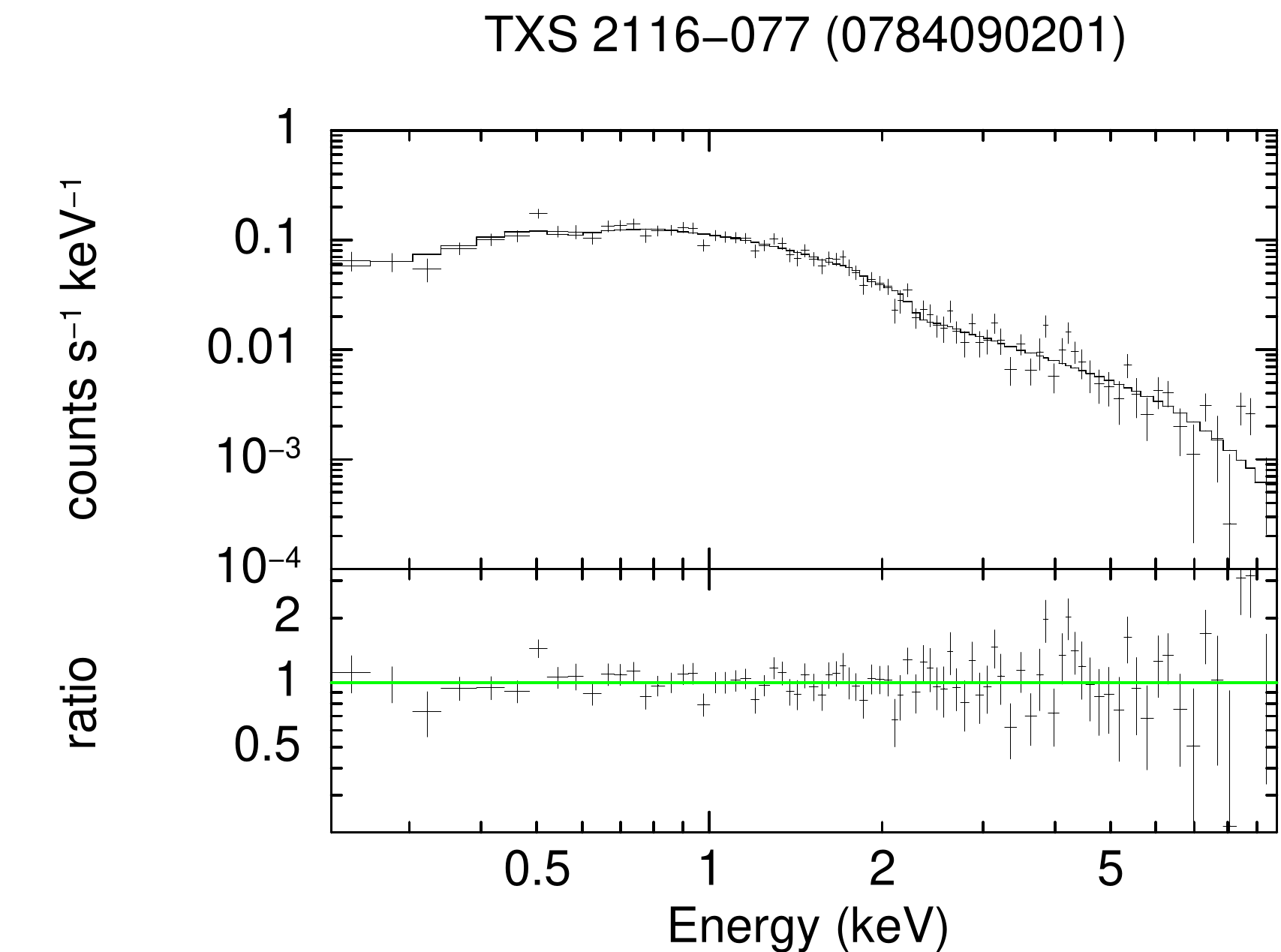}
	\end{minipage}

    \begin{minipage}{.3\textwidth} 
		\centering 
		\includegraphics[width=.99\linewidth]{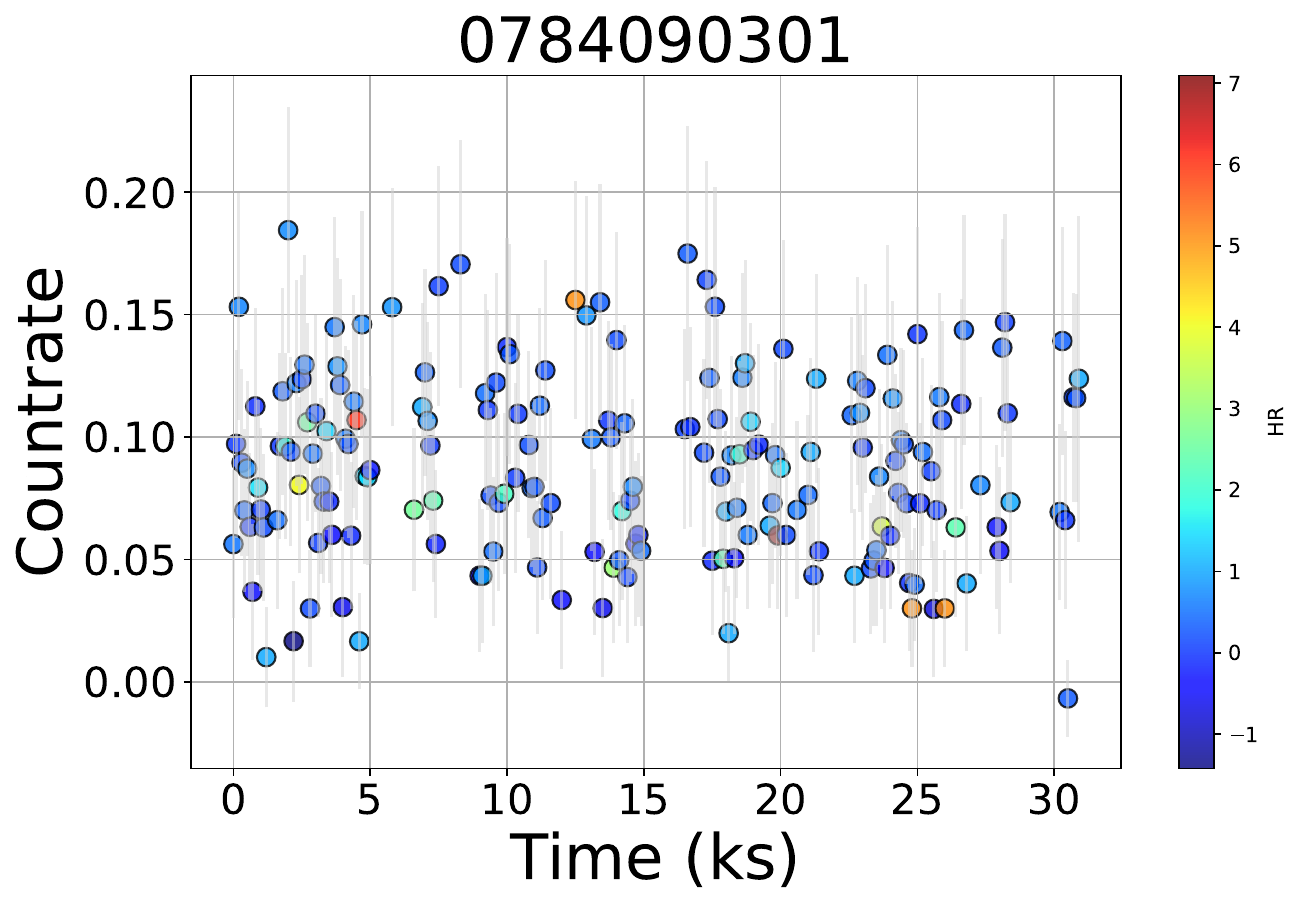}
	\end{minipage}
	\begin{minipage}{.3\textwidth} 
		\centering 
		\includegraphics[width=.99\linewidth]{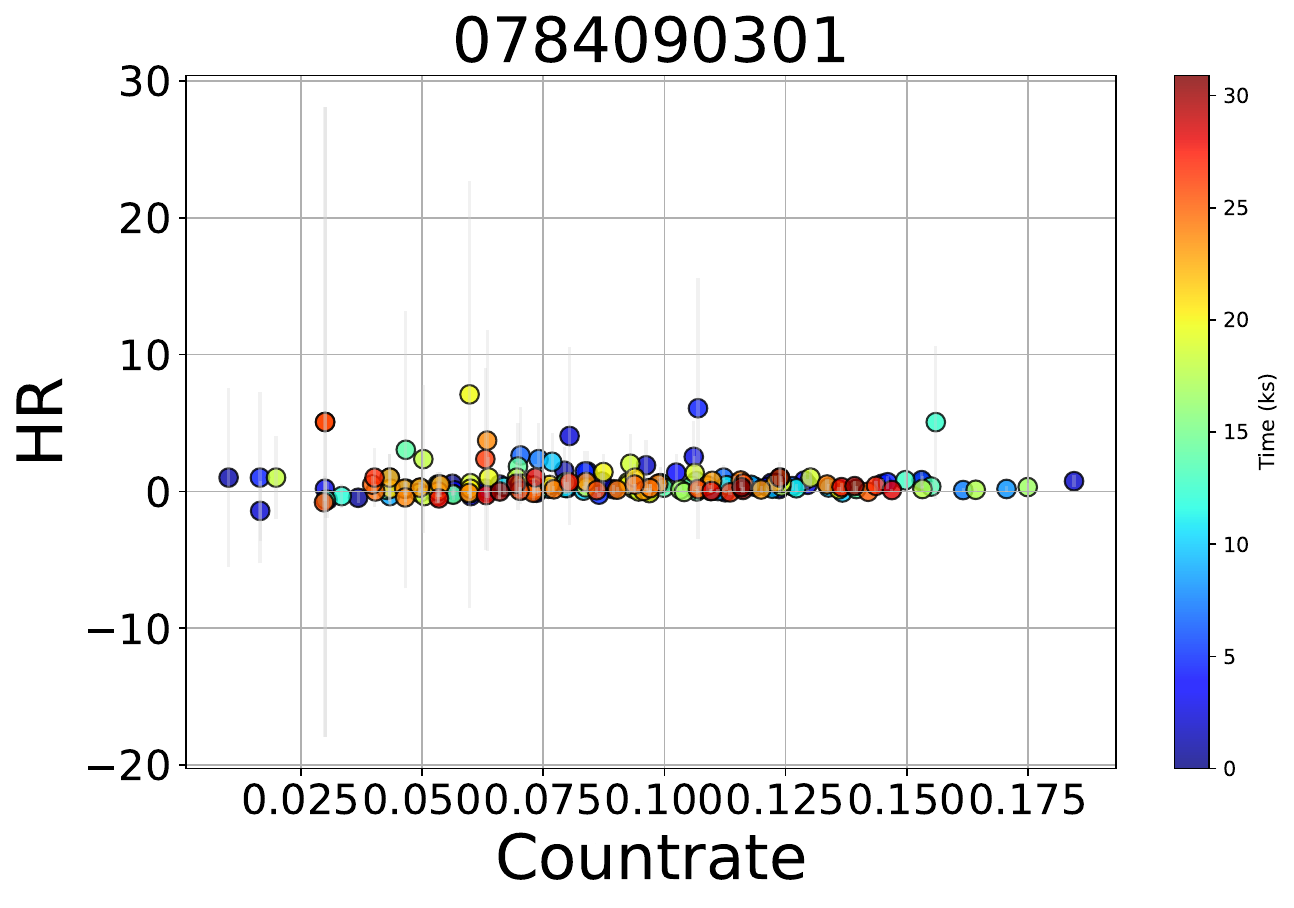}
	\end{minipage}
	\begin{minipage}{.3\textwidth} 
		\centering 
		\includegraphics[width=.99\linewidth, angle=0]{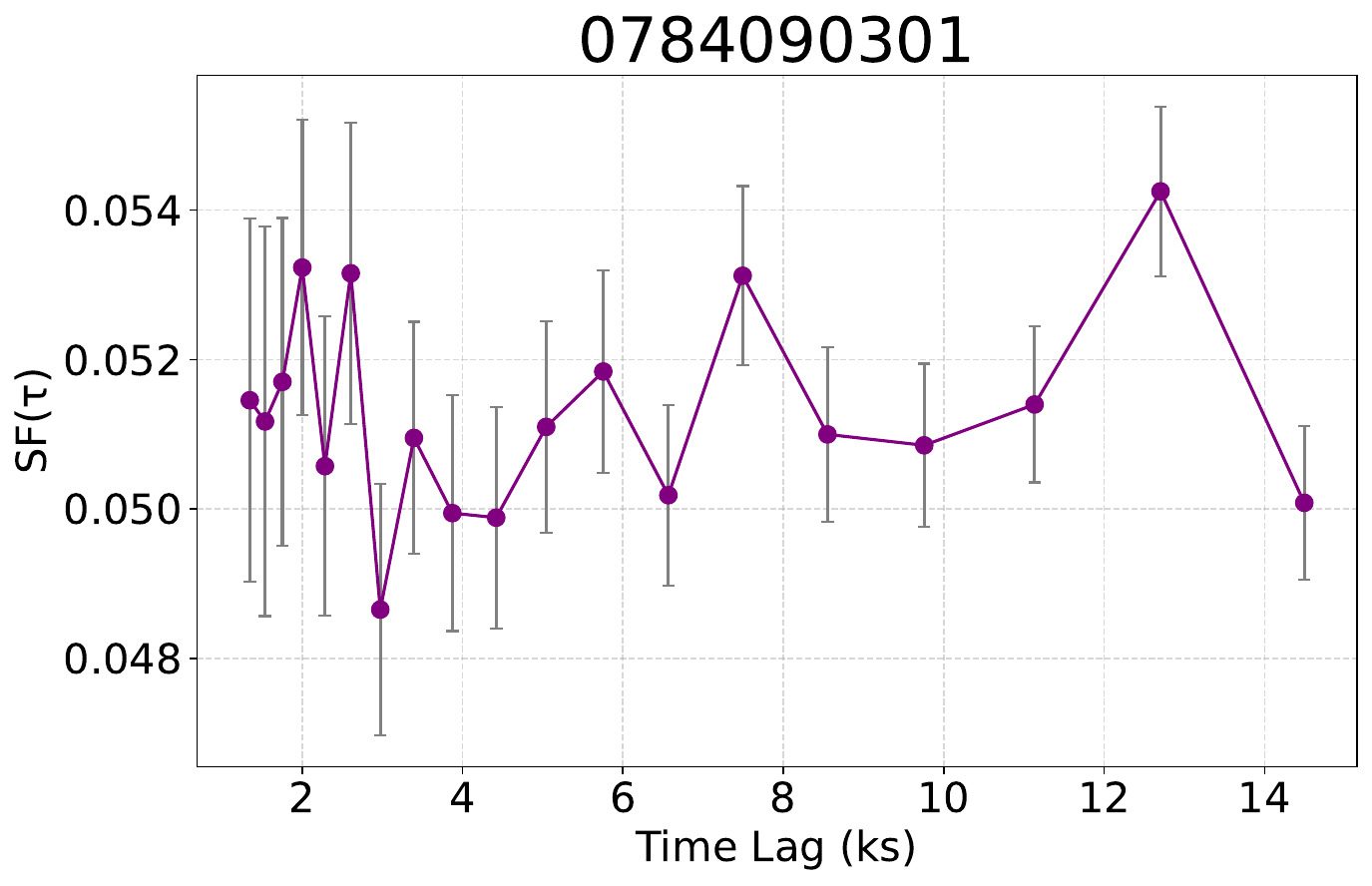}
	\end{minipage}
	\begin{minipage}{.3\textwidth} 
		\centering 
		\includegraphics[width=.99\linewidth]{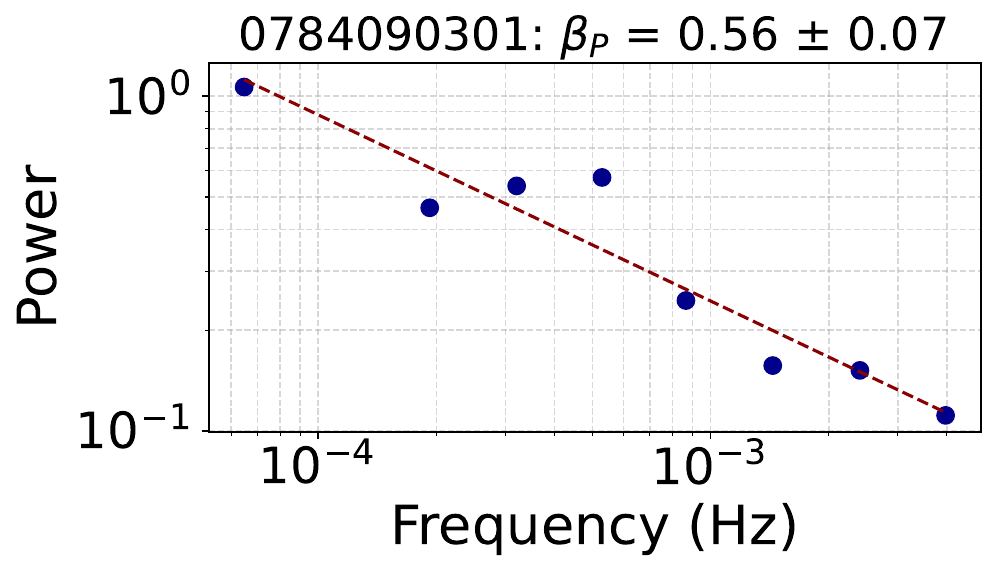}
	\end{minipage}
	\begin{minipage}{.3\textwidth} 
		\centering 
		\includegraphics[width=.99\linewidth]{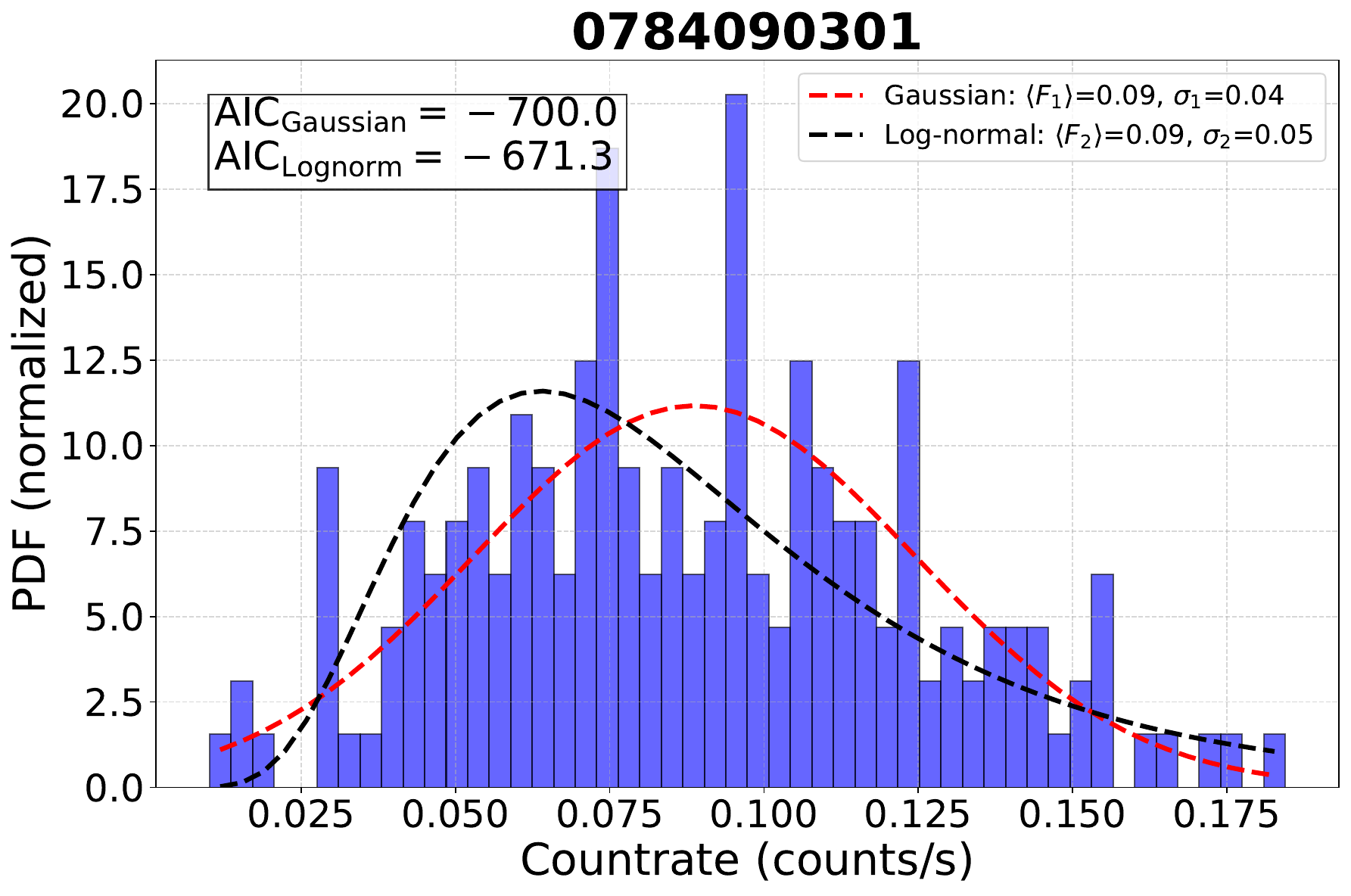}
	\end{minipage}
	\begin{minipage}{.3\textwidth} 
		\centering 
		\includegraphics[height=.99\linewidth, angle=-90]{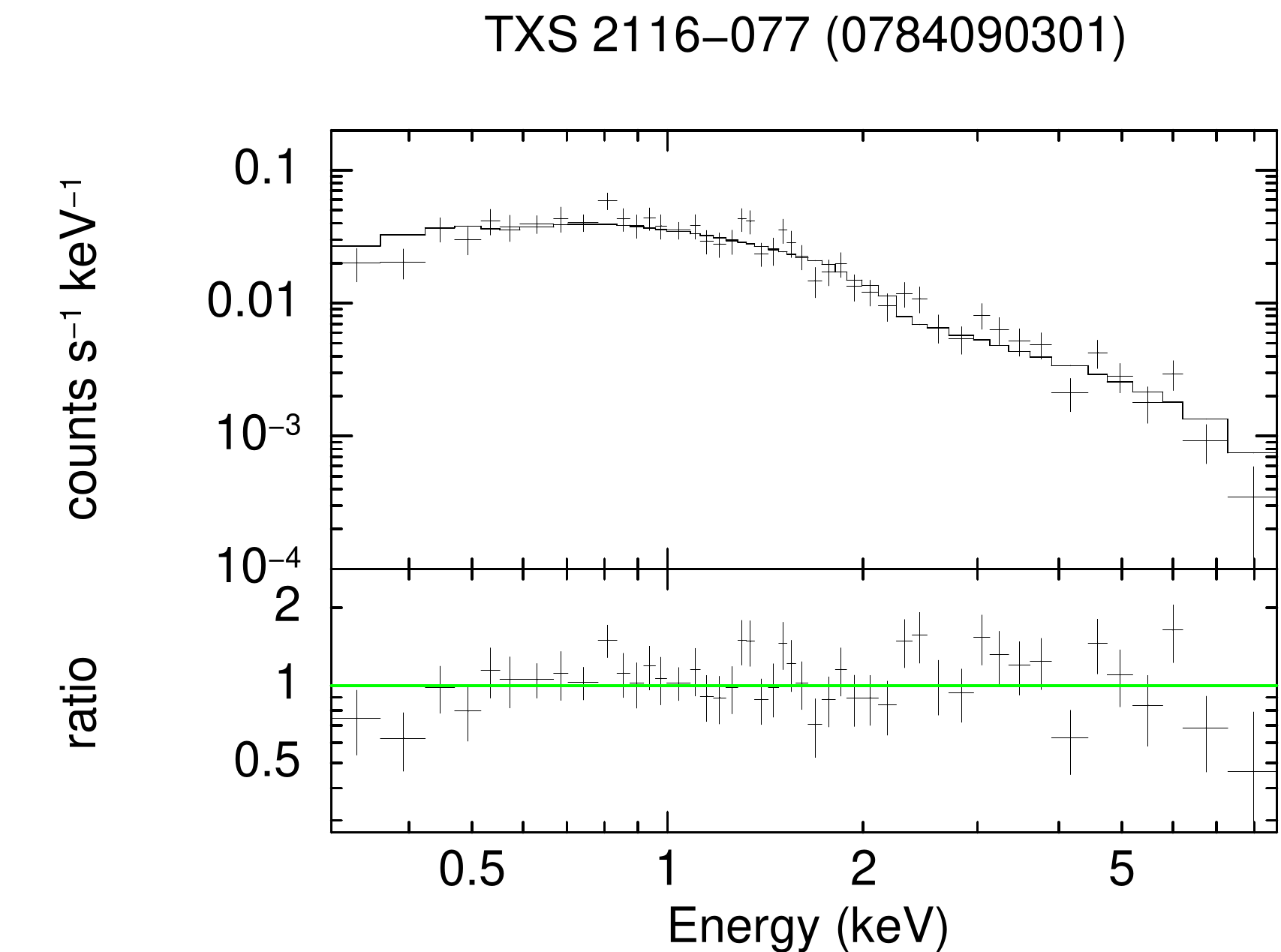}
	\end{minipage}
\end{figure*}

\begin{figure*}\label{app:J1102+2239}
	\centering
	\caption{LCs, HR plots, Structure Function, PSD, PDF, and spectral fits derived for FBQS J1102+2239.}
	\begin{minipage}{.3\textwidth} 
		\centering 
		\includegraphics[width=.99\linewidth]{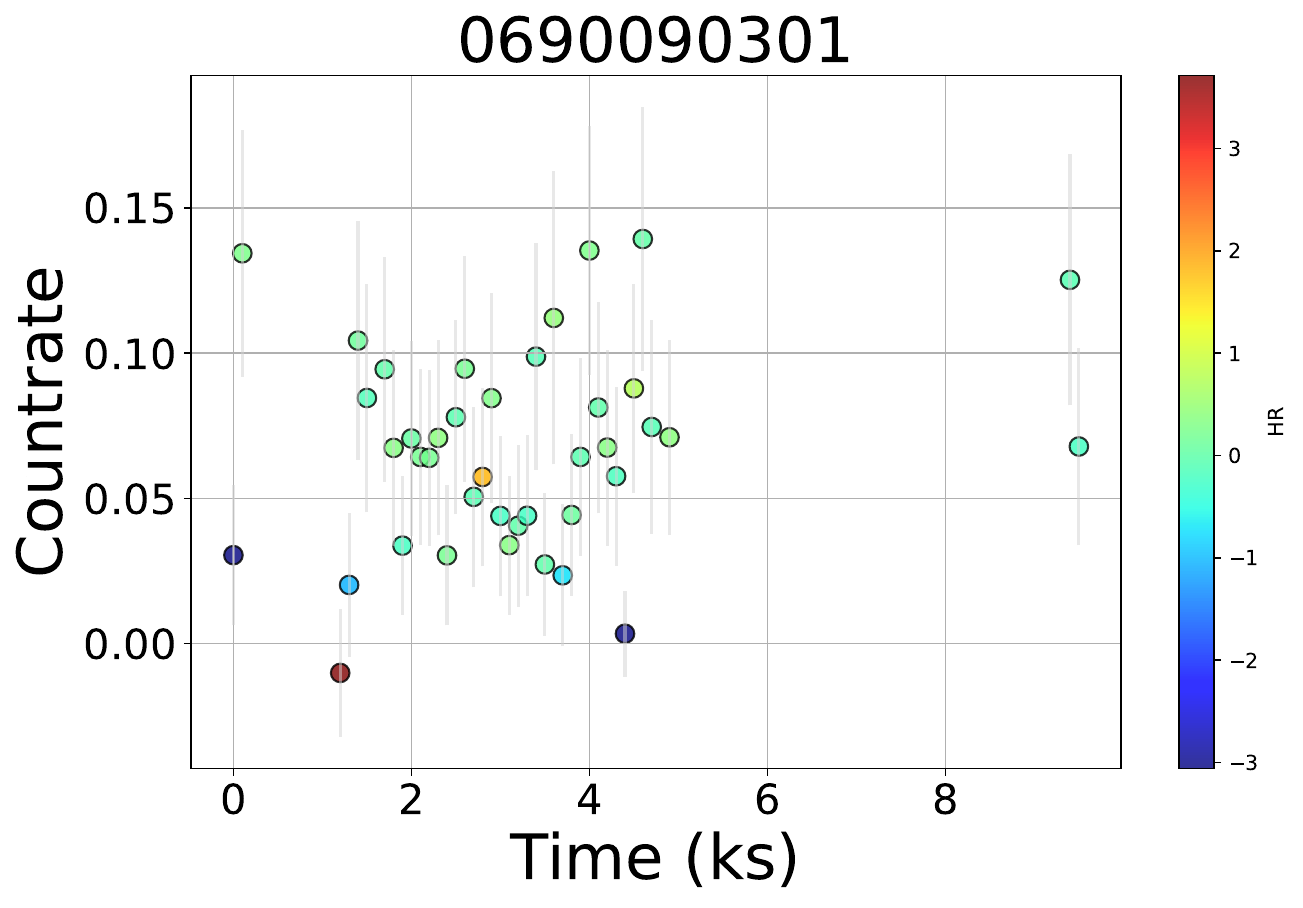}
	\end{minipage}
	\begin{minipage}{.3\textwidth} 
		\centering 
		\includegraphics[width=.99\linewidth]{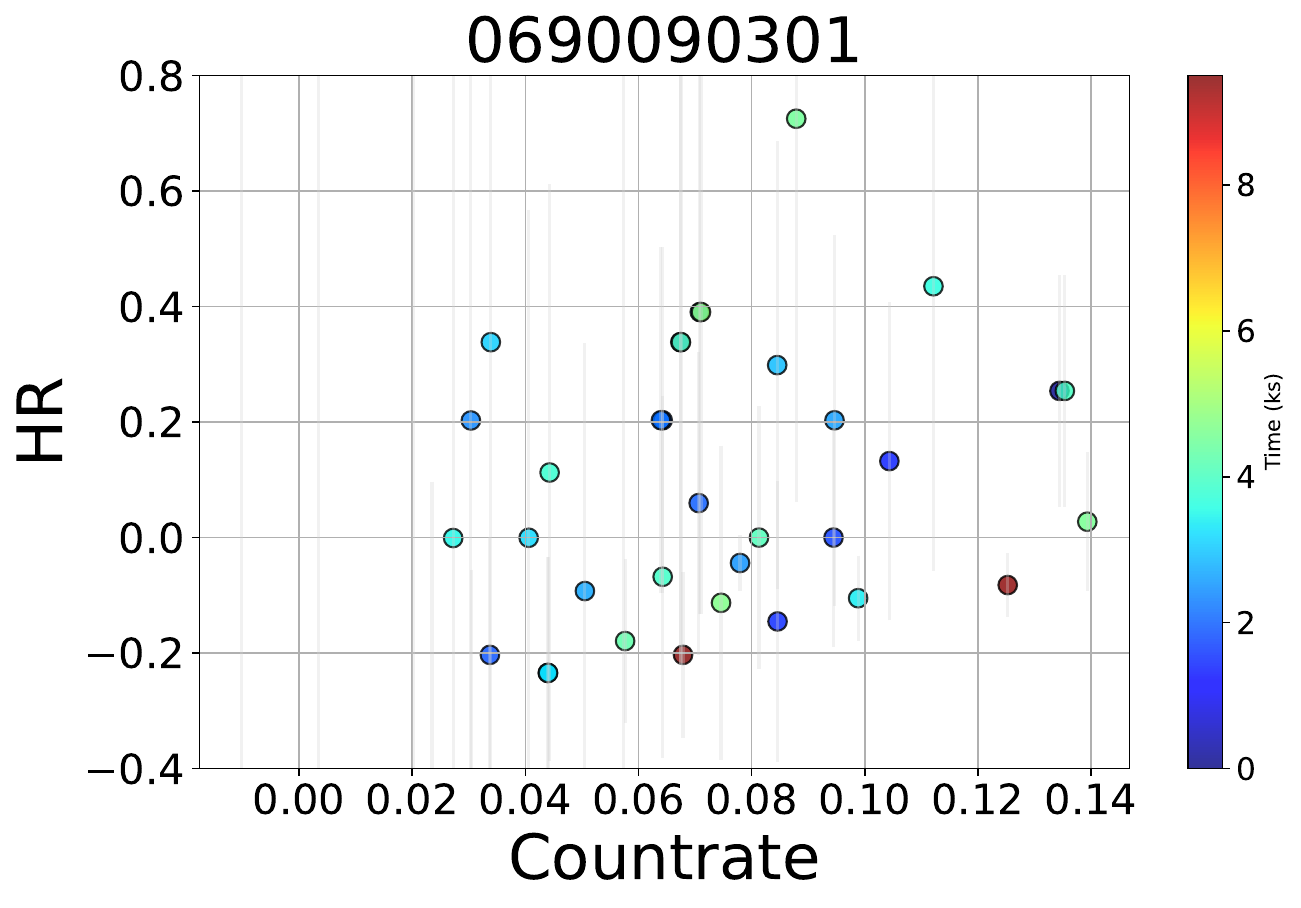}
	\end{minipage}
	\begin{minipage}{.3\textwidth} 
		\centering 
		\includegraphics[width=.99\linewidth, angle=0]{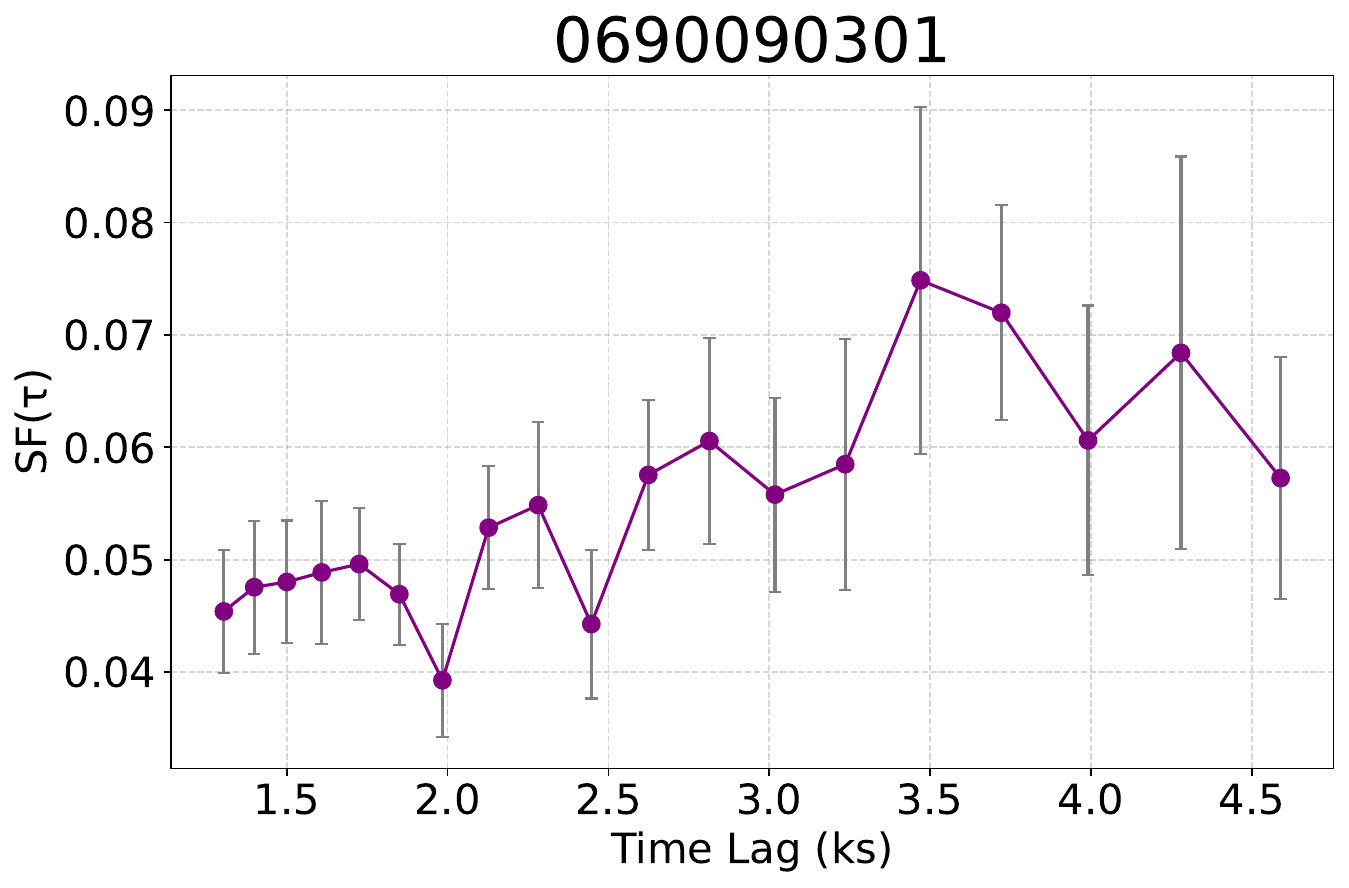}
	\end{minipage}
	\begin{minipage}{.3\textwidth} 
		\centering 
		\includegraphics[width=.99\linewidth]{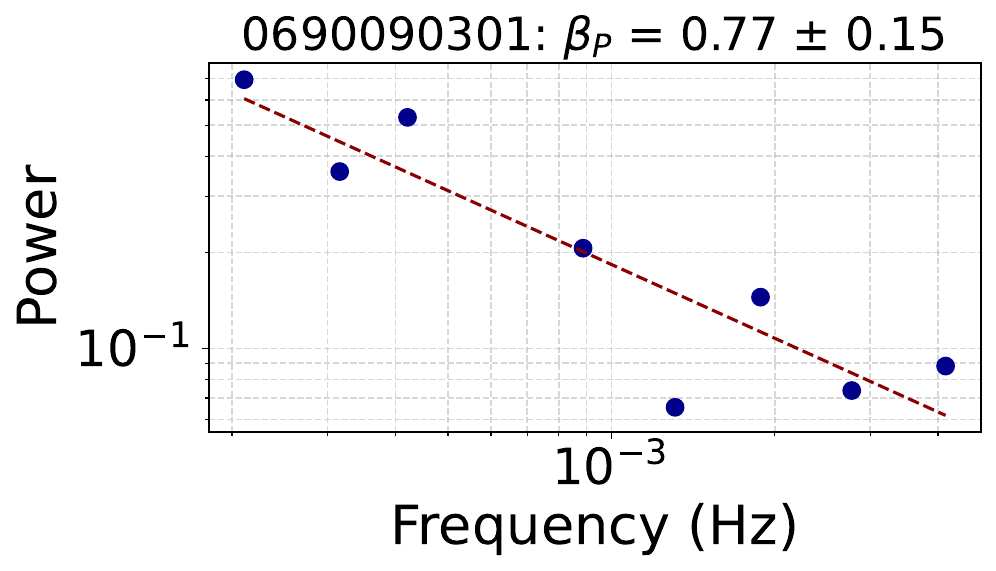}
	\end{minipage}
	\begin{minipage}{.3\textwidth} 
		\centering 
		\includegraphics[width=.99\linewidth]{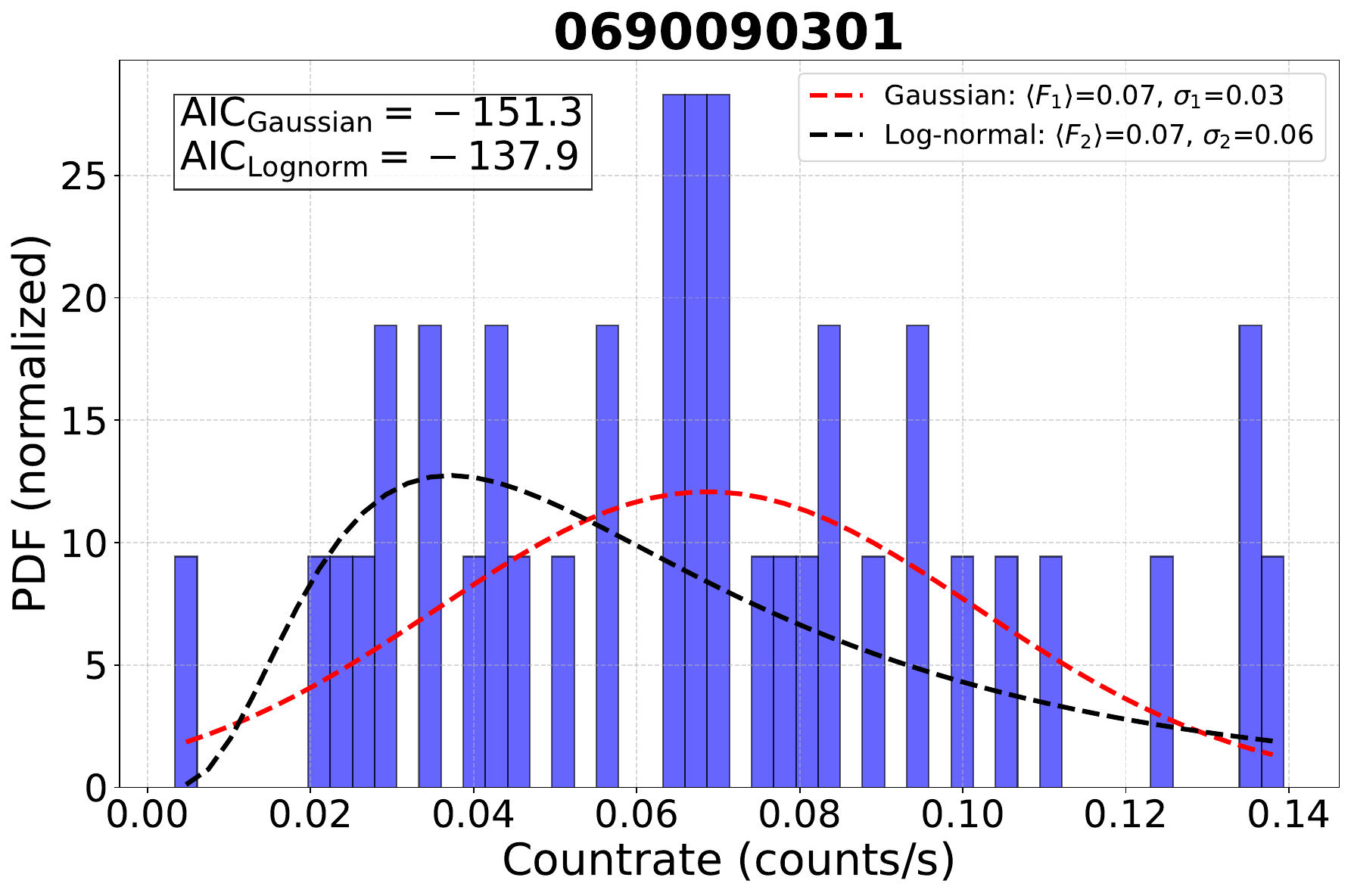}
	\end{minipage}
	\begin{minipage}{.3\textwidth} 
		\centering 
		\includegraphics[height=.99\linewidth, angle=-90]{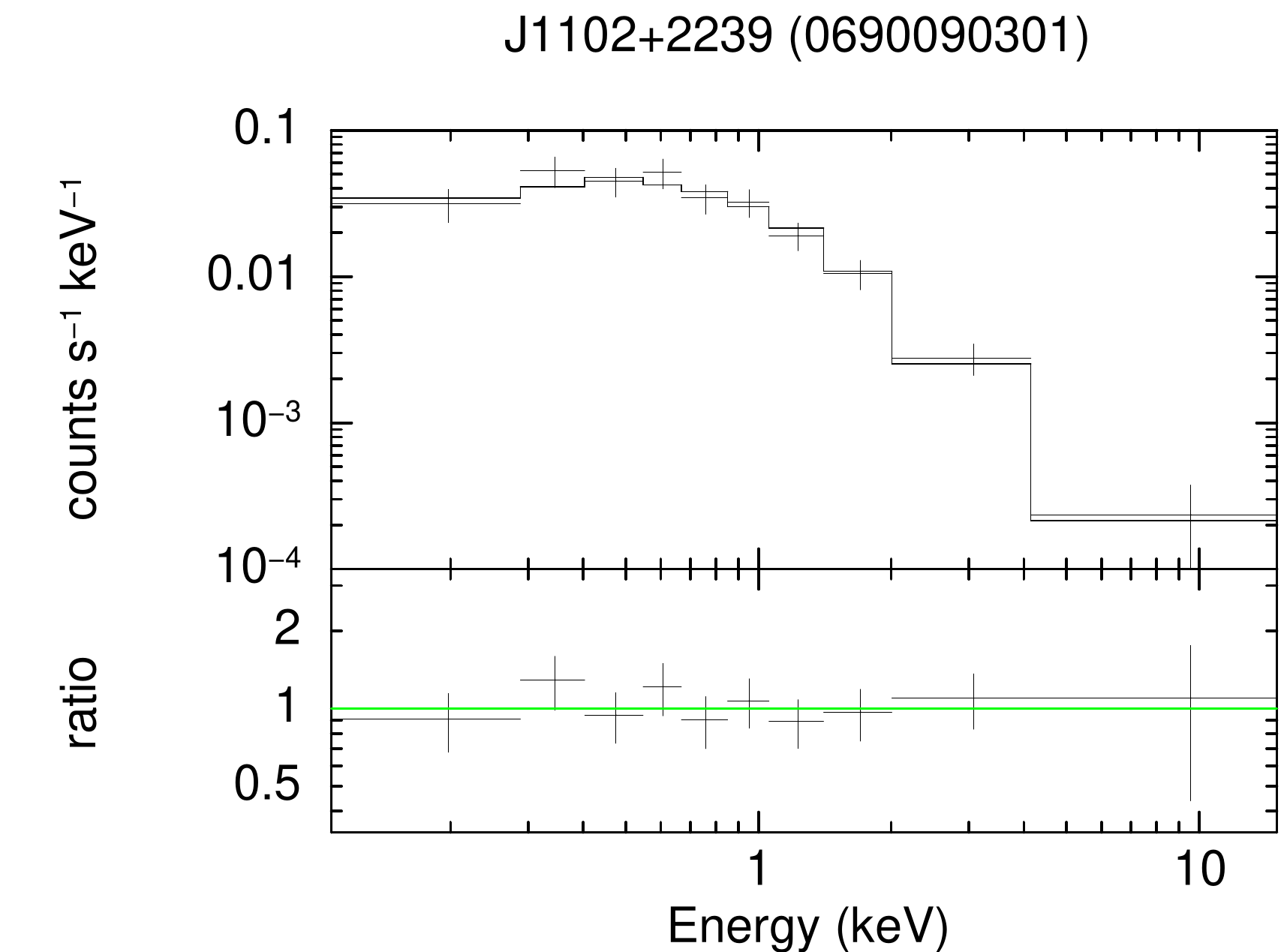}
	\end{minipage}
\end{figure*}

\begin{figure*}\label{app:J1222+0413}
	\centering
	\caption{LCs, HR plots, Structure Function, PSD, PDF, and spectral fits derived for J1222+0413.}
	\begin{minipage}{.3\textwidth} 
		\centering 
		\includegraphics[width=.99\linewidth]{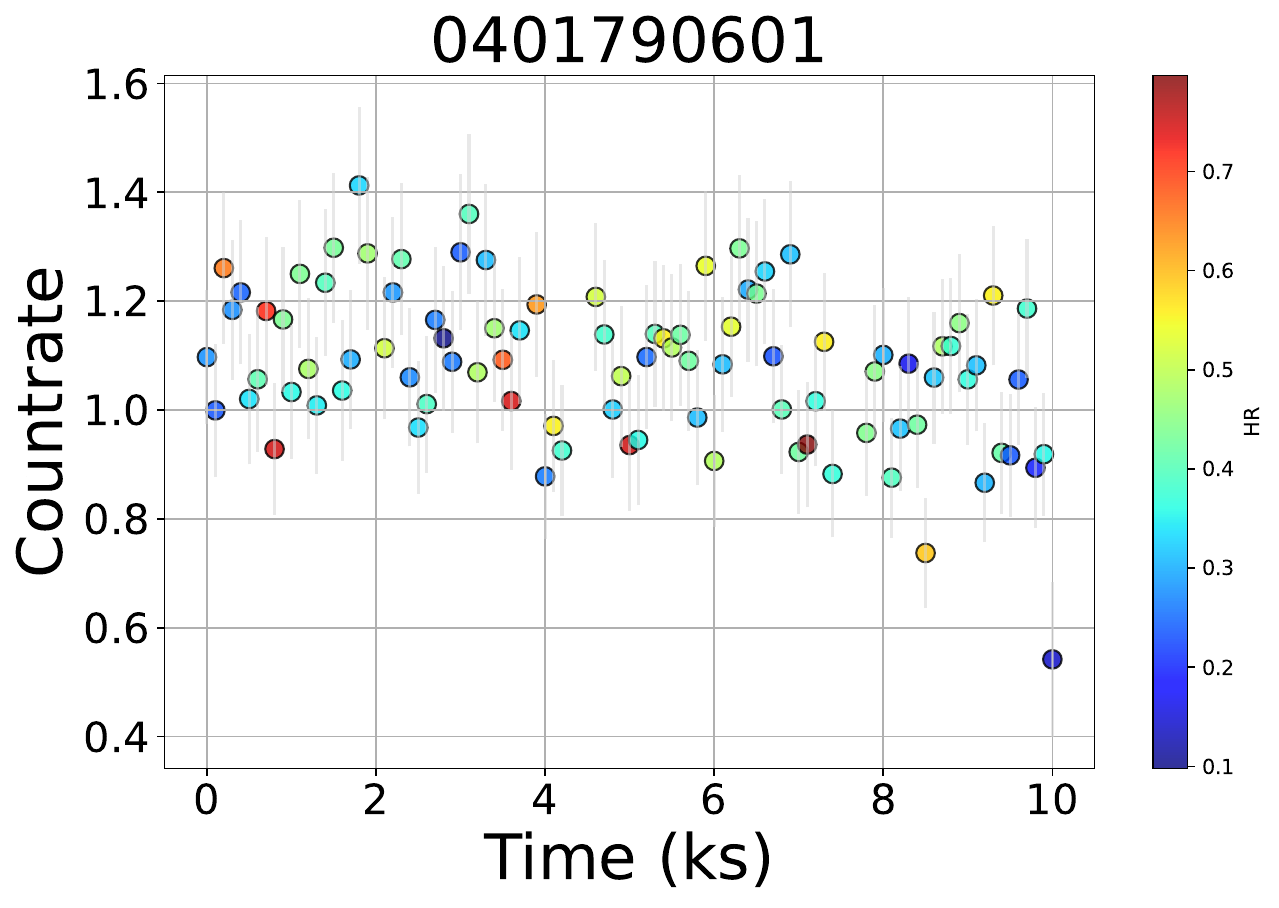}
	\end{minipage}
	\begin{minipage}{.3\textwidth} 
		\centering 
		\includegraphics[width=.99\linewidth]{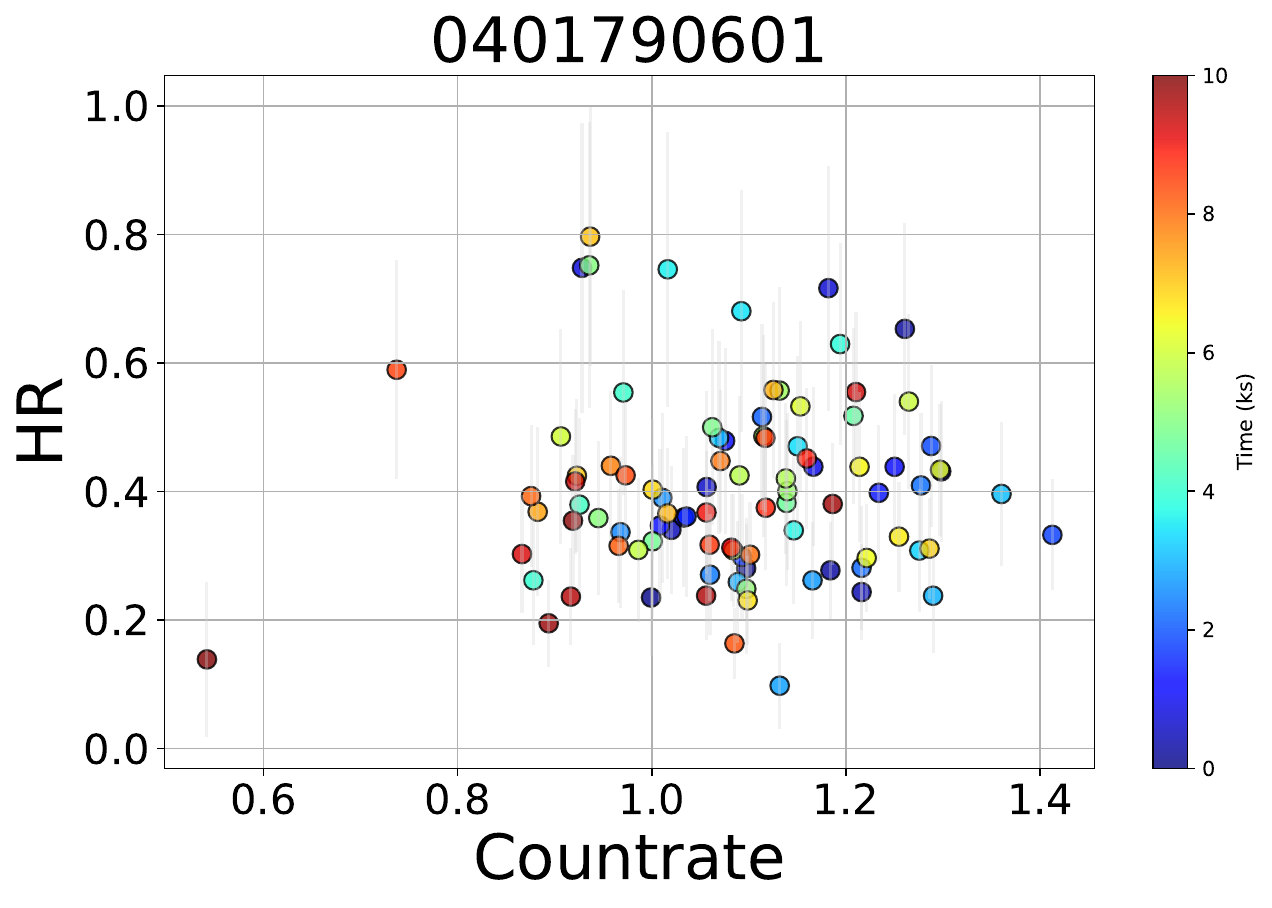}
	\end{minipage}
	\begin{minipage}{.3\textwidth} 
		\centering 
		\includegraphics[width=.99\linewidth, angle=0]{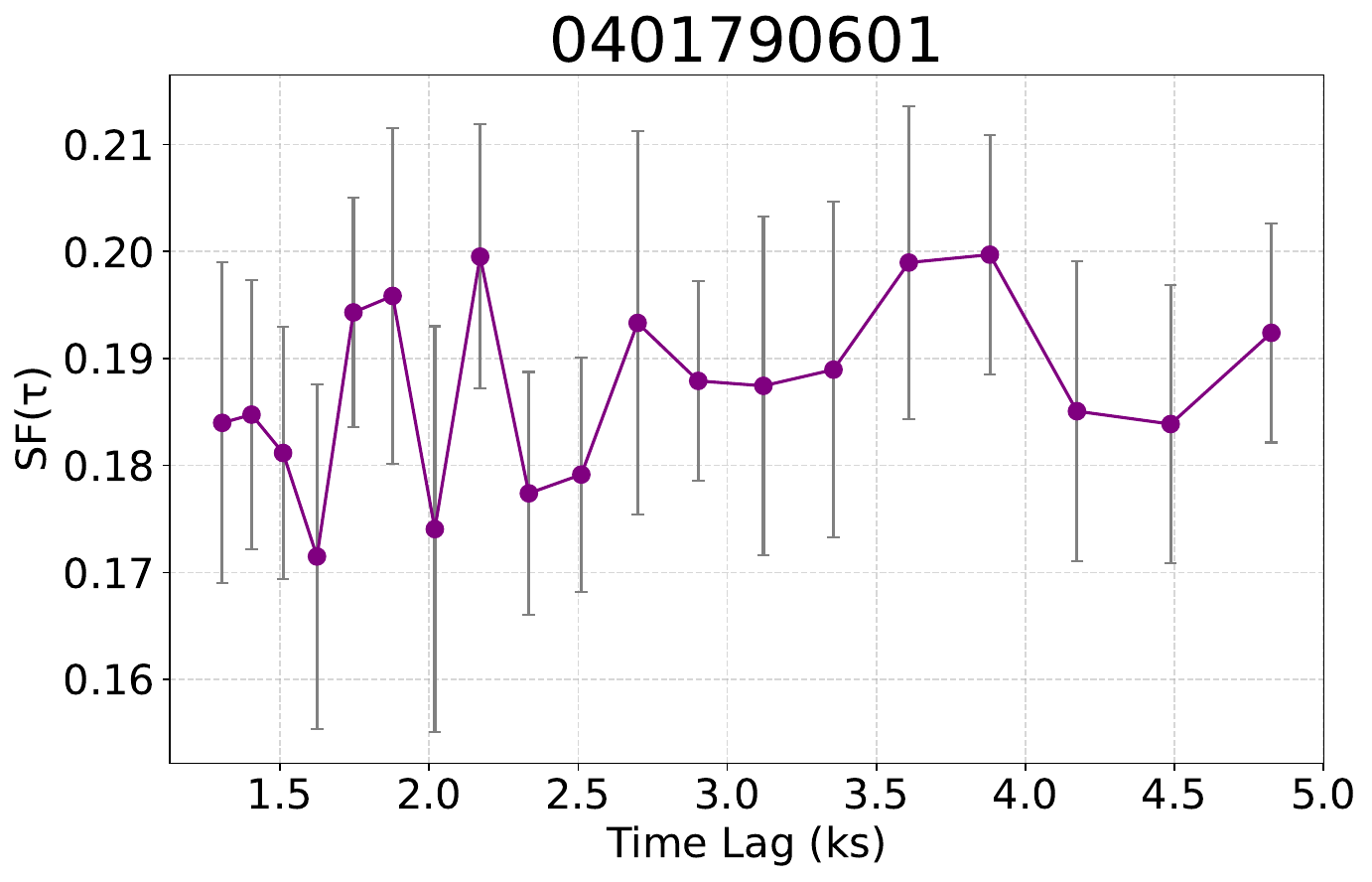}
	\end{minipage}
	\begin{minipage}{.3\textwidth} 
		\centering 
		\includegraphics[width=.99\linewidth]{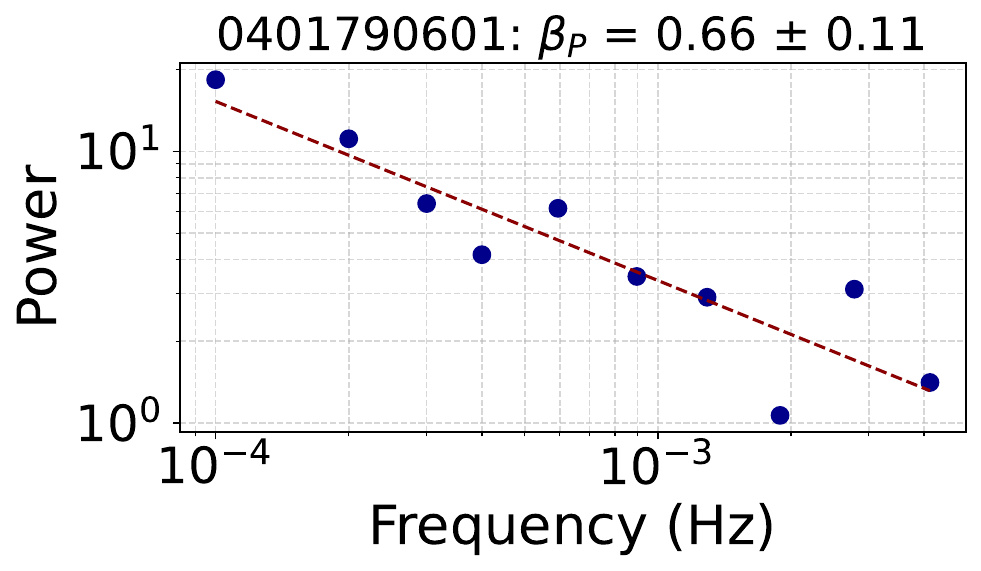}
	\end{minipage}
	\begin{minipage}{.3\textwidth} 
		\centering 
		\includegraphics[width=.99\linewidth]{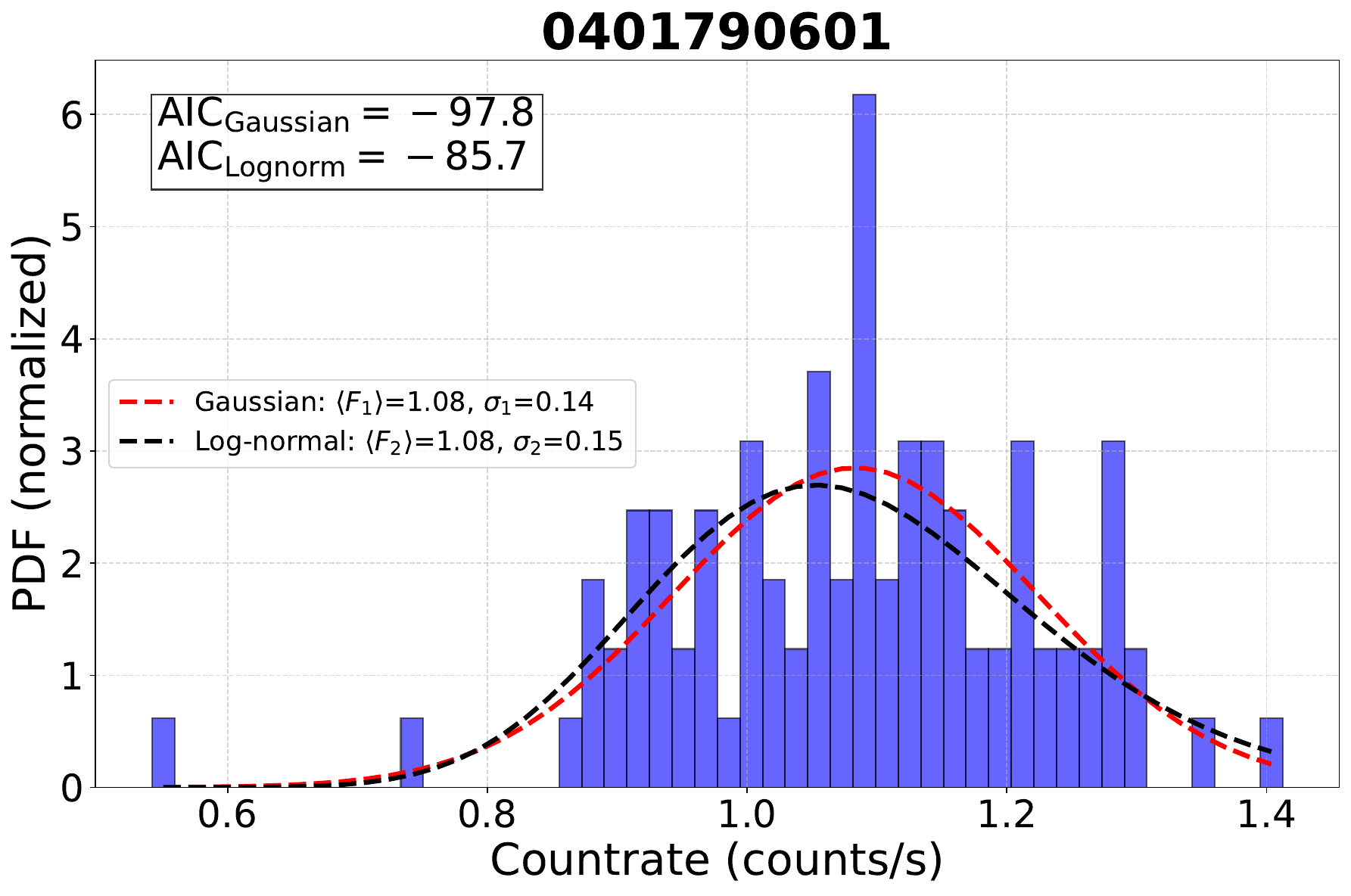}
	\end{minipage}
	\begin{minipage}{.3\textwidth} 
		\centering 
		\includegraphics[height=.99\linewidth, angle=-90]{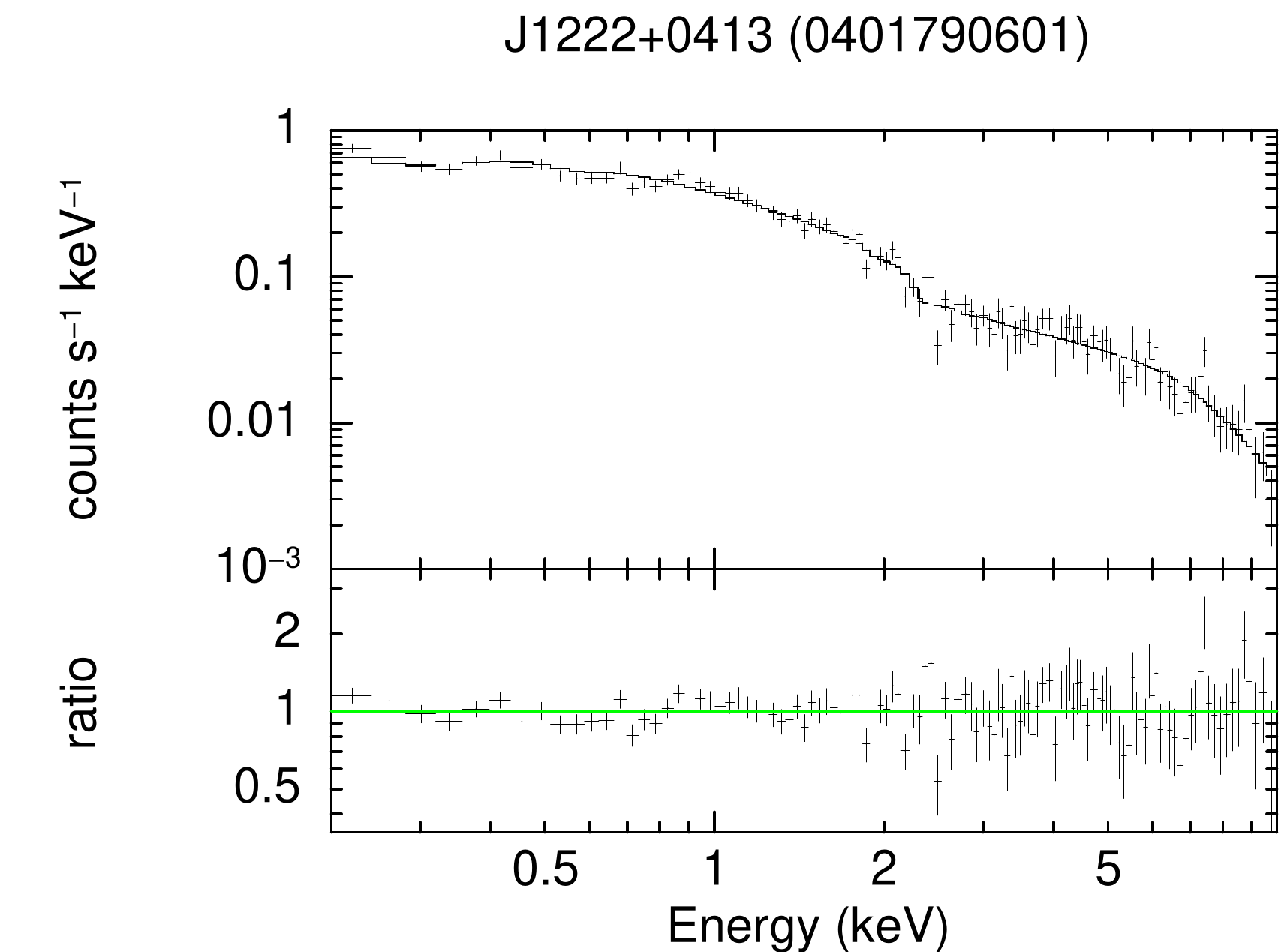}
	\end{minipage}
\end{figure*}

\begin{figure*}\label{app:TXS 0103+189}
	\centering
	\caption{LCs, HR plots, Structure Function, PSD, PDF, and spectral fits derived for TXS 0103+189.}
	\begin{minipage}{.3\textwidth} 
		\centering 
		\includegraphics[width=.99\linewidth]{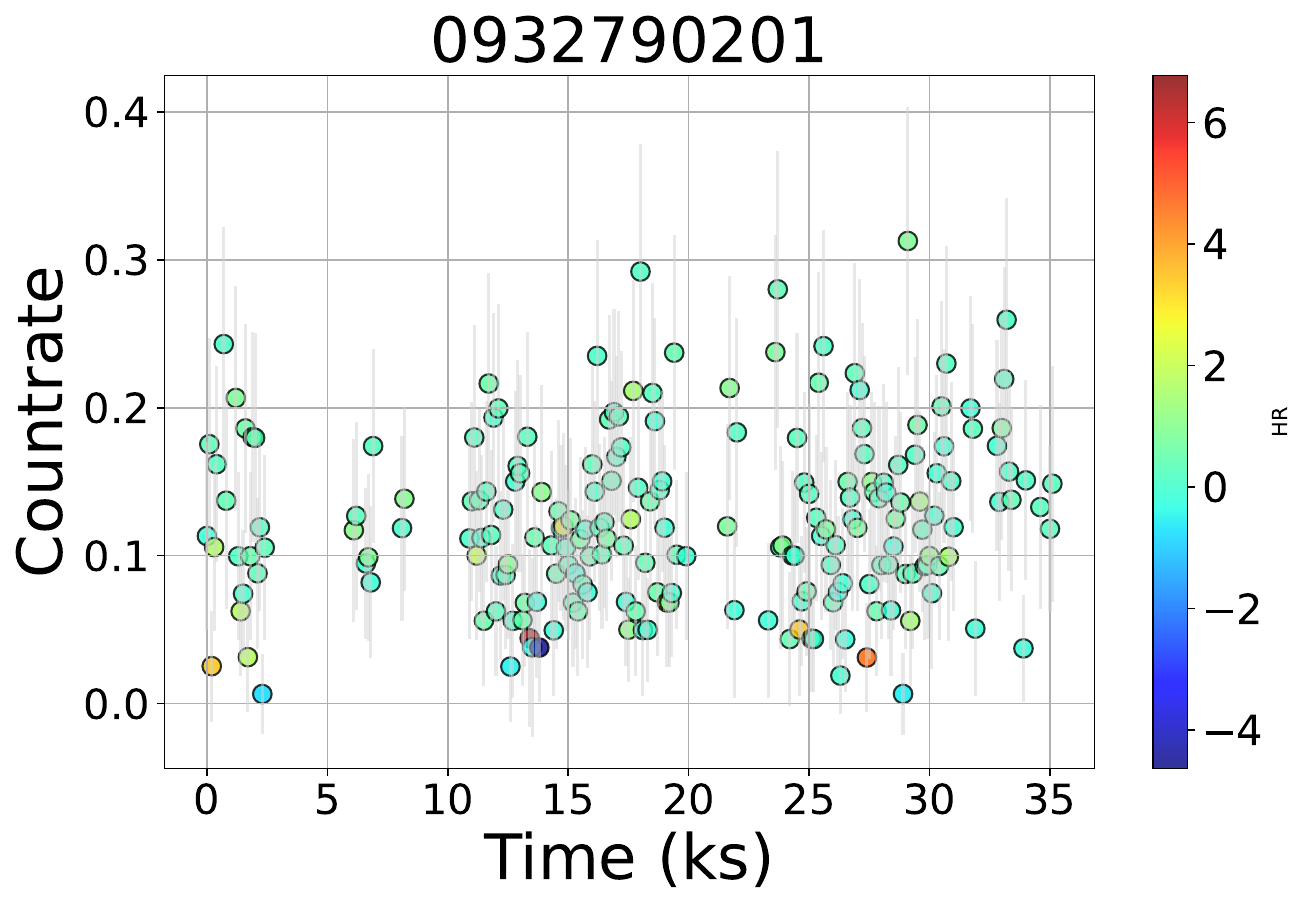}
	\end{minipage}
	\begin{minipage}{.3\textwidth} 
		\centering 
		\includegraphics[width=.99\linewidth]{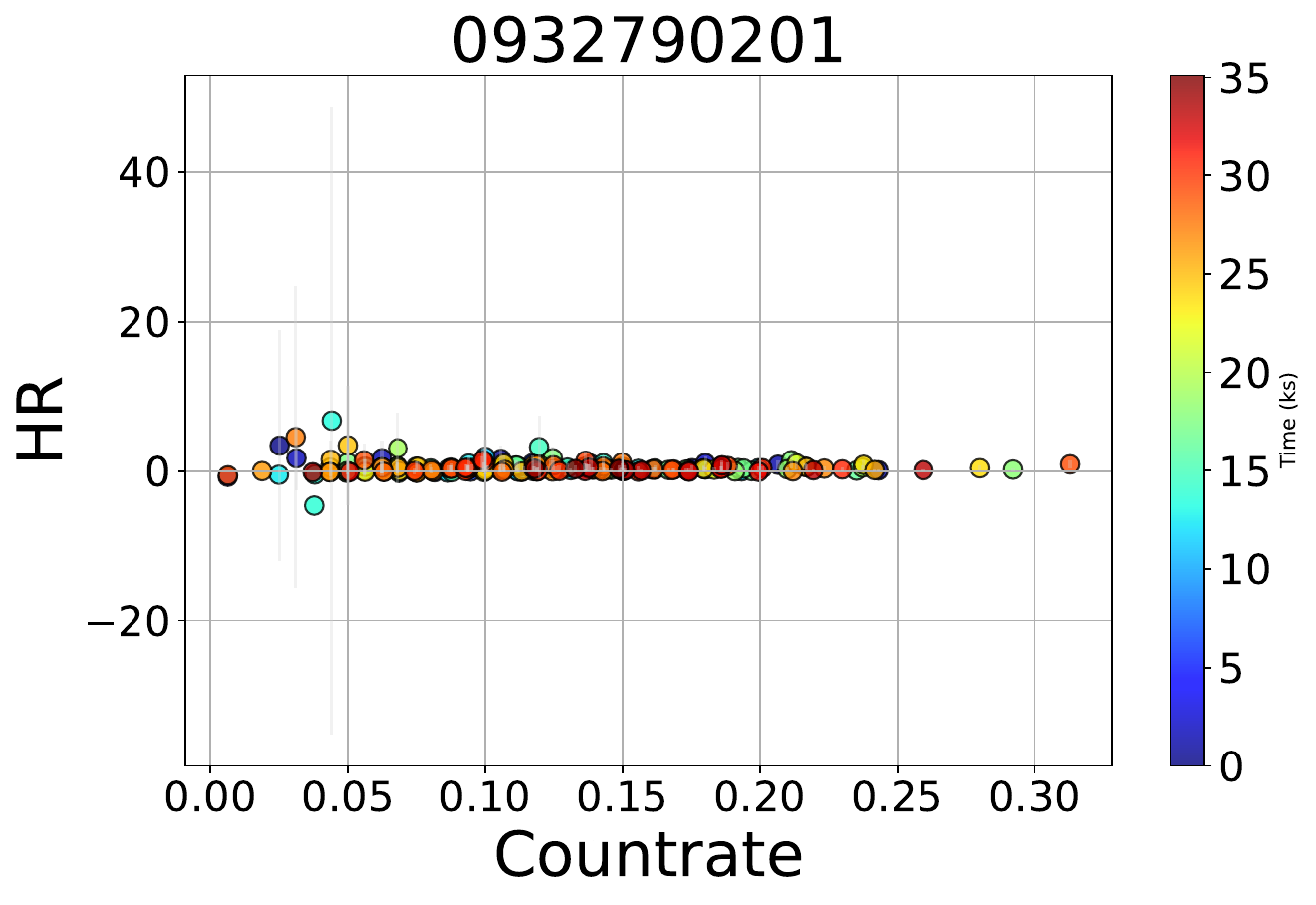}
	\end{minipage}
	\begin{minipage}{.3\textwidth} 
		\centering 
		\includegraphics[width=.99\linewidth, angle=0]{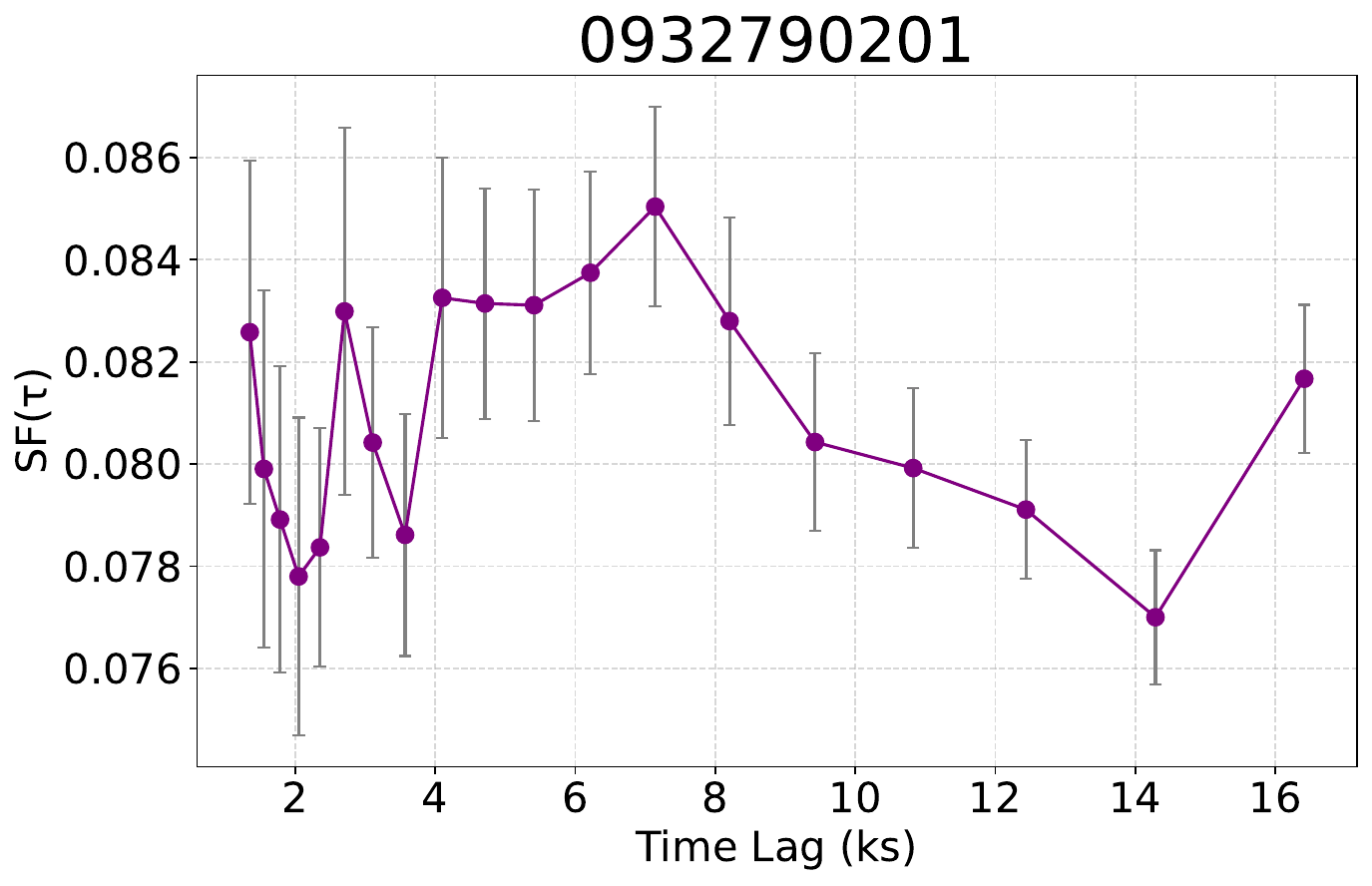}
	\end{minipage}
	\begin{minipage}{.3\textwidth} 
		\centering 
		\includegraphics[width=.99\linewidth]{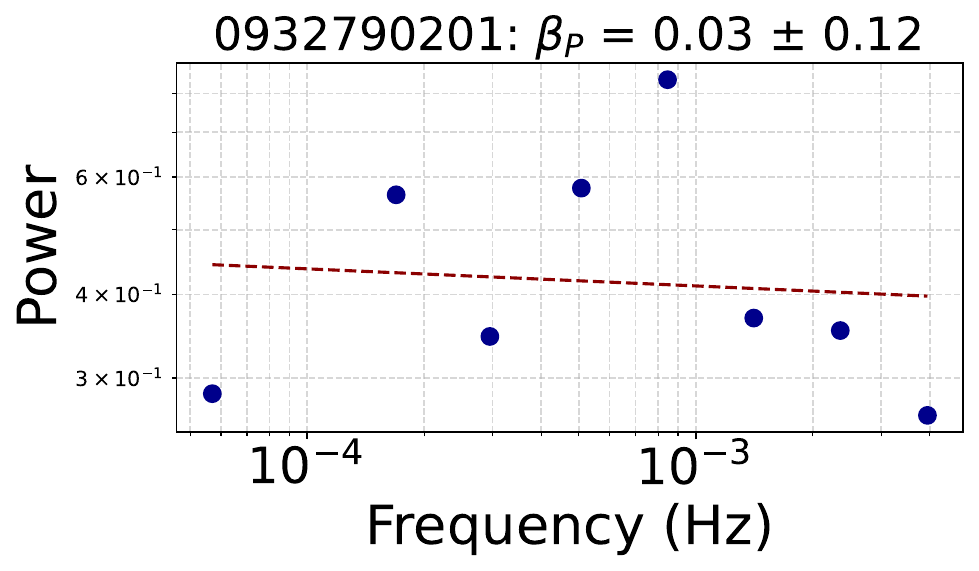}
	\end{minipage}
	\begin{minipage}{.3\textwidth} 
		\centering 
		\includegraphics[width=.99\linewidth]{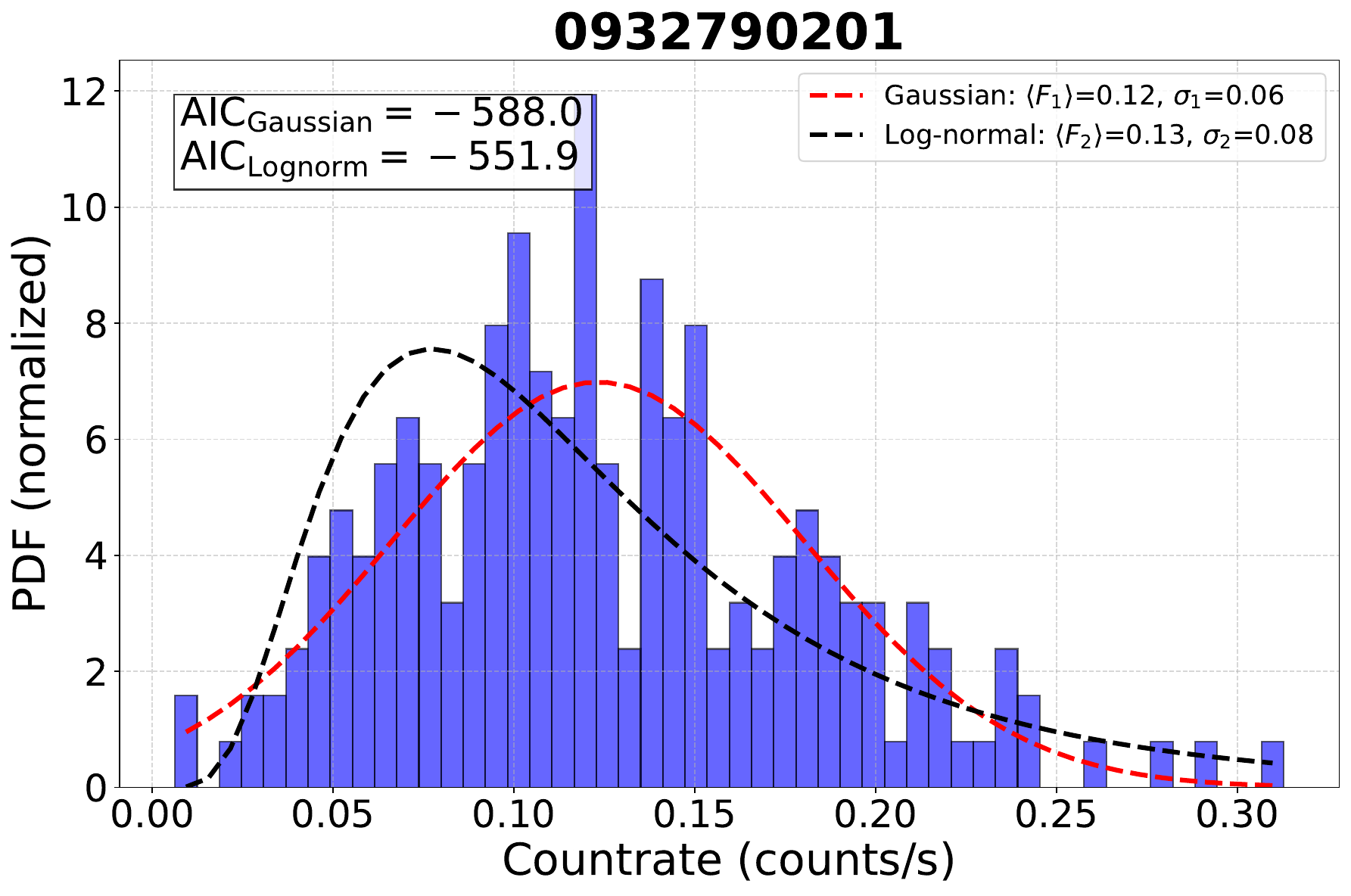}
	\end{minipage}
	\begin{minipage}{.3\textwidth} 
		\centering 
		\includegraphics[height=.99\linewidth, angle=-90]{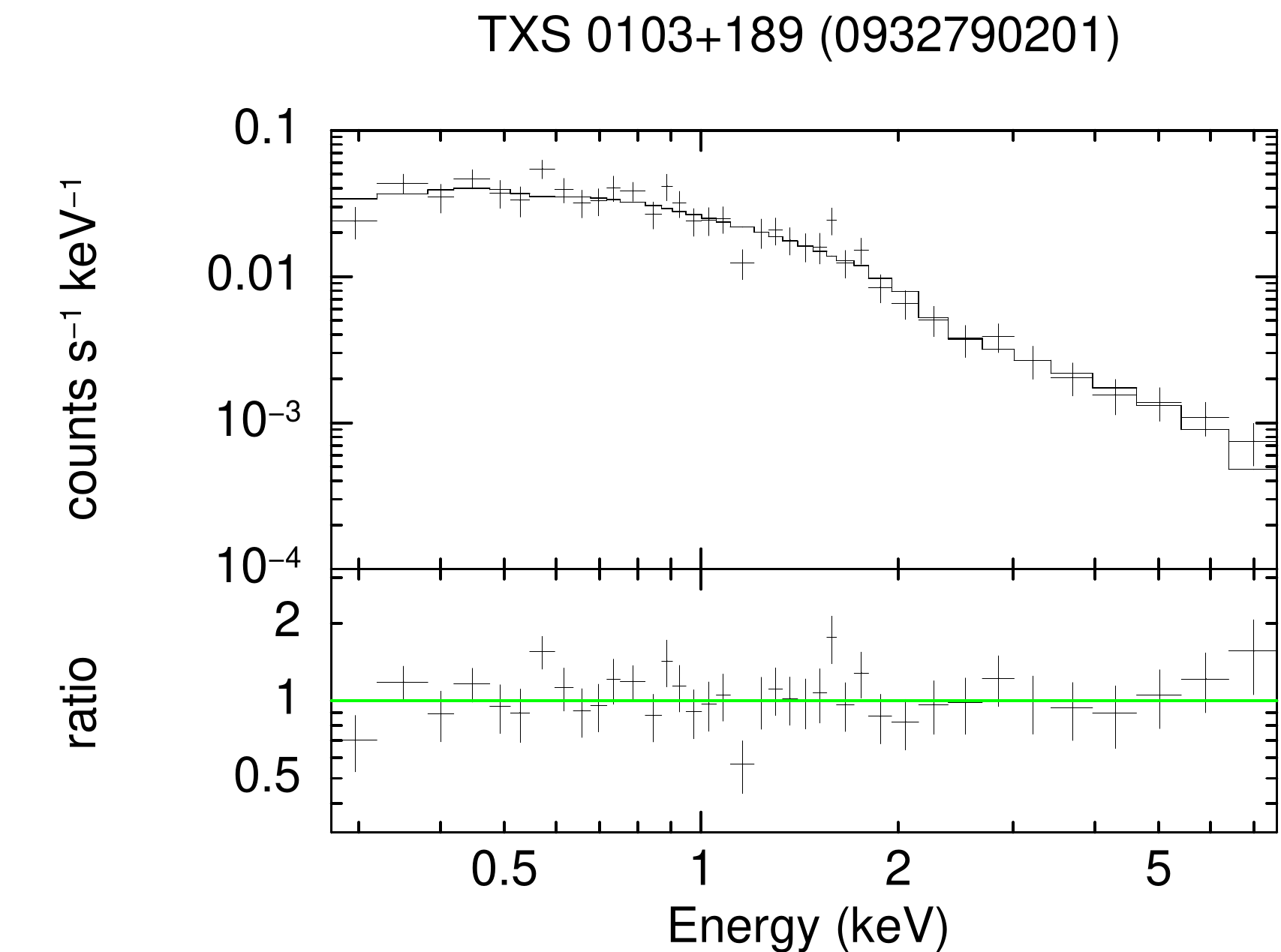}
	\end{minipage}
\end{figure*}

\begin{figure*}\label{app: PKS 1244-255}
	\centering
	\caption{LCs, HR plots, Structure Function, PSD, PDF, and spectral fits derived for PKS 1244-255.}
	\begin{minipage}{.3\textwidth} 
		\centering 
		\includegraphics[width=.99\linewidth]{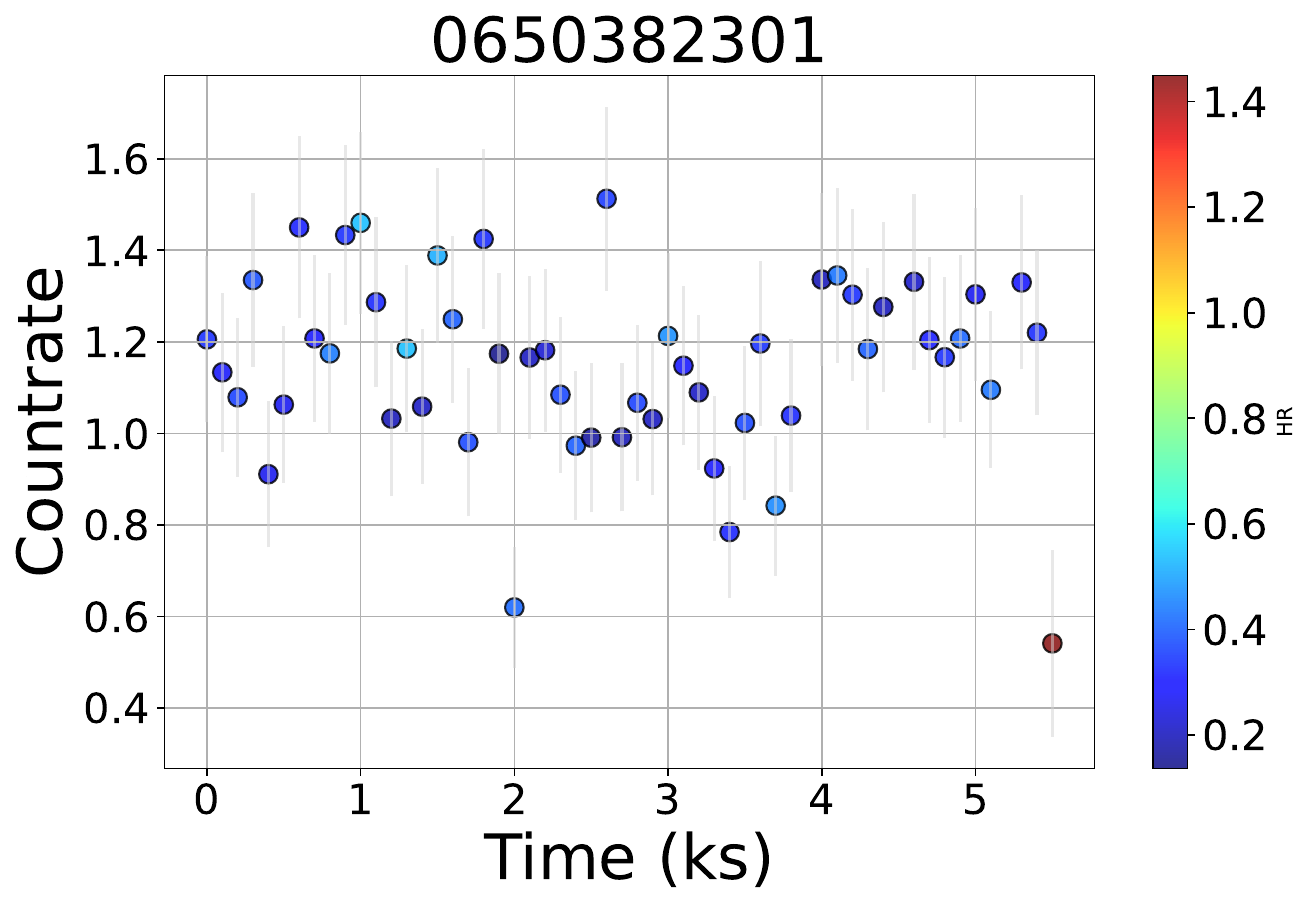}
	\end{minipage}
	\begin{minipage}{.3\textwidth} 
		\centering 
		\includegraphics[width=.99\linewidth]{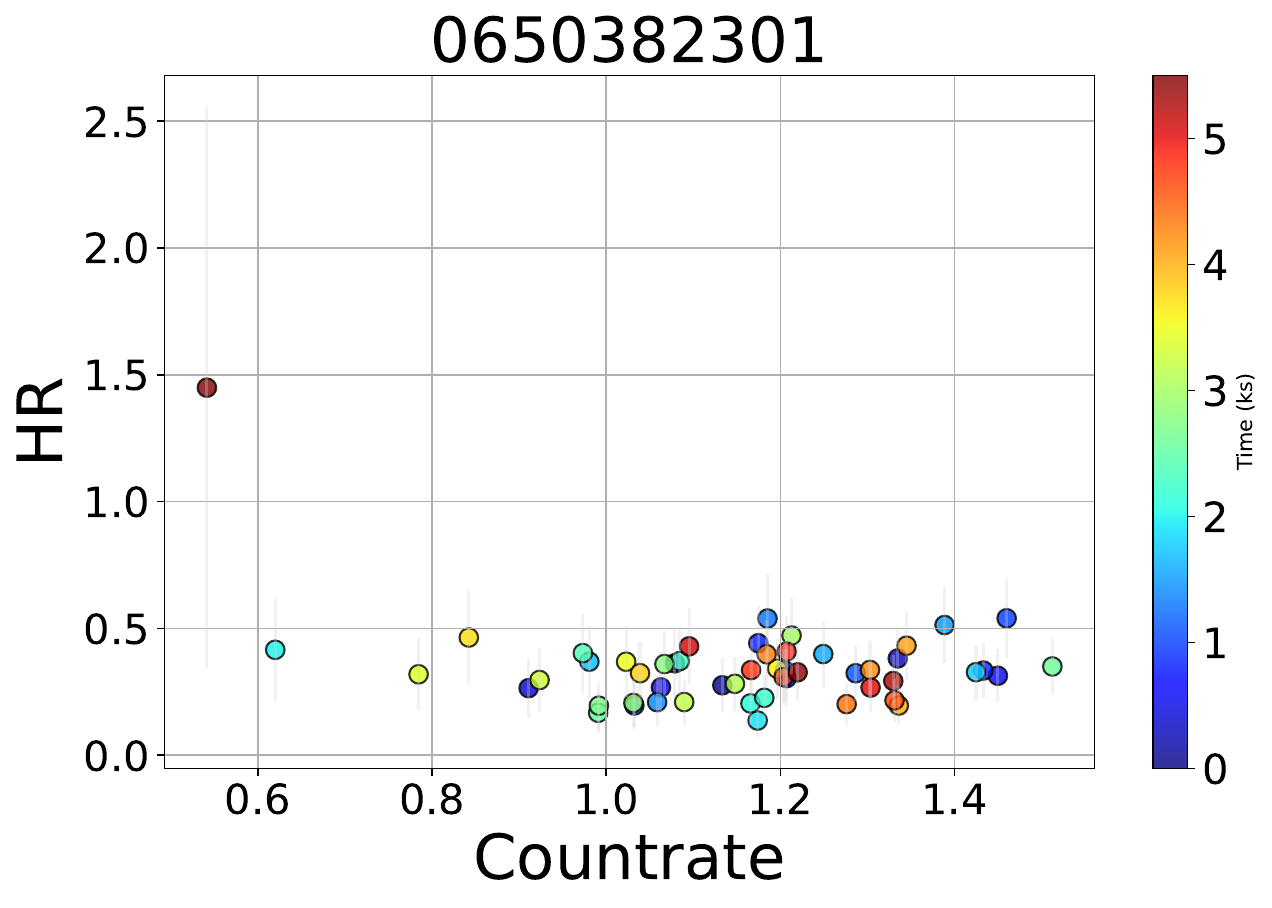}
	\end{minipage}
	\begin{minipage}{.3\textwidth} 
		\centering 
		\includegraphics[width=.99\linewidth, angle=0]{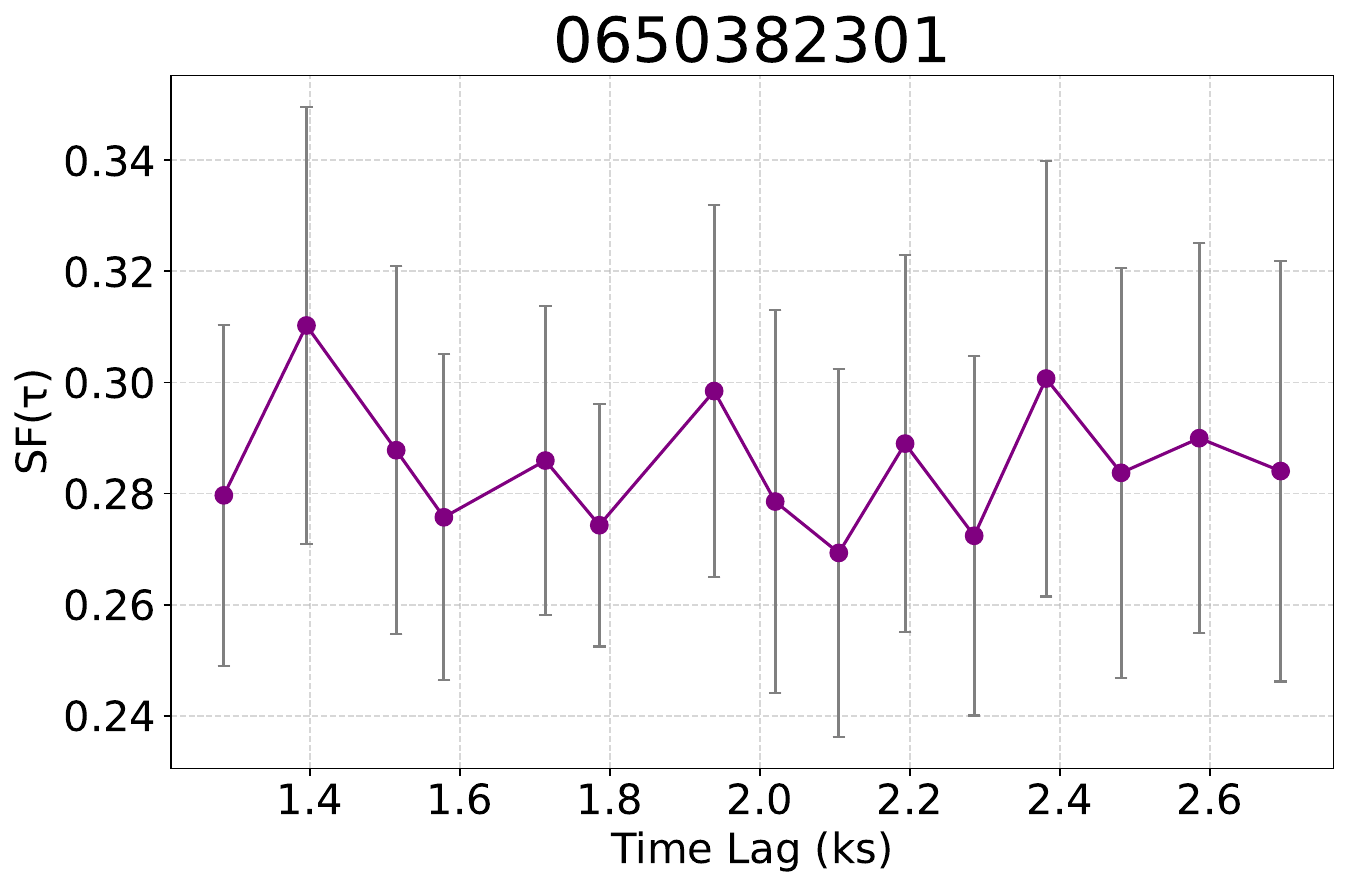}
	\end{minipage}
	\begin{minipage}{.3\textwidth} 
		\centering 
		\includegraphics[width=.99\linewidth]{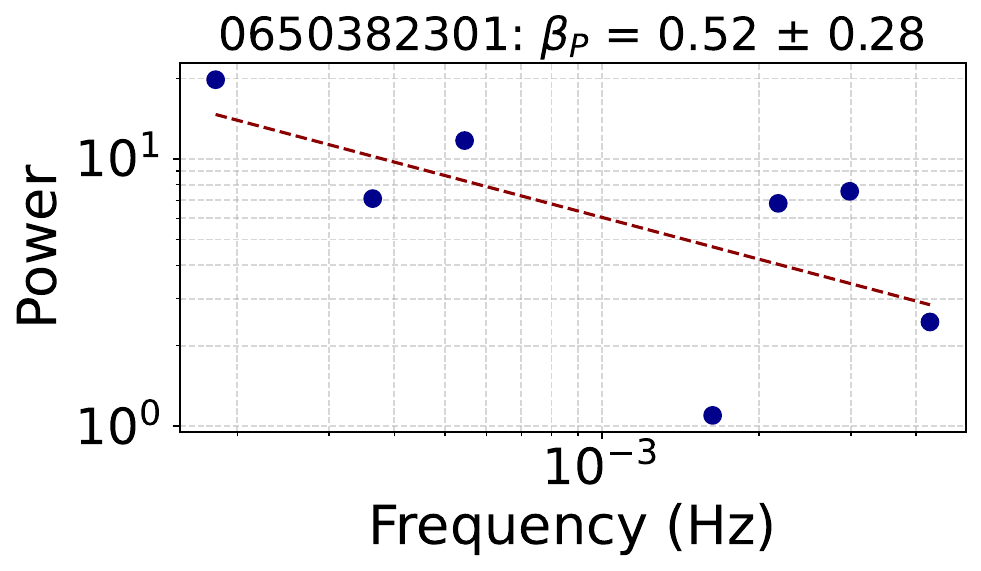}
	\end{minipage}
	\begin{minipage}{.3\textwidth} 
		\centering 
		\includegraphics[width=.99\linewidth]{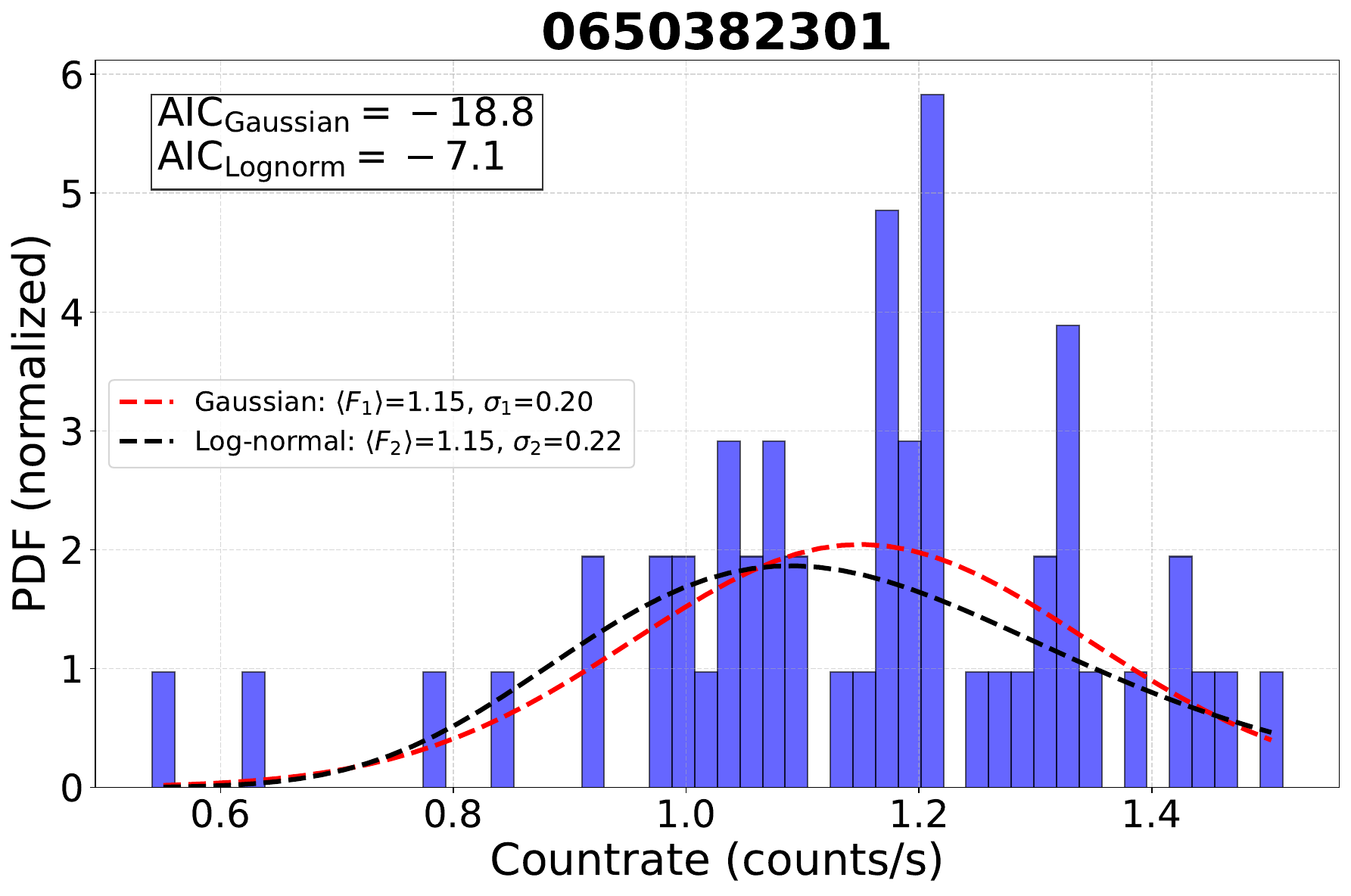}
	\end{minipage}
	\begin{minipage}{.3\textwidth} 
		\centering 
		\includegraphics[height=.99\linewidth, angle=-90]{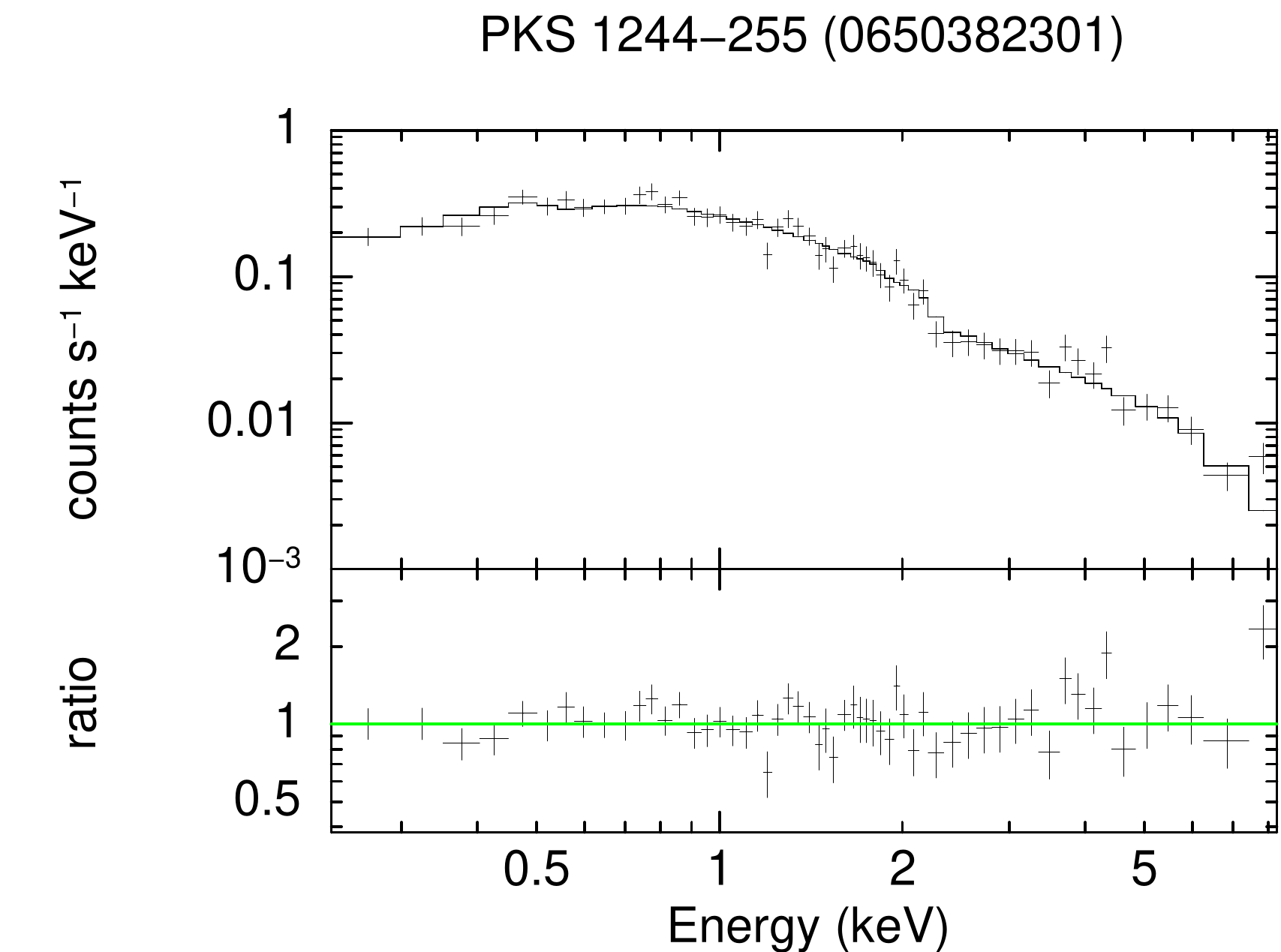}
	\end{minipage}
\end{figure*}

\begin{figure*}\label{app: TXS 1308+554}
	\centering
	\caption{LCs, HR plots, Structure Function, PSD, PDF, and spectral fits derived for  TXS 1308+554.}
	\begin{minipage}{.3\textwidth} 
		\centering 
		\includegraphics[width=.99\linewidth]{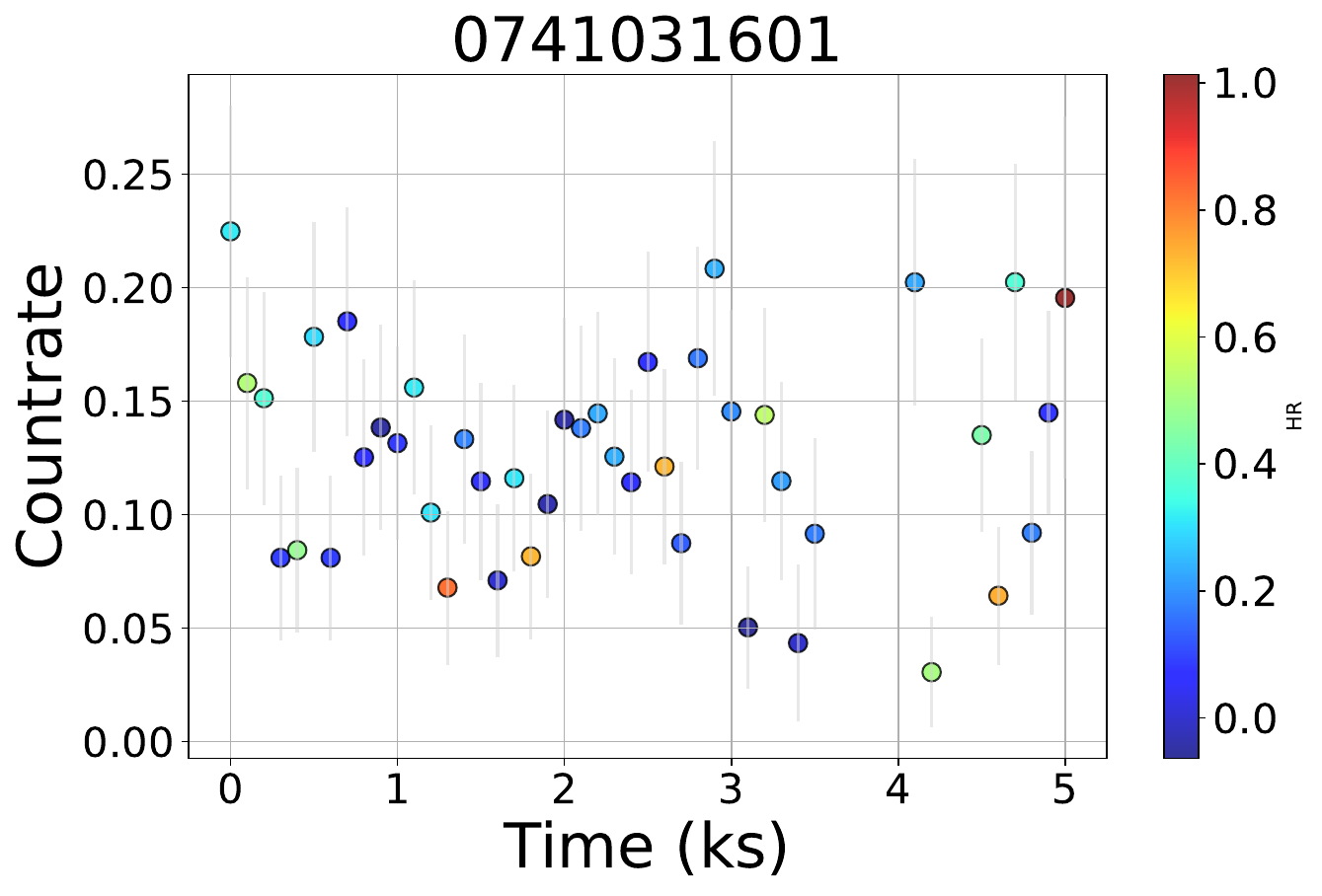}
	\end{minipage}
	\begin{minipage}{.3\textwidth} 
		\centering 
		\includegraphics[width=.99\linewidth]{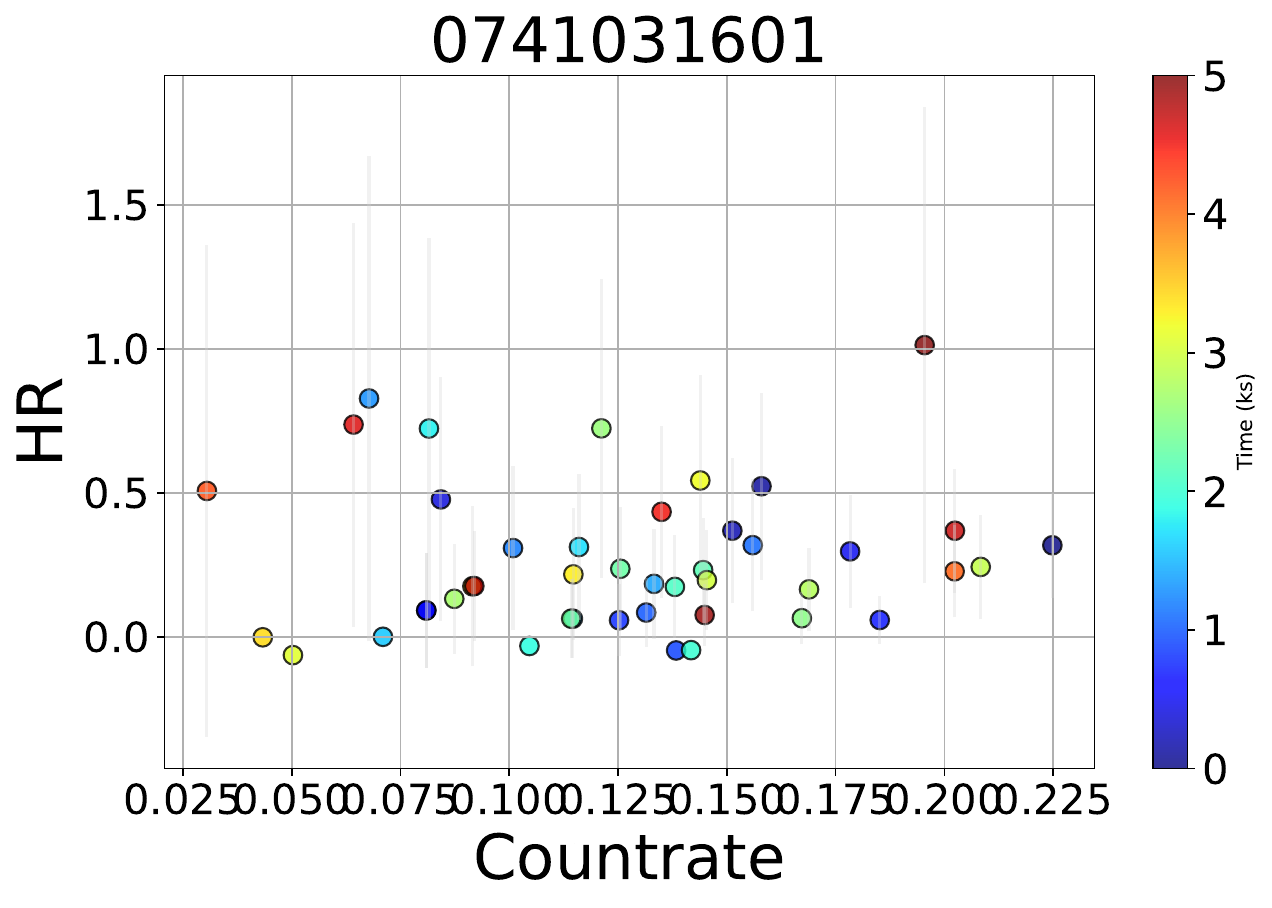}
	\end{minipage}
	\begin{minipage}{.3\textwidth} 
		\centering 
		\includegraphics[width=.99\linewidth, angle=0]{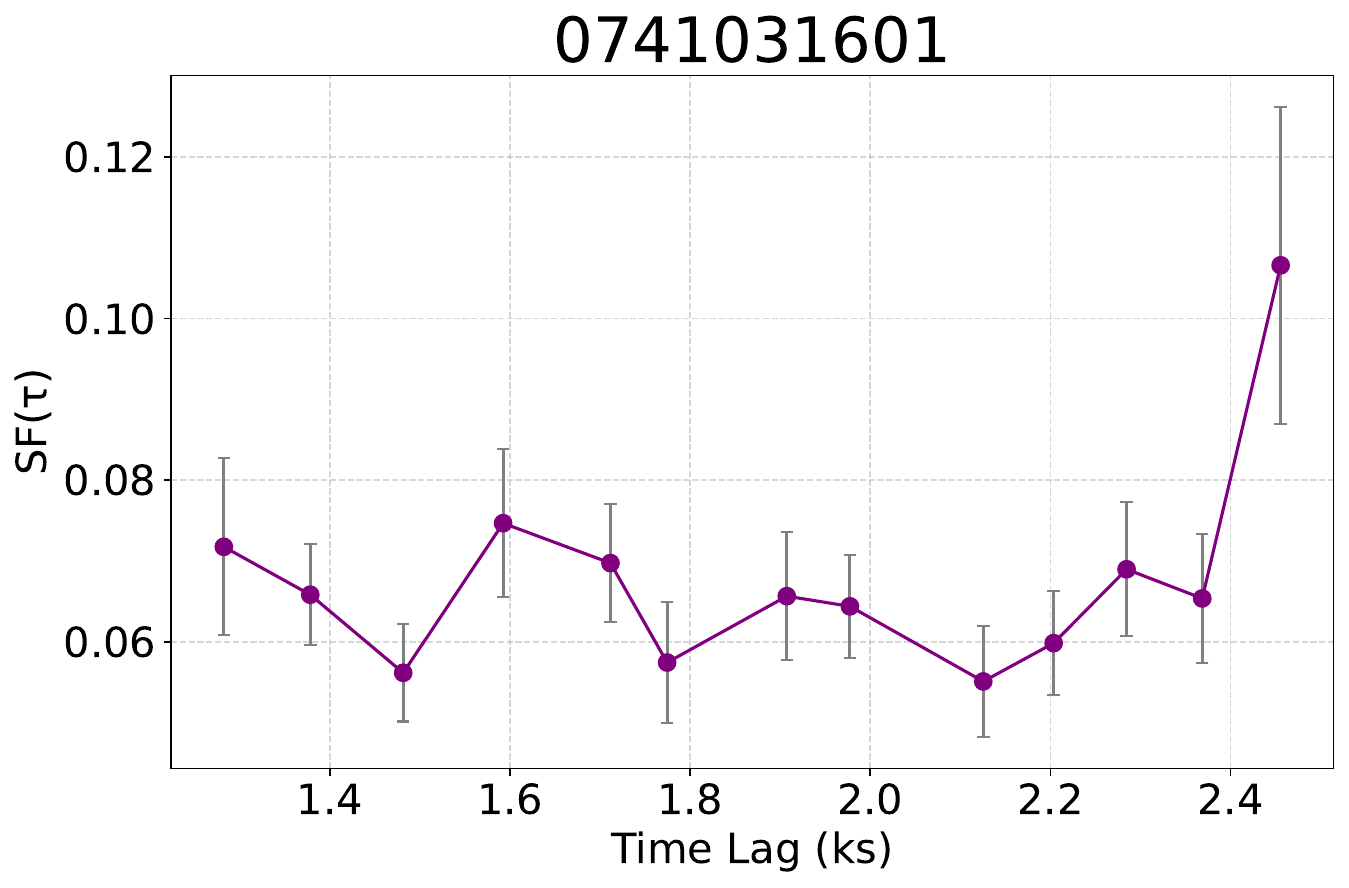}
	\end{minipage}
	\begin{minipage}{.3\textwidth} 
		\centering 
		\includegraphics[width=.99\linewidth]{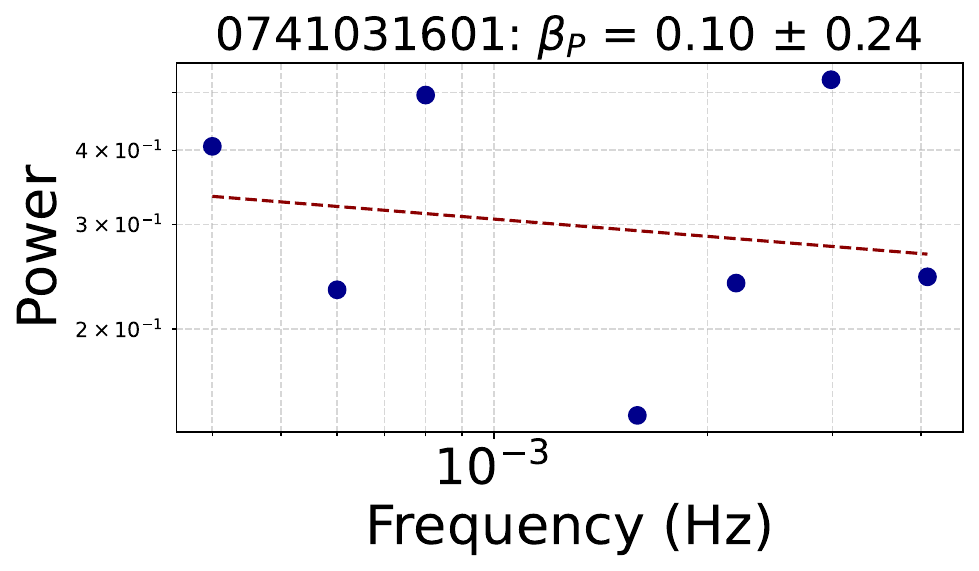}
	\end{minipage}
	\begin{minipage}{.3\textwidth} 
		\centering 
		\includegraphics[width=.99\linewidth]{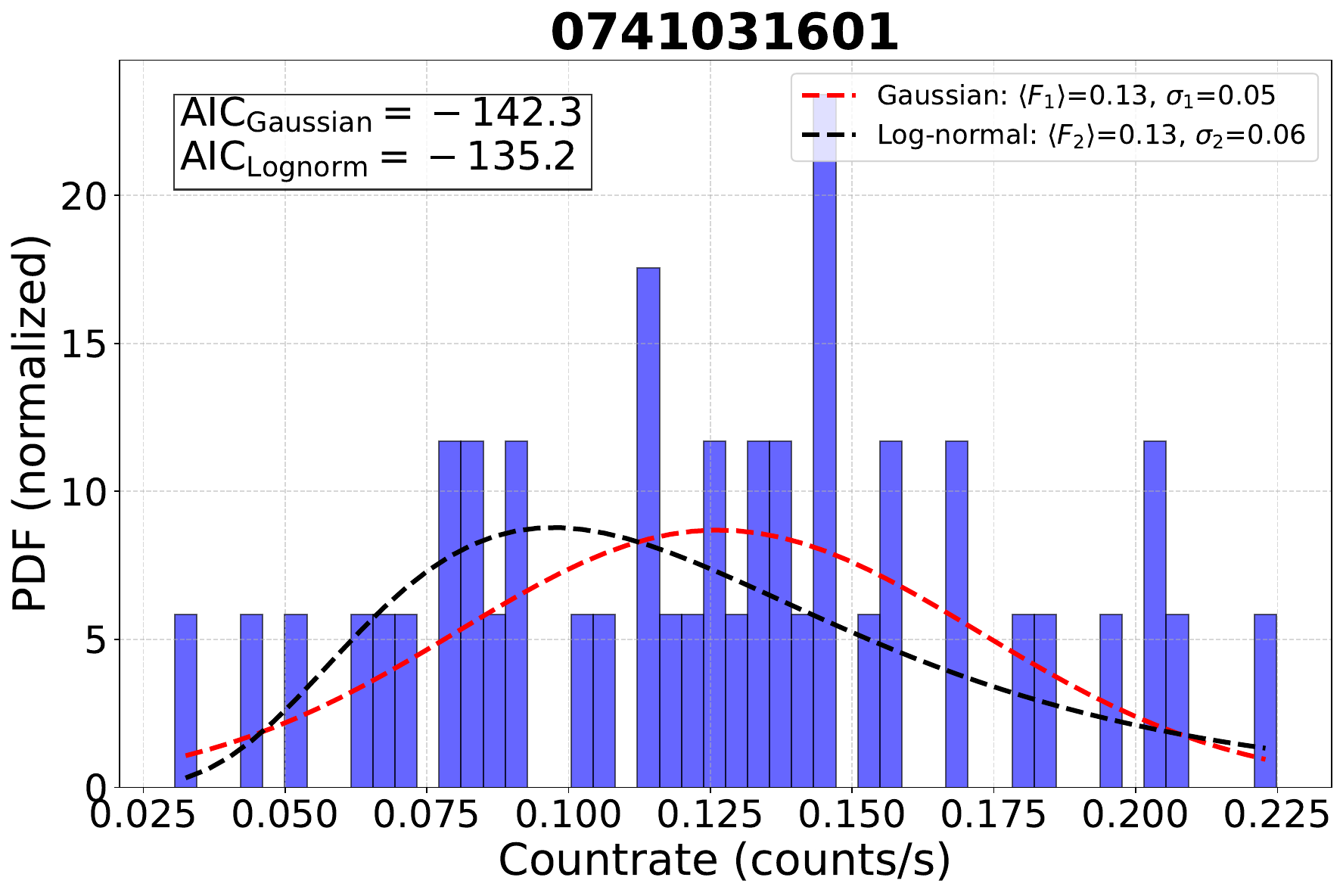}
	\end{minipage}
	\begin{minipage}{.3\textwidth} 
		\centering 
		\includegraphics[height=.99\linewidth, angle=-90]{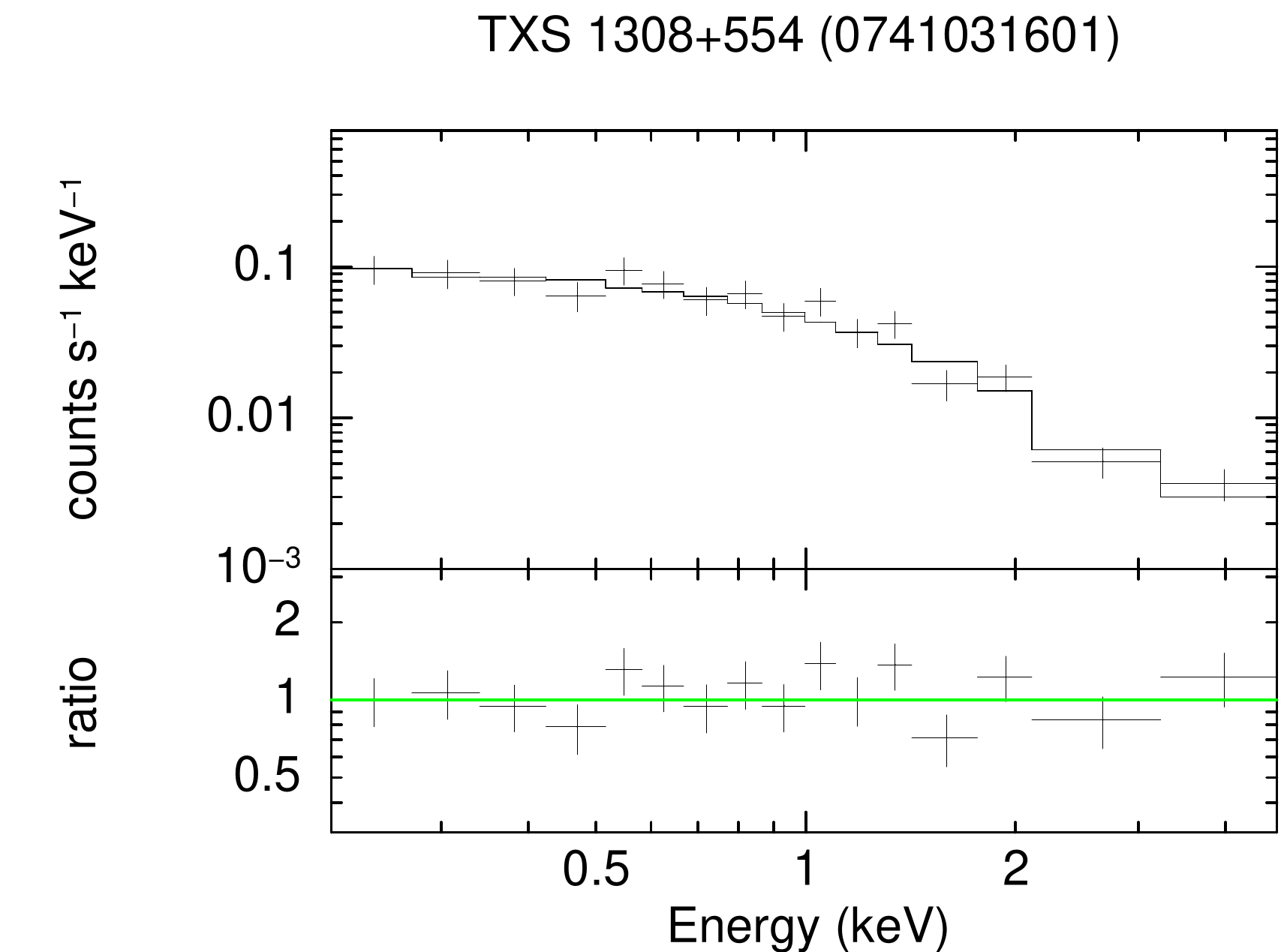}
	\end{minipage}
\end{figure*}


\bsp	
\label{lastpage}
\end{document}